\documentclass[final,journal, twoside, 10pt]{IEEEtran}                     
\usepackage[hyperref]{Bamble}
\usepackage{Bmacros}
\usepackage[noadjust]{cite}
\usepackage{enumitem}
\usetikzlibrary{plotmarks}
\newcommand{\rcdiff}{\delta}
\newcommand{\rcdiffprime}{\delta'}
\newcommand{\candidate}{\mathsf{D}}
\newcommand{\candidateA}{\mathsf{A}}
\newcommand{\candidateB}{\mathsf{B}}
\newcommand{\candidateC}{\mathsf{C}}

\begin{document}
\title{Universal Polarization for Processes with Memory}
\date{}
\author{\IEEEauthorblockN{Boaz~Shuval, Ido~Tal,~\IEEEmembership{Senior Member,~IEEE}}}
\maketitle

\begin{abstract}
    A transform that is universally polarizing over a set of channels with memory is presented. 
    Memory may be present in both the input to the channel and the channel itself. 
    Both the encoder and the decoder are aware of the input distribution, which is fixed. 
    However, only the decoder is aware of the actual channel being used. 
    The transform can be used to design a universal code for this scenario.
    The code is to have vanishing error probability when used over any channel in the set, and achieve the infimal information rate over the set. 
    The setting considered is, in fact, more general: we consider a set of processes with memory.
    Universal polarization is established for the case where each process in the set: (a) has memory in the form of an underlying hidden Markov state
    sequence that is aperiodic and irreducible, and (b) satisfies a `forgetfulness' property. 
    Forgetfulness, which we believe to be of independent interest, occurs when two hidden Markov states become approximately independent of each other given
    a sufficiently long sequence of observations between them. 
    We show that aperiodicity and irreducibility of the underlying Markov chain is not sufficient for forgetfulness, and develop a sufficient
    condition for a hidden Markov process to be forgetful.
\end{abstract}
\begin{IEEEkeywords}
	Polar codes, universal polarization, universal codes, channels with memory, hidden Markov processes
\end{IEEEkeywords}

\section{Introduction}
\IEEEPARstart{I}{mperfect} channel knowledge characterizes many practical communication scenarios. 
There are various models for imperfect channel knowledge; see~\cite{fadingChannels} for a comprehensive discussion. 
We consider the scenario where the decoder has full channel information, but the encoder is only aware of a \emph{set} to which the actual
channel belongs. 
    Both the encoder and the decoder are aware of the input distribution, which is fixed. 
We wish to build a polarization-based code that is universal over the set: it achieves vanishing error probability for any channel in the set, and its
rate approaches the infimal information rate over all channels in the set. 

In fact, this work tackles a more general setting. 
The universal construction in this paper applies both to channel coding and source coding scenarios. 
However, to keep the introduction focused, we concentrate on a channel-coding scenario. 
We wish to design polarization-based codes that achieve vanishing error probability over a set of channels \emph{with memory}. 
The input distribution to all channels in the set is fixed and known at the encoder and decoder. 
The encoder only knows that the channel belongs to the set, while the decoder is aware of the actual channel used. 
Examples of channels with memory are finite-state channels, input-constrained channels, and intersymbol-interference
channels. 
We show a polar coding construction that approaches the infimal information rate among the set of channels under successive-cancellation decoding,
provided that every input-output process in the set satisfies some mild technical constraints. 
This construction achieves vanishing error probability over all processes in this set with the same exponent as \arikan's polar
codes~\cite{Arikan_2009,Arikan_Telatar_2009}. That is, roughly $2^{-\sqrt{\Lambda}}$, where $\Lambda$ is the codeword length,

The informal statements of the previous paragraph are stated formally in our main theorem, below. The theorem contains several terms that will be defined throughout the paper.
\begin{theorem}
	\label{thm_main}
Consider a family of FAIM \soprocesses with an upper bound on forgetfulness and mixing, where all \soprocesses share the same input distribution. We consider the case in which the encoder does not know which $\soprocess$ was used, but the decoder does know. Let $I^*$ be the infimal information rate over the family of \soprocesses. Then, for any $R < I^*$ and $\beta < 1/2$, there exists a sequence of codes with growing lengths such that the following holds for each code:
\begin{enumerate}
	\item \label{thm_main:rate} The code rate is at least $R$.
	\item \label{thm_main:error probability} The probability of error is upper-bounded by $2^{-\Lambda^\beta}$, where $\Lambda$ is the codeword length. This bound holds universally for any $\soprocess$ in the family.
	\item \label{thm_main:complexity}  Both the encoding and decoding complexities are $O(|\mathcal{S}|^3 \Lambda \log \Lambda)$, where $|\mathcal{S}|$ is an upper bound on the number of states of an \soprocess in the family.
\end{enumerate}
\end{theorem}
For the proof of the theorem, see the end of \Cref{sec_decoding}.
\subsection{Prior work on universal polar codes}

The study of polar coding for a class of memoryless channels with full channel knowledge at the decoder was first considered in~\cite{Hassani_compound2009}. 
Hassani et al.\ showed that \arikan's  polar codes~\cite{Arikan_2009}, under successive-cancellation decoding, cannot achieve the compound capacity~\cite{blackwell1959}
of a set of binary-input, memoryless, and symmetric (BMS) channels.   
 In~\cite[Proposition 7.1]{sasoglu_thesis} it was shown that polar codes are universal over a set of BMS channels if optimal decoding is employed. 
Thus, the non-universality exhibited in~\cite{Hassani_compound2009} is an artifact of using successive-cancellation decoding. 
Nevertheless, as described below, coding methods that are based on polarization and successive-cancellation decoding have been shown to yield universal codes. 

In~\cite{Hassani_universal}, Hassani and Urbanke present two designs based on \arikan's polar codes that achieve universality over a set of BMS channels.
Their first construction combines \arikan's polar codes and Reed-Solomon codes designed for an erasure channel. 
Their second construction may be viewed as a two-stage method.
In the first stage, one forms several \arikan polar codes, in which identical channels are combined recursively. 
In the second stage, different channels are combined to obtain universality.

\sasoglu and Wang~\cite{sasoglu_2016_universal} presented another universal polar coding construction for BMS channels. 
Their construction is also a recursive two-stage method. 
The first stage, called the slow stage, transforms multiple channel-uses into ones that universally have high-entropy and ones that universally have low-entropy. 
The second stage, invoked once sufficient polarization is obtained, combines the channels that are universally low-entropy using \arikan's polar codes to yield
vanishing error probability. 
The construction presented in this paper is a simplified variation of the \sasoglu-Wang construction. 

We briefly mention other works concerning universality of polar codes. 
Universal polar codes for families of ordered BMS channels or memoryless sources, with full decoder side information, was considered in
\cite{Sutter_universal}. 
See also~\cite{Abbe_universal} for the case of universal polar source codes, with specialization to the binary case. 
Universal source polarization was studied in~\cite{Ye_Barg_universal}, in which polar-based codes were used to compress a memoryless source
to be losslessly recovered by multiple users, each observing different local side information on the source sequence. 
Finally, universal polar coding for certain classes of BMS channels with channel knowledge at the encoder was considered in~\cite{Alsan_2014}.  

\subsection{Overview of this paper}
We present our universal construction in \Cref{sec_universal construction}. 
It consists of two stages, a slow stage, described in \Cref{sec_slow polarization stage},  followed by a fast stage, described in \Cref{sec_fast stage}. 
Both stages are recursive and use \arikan transforms as building blocks. 
The fast stage consists of multiple applications of \arikan transforms as in the seminal paper~\cite{Arikan_2009}.
The slow stage uses \arikan transforms in a different manner.
Properties of the slow stage, as well as a variation of it that will be useful for our proof of universality, are presented in \Cref{sec_properties of
BST and a variation}. 
When used over a set of BMS channels and specialized appropriately, this universal construction is functionally equivalent to the one presented in~\cite{sasoglu_2016_universal}. 
Our goal, however, is to use it over a set of processes with memory. 

Polar codes were shown to achieve vanishing error probability for processes with memory in~\cite{sasoglu_Tal_mem} and~\cite{Shuval_Tal_Memory_2017}. 
It was shown in~\cite{sasoglu_Tal_mem} that a large class of processes with memory polarizes under \arikan's polar transform. 
This result extended \sasoglu's earlier findings in~\cite[Chapter 5]{sasoglu_thesis}. 
It was further shown in~\cite{sasoglu_Tal_mem} that the Bhattacharyya parameter polarizes fast to $0$ for this class. 
Later, it was shown in~\cite{Shuval_Tal_Memory_2017} that for processes with an underlying hidden Markov structure, the Bhattacharyya parameter also polarizes
fast to $1$. 
Combined, the results of~\cite{sasoglu_Tal_mem} and~\cite{Shuval_Tal_Memory_2017} enable information-rate-achieving polar codes for such processes
with memory.
A practical, low-complexity, decoding algorithm for processes with memory with an underlying hidden Markov structure was described
in~\cite{wang2014joint} and~\cite{Wang_2015}. 
This algorithm is a variation of successive-cancellation decoding that takes into account the hidden state. 

One drawback of polar codes for processes with memory using the strategy in the previous paragraph is that they must be tailored for the process. 
For example, to design a polar code for a channel with intersymbol interference, one must know the exact transfer function of the channel.  
In a practical scenario, it is reasonable to assume that the decoder has full channel knowledge, obtained, for example, by channel estimation based on
a reference sequence~\cite{cellMobileRadio}. 
However, the assumption that the encoder also has full channel knowledge \emph{before} transmission may be unrealistic. 
This is where universal polar codes come into play. 

In the universal setting we consider, the encoder has partial information: it knows that the process belongs to some set of processes with memory. 
The exact process is known only to the decoder, at the time of decoding.
The encoder must employ a code that will enable vanishing error probability no matter which process in the set is used. 
We wish to design a universal code with the highest possible rate over the entire set. 
Thus, the code is to approach the infimal information rate over the entire set. 

This is indeed what we achieve in this work. 
We show that our polarization-based construction is universal over sets of processes with memory. 
We prove universality when the sets contain processes with memory that satisfy two technical constraints, presented in detail in \Cref{sec_probabilistic model with memory}.
Briefly, the processes have an underlying hidden finite-state Markov structure that is regular (aperiodic and irreducible); and they have a property we call
\emph{forgetfulness}, which we believe is of independent interest.

Forgetfulness is a property we now describe informally. 
In a hidden Markov process, we are given a sequence of observations that are known to be probabilistic functions of some Markov chain called the state process. 
The process is called forgetful if, given a long-enough sequence of observations, the state at the time of the first observation and the state at the time of the
last observation become approximately independent. 
Surprisingly, regularity of the underlying Markov chain is not sufficient to ensure forgetfulness. 
We note that forgetfulness was not required in the non-universal setting of~\cite{sasoglu_Tal_mem,Shuval_Tal_Memory_2017}, yet in our proof of the universal case it plays a key role. 

Hochwald and Jelenkovi\'c~\cite{Hochwald_Jelenkovic_Markov_1999} considered a property similar to forgetfulness under the restrictive assumption that 
there is a positive probability of transitioning between any two states in one step. 
Leveraging ideas from Kaijser~\cite{kaijser1975}, who considered a related setting for hidden Markov processes, we lift this restrictive assumption and prove, in \Cref{sec_contraction,sec_HMM}, a sufficient condition 
 for forgetfulness of a hidden Markov model. 
This condition, which we call \Cref{cond_kaijser}, takes into account both the transition matrix of the state process as well as the probabilistic function that generates the
observations. 
Specifically, we use mutual information as a measure for independence, and show that under \Cref{cond_kaijser}, the mutual information between the states at the beginning and
end of a block, given the observations in between, vanishes with the length of the block. 

The slow stage of the construction is the one responsible for its universality.
The proof of universality is given in \Cref{sec_biprocess,sec_monopolarization for FAIM derived processes}.
Low complexity decoding of the universal polar codes is based on the successive-cancellation trellis decoding of~\cite{Wang_2015}; details are given
in \Cref{sec_decoding}. 
In~\Cref{sec_construction} we explain how to construct universal polar codes for a given family of processes. Numerical results for a particular universal polar code, constructed using the method of~\Cref{sec_construction} and used over several different channels with and without memory, can be found in~\Cref{sec_numerical results}. 

\begin{figure}
    \begin{center}
        \begin{tikzpicture}[rrect/.style={draw,rounded corners, minimum width = 1.5cm, minimum height = 1cm}, 
                            >=latex]
            \node[rrect,label=above:start] (notation) at (0,0) {\ref{sec_notation conventions and reminders}};
            \node[rrect, below = 0.5 of notation] (construction) {\ref{sec_universal construction}};
            \node[rrect, below = 0.5 of construction] (model) {\ref{sec_probabilistic model with memory}};

            \node at ($(model)-(0,0.35)$) {\scriptsize to Example~\ref{ex_function counter example}};

            \node[rrect, below = 0.5 of model] (decoding) {\ref{sec_decoding}}; 
            \node[rrect, below = 0.5 of decoding] (howto) {\ref{sec_construction}}; 
            \node[rrect, below = 0.5 of howto,label=below:end] (numerical) {\ref{sec_numerical results}}; 

            \draw[->] (notation) -- (construction);
            \draw[->] (construction) -- (model);
            \draw[->] (model) -- (decoding); 

            \node[rrect, right = of construction] (properties) {\ref{sec_properties of BST and a variation}}; 
            \node[rrect, below = 0.5 of properties] (proof1) {\ref{sec_probabilistic model with memory}}; 

            \draw[->] (construction) -- (properties);
            \draw[->] (properties) -- (proof1);

            \node[rrect, right = of proof1] (contraction) {\ref{sec_contraction}}; 
            \node[rrect, below = 0.5 of proof1] (proof2) {\ref{sec_biprocess},\ref{sec_monopolarization for FAIM derived processes}}; 
            \node[rrect, below = 0.5 of contraction] (hmm) {\ref{sec_HMM}}; 

            \draw[->] (proof1) -- (proof2); 
            \draw[->] (proof2) -- (decoding); 
            \draw[->] (proof1) -- (contraction); 
            \draw[->] (contraction) -- (hmm); 
            \draw[->] (hmm) -- (proof2); 
            \draw[->] (decoding) -- (howto); 
            \draw[->] (howto) -- (numerical); 
        
        \end{tikzpicture}
    \end{center}
    \caption{Roadmap of the various ways to read this paper. All paths start at \Cref{sec_notation conventions and reminders} and end at
    \Cref{sec_numerical results}.} \label{fig_roadmap}
\end{figure}

\subsection{Paper Roadmap}
There are several ways to read this paper, with increasing levels of detail. 
A map of the various paths is shown in \Cref{fig_roadmap}.  
All readers are advised to familiarize themselves with the notations and definitions of \Cref{sec_notation conventions and reminders}. 
In it, we introduce the notion of a symbol/observation pair, which generalizes the concept of a channel and allows for simultaneous description of
channel and source coding.
\Cref{sec_universal construction} is also recommended for all readers, for it introduces the details of the universal construction.
At this point, there are several options. 
\begin{itemize}
    \item A practitioner who wishes to understand and implement the construction, without getting bogged down with the proofs, is advised to skip to
        \Cref{sec_probabilistic model with memory}, and read it up to \Cref{ex_function counter example}. 
        This introduces the assumptions on the processes for which we can prove universality. 
        \Cref{ex_kaijser counter example,ex_function counter example} are important as they illustrate that forgetfulness does not follow from regularity (aperiodicity and irreducibility) of the underlying Markov chain. 
        Then, the practitioner may skip straight to the decoding process in \Cref{sec_decoding}.
        The practitioner is also well-advised to read~\Cref{sec_construction,sec_numerical results} to understand how to construct universal polar codes in practice, and to realize the benefits of using list decoding to decode these universal polar codes. We note that some definitions from~\Cref{sec_monopolarization for FAIM derived processes} are required to follow~\Cref{sec_construction}, but we refer to the relevant equations as the need arises.  
    \item A reader who is interested in understanding why the construction is universal is advised to turn to \Cref{sec_properties of BST and
        a variation,sec_polarizable soprocesses} after \Cref{sec_universal construction}. 
        These sections contain a detailed proof of universality of the construction, provided that one takes on faith that forgetful processes exist. 
    \item A sufficient condition for the existence of forgetful processes is developed in \Cref{sec_contraction,sec_HMM}. 
        The interested reader is advised to read them following \Cref{sec_probabilistic model with memory}. 
        \Cref{sec_contraction,sec_HMM} are written for a general hidden Markov model and may be read independently.
\end{itemize}

\section{Notation and Basic Definitions}\label{sec_notation conventions and reminders}
A discrete set of elements is denoted as a list in braces, e.g., $\{1,2,\ldots,L\}$, usually denoted with a calligraphic letter, e.g., $\mathcal{A}$. 
The number of elements in a discrete set $\mathcal{A}$ is denoted by $|\mathcal{A}|$. 
We denote $y_j^k = \begin{bmatrix} y_j &y_{j+1}& \cdots & y_k \end{bmatrix}$ for $j < k$. 
If $ j = k $ then $y_j^k = y_j$ and if $ j > k$ then $y_j^k$ is a null vector.

We use boldface to denote vectors, and, unless stated otherwise, vectors are assumed to be column vectors. 
The transpose of a column vector $\bv{x}$ is the row vector $\trp{\bv{x}}$. 
The $i$th element of a vector $\bv{x}$ is denoted by $\matel{\bv{x}}{i}$ (usually, and unless stated
otherwise, 
$\matel{\bv{x}}{i} = x_i$). 
Special vectors are the all-ones vector $\bv{1}$, all-zeros vector $\bv{0}$, and the unit vector $\bv{e}_i$, which has $1$ in its $i$th entry and zero 
in all other entries. 
We further define the norm 
\[ 
    \norm{\bv{x}}_1 = \sum_i | x_i |.
\]
An inequality involving vectors is assumed to be element-wise. 
Therefore, if $a$ is a scalar and $\bv{b}$ is a vector, $\bv{x} \geq a$ implies that $x_i \geq a$ for all $i$, and $\bv{x} \geq \bv{b}$ implies that 
$x_i \geq b_i$ for all $i$.
For two vectors (possibly of different lengths) $\bv{a}$ and $\bv{b}$ we write $\bv{a} \stackrel{f}{\equiv} \bv{b}$ if there is a one-to-one mapping $f$
between  $\bv{a}$ and $\bv{b}$; usually, $f$ is clear from the context, so we omit it and simply write
$\bv{a} \equiv \bv{b}$. 
The \emph{support} $\support{\bv{x}}$ of a vector $\bv{x}$ is the set of indices $i$ such that $x_i \neq 0$.
A vector is said to be nonzero if it has a non-empty support. 

Matrices are denoted using capital letters in sans-serif font, e.g., $\mat{M}$.
The $i,j$ element of a matrix $\mat{M}$ is denoted by $\matel{\mat{M}}{i,j}$.
The $i$th row of $\mat{M}$ is denoted by $\matel{\mat{M}}{i,:}$ and the $j$th column of $\mat{M}$ is denoted by $\matel{\mat{M}}{:,j}$.   
The identity matrix is denoted by $\Idmat$.
For matrix $\mat{M}$, we denote its set of nonzero rows\footnote{A row or column is nonzero if it has at least one nonzero element.} by
$\nzrows{\mat{M}}$ and its set of nonzero columns by $\nzcols{\mat{M}}$.  
The support $\support{\mat{M}}$ of a matrix $\mat{M}$ is the set of index pairs $(i,j)$ such that $i \in \nzrows{\mat{M}}$ and $j \in \nzcols{\mat{M}}$. 

The probability of an event $A$ is denoted by $\Probi{A}$.
Random variables are usually denoted using upper-case letters, e.g., $X$,  and their realizations using lower-case letters, e.g., $x$. 
The distribution of random variable $X$ is denoted by $P_X$.
The expectation of $X$ is denoted by $\Exp{X}$.
When $X_n$ is a sequence of random variables and $\bv{b} = \begin{bmatrix} b_1 & b_2 & \cdots & b_m \end{bmatrix}$ is a vector of indices, then
$X_{\bv{b}} = (X_{b_1}, X_{b_2}, \ldots, X_{b_m})$. 

Let $X$ and $Y$ be two discrete random variables taking values in alphabets $\mathcal{X}$ and $\mathcal{Y}$, respectively.
We define $H(X)$, the entropy of $X$, and $H(X|Y)$, the conditional entropy of $X$ given $Y$, by 
\begin{align*}
    H(X) &= -\sum_{x\in \mathcal{X}} \prrv{X}{x}\log \prrv{X}{x}, \\ 
    H(X|Y) &= -\sum_{y \in \mathcal{Y}}\sum_{x\in\mathcal{X}} \prrv{X,Y}{x,y}\log \prrv{X|Y}{x|y},
\end{align*}
where we follow the usual convention that $0\cdot \log 0 = 0$. 
Logarithms are base $2$ unless stated otherwise. 
The binary entropy function $h_2:[0,1]\to[0,1]$ is defined by 
\begin{equation} \label{eq_def of h2}
    h_2(x) = -x \log x - (1-x) \log(1-x).
\end{equation} 

The mutual information between $X$ and $Y$, denoted $I(X;Y)$ is defined by 
\[ I(X;Y) = H(X) - H(X|Y).\] 
Let $Q$ be an additional discrete random variable; the conditional mutual information of $X$ and $Y$ given $Q$ is  $I(X;Y|Q) = H(X|Q) - H(X|Y,Q)$. 

    The following variation of the data processing inequality will be useful. 
    Let $X,Y,Q,W$ be four random variables. 
    We introduce the notation $X \markov (Y,Q) \markov W$ whenever $X$ and $W$ are independent given $Y$ and $Q$.
    We then have the following variation of the data processing inequality:
    \begin{equation} \label{eq_DPI}
        X \markov (Y,Q) \markov W \Rightarrow I(X;Y|Q) \geq I(X;W|Q).
    \end{equation}  
    Indeed, on the one hand, $I(X;(Y,W) | Q) = I(X;Y|Q) + I(X;W|Y,Q) = I(X;Y|Q)$, where the last equality is by conditional independence.
    On the other hand $I(X;(Y,W)|Q) = I(X;W|Q) + I(X;Y|W,Q) \geq I(X;W|Q)$, since mutual information is nonnegative.
    
The following definition generalizes the concept of a channel. 
This generalization allows us to describe polarization transforms for channel coding and source coding in one fell swoop. 
\begin{definition}[\sopair]\label{def_so pair}
    A \emph{symbol-observation pair}, or \emph{\sopair} in short, is a pair of dependent random variables $X$ and $Y$.
    The random variable $X$ is called the \emph{symbol} and the random variable $Y$ is called the \emph{observation}.
    We use the notation $X\sarrow Y$ to denote an \sopair whose symbol is $X$ and whose observation is $Y$.
    The joint distribution of the \sopair is given by $\prrv{X,Y}{x,y}=\prrv{X}{ x }\prrv{Y|X}{ y|x }$. 
    The conditional entropy of an \sopair $X\sarrow Y$ is $H(X|Y)$.
\end{definition}
We emphasize that an \sopair is specified using the \emph{joint} distribution of $X$ and $Y$.
This is in contrast to a channel that is specified using only the conditional distribution of the output given its input.  
A channel with input $X$ and output $Y$ becomes an \sopair once the input distribution is specified. 
Another example of an \sopair is a source $X$ with distribution $\prrv{X}{x}$ to be estimated based on dependent observation $Y$ distributed according
to $\prrv{Y|X}{y|x}$. 

\begin{definition}[\soprocess] \label{def_so process}
    A sequence of \sopairs $X_i\sarrow Y_i$, $i=1,2,\ldots$ is called a \emph{symbol-observation process}, or \emph{\soprocess} in short.
    We use the notation $X_{\star} \sarrow Y_{\star}$. 
\end{definition}
\begin{definition}[\soblock] \label{def_so block}
    A sequence of $N$ consecutive \sopairs of an \soprocess is called an \emph{\soblock}. 
    We use the notation $X_1^N \sarrow Y_1^N$.  
    An \soblock has a natural indexing: $X_j \sarrow Y_j$ is \sopair $j$ of  \soblock $X_1^N \sarrow Y_1^N$. 
    The joint distribution of an \soblock is given by $\prrv{X_1^N, Y_1^N}{x_1^N,y_1^N} = \prrv{X_1^N}{x_1^N}\prrv{Y_1^N|X_1^N}{y_1^N|x_1^N}$.
\end{definition}

Generally, the \sopairs in an \soblock are dependent; that is, there is memory in the process. 
In this paper, we assume that \soprocesses are stationary.
In particular, this implies that for an \soblock $X_1^N \sarrow Y_1^N$, the \sopairs $X_i \sarrow Y_i$ are identically distributed for all $i$.

The \emph{conditional entropy rate} of a stationary \soprocess $X_{\star}\sarrow Y_{\star}$ is 
\begin{align*}
    \ENT{X_{\star}|Y_{\star}} &\triangleq \lim_{N\to\infty} \frac{1}{N}H(X_1^N|Y_1^N) \\ 
    &= \lim_{N\to\infty} \frac{1}{N}H(X_1^N,Y_1^N) - \lim_{N\to\infty} \frac{1}{N}H(X_1^N).
\end{align*} 
The limits on the right-hand side exist due to stationarity (see, e.g., \cite[Theorem 4.2.1]{cover_thomas}).

For simplicity, we assume throughout that \sopairs have binary symbols and that their observations are over a finite alphabet. 
Extension to the case where symbols are non-binary over an alphabet of prime size is possible using the techniques of~\cite[Chapter
3]{sasoglu_thesis}.
This entails replacing modulo-$2$ addition with modulo-$|\mathcal{X}|$ addition, where $|\mathcal{X}|$ is the symbol alphabet size, and replacing binary
entropies with non-binary entropies.

\section{Universal Polar Transform} \label{sec_universal construction}

In this section we describe the universal polar transform, which is based on~\cite{sasoglu_2016_universal}.
The transform described in~\cite{sasoglu_2016_universal} was used to construct a universal code over memoryless symmetric channels subject to
a capacity constraint. 
In this work, we extend the transform of~\cite{sasoglu_2016_universal} for \soprocesses with memory. 

This section is focused on describing the transform.
Properties of the transform and proof of its universality are presented in \Cref{sec_properties of BST and a variation,sec_polarizable soprocesses}. 
The decoding operation is described in \Cref{sec_decoding}. 

\subsection{Overview of the Transform}
In this section, we provide a general overview of the universal polar transform. 
It is a type of \htransform, a concept that we now define. 
\begin{definition}[\htransform] \label{def_H transform}
    A one-to-one and onto mapping $f$ between two symbol vectors of length $N$  is called an \emph{\htransform}. 

    Moreover, when we say that \soblock $X_1^N \sarrow Y_1^N$ is transformed to \soblock  $F_1^N \sarrow G_1^N$ by \htransform $f$, 
    we mean that:
    \begin{enumerate}
        \item $F_1^N = f(X_1^N)$; 
        \item $G_i = (F_1^{i-1}, Y_1^N)$, for any $i$. 
    \end{enumerate}
\end{definition}

\begin{example} \label{ex_arikan}
    \arikan's polar codes~\cite{Arikan_2009} are based on \htransforms. In this case, the mapping $f$ is given by $F_1^N = f(X_1^N)= \mat{B}_N
    \mat{G}_2^{\otimes n} X_1^N$, where $N = 2^n$, $\mat{B}_N$ is the $N\times N$ bit-reversal matrix, $\mat{G}_2 = \begin{bsmallmatrix} 1 & 0 \\
    1 & 1\end{bsmallmatrix}$, and $\otimes$ denotes a Kronecker product. 
\end{example}

The name ``\htransform'' is motivated by the equality
\begin{equation}       \label{eq_entropy equality}
    H(X_1^N|Y_1^N) = H(F_1^N | Y_1^N) = \sum_{i=1}^N H(F_i | G_i).
\end{equation} 
The right-most equality follows from the chain rule for entropies and the definition of $G_i$. 
Typically, the $f$ of an \htransform is defined recursively. 

Consider an \soblock $X_1^N \sarrow Y_1^N$, with \htransform $F_1^N \sarrow G_1^N$.  
We wish to recover the symbols $X_1^N$ from the observations $Y_1^N$. 
We denote the recovered symbols with a hat, $(\,\hat{\cdot}\,)$.
That is, $\hat{X}_1^N = \Phi(Y_1^N)$, where $\Phi(\cdot)$ is the algorithm for recovery.
We assess $\Phi$ by its error probability, $\Probi{\hat{X}_1^N \neq X_1^N}$. 
\htransforms, thanks to~\eqref{eq_entropy equality}, naturally give rise to a sequential algorithm called \emph{successive cancellation}.

Rather than computing $\hat{X}_1^N$ from $Y_1^N$ directly, we may compute $\hat{F}_1^N$ from $Y_1^N$.
By the properties of the \htransform, there exists a mapping $f$, with inverse $f^{-1}$, such that $X_1^N = f^{-1}(F_1^N)$. 
Any algorithm for recovering $F_1^N$ from $Y_1^N$ is equivalent to an algorithm for recovering $X_1^N$ from $Y_1^N$. 
For, if $\hat{F}_1^N = \Phi(Y_1^N)$ we can define $\hat{X}_1^N = f^{-1}(\hat{F}_1^N) = f^{-1}(\Phi(Y_1^N))$ and vice versa. 
Since $\Probi{\hat{F}_1^N \neq F_1^N} = \Probi{\hat{X}_1^N \neq X_1^N}$, we concentrate on an algorithm to recover $F_1^N$. 

One approach is to compute $\hat{F}_1^N$ sequentially as follows. 
Let $\Phi_i$ be a maximum-likelihood decoder of $F_i$ from $G_i$.
Compute $\hat{F}_1 = \Phi_1(\hat{G}_1)$, where $\hat{G}_1 = G_1 = Y_1^N$; then, assuming that $\hat{F}_1 = F_1$, form $\hat{G}_2 = (\hat{F}_1,
Y_1^N)$ and compute $\hat{F}_2 = \Phi_2(\hat{G}_2)$, and so on, culminating with $\hat{F}_N=\Phi_N(\hat{G}_N)$. 
This is tantamount to the successive-cancellation decoding described in~\cite{Arikan_2009}, and we will use the name ``successive cancellation'' to
describe this algorithm. 

It is well known~\cite[Proposition 2.1]{sasoglu_thesis} that the error probability of recovering $\hat{F}_1^N$ sequentially from $\hat{G}_1^N$ using
successive cancellation as described above is the same as if a genie had replaced $\hat{G}_i$ with $G_i$ at every step. 
That is, 
\[ 
    \Prob{\big(\Phi_i(\hat{G}_i)\big)_{i=1}^N \neq \big(F_i\big)_{i=1}^N} = \Prob{\big(\Phi_i(G_i)\big)_{i=1}^N \neq \big(F_i\big)_{i=1}^N}.
\]
(To see this, observe that if $\Phi_i(G_i) = F_i$ for all $i < i_0$ and $\Phi_{i_0}(G_{i_0}) \neq F_{i_0}$ then we must also have $\Phi_i(\hat{G}_i)
= F_i$ for all $i < i_0$ and $\Phi_{i_0}(\hat{G}_{i_0}) \neq F_{i_0}$.)
Therefore, when assessing the performance of successive cancellation, we may assume that at step $i$, $G_i$ (in contrast to $\hat{G}_i$) is known. 

\begin{definition}[Monopolarizing \htransform]
    Let $\eta > 0$ and let $ \setmonodown, \setmonoup \subseteq \{1,2, \ldots, N\}$ be two index sets. 
    An \htransform $f$ is ($\eta$, $\setmonodown$, $\setmonoup$)-\emph{monopolarizing} for a family of \soprocesses if
    for any  \soblock $X_1^{N} \sarrow Y_1^{N}$ in the family,
    either $H(F_i | G_i) \leq \eta$ for all $i \in \setmonodown$ or $H(F_i | G_i) \geq 1-\eta$ for all $i \in \setmonoup$, where  \soblock
    $F_1^N\sarrow G_1^N$   
    denotes the transformed \soblock.
\end{definition}

Monopolarizing \htransforms are useful because they make the process of recovering $\hat{F}_i$ from $G_i$ very easy whenever $H(F_i|G_i) \approx 0$,
because then $F_i$ is approximately a deterministic function of $G_i$. 
On the other hand, if $H(F_i|G_i) \approx 1$ we know that $F_i$ is essentially a result of a uniform coin flip, independent of $G_i$.

The universal transform is a moniker for a family of \htransforms with increasing lengths.
It comprises two stages: a slow polarization stage and a fast polarization stage. 
Each is an \htransform that is constructed recursively. 
Our goal is to show that, as the blocklength increases, they become monopolarizing.

Recursive construction of an \htransform begins with an initial \htransform $f_0$ of length $N_0$. 
Then, at step $n+1$ we take step-$n$ \htransforms of consecutive symbol vectors to generate a step-$(n+1)$ \htransform of a single, larger, symbol
vector.
A typical case is as follows.                                                 
Let $f_n$ be an \htransform of length $N_n$ that results from step $n$, 
and let $\varphi_{n+1}$ be a one-to-one and onto mapping from two length $N_n$ vectors to a vector of length $N_{n+1} = 2N_n$. 
Apply $f_n$ to two consecutive  symbol vectors: $U_1^{N_n}= f_n( X_1^{N_n})$ and $V_1^{N_n} = f_n( X_{N_n+1}^{2N_n})$.
Then, form $F_1^{N_{n+1}} = \varphi_{n+1}(U_1^{N_n}, V_1^{N_n}) = f_{n+1}(X_1^{N_{n+1}})$.

A basic building block is the \arikan transform~\cite{Arikan_2009}, illustrated in \Cref{fig_arikan transform}.  
It operates on two input symbols: input-$\seta$: $U$ (with observation $Q$) and input-$\setb$: $V$ (with observation $R$)
and transforms them to two new symbols: 
a `$\setMinus$' symbol $F_1$ (with observation $G_1$) and a `$\setPlus$' symbol $F_2$ (with observation $G_2$), where $F_1 = U + V$, $G_1 = (Q,R)$ and $F_2 = V$, $G_2 = (F_1, Q, R)$.  
Schematically, the \arikan transform is as follows:
\[ 
    \left\{
        \begin{IEEEeqnarraybox}[][c]{rCl}
            \IEEEstrut
                \seta: U   &\sarrow& Q    \\
                \setb: V   &\sarrow& R
            \IEEEstrut
        \end{IEEEeqnarraybox}
    \right. 
    \Rightarrow                                    
    \left\{
        \begin{IEEEeqnarraybox}[][c]{lCc}
            \IEEEstrut
            \text{`$\setMinus$'}: \underbrace{U+V \vphantom{)}}_{F_1} &\sarrow& \underbrace{    (Q,R)}_{G_1}\phantom{.}    \\
            \text{`$\setPlus$'}: \underbrace{\vphantom{)}\hphantom{\:{}+}V\hphantom{+{}}}_{F_2}    &\sarrow& \underbrace{(F_1,Q,R)}_{G_2}.
            \IEEEstrut
        \end{IEEEeqnarraybox}
    \right. 
\]
It is evident that an \arikan transform is an \htransform of length $2$. 

\begin{figure}[t]
    \begin{center}
    \begin{tikzpicture}[rect/.style={rectangle,minimum width = 0.97cm, minimum height = 0.6cm},
        circ/.style={fill=white,draw,circle,minimum size = 8pt, inner sep = 0pt}, 
                        dot/.style={draw,circle, fill, minimum size = 3 pt, inner sep = 0pt}, >=latex,
                        label distance = 2pt ]
        
        \def\YDIST{1.4}
        \def\XDIST{3.4}

        \foreach \i/\j/\k in {1/U/Q,2/V/R}
        {
            \node (F\i) at (0, -{\i-1}*\YDIST) {$F_{\i}$}; 
            \node[right = \XDIST of F\i] (U\i) {$\j$};
            \node[right = \XDIST/5 of U\i] (Q\i) {$\k$};
            \draw[dotted] (U\i) -- (Q\i);
        }

        \foreach \i/\j/\k/\l/\h in {1/circ/{+}/{$\seta$}/{$\setMinus$}, 2/dot/{}/{$\setb$}/{$\setPlus$}}
        {
            \coordinate (D\i) at ($(F\i.center)!0.5!(U\i.center)$);
            \draw (F\i)--node[above, pos=0.2]{\scriptsize `\h'} (D\i)-- node[above, pos=0.8]{\scriptsize \l}  (U\i);
            \node[\j] (E\i) at (D\i) {$\k$}; 
        }
        \draw (E1)--(E2); 

        \node[draw, rounded corners, dashed,  label=above:{\footnotesize \arikan Transform}, fit = {(D1) (D2)}, minimum width = 2.0cm, minimum height
            = 1.6cm] {}; 

    \end{tikzpicture}
\end{center}
\caption{Illustration of an \arikan transform. It transforms two input symbols, $U$ (input-$\seta$) and $V$ (input-$\setb$) to two output symbols,
    $F_1$ (output `$\setMinus$') and $F_2$ (output `$\setPlus$').}
    \label{fig_arikan transform}
\end{figure}

For an \arikan transform, we obtain
\begin{align*} 
    H(F_1 | G_1) + H(F_2 | G_2) &= H(F_1^2 | Q,R) \\ 
    &= H(U,V |Q,R) \leq H(U|Q) +H(V|R).
\end{align*} 
The inequality is because the \sopairs $U\sarrow Q$ and $V\sarrow R$ are generally dependent. 
Informally, \arikan transforms facilitate polarization if one can show that $H(F_1|G_1) \geq \max\{H(U|Q), H(V|R)\}$ 
and that the inequality is strict unless either $H(U|Q)$ or $H(V|R)$ is extremal. 
This was the strategy of obtaining polarization for standard (\arikan's) polar codes, with and without memory. 
See, for example,~\cite{Arikan_2009,sasoglu_thesis,sasoglu_Tal_mem}. 
We will also pursue such a strategy. 

\subsection{Slow Polarization Stage}\label{sec_slow polarization stage}
In this subsection we describe the slow polarization stage. 
We will focus on describing a slow stage transform called a \emph{basic slow transform} (\BST). 
It is an extension of the transform shown in~\cite[Section II]{sasoglu_2016_universal}.

The basic slow transform is constructed recursively. 
We call each step in the construction a \emph{level}. 
Each level is an \htransform of length $N_n = 2L_n+M_n$. 
We will specify how to compute $L_n$ and $M_n$ later in~\eqref{eq_computation of LN1 MN1}. 
We call the transformed \soblock a \emph{level-$n$ block}. 

We define the following index sets for a level-$n$ block, $n \geq 0$.   
See \Cref{fig_level n block} for an illustration.
\begin{IEEEeqnarray}{rCl} \IEEEyesnumber \label{eq_defs of lat-med sets}
    \latsettop{n} &\triangleq& \{i \ | \ \hphantom{L_n+M_n+{}} 1 \leq i \leq L_n\},  \IEEEyessubnumber \label{eq_def of lat1}\\
    \latsetbot{n} &\triangleq& \{i \ | \  L_n + M_n + 1 \leq i \leq N_n\},  \IEEEyessubnumber \label{eq_def of lat2} \\
    \latset{n}    &\triangleq& \latsettop{n} \cup \latsetbot{n},  \IEEEyessubnumber \label{eq_def of latset} \\
    \medsetMinus{n}   &\triangleq& \{i \ | \ i = L_n + 2k-1, \ 1 \leq k \leq M_n/2\},  \IEEEeqnarraynumspace \IEEEyessubnumber \label{eq_def of medA} \\    
    \medsetPlus{n}   &\triangleq& \{i \ | \ i = L_n + 2k,\hphantom{{}-1} \ 1 \leq k \leq M_n/2\},  \IEEEyessubnumber \label{eq_def of medB}\\
    \medset{n}    &\triangleq& \medsetMinus{n} \cup \medsetPlus{n}.  \IEEEyessubnumber \label{eq_def of medset}
\end{IEEEeqnarray}                              
In words, the sets $\latsettop{n}$ and $\latsetbot{n}$ are, respectively, the first $L_n$ and last $L_n$ indices in a level-$n$ block. 
Then, the remaining $M_n$ indices alternate between $\medsetMinus{n}$ and $\medsetPlus{n}$, starting with $\medsetMinus{n}$ and ending with
$\medsetPlus{n}$. 

We  classify symbols in an \soblock according to their indices as follows: 
\begin{itemize}
    \item $i \in \latset{n}$ \tabto{2.2cm} $\Rightarrow $ symbol $i$ is lateral; 
    \item $i \in \medset{n}$ \tabto{2.2cm} $\Rightarrow $ symbol $i$ is medial; 
\end{itemize}
We will sometimes classify \sopairs based on the classification of the indices. 
For example, we say that \sopair $i$ is lateral if symbol $i$ is lateral.

The construction is initialized with integer parameters $L_0$ and $M_0$. 
We assume that $M_0$ is even.\footnote{This is not necessary, and it is possible to initialize the construction with odd $M_0$.
However, assuming that $M_0$ is even  ensures that the index sets defined in~\eqref{eq_defs of lat-med sets} hold also for $n=0$.} 
\begin{itemize}
    \item The parameter $L_0$ determines, informally, ``how much memory'' in the \soprocess the transform can handle; see \Cref{sec_polarizable
        soprocesses} for more details. 
        For a memoryless process, it may be set to $0$. 
    \item The parameter $M_0$ has a dual role: 
        \begin{itemize}
            \item Informally, it is set large enough so that two \sopairs that are $M_0$ time-indices apart may be considered almost independent. See
                \Cref{sec_polarizable soprocesses} for more details. 
            \item It controls the fraction of medial symbols in an \soblock.
            See \Cref{lem_proportion of lateral sopairs} for details.  
        \end{itemize}
\end{itemize}

The initial step $f_0$, which generates a level-$0$ block, is an \htransform of length $N_0 = 2 L_0 + M_0$. 
We set $f_0$ as the identity mapping. 
Thus, the initial step transforms an \soblock $X_1^{N_0} \sarrow Y_1^{N_0}$ into an \soblock $F_1^{N_0} \sarrow G_1^{N_0}$, where, for $1 \leq i  \leq N_0$, 
\begin{IEEEeqnarray}{rCl} \IEEEyesnumber \label{eq_level0}
    F_i &=& X_i, \IEEEyessubnumber\\ 
    G_i &=& (F_1^{i-1}, Y_1^{N_0}). \IEEEyessubnumber
\end{IEEEeqnarray}

\begin{figure}[t]
    \begin{center}
        \begin{tikzpicture}[rect/.style={draw,rectangle,minimum width = 0.97cm, minimum height = 0.6cm},
                            circ/.style={draw,circle,minimum size = 8pt, inner sep = 0pt}, 
				         	dot/.style={draw,circle, fill, minimum size = 1 pt, inner sep = 0pt}, >=latex]
         
            \node[rectangle, minimum width = 3.5cm, minimum height = 5cm, label={Level-$n$ block}] (R) at (1.75,2.5) {}; 

            \foreach \i/\j in {1/0.025 , 2/0.175 , 3/0.225, 4/0.775 , 5/0.825 , 6/0.975 }
            {
                \node (X\i) at ($(R.north west)!\j!(R.south west)$) {};  
                \node (Y\i) at ($(R.north east)!\j!(R.south east)$) {};  
            }
            
            \foreach \i/\j in {1/0.225 , 2/0.2750 , 3/0.325, 4/0.375 , 5/0.625, 6/0.675, 7/0.7250, 8/0.775 }
            {
                \node (G\i) at ($(R.north east)!\j!(R.south east)$) {};  
                \draw (G\i.center) -- +(0.2,0); 
            }

            \foreach \i in {0.25,0.5,0.75}
            {
                \node[dot] at ($(G4)!\i!(G5) + (0.1,0)$) {}; 
            }

            \path[fill = orange!25!white] (0,0) rectangle ($(Y4)!0.5!(Y5)$) node[pos=0.5]{\footnotesize lateral \sopairs} ; 
            \path[fill = orange!25!white] ($(X2)!0.5!(X3)$) rectangle (3.5,5) node[pos=0.5]{\footnotesize lateral \sopairs} ; 
            \path[fill = green!25!white] ($(X2)!0.5!(X3)$) rectangle ($(Y4)!0.5!(Y5)$) node[pos=0.5]{\footnotesize medial \sopairs} ; 
            
            \foreach \i/\j in {1/{1}, 2/{L_n}, 3/{L_n+1}, 4/{L_n+M_n}, 5/{L_n + M_n+1}, 6/{N_n} }   
            {
                \draw ($(X\i.center) - (0.2,0)$) -- (X\i.center);  
                \node[outer sep = 0pt, inner sep = 0pt] (M\i) at ($(X\i.center)!0.5!(Y\i.center)$) {  \tiny $F_{\j} \sarrow G_{\j}$};
                \draw[dotted] (X\i.center) -- (M\i) --    (Y\i.center); 
                \draw (Y\i.center) -- ($(Y\i.center) + (0.2,0)$); 
            }

            \node[magenta] (MA) at ($(G1) + (2,0)$) {$\medsetMinus{n}$}; 
            \node[blue] (MB) at ($(G8) + (2,0)$) {$\medsetPlus{n}$}; 
            
            \foreach \i/\j in {1/2,3/4,5/6,7/8}
            {
                \draw[magenta] (G\i.center) -- +(0.2,0) -- (MA.west); 
                \draw[blue]    (G\j.center) -- +(0.2,0) -- (MB.west); 
            }

            \node[draw, rectangle, minimum width = 3.5cm, minimum height = 5cm] at (R) {}; 

            \draw[orange, decorate, thick, decoration={brace,mirror}] ($(R.north west)-(0.5,0)$) -- node[left]{$\latsettop{n}$} ($(X2)!0.5!(X3) - (0.5,0)$) ;
            \draw[orange, decorate, thick, decoration={brace,mirror}] ($(X4)!0.5!(X5) - (0.5,0)$)  -- node[left]{$\latsetbot{n}$} ($(R.south west)-(0.5,0)$);
            \draw[green!50!black, decorate, thick, decoration={brace,mirror}]  ($(X2)!0.5!(X3) - (0.5,0)$) -- node[left]{$\medset{n}$}  ($(X4)!0.5!(X5) - (0.5,0)$);

            \draw[dotted, thin] ($(X2)!0.5!(X3) - (1.5,0)$) -- +(1.5,0); 
            \draw[dotted, thin] ($(X4)!0.5!(X5) - (1.5,0)$) -- +(1.5,0); 
        \end{tikzpicture}
    \end{center}
    \caption{Index sets in level $n$ of the basic slow transform. A Level-$n$ block comprises $N_n = 2L_n + M_n$ \sopairs. 
             The first $L_n$ and the last $L_n$ \sopairs are lateral \sopairs and the remaining $M_n$ \sopairs are medial \sopairs.}
        \label{fig_level n block}
    \end{figure}

    We now construct a level-$(n+1)$ \BST from two level-$n$ \BSTs. 
Denote by $f_n$ a \BST of length $N_n$. 
We will define $f_{n+1}$ using a one-to-one and onto mapping $\varphi_{n+1}$ from two length-$N_n$ vectors to a single length-$N_{n+1} = 2N_n$ vector. 
The mapping $\varphi_{n+1}$ is defined in~\eqref{eq_Fj for lateral indices} and~\eqref{eq_Fj for medial indices} below. 

The \BSTs of the two consecutive level-$n$ \soblocks are 
\begin{IEEEeqnarray}{rCl'rCl'rCl} \IEEEyesnumber \label{eq_defs of UVQR}
    U_1^{N_n} &=& f_n(X_1^{N_n}),        & Q_i &=& (U_1^{i-1}, Y_1^{N_n}),       & 1 &\leq &i \leq N_n, \IEEEyessubnumber \label{eq_defs of UQ}
    \\[0.15cm] 
    V_1^{N_n} &=& f_n(X_{N_n+1}^{2N_n}), & R_i &=& (V_1^{i-1}, Y_{N_n+1}^{2N_n}),& 1 &\leq &i \leq N_n.  \IEEEyessubnumber \IEEEeqnarraynumspace  \label{eq_defs of VR}  
\end{IEEEeqnarray}
Denoting $N_{n+1} = 2N_n$, we obtain the level-$(n+1)$ transformed \soblock 
\begin{IEEEeqnarray}{rCl} \IEEEyesnumber\label{eq_varphi level n+1}
    F_1^{N_{n+1}} &=& \varphi_{n+1}(U_1^{N_n}, V_1^{N_n}) = f_{n+1}(X_{1}^{N_{n+1}}), \IEEEyessubnumber\\ 
    G_i           &=& (F_1^{i-1}, Y_1^{N_{n+1}}), \quad 1 \leq i \leq N_{n+1}. \IEEEyessubnumber 
\end{IEEEeqnarray}

The level-$(n+1)$ block is of length $N_{n+1} = 2L_{n+1} + M_{n+1}$, where 
\begin{IEEEeqnarray}{rCl} \IEEEyesnumber \label{eq_computation of LN1 MN1}
    L_{n+1} &=& 2L_n + 1 \IEEEyessubnumber    \label{eq_computation of LN1} \\ 
    M_{n+1} &=& 2(M_n - 1). \IEEEyessubnumber \label{eq_computation of MN1}
\end{IEEEeqnarray}
Indeed, $N_{n+1} = 2L_{n+1} + M_{n+1} = 2(2L_n + M_n) = 2N_n$. 
\begin{remark} \label{rem_sets alt}
Observe that $L_n$ is odd and $M_n$ is even for any $n \geq 1$. 
Therefore, by \eqref{eq_defs of lat-med sets}, for any $n\geq 1$,  the set $\medsetMinus{n}$ is the set of even indices of $\medset{n}$ and the set $\medsetPlus{n}$ is the set of odd
indices of $\medset{n}$. 
\end{remark}

    Lateral symbols of a level-$(n+1)$ block are formed by renaming symbols of level-$n$ \sopairs, as follows: 
\begin{equation}  \label{eq_Fj for lateral indices}
    i \in \latset{n+1} \Rightarrow    
     F_i = \begin{cases}
                U_j, & i = 2j-1, \\  
                V_j, & i = 2j.
            \end{cases}  
\end{equation}
This is illustrated in \Cref{fig_schematic ln to ln+1}. 
Observe that all lateral symbols of the level-$n$ blocks become lateral symbols of the level-$(n+1)$ block.  
Additionally, note that, by~\eqref{eq_defs of lat-med sets},~\eqref{eq_computation of LN1 MN1}, and~\eqref{eq_Fj for lateral indices}, two medial symbols of the level-$n$ blocks become
lateral symbols of the level-$(n+1)$ block: 
\[ 
    F_{L_{n+1}} = F_{2(L_n+1) - 1} = U_{L_n+1}
\] 
and  
\[ 
    F_{L_{n+1}+M_{n+1}+1} = F_{2(L_n+M_n)} = V_{L_n+M_n}. 
\] 

\begin{figure}[t]
    \begin{center}
        \begin{tikzpicture}[>=latex,rect/.style={rectangle,minimum width = 1.5cm, minimum height = 2.5cm},
            bigrect/.style={rectangle,minimum width = 1.5cm, minimum height = 5cm}]

            \node[rect] (R1) at (0,0) {}; 
            \node[rect, below = 0.3 of R1] (R2) {}; 

            \coordinate (Rmid) at ($(R1)!0.5!(R2)$); 
            \node[bigrect, left = 2.5 of Rmid] (R3) {}; 

            \foreach \i/\j in {1/{U\sarrow Q},2/{V\sarrow R},3/{F\sarrow G}}
            {
                \fill[orange!25!white]   ($(R\i.north west)!0.1!(R\i.south west)$) rectangle (R\i.north east) node[black, pos = 0.5] {\scriptsize lateral};
                \fill[orange!25!white]   ($(R\i.north west)!0.9!(R\i.south west)$) rectangle (R\i.south east) node[black, pos = 0.5] {\scriptsize lateral};
                \fill[green!25!white] ($(R\i.north west)!0.9!(R\i.south west)$) rectangle ($(R\i.north east)!0.1!(R\i.south east)$) node[black, pos = 0.5] {$\j$};
            }

            \foreach \i in {1,2}
            {
                \draw[orange, decorate, decoration={brace, amplitude=1pt, mirror}] ($(R\i.north west)-(0.05,0)$) -- node (T\i){}( $(R\i.north west)!0.1!(R\i.south west) - (0.05,0)$) ;
                \draw[orange, decorate, decoration={brace, amplitude=1pt, mirror}] ($(R\i.north west)!0.9!(R\i.south west)-(0.05,0)$) -- node (B\i){} ($(R\i.south west) - (0.05,0)$) ;
            }

            \draw[orange, decorate, decoration={brace}] ($(R3.north east)!0.9!(R3.south east)+(0.05,0)$) -- node (B3){} ($(R3.south east) + (0.05,0)$) ;
            \draw[orange, decorate, decoration={brace, mirror}]         ($(R3.north east)!0.1!(R3.south east)+(0.05,0)$) -- node (T3){} ($(R3.north east) + (0.05,0)$) ;

            \foreach \i/\j in {1/0,2/0}
            {
                \draw[-{Latex[open, round, width = 3.5pt]},orange, double] (T\i.west) -- (T3.\j); 
                \draw[-{Latex[open, round, width = 3.5pt]},orange, double] (B\i.west) -- (B3.\j); 
            }

            \draw[orange,ultra thin, -{Latex[length=2pt]}] ($(R1.north west)!0.12!(R1.south west)$) -- ($(T3.0) + (-0.085,-0.225)$); 
            \draw[orange,ultra thin, -{Latex[length=2pt]}] ($(R2.north west)!0.88!(R2.south west)$) -- ($(B3.0) + (-0.085, 0.225)$);

            \foreach \i in {1,2}
            {
                \draw[Bar-, ultra thin]    ($(R\i.north east)+(0.15,0)$) -- node[right]{\scriptsize $L_n$} ( $(R\i.north east)!0.1!(R\i.south east) + (0.15,0)$) ;
                \draw[Bar-Bar, ultra thin] ($(R\i.north east)!0.1!(R\i.south east)+(0.15,0)$) -- node[right]{\scriptsize $M_n$} ($(R\i.north east)!0.9!(R\i.south east)+(0.15,0)$);
                \draw[-Bar, ultra thin]    ($(R\i.north east)!0.9!(R\i.south east)+(0.15,0)$) -- node[right] {\scriptsize $L_n$} ($(R\i.south east) + (0.15,0)$) ;
            }

            \draw[Bar-, ultra thin]    ($(R3.north west)-(0.15,0)$) -- node[left]{\scriptsize $L_{n+1} = 2L_n + 1$} ( $(R3.north west)!0.1!(R3.south west) - (0.15,0)$) ;
            \draw[Bar-Bar, ultra thin] ($(R3.north west)!0.1!(R3.south west)-(0.15,0)$) -- node[left]{\scriptsize $M_{n+1} = 2(M_n-1)$} ($(R3.north west)!0.9!(R3.south
                west)-(0.15,0)$);
            \draw[-Bar, ultra thin]    ($(R3.north west)!0.9!(R3.south west)-(0.15,0)$) -- node[left] {\scriptsize $L_{n+1} = 2L_n + 1$} ($(R3.south west) - (0.15,0)$) ;

            \node[draw, rect, label={\small Level-$n$ block}] at (R1) {} ; 
            \node[draw, rect, label=below:{\small Level-$n$ block}] at (R2) {}; 
            \node[draw, bigrect, label={\small Level-$(n+1)$ block}] at (R3) {}; 

        \end{tikzpicture}
        \end{center}
        \caption{A schematic description of forming lateral \sopairs of a level-$(n+1)$ block from two level-$n$ blocks.}
        \label{fig_schematic ln to ln+1}
    \end{figure}

The medial symbols of a level-$(n+1)$ block are formed using \arikan transforms, as illustrated in \Cref{fig_level n+1 of universal construction}.
That is, medial symbols of a level-$(n+1)$ block are computed according to: 
\begin{align} \label{eq_Fj for medial indices}
    &i \in \medset{n+1}  \Rightarrow \nonumber \\ 
    &\qquad F_i = \begin{cases} 
        U_{j+1} + V_j, & i = 2j, \\ 
        V_j,           & i = 2j+1,\ j \in \medsetMinus{n}, \\ 
        U_{j+1},       & i = 2j+1,\ j \in \medsetPlus{n}.
        \end{cases}
\end{align}
We emphasize that by \eqref{eq_defs of lat-med sets} and \eqref{eq_computation of LN1 MN1},
\begin{equation} \label{eq_j is medial}
    i \in \medset{n+1} \Leftrightarrow \left\{\left\lfloor \frac{i}{2}\right\rfloor, \left\lfloor \frac{i}{2}\right\rfloor + 1\right\}\in \medset{n}. 
\end{equation}
That is, medial symbols of a level-$(n+1)$ \BST are generated by combining medial symbols of level-$n$ \BSTs. 
This can be seen either from \Cref{fig_level n+1 of universal construction} or from~\eqref{eq_defs of lat-med sets},~\eqref{eq_computation of LN1
MN1}, and~\eqref{eq_Fj for medial indices}. 
In particular,~\eqref{eq_Fj for medial indices} and \eqref{eq_j is medial} imply that for any $n \geq 0$, 
\begin{equation*} \label{eq_j is 2i}
    \begin{split}
        i \in \medsetMinus{n+1} &\Leftrightarrow i = 2j,\hphantom{{}+1}\ j \in \medset{n}, \  j \neq N_n - L_n, \\
        i \in \medsetPlus{n+1} &\Leftrightarrow i = 2j+1, \ j \in \medset{n}, \  j \neq N_n - L_n. 
    \end{split}
    \end{equation*}

    \begin{figure}[t]
    \begin{center}
        \begin{tikzpicture}[rect/.style={draw,rectangle,minimum width = 0.97cm, minimum height = 0.6cm},
                            smallrect/.style={draw,rectangle,minimum width = 0.32cm, minimum height
                            = 0.2cm},
                            largerect/.style={rectangle,minimum width = 2.4cm, minimum height = 5.0cm},
                            largerect1/.style={rectangle,minimum width = 2.4cm, minimum height = 4.35cm},
                            circ/.style={draw,fill=white,circle,minimum size = 4pt, inner sep = 0pt}, 
				         	dot/.style={draw,circle, fill, minimum size = 2 pt, inner sep = 0pt},
				         	ldot/.style={draw,circle, fill, minimum size = 3 pt, inner sep = 0pt},
				         	sdot/.style={draw,circle, fill, minimum size = 1 pt, inner sep = 0pt}, >=latex]

            \def\YDIST{1.0}

            \node[largerect] (R1) at (0,0) {}; 
            \node[largerect, below = -0.5 of R1] (R2) {}; 

            \foreach \i/\j in {1/{U\sarrow Q},2/{V\sarrow R}}
            {
                \fill[orange!25!white]   ($(R\i.north west)!0.155!(R\i.south west)$) rectangle ($(R\i.north east)!0.065!(R\i.south east)$) node[black, pos = 0.5] {\scriptsize lateral};
                \fill[orange!25!white]   ($(R\i.north west)!0.845!(R\i.south west)$) rectangle ($(R\i.north east)!0.935!(R\i.south east)$) node[black, pos = 0.5] {\scriptsize lateral};
                \fill[green!25!white] ($(R\i.north west)!0.845!(R\i.south west)$) rectangle ($(R\i.north east)!0.155!(R\i.south east)$) node[black, pos = 0.5] {$\j$};
            }

            \foreach \i [evaluate=\i as \x using \i/17] in  {0,1,...,17}
            {
                \coordinate (TW\i) at ($(R1.north west)!\x!(R1.south west)$);
                \coordinate (BW\i) at ($(R2.north west)!\x!(R2.south west)$);
                \coordinate (TE\i) at ($(R1.north east)!\x!(R1.south east)$);
                \coordinate (BE\i) at ($(R2.north east)!\x!(R2.south east)$);
            } 

            \draw (TW3)  -- node[above = -0.075]{\tiny $U_{L_n+1}$}     +(-1.35,0); 
            \draw[thick] (TW4)  -- node[above = -0.075]{\tiny $U_{L_n+2}$}     +(-1.35,0); 
            \draw (TW5)  -- node[above = -0.075]{\tiny $U_{L_n+3}$}     +(-1.35,0); 
            \draw (TW13) -- node[above = -0.075]{\tiny $U_{L_n+M_n-1}$} +(-1.35,0); 
            \draw (TW14) -- node[above = -0.075]{\tiny $U_{L_n+M_n}$}   +(-1.35,0); 
            \draw (BW3)  -- node[above = -0.075]{\tiny $V_{L_n+1}$}     +(-1.35,0); 
            \draw (BW4)  -- node[above = -0.075]{\tiny $V_{L_n+2}$}     +(-1.35,0); 
            \draw (BW13) -- node[above = -0.075]{\tiny $V_{L_n+M_n-1}$} +(-1.35,0); 
            \draw (BW14) -- node[above = -0.075]{\tiny $V_{L_n+M_n}$}   +(-1.35,0); 

            \foreach \x in {TW,BW}
            {
                \foreach \i [evaluate=\i as \j using int(\i+1)] in {3,5,11,13}
                {
                    \draw (\x\i) -- +(0.1,0) node[right, magenta, pos = -0.075]  (C\x\i) {\scriptsize $\bs{\setMinus}$}; 
                    \draw (\x\j) -- +(0.1,0) node[right, blue, pos = -0.075] (C\x\j) {\scriptsize $\bs{\setPlus}$}; 
                }

                \foreach \y in {0.25,0.5,0.75}
                {
                    \node[sdot] at ($(C\x6)!\y!(C\x11)$){}; 
                    \node[sdot] at ($(C\x6)!\y!(C\x11) + (-0.95,0)$){}; 
                    \node[sdot] at ($(C\x6)!\y!(C\x11) + (-2.35,0)$){}; 
                }

            }

            \node[circ,thick] (X1) at ($(CTW4) + (-1.7,0)$) {\scriptsize $\bs{+}$}; 
            \node[ldot]  (Y1) at ($(CBW3) + (-1.7,0)$) {}; 
            \node[circ] (X2) at ($(CBW4) + (-2.1,0)$) {\scriptsize $+$}; 
            \node[dot]  (Y2) at ($(CTW5) + (-2.1,0)$) {}; 

            \node[circ] (X3) at ($(CTW14) + (-3.0,0)$) {\scriptsize $+$}; 
            \node[dot]  (Y3) at ($(CBW13) + (-3.0,0)$) {}; 
            \node[circ] (X4) at ($(CBW12) + (-2.6,0)$) {\scriptsize $+$}; 
            \node[dot]  (Y4) at ($(CTW13) + (-2.6,0)$) {};   

            \draw[very thick] (CTW4) -- (X1) -- ($(CTW4)-(3.5,0)$) node[left] (Ft) {\scriptsize $F_{2L_n+2}$} 
                  (X1) -- (Y1)
                  (CBW3) -- (Y1) -- ($(CBW3)-(3.5,0)$) node[left] {\scriptsize $F_{2L_n + 3}$}; 
            \draw (CBW4) -- (X2) -- ($(CBW4)-(3.5,0)$) node[left] {\scriptsize $F_{2L_n+4}$} 
                  (X2) -- (Y2)
                  (CTW5) -- (Y2) -- ($(CTW5)-(3.5,0)$) node[left] {\scriptsize $F_{2L_n + 5}$}; 

            \draw (CBW12) -- (X4) -- ($(CBW12)-(3.5,0)$) node[left] {\scriptsize $F_{2L_n+2M_n-4}$} 
                  (X4) -- (Y4)
                  (CTW13) -- (Y4) -- ($(CTW13)-(3.5,0)$) node[left] {\scriptsize $F_{2L_n+2M_n-3}$};
            \draw (CTW14) -- (X3) -- ($(CTW14)-(3.5,0)$) node[left] {\scriptsize $F_{2L_n+2M_n-2}$} 
                  (X3) -- (Y3)
                  (CBW13) -- (Y3) -- ($(CBW13)-(3.5,0)$) node[left] (Fb) {\scriptsize $F_{2L_n+2M_n - 1}$};

            \node[largerect1, draw, label = above:{Level-$n$ block}] at (R1) {};
            \node[largerect1, draw, label = below:{Level-$n$ block}] at (R2) {};

            \node[draw,green!50!black, thick, rounded corners, dashed, fit = {(Ft) (Fb)}, label ={[green!50!black]above:{$\medset{n+1}$}}] {};

        \end{tikzpicture}
    \end{center}
    \caption{ Forming the medial symbols of level $n+1$ of the basic slow transform. 
        \arikan transforms are used with a symbol from $\medsetPlus{n}$ of one block as their input-$\seta$ and a symbol from $\medsetMinus{n}$ of the other
        block as their input-$\setb$. One \arikan transform is highlighted using thicker edges.} 
    \label{fig_level n+1 of universal construction}
\end{figure}

\Cref{fig_level n+1 of universal construction} makes it clear that the medial symbols of a level-$(n+1)$ block are formed in pairs. 
Overall, $M_n-1$ \arikan transforms are performed in forming the medial symbols of a level-$(n+1)$ block. 
Recall that an \arikan transform has two inputs, $\seta$ and $\setb$, see \Cref{fig_arikan transform}. 
In each \arikan transform, input-\seta is a symbol from $\medsetPlus{n}$ of one level-$n$ block and input-\setb is a symbol from $\medsetMinus{n}$ of the
other level-$n$ block. 
The blocks \emph{alternate} between successive \arikan transforms: look at $F_{2L_n+2}, F_{2L_n+3}, F_{2L_n+4},$ and $F_{2L_n+5}$ in \Cref{fig_level
n+1 of universal construction}. 

We saw above that the first medial symbol of the first level-$n$ block and the last medial symbol of the second level-$n$ block become lateral
symbols of the level-$(n+1)$ block;  they do not participate in forming medial symbols of the level-$(n+1)$ block. 
This explains why the index of $U$ leads by one the index of $V$ in~\eqref{eq_Fj for medial indices}. 

By \eqref{eq_Fj for medial indices}, when $2j \in \medset{n+1}$, $F_{2j}$ and $F_{2j+1}$ are the outputs of an \arikan transform of $U_{j+1}$ and $V_j$. 
 The expression for $F_{2j}$ is always the same: $F_{2j} = U_{j+1} + V_j$. 
The expression for $F_{2j+1}$ depends on which of $U_{j+1}$ or $V_j$ is input-$\setb$ of the \arikan transform. 
One of $j$ and $j+1$ is in $\medsetMinus{n}$ and the other is in $\medsetPlus{n}$. 
Since we form medial symbols using \arikan transforms with input-$\setb$ symbols from $\medsetMinus{n}$ of a level-$n$ block, 
$F_{2j+1}$ is assigned according to  the classification of $j$. 
Observe that for any $n \geq 1$, by \Cref{rem_sets alt}, the condition ``$j \in \medsetMinus{n}$'' is the same as ``$j$ is even'', and the condition ``$j \in
\medsetPlus{n}$'' is the same as ``$j$ is odd.''

We pause momentarily to introduce some terminology that will be useful in the sequel. 
\begin{definition}[Ancestors and Base-ancestors] \label{def_ancestors}
    An \arikan transform --- see \Cref{fig_arikan transform} --- maps two symbols, $U$ and $V$, into two transformed symbols, $F_1$ and $F_2$. 
    Medial symbols are generated by \arikan transforms, as evident by \Cref{fig_level n+1 of universal construction} and~\eqref{eq_Fj for medial indices}. 
    Let $i = 2 j \in \medset{n+1}$. 
    Then, $i+1 \in \medset{n+1}$ as well, see~\eqref{eq_defs of lat-med sets} and \Cref{rem_sets alt}.  
    Medial symbols $F_{i}$ and $F_{i+1}$, by~\eqref{eq_Fj for medial indices},  are generated by an \arikan transform of $U_{j+1}$ and $V_j$. 
    Symbol $U_{j+1}$ is in the first level-$n$ block 
    and symbol $V_j$ is in the second level-$n$ block.
    Hence, we define the (immediate) \emph{ancestors} of both medial symbols $F_{i}$ and $F_{i+1}$ as $U_{j+1}$ and $V_j$. 
Since the immediate ancestors are of level $n$, we may also call them level-$n$ ancestors. 
    
Each medial symbol of level $n$, in turn, has two level-$(n-1)$ medial symbols as its immediate ancestors, see the discussion following~\eqref{eq_j is
medial}.  
Thus, we say that a medial symbol in level $n+1$ has four level-$(n-1)$ ancestors, all medial symbols from four different level-$(n-1)$ blocks. 
Continuing in this manner, a level-$(n+1)$ symbol has $2^{n+1}$ level-$0$ ancestors, all medial symbols from $2^{n+1}$ different level-$0$
blocks. 
The level-$0$ ancestors of a symbol are called \emph{base-ancestors}. 
\end{definition}

Equations~\eqref{eq_Fj for lateral indices} and~\eqref{eq_Fj for medial indices} form a one-to-one and onto mapping from $(U_1^{N_n}, V_1^{N_n})$ to
$F_1^{N_{n+1}}$. 
We define the function $\varphi_{n+1}$ of~\eqref{eq_varphi level n+1} using these equations.
While the level-$(n+1)$ \BST is completely specified by~\eqref{eq_varphi level n+1}, the following lemma provides a direct method of computing
$G_1^{N_{n+1}}$ from $Q_1^{N_n}$ and $R_1^{N_n}$. 
\begin{lemma} \label{lem_G computation level n+1}
    Consider the \BST defined by~\eqref{eq_varphi level n+1}, where $\varphi_{n+1}$ is defined according to~\eqref{eq_Fj for lateral indices}
    and~\eqref{eq_Fj for medial indices}. Then, for any $n \geq 0$, 
    \begin{equation} \label{eq_Gj for lateral indices} 
     i \in \latset{n+1} \Rightarrow 
       G_i \equiv \begin{cases} 
            (Q_j, R_j), & i = 2j-1, \\ 
        (Q_{j+1}, R_j), & i = 2j \neq 2N_n, \\
            (F_{i-1}, Q_{N_n}, R_{N_n}), & i = 2N_n
        \end{cases}
    \end{equation}
    and
    \begin{equation} \label{eq_Gj for medial indices}
        i \in \medset{n+1} \Rightarrow 
         G_i  \equiv \begin{cases} 
            (Q_{j+1}, R_j), & i = 2j, \\ 
            (F_{i-1}, Q_{j+1}, R_j), & i = 2j+1.  
        \end{cases} 
    \end{equation}
\end{lemma}

\begin{IEEEproof}
    By construction, for $1 \leq j \leq N_n$, we have 
    \[ 
        Q_j = (U_1^{j-1}, Y_1^{N_n}), \quad R_j = (V_1^{j-1}, Y_{N_n+1}^{2N_n}). 
    \]
    Since 
    \[ 
        G_i = (F_1^{i-1},Y_1^{2N_n}),
    \] 
    we need only show that there is a one-to-one  mapping between the non-$Y$ portions of the right-hand-sides 
    of~\eqref{eq_Gj for lateral indices} and~\eqref{eq_Gj for medial indices} to $F_1^{i-1}$. 
    We proceed in cases, based on the index $i$ in the level-$(n+1)$ block.

    \emph{Case 1:} $i \in \latsettop{n+1}$ --- the first half of the lateral set, see~\eqref{eq_def of lat1}.  \\  
    In this case, to show~\eqref{eq_Gj for lateral indices} it suffices to establish 
    \begin{equation} \label{eq_Gj lateral helper 1}
        F_1^{i-1} \equiv \begin{cases} 
            (U_1^{j-1}, V_1^{j-1}), & i = 2j-1, \\ 
            (U_1^j, V_1^{j-1}), & i = 2j.
        \end{cases}
    \end{equation} 
By~\eqref{eq_Fj for lateral indices}, if $i = 2j-1$ we have $F_1^{i-1} \equiv (U_1^{j-1}, V_1^{j-1})$. If $i= 2j$ then $F_1^{i-1} \equiv (U_1^j, V_1^{j-1})$. 
      Thus,~\eqref{eq_Gj lateral helper 1} holds for any $i \in \latsettop{n+1}$. 

      \emph{Case 2:} $i \in \medset{n+1}$ --- the medial set, see~\eqref{eq_def of medset}. \\  
    In this case, to show~\eqref{eq_Gj for medial indices} it suffices to establish 
    \begin{equation}  \label{eq_Gj medial helper}
        F_1^{i-1} \equiv \begin{cases} 
            (U_1^j, V_1^{j-1}), & i = 2j, \\ 
            (F_{i-1}, U_1^j, V_1^{j-1}), & i = 2j+1.
        \end{cases}
    \end{equation} 
    By~\eqref{eq_computation of LN1}, if $i$ is the first medial index, $i = L_{n+1}+1 = 2(L_n+1)$.
    Hence, $i-1$ is odd and lateral, so by~\eqref{eq_Fj for lateral indices}, $F_1^{i-1} \equiv (U_1^{L_n+1}, V_1^{L_n})$, and
    trivially $F_1^i \equiv (F_i, U_1^{L_n+1}, V_1^{L_n})$. 
    This implies~\eqref{eq_Gj medial helper} for the first two medial indices. 
    We continue by induction.
    Assume that for $i= 2j\in \medset{n+1}$ we have $F_1^{2j-1} \equiv (U_1^j, V_1^{j-1})$. 
    Trivially, $F_1^{2j} \equiv (F_{2j}, U_1^j, V_1^{j-1})$; hence~\eqref{eq_Gj medial helper} holds for $i+1$ as well. 
    By~\eqref{eq_Fj for medial indices}, 
    \begin{align}
        F_1^{2(j+1)-1} &\equiv (F_1^{2j-1}, F_{2j}, F_{2j+1}) \nonumber \\
        &\equiv (F_1^{2j-1}, U_{j+1},V_j)\nonumber  \\ 
        &\equiv (U_1^{j+1}, V_1^j),  \label{eq_Gj medial i2}
    \end{align}
    where for the last equivalence we used the induction assumption. 
    This implies~\eqref{eq_Gj medial helper} for $i+2$. 

    Observe that when $i = 2(L_n+M_n-1) \in \medset{n+1}$, that is, when $i$ is the last even index in $\medset{n+1}$, then $i+2$ is the first lateral
    index in $\latsetbot{n+1}$. 
    Equation~\eqref{eq_Gj medial i2} still holds for $i+2 = 2(L_n+M_n)$. 

    \emph{Case 3:} $i \in \latsetbot{n+1}$ --- the second half of the lateral set, see~\eqref{eq_def of lat2}. \\ 
    In this case, to show~\eqref{eq_Gj for lateral indices} it suffices to establish 
    \begin{equation} \label{eq_Gj lateral helper 2}
        F_1^{i-1} \equiv \begin{cases} 
            (U_1^{j-1}, V_1^{j-1}), & i = 2j-1, \\ 
            (U_1^j, V_1^{j-1}), & i = 2j \neq 2N_n, \\  
            (F_{i-1}, U_1^{N_n-1}, V_1^{N_n-1}), & i = 2N_n.  
        \end{cases}
    \end{equation} 
    If $i$ is the first lateral index in $\latsetbot{n+1}$, by~\eqref{eq_computation of LN1 MN1} we have $i = L_{n+1} + M_{n+1} +1 = 2(L_n+M_n)$. 
    Thus, by the observation at the end of case 2, $F_1^{2(L_n+M_n)-1} \equiv (U_1^{L_n+M_n},
    V_1^{L_n+M_n-1})$. 
    For any other index $i \in \latsetbot{n+1}$, by~\eqref{eq_Fj for lateral indices} indeed~\eqref{eq_Gj lateral helper 2} holds, similar to case 1. 
\end{IEEEproof}

We conclude this section by computing the fraction of medial symbols out of all symbols in a level-$n$ block. 
To this end, denote
\begin{equation} 
    \alpha_n \triangleq \frac{M_n}{2L_n+M_n}. \label{eq_def of alphan}
\end{equation}

\begin{lemma} \label{lem_proportion of lateral sopairs}
    Consider a \BST initialized with parameters $L_0 \geq 0$ and $M_0$, and let $0 < \alpha < 1$. 
    If   
    \[ 
        M_0 \geq \left\lceil\frac{2(1 + \alpha L_0)}{1-\alpha} \right \rceil,
    \] 
    then $\alpha_n \geq \alpha$ for any $n \geq 0$. 
\end{lemma}
\begin{IEEEproof}
    Plugging $n=0$ in~\eqref{eq_def of alphan} yields $\alpha_0 = M_0/(2L_0+M_0)$. 
    It is straightforward to show from~\eqref{eq_computation of LN1 MN1} that for any $n\geq 0$, 
    \begin{IEEEeqnarray}{rCl} \IEEEyesnumber \label{eq_number of lateral and medial indices of a BST} 
	    L_n &=& 2^n (L_0 + (1 - 2^{-n})) \IEEEyessubnumber \label{eq_number of lateral and medial indices of a BST lateral} \\ 
        M_n &=& 2^n (M_0 - 2(1 - 2^{-n})). \IEEEyessubnumber \label{eq_number of lateral and medial indices of a BST medial}
    \end{IEEEeqnarray}
    Therefore, recalling that $N_0 = 2L_0+M_0$,  
    \[
        \alpha_n = \frac{M_n}{2L_n + M_n}  
                 = \frac{M_0 - 2(1 - 2^{-n})}{2L_0 +M_0}
                 = \alpha_0 - \frac{2(1-2^{-n})}{N_0}. 
    \]
    This implies that 
    \[ 
        \alpha_n \geq \alpha_0 - \frac{2}{N_0} = \frac{M_0 - 2}{M_0 + 2 L_0}. 
    \] 
    The right-hand side is an increasing function of $M_0$, since its derivative with respect to $M_0$ is $2(1+L_0)/(2L_0 + M_0)^2 >0$. 
    It remains to find $m_0$ such that $(m_0-2)/(m_0+2L_0) = \alpha$. 
    Then, for any $M_0 \geq \lceil m_0 \rceil$, we will have $\alpha_n \geq \alpha$. 
    The proof is complete by noting that $m_0=2(1+\alpha L_0)/(1-\alpha)$.  
\end{IEEEproof}
\begin{discussion}
    The transform presented in~\cite{sasoglu_2016_universal}, henceforth referred to as the \sasoglu-Wang transform (SWT), is the basis for the \BST. 
    The first two levels of the SWT (levels $1$ and $2$ in~\cite{sasoglu_2016_universal}) differ from the first two levels of the \BST (levels $0$
    and $1$ here). After that, the construction of the two transforms coincide (compare our \Cref{fig_level n+1 of universal construction} with
    \cite[Figure 5]{sasoglu_2016_universal}).
    The \BST is simpler and more streamlined than the SWT, since all levels of the \BST share the same construction. 
    In the memoryless case one can verify that the SWT and \BST (with $L_0 = 0$) have the same performance. 

    We will see in \Cref{sec_polarizable soprocesses} that the \BST is effective also for processes with memory, by taking $L_0 > 0$.  

    In \Cref{sec_polarizable soprocesses} we will show that for an appropriate $\eta$ and family of \soprocesses, the \BST is $(\eta, \setmonodown, \setmonoup)$-monopolarizing, with $\setmonodown
= \medsetPlus{n}$ and $\setmonoup = \medsetMinus{n}$, where $n$ is the level number of the \BST. In particular, this implies that $|\setmonodown| = |\setmonoup|$, which limits to $1/2$ the
achievable rates the universal code can yield. It is possible to generate slow stage transforms for which $\setmonodown$ and $\setmonoup$ are of
different sizes.  
One way to achieve this is by cascading multiple \BSTs. 
This idea originates in~\cite[Section III]{sasoglu_2016_universal}; a brief description on how this is accomplished follows. 
After a \BST, all symbols in $\medsetMinus{n}$ have approximately the same conditional entropy; the same is true for all symbols in $\medsetPlus{n}$. 
If $n$ is sufficiently large, one set will have polarized (e.g., the conditional entropies of \sopairs in $\medsetMinus{n}$ are all very close to $1$). 
By applying a \BST to multiple copies of the other set, we divide its \sopairs into two new sets of equal size, one of which will have polarized. 
This operation can be repeated to tailor the size of the polarized set. For further details, see \Cref{sec_monopolarization for FAIM derived processes cascade}.

An alternative strategy to modify the sizes of $\setmonodown$ and $\setmonoup$ is to form medial symbols with kernels other than the \arikan transform. 
A family of kernels are introduced in~\cite[Section III]{sasoglu_2016_universal}. 
They can also be adapted to our construction, and we leave this to the interested reader. 
\end{discussion}

\subsection{Fast Polarization Stage}\label{sec_fast stage}
We will show in \Cref{sec_polarizable soprocesses} that the \BST is $(\eta, \mathcal{L}, \mathcal{H})$-monopolarizing for
a suitable family of \soprocesses with memory.
This is also true for a cascade of \BSTs, see \Cref{sec_monopolarization for FAIM derived processes cascade}.
Moreover, the sets $\mathcal{L}$ and $\mathcal{H}$ are predetermined and independent of the \soprocess. 
However, even in the memoryless case \cite{sasoglu_2016_universal}, the speed of polarization is too slow to enable a successive-cancellation decoder
to succeed. 
Therefore, as in~\cite{sasoglu_2016_universal}, we append a fast  polarization stage to the \BST cascade that facilitates  error-free successive-cancellation decoding.

The fast polarization stage is based on \arikan's seminal transform~\cite{Arikan_2009}, which is known to polarize fast also under
memory~\cite{sasoglu_Tal_mem,Shuval_Tal_Memory_2017}. 
One strategy to incorporate a fast polarization stage, suggested in~\cite{sasoglu_2016_universal}, is as follows. 

As in the proof of~\Cref{thm_main}, we fix a sufficiently small $\eta$, which determines the back-off from extremality that the \BST cascade will achieve. 
This value, as shown in Appendix~\ref{ap_fast polarization}, is set small enough to ensure fast polarization of this stage. 
Further following the proof of~\Cref{thm_main}, we set the \BST cascade parameters. These include the \BST parameters $L_0$, $M_0$, $n$, as well parameters $t$ and $\bv{c}$ defining the cascade, to be discussed in \Cref{sec_monopolarization for FAIM derived processes cascade}. 
These parameters ensure that the BST cascade in $(\eta, \mathcal{L}, \mathcal{H})$-monopolarizing and that $|\mathcal{H}|/N$ is sufficiently close to the infimal conditional entropy rate of the family of \soprocesses. 

For the fast stage, take $\hat{N}=2^{\hat{n}}$ copies of the \BST cascade of length $N$.
Apply multiple copies of \arikan's seminal transform (``fast transform'') of length $\hat{N}$, as illustrated in \Cref{fig_fast stage}. 
In words, the $j$th fast transform operates on the $j$th \sopair from each copy of the \BST cascade.

\begin{figure} 
    \begin{center}
\begin{tikzpicture}[>=latex]
    \foreach \x/\y in {0/0,1.4cm/1,2.8cm/2,5.0cm/3}
    {
        \begin{scope}[yshift = -\x, scale = 0.5, every node/.style={scale=0.5}]
            \node[draw, rectangle, minimum width = 1.6cm, minimum height = 2cm] (X\y) at (3.0cm,0) {}; 
            \node[draw, rectangle, minimum width = 1.6cm, minimum height = 2cm, align=center, text width = 1.2cm, fill=green!30!white] (F\y) at (0,0) {BST cascade}; 
	    \draw[->] (X\y.135) -- (F\y.45); 
	    \draw[->] (X\y.180) -- (F\y); 
	    \draw[->] (X\y.225) -- (F\y.315); 

        \end{scope}
    }
    \coordinate (XF2) at ($(F2)!0.5!(X2)$); 
    \coordinate (XF3) at ($(F3)!0.5!(X3)$); 
    
    \node[above = 0.2 of X0, text width = 2.0cm, inner sep = 0, align = center] { \soprocess blocks }; 

    \node[circle,draw,fill, minimum size = 2pt, inner sep = 0] (C) at ($(XF2)!0.5!(XF3)$){}; 
    \node[circle,draw,fill, minimum size = 2pt, inner sep = 0] at ($(XF2)!0.7!(XF3)$){}; 
    \node[circle,draw,fill, minimum size = 2pt, inner sep = 0] at ($(XF2)!0.3!(XF3)$){}; 
    \node[below = 0.5 of XF3] {$\hat{N}$ copies}; 

    \draw[<->] ($(X0.north east) + (0.2,0)$) -- node[right]{$N$}($(X0.south east) + (0.2,0)$) ; 

    \node[draw, rectangle, fill=orange!30!white, minimum height = 1cm, text width = 1.0cm, left = 3 of X0, align=center, font = \scriptsize ] (fa) {Fast transform}; 
    \foreach \x/\y in {0/20,1/10,2/0,3/-10}
    {
        \draw[->] ($(F\x.north west)!0.1!(F\x.south west)$) --  (fa.\y); 
        \draw[->] (fa.180) -- +(-0.4,0); 
        \draw[->] (fa.210) -- +(-0.4,0); 
        \draw[->] (fa.150) -- +(-0.4,0); 
    }
    \draw[<->] ($(fa.north west) - (0.6,0)$) -- node[left]{$\hat{N}$}($(fa.south west) - (0.6,0)$) ; 
    \node[draw, very thick, fill=orange!30!white, rectangle, minimum height = 1cm, text width = 1.0cm, below = 1.2 of fa, align=center, font = \scriptsize] (fb) {Fast transform}; 
    \foreach \x/\y in {0/15,1/5,2/-5,3/-15}
    {
        \draw[->,very thick] ($(F\x.north west)!0.3!(F\x.south west)$)node[right=-0.1] {\tiny $j$} -- (fb.\y); 
        \draw[->] (fb.180) -- +(-0.4,0); 
        \draw[->] (fb.210) -- +(-0.4,0); 
        \draw[->] (fb.150) -- +(-0.4,0); 
    }
    \node[draw, rectangle, fill=orange!30!white, minimum height = 1cm, text width = 1.0cm, below = 3.4 of fa, label=below:{$N$ copies}, align=center, font = \scriptsize] (fc) {Fast transform}; 
    \foreach \x/\y in {0/10,1/0,2/-10,3/-20}
    {
        \draw[->] ($(F\x.north west)!0.94!(F\x.south west)$) -- (fc.\y); 
        \draw[->] (fc.180) -- +(-0.4,0); 
        \draw[->] (fc.210) -- +(-0.4,0); 
        \draw[->] (fc.150) -- +(-0.4,0); 
    }
    \foreach \x/\y in {fa/fb,fb/fc}
    {
    \node[circle,draw,fill, minimum size = 2pt, inner sep = 0] (C) at ($(\x)!0.5!(\y)$){}; 
    \node[circle,draw,fill, minimum size = 2pt, inner sep = 0] at ($(\x)!0.65!(\y)$){}; 
    \node[circle,draw,fill, minimum size = 2pt, inner sep = 0] at ($(\x)!0.35!(\y)$){}; 
    }
\end{tikzpicture}
\end{center}
\caption{Illustration of the slow and fast stages. 
    First, $\hat{N}$ length-$N$ blocks of the \soprocess are transformed using \BST cascades of length $N$ (denoted in green \protect \tikz{ \protect \node[fill=green!30!white] at (0,0) {\protect \pgfuseplotmark{square}}  ;}).
    Then, $N$ fast transforms of length $\hat{N}$ are applied (denoted in orange \protect \tikz{ \protect \node[fill=orange!30!white] at (0,0) {\protect \pgfuseplotmark{square}}  ;}). 
    The $j$th fast transform (in bold) operates on the $j$th \sopair in each \BST cascade. 
}
\label{fig_fast stage}
\end{figure}

As shown in the proof of~\Cref{thm_main}, this construction is universal over the family of \soprocesses. That is, the code is the 
same for any \soprocess in the family. Recall that when decoding, we assume that the \soprocess is known at the decoder side.

\section{Properties of the \BST and a Variation} \label{sec_properties of BST and a variation}
In this section we explore some of the properties of the \BST. 
We also introduce a variation of the \BST, the Observation-truncated \BST. 
We will call upon these when analyzing the \BST in \Cref{sec_monopolarization for FAIM derived processes}. 

\subsection{Properties of the \BST}\label{sec_properties of BST}
We now explore some properties of the \BST that will be useful in the sequel. 
To this end, throughout this section we assume that \BSTs are initialized with parameters $L_0$ and $M_0$. 
A level-$0$ \BST is thus of length $N_0 = 2L_0 + M_0$, and a level-$n$ \BST is of length $N_n = 2^n N_0$. 

Since $N_n = 2^n N_0$, we say that a level-$n$ \BST is formed from $2^n$ level-$0$ \BSTs. 
We call each level-$0$ \BST a \bblock,\footnote{The letter `b' here and also in the name \bindex below stands for
`base,' as the \BST may be thought of as consisting of $2^n$ ``base blocks'' of length $N_0$.}  and we number them sequentially. 
The \bblock numbered $\ell$ contains \sopairs with indices $(\ell-1)N_0 +k$, $1 \leq k \leq N_0$. 
The following definition names both $\ell$ and $k$. 

\begin{definition}[\bblock number and \bindex]  \label{def_bblock bindex}
    In a level-$n$ \BST, an index $j$ is a number between $1$ and $N_n$.
We write it in the form
\begin{equation} \label{eq_bblock bindex representation}
    j = (\ell-1)N_0 + k, \quad 1 \leq \ell \leq 2^n, \quad 1 \leq k \leq N_0. 
\end{equation}
We call $\ell$ the \emph{\bblock number} and $k$ the \emph{\bindex} that correspond to index $j$.
\end{definition}

Recall from \Cref{def_ancestors} that each medial level-$n$ symbol has $2^n$ medial level-$0$ indices as its base-ancestors. 
These base-ancestors are a subvector of $X_1^{N_n}$. 
Each of these level-$0$ indices has a different \bblock number, computed via~\eqref{eq_bblock bindex representation}. 
We collect the sorted indices of these symbols in a vector as follows. 
From this point onwards, we use the term `ancestor' to apply to both the symbol and its index; it will be clear from the context if we refer to the
symbol or to its index. 

\begin{definition}[Base-vector and modulo-base-vector] \label{def_base vector}
    The \emph{base-vector} $\bv{b}$ of a medial index $i$ is a \emph{row} vector whose $\ell$th entry is the base-ancestor of 
    $i$ from \bblock $\ell$. 
    Therefore, 
    \begin{equation} \label{eq_bell of base-vector}
        \matel{\bv{b}}{\ell} = (\ell-1)N_0 + k 
    \end{equation}                              
    for some $L_0+1 \leq k \leq N_0-L_0$.

    The \emph{modulo-base-vector} $\bar{\bv{b}}$ of $i$ is defined by 
    \begin{equation} \label{eq_modulo determining}
        \matel{\bar{\bv{b}}}{\ell} = \matel{\bv{b}}{\ell} - (\ell-1)N_0, \quad 1 \leq \ell \leq 2^n, 
    \end{equation} 
    where $n$ is the level of $i$. 
    This vector contains in its $\ell$th entry the \bindex  of $i$'s base-ancestor in the $\ell$th \bblock.
    That is, $k$ in~\eqref{eq_bell of base-vector}. 
\end{definition}

\begin{remark} \label{rem_medial only}
    We only define base-vectors for medial indices. 
    While it is possible to extend the definition to apply to lateral indices, this will not be of interest to us. 
    This is afforded because the ancestors of medial indices can only be medial indices, so we will not need to consider lateral indices. 
    In particular, equation~\eqref{eq_base-vector recursion med} below is well-defined because each vector on the right-hand side is
    a modulo-base-vector of a medial index. 
\end{remark}

To motivate the definition of the base-vector, assume momentarily that the \soprocess being transformed were memoryless. 
If we tried to recover some transformed symbol $F_i$ using successive-cancellation decoding, we could discard all observations except for
those whose indices are in the base-vector.  
That is, in the memoryless case
\[
    \Prob{F_i = 0\  \big| \ F_1^{i-1}, Y_1^{N_n}} = \Prob{F_i =0 \ \big|\  F_1^{i-1}, Y_{\bv{b}} },
\]
where $Y_{\bv{b}} = \{Y_{\matel{\bv{b}}{1}}, Y_{\matel{\bv{b}}{2}}, \ldots, Y_{\matel{\bv{b}}{2^n}}\}$. 
We emphasize that the aforementioned assumption of a memoryless process was made solely for the 
purpose of motivating the base-vector. 
In fact, the base-vector is a product of the \BST itself, and has nothing to do with the \soprocess being transformed. 
Henceforth, in this section we look at a \BST as a transformation between two vectors, and study some of its properties. 

To compute the base-vector of an index, we first compute its modulo-base-vector, and then use~\eqref{eq_modulo determining}.  
The modulo-base-vectors are constructed recursively. 
To this end, we augment the notation for base- and modulo-base-vectors with the index and level specification. 
Thus, for $i \in \medset{n}$, we use $\bv{b}_i^{(n)}$ and $\bar{\bv{b}}_i^{(n)}$ to denote the base-vector and modulo-base-vector,
respectively.

For a level-$0$ \BST, the modulo-base-vector for medial index $L_0+1 \leq i \leq N_0-L_0$ contains just one index: 
\[ 
    \bar{\bv{b}}^{(0)}_i = \begin{bmatrix} i \end{bmatrix}. 
\] 
For higher levels, by \Cref{def_ancestors}, the modulo-base-vectors are constructed by 
\begin{equation}\label{eq_base-vector recursion med}
    \bar{\bv{b}}_i^{(n+1)} = \begin{bmatrix} \bar{\bv{b}}_{j+1}^{(n)} & \bar{\bv{b}}_{j}^{(n)} \end{bmatrix}, \quad j = \left\lfloor \frac{i}{2}
        \right\rfloor. 
\end{equation}
Recall from \Cref{rem_sets alt} that if $i \in \medsetMinus{n+1}$, then $i$ is even, so  $i$ and $i+1$ share the same base-vector.

\begin{example}  \label{ex_level 3 BST}
    Consider a \BST initialized with $L_0 = 3, M_0 = 6$. 
    A level-$0$ \BST is of length $N_0 = 2L_0 + M_0 = 12$. 
    A level-$1$ \BST is of length $N_1 = 2N_0 = 24$. 
    The first medial index is $L_1 + 1 = (2L_0+1)+1 = 8$. 
    We have 
    \begin{equation*}
        \bar{\bv{b}}^{(1)}_{8} =  \bar{\bv{b}}^{(1)}_{9}  =  \begin{bmatrix} 5 & 4 \end{bmatrix}, \quad 
        \bar{\bv{b}}^{(1)}_{10} = \bar{\bv{b}}^{(1)}_{11} = \begin{bmatrix} 6 & 5 \end{bmatrix},
    \end{equation*}
    and so on. 
    A level-$2$ \BST is of length $N_2 = 2N_1 = 48$, and its first medial index is $L_2 + 1 = (2L_1 + 1) + 1 = 16$. 
    Thus, 
    \begin{equation*}
        \bar{\bv{b}}^{(2)}_{16}= \bar{\bv{b}}^{(2)}_{17} = \begin{bmatrix} 5 & 4 & 5 & 4 \end{bmatrix}, \quad 
        \bar{\bv{b}}^{(2)}_{18} = \bar{\bv{b}}^{(2)}_{19}  = \begin{bmatrix} 6 & 5 & 5 & 4 \end{bmatrix}.  
    \end{equation*}
    A level-$3$ \BST is of length $N_3 = 2N_2 = 96$, its first medial index is $L_3 + 1 = (2L_2 + 1) + 1 = 32$, and
    \begin{align*}
        \bar{\bv{b}}^{(3)}_{32}= \bar{\bv{b}}^{(3)}_{33} &= \begin{bmatrix} 5 & 4 & 5 & 4 & 5 & 4 & 5 & 4 \end{bmatrix}, \\ 
        \bar{\bv{b}}^{(3)}_{34}= \bar{\bv{b}}^{(3)}_{35} &= \begin{bmatrix} 6 & 5 & 5 & 4 & 5 & 4 & 5 & 4 \end{bmatrix}.
    \end{align*}
    Computing a base-vector, say $\bv{b}^{(3)}_{35}$, is easily done using~\eqref{eq_modulo determining}:
    \[ 
        \bv{b}_{35}^{(3)} = \begin{bmatrix} 6 & 17 & 29 & 40 & 53 & 64 & 77 & 88 \end{bmatrix}. 
    \] 
    In \Cref{fig_level 3 BST} we illustrate a portion of a level-$3$ \BST and show the base-vector $\bv{b}^{(3)}_{34} = 
    \bv{b}^{(3)}_{35}$.

    \begin{figure}[t]
        \begin{center}
            \begin{tikzpicture}[rect/.style={draw,rectangle,minimum width = 0.3cm, minimum height = 0.6cm},
                                circ/.style={draw,fill=white,circle,minimum size = 4pt, inner sep = 0pt}, 
                                dot/.style={draw,circle, fill, minimum size = 2 pt, inner sep = 0pt},
                                sdot/.style={draw,circle, fill, minimum size = 1 pt, inner sep = 0pt}, >=latex]

                \def\YDIST{1.1}
                \def\XDIST{0.5}
                \foreach \x in {1,2,...,8}
                {
                    \node[rect] (R\x) at (0,-\x*\YDIST) {}; 
                }

                \node[circ, left = \XDIST of R2] (C2) {$+$}; 
                \node[dot] (C1) at (C2|-R1) {}; 

                \foreach \i [evaluate=\i as \j using int(\i+1)] in {3,5,7}
                {
                    \node[circ, left = \XDIST of R\i] (C\i) {$+$};
                    \node[dot] (C\j) at (C\i |- R\j) {};
                }

                \draw (R1) -- node[above](L0){\tiny $6$} node[below,green!50!black]{\tiny $(6)$} (C1) -- (C2) -- node[above,pos=0.4]{\tiny $5$}
                    node[below,pos=0.4, green!50!black]{\tiny $(17)$} (R2);
                \draw (R3) -- node[above, pos=0.6]{\tiny $5$}  node[below,pos=0.6, green!50!black]{\tiny $(29)$}(C3) -- (C4) -- 
                    node[above]{\tiny $4$} node[below,green!50!black]{\tiny $(40)$} (R4);
                \draw (R5) -- node[above, pos=0.6]{\tiny $5$}  node[below,pos=0.6, green!50!black]{\tiny $(53)$}(C5) -- (C6) -- 
                    node[above]{\tiny $4$} node[below,green!50!black]{\tiny $(64)$} (R6);
                \draw (R7) -- node[above, pos=0.6]{\tiny $5$}  node[below,pos=0.6, green!50!black]{\tiny $(77)$}(C7) -- (C8) -- 
                    node[above]{\tiny $4$} node[below,green!50!black]{\tiny $(88)$} (R8);

                \node[circ, left = 2*\XDIST of C4] (D4) {$+$}; 
                \node[dot] (D2) at (D4|-R2) {}; 
                \node[circ] (D6) at (D4|-R6) {$+$}; 
                \node[dot] (D7) at (D4|-R7) {}; 

                \draw (C2) -- node[above,pos=0.5](L1){\tiny $10$} (D2) -- (D4) -- node[above, pos = 0.4]{\tiny $9$} (C4); 
                \draw (C7) -- node[above,pos=0.5]{\tiny $8$} (D7) -- (D6) -- node[above, pos =0.4]{\tiny $9$} (C6); 

                \node[circ, left = 2*\XDIST of D7] (E7) {$+$}; 
                \node[dot] (E4) at (E7|-R4) {}; 

                \draw (D7) -- node[above,pos=0.6]{\tiny $17$} (E7) -- (E4) -- node[above](L2){\tiny $18$} (D4); 

                \coordinate[left = \XDIST of E7] (F7);
                \coordinate[left = \XDIST of E4] (F4);

                \draw (F4) -- node[above, pos=0.4](L3){\tiny $35$} (E4); 
                \draw (F7) -- node[above]{\tiny $34$} (E7); 

                \node[red, rounded corners, dashed, fit = {(R1) (R2) (C1) (C2)}, draw] (A12) {};  
                \node[red, rounded corners, dashed, fit = {(R3) (R4) (C3) (C4)}, draw] (A34) {};  
                \node[red, rounded corners, dashed, fit = {(R5) (R6) (C5) (C6)}, draw] (A56) {};  
                \node[red, rounded corners, dashed, fit = {(R7) (R8) (C7) (C8)}, draw] (A78) {};  
                
                \node[blue, rounded corners, dashdotted, fit = {(A12) (A34) (D4)}, draw, inner sep = 2pt] (A1234) {}; 
                \node[blue, rounded corners, dashdotted, fit = {(A56) (A78) (D6)}, draw, inner sep = 2pt] (A5678) {};

                \foreach \x in {0,1,2,3}
                {
                    \node[above = 0.2 of L\x, gray] {\scriptsize level-$\x$};
                }
            \end{tikzpicture}
        \end{center}
        \caption{A portion of a level-$3$ \BST, initialized with $L_0 = 3, M_0 = 6$. The base-vector $\bv{b}^{(3)}_{34} = 
        \bv{b}^{(3)}_{35}$ is illustrated. The rectangles denote level-$0$ \BSTs.
        Level-$1$ \BSTs are delimited with dashed lines (in red) and level-$2$ \BSTs are delimited with dash-dotted lines (in blue).
        Above each line, we show its index with respect to its relevant-level \BST (the rightmost are level-$0$). 
    The level-$0$ indices are also \bindices; below them we noted in parentheses (in green) their respective indices in a level-$3$ \BST.} 
        \label{fig_level 3 BST}
    \end{figure}
\end{example}

    Let $n \leq m$. 
    Fix some $i \in \medset{m}$ and apply~\eqref{eq_base-vector recursion med} recursively $m-n$ times. 
    This expresses the modulo-base-vector of $i$ as a concatenation of $2^{m-n}$ level-$n$ modulo-base-vectors. 
    These are the modulo-base-vectors of the level-$n$ ancestors of this level-$m$ index. 
    In particular, the modulo-base-vector of any level-$n$ ancestor of $i$ is a sub-vector of $i$'s modulo-base-vector.

\begin{example}[continues = ex_level 3 BST] 
    We can express the modulo-base-vector of level-$3$ index $34$ as a concatenation of the modulo-base-vectors of its level-$1$ ancestors: 
    \begin{align*}
    \bar{\bv{b}}_{34}^{(3)} 
        &= \begin{bmatrix} \begin{bmatrix} 6 & 5 \end{bmatrix} & \begin{bmatrix} 5 & 4 \end{bmatrix} 
        & \begin{bmatrix} 5 & 4 \end{bmatrix} & \begin{bmatrix} 5 & 4 \end{bmatrix} \end{bmatrix} \\ 
    &= \begin{bmatrix} \bar{\bv{b}}_{10}^{(1)} & \bar{\bv{b}}_9^{(1)} & \bar{\bv{b}}_9^{(1)} & \bar{\bv{b}}_8^{(1)}
        \end{bmatrix}. 
    \end{align*}                                                             
\end{example}

Observe that in \Cref{ex_level 3 BST}, the modulo-base-vectors of medial indices contain at least two and at most three distinct \bindices, and these
\bindices are consecutive. 
This is not a coincidence, as the corollary to the following two lemmas will show. 

\begin{lemma} \label{lem_BST maxmin}
    For any $i, i+1 \in \medset{n}$ and any $1 \leq \ell \leq 2^n$ we have 
    \begin{equation} \label{eq_detvec max inequality}
        \matel{\bar{\bv{b}}_{i+1}^{(n)}}{\ell} \geq \matel{\bar{\bv{b}}_i^{(n)}}{\ell}.
    \end{equation}
\end{lemma}
\begin{IEEEproof}
    This follows from~\eqref{eq_base-vector recursion med} by straightforward induction. 
    Specifically, note that if the index $i$ on the left-hand-side of ~\eqref{eq_base-vector recursion med} increases, the indices $j$ and $j+1$ on the
    right-hand-side cannot decrease. 
\end{IEEEproof}

\begin{lemma} \label{lem_BST base-vector}
    For any $i \in \medset{n}$ and any $1 \leq \ell \leq 2^n$ we have 
    \begin{equation}  \label{eq_maxmin dj}
        \left\lfloor \frac{i}{2^n} \right\rfloor = \matel{\bar{\bv{b}}^{(n)}_i}{2^n} 
        \leq \matel{\bar{\bv{b}}^{(n)}_i}{\ell}  \leq  \matel{\bar{\bv{b}}^{(n)}_i}{1} = 1 + \left\lceil \frac{i-1}{2^n} \right\rceil. 
   \end{equation}
\end{lemma}
In words, for any medial index $i$, the first element of $\bar{\bv{b}}_i^{(n)}$ is its maximal, which equals $1+\lceil(i-1)\cdot2^{-n} \rceil$, 
and the last element of $\bar{\bv{b}}_i^{(n)}$ is its minimal, which equals $\lfloor i\cdot2^{-n} \rfloor$. 

\begin{IEEEproof}
    The proof consists of several steps, all proved using induction. 
    First, we prove
    \emph{claim 1}: $\matel{\bar{\bv{b}}_i^{(n)}}{1} \geq \matel{\bar{\bv{b}}_i^{(n)}}{\ell} \geq \matel{\bar{\bv{b}}_i^{(n)}}{2^n}$ for any $i \in
    \medset{n}$ and $1 \leq \ell \leq 2^n$. 
    Then, we will establish the formulas for computing the values of these elements. 

    \emph{Proof of Claim 1:}
    For $n=0$ claim 1 is trivially true, as for any $i \in \medset{0}$, $\bar{\bv{b}}_i^{(0)}$ is a singleton. 
    Assume that claim 1 holds for some  $n\geq 0$; we will establish that it is true also for $n+1$. 
    Let $i \in \medset{n+1}$. 
    Then, by~\eqref{eq_base-vector recursion med}, $\bar{\bv{b}}_i^{(n+1)} = \begin{bmatrix} \bar{\bv{b}}_{j+1}^{(n)} & \bar{\bv{b}}_{j}^{(n)} \end{bmatrix}$,
    where $j = \lfloor i/2 \rfloor$.
    By the induction hypothesis, 
    \begin{IEEEeqnarray*}{rcCcCl}
        \matel{\bar{\bv{b}}_{i}^{(n+1)}}{1} ={}& \matel{\bar{\bv{b}}_{j+1}^{(n)}}{1} &\geq& \matel{\bar{\bv{b}}_{j+1}^{(n)}}{\ell} &\geq&
        \matel{\bar{\bv{b}}_{j+1}^{(n)}}{2^n}, \\
        &\matel{\bar{\bv{b}}_{j}^{(n)}}{1} &\geq& \matel{\bar{\bv{b}}_{j}^{(n)}}{\ell} &\geq& \matel{\bar{\bv{b}}_{j}^{(n)}}{2^n} = \matel{\bar{\bv{b}}_{i}^{(n+1)}}{2^{n+1}}
    \end{IEEEeqnarray*}
for any $1 \leq \ell \leq 2^n$. 
By \Cref{lem_BST maxmin}, $\matel{\bar{\bv{b}}_{j+1}^{(n)}}{\ell} \geq \matel{\bar{\bv{b}}_j^{(n)}}{\ell}$ for any $1 \leq \ell \leq 2^n$. 
    Therefore, 
    \[
        \matel{\bar{\bv{b}}_{i}^{(n+1)}}{1} \geq \matel{\bar{\bv{b}}_{i}^{(n+1)}}{\ell} \geq \matel{\bar{\bv{b}}_{i}^{(n+1)}}{2^{n+1}}
    \]
    for any $1\leq \ell\leq 2^{n+1}$, thereby proving claim 1. 

    \emph{Proof of the right-hand side of~\eqref{eq_maxmin dj}:}
    For $n=0$ and any $i \in \medset{0}$, trivially $\matel{\bar{\bv{b}}_i^{(0)}}{1} = 1 + \lceil(i-1)\cdot 2^{-0}\rceil = i$. 
    Assume that the right-hand side of~\eqref{eq_maxmin dj} holds for some $n \geq 0$; we will show it holds for $n+1$ as well.  
    Let $i \in \medset{n+1}$; by~\eqref{eq_base-vector recursion med}, $\matel{\bar{\bv{b}}_i^{(n+1)}}{1} = \matel{\bar{\bv{b}}_{j+1}^{(n)}}{1}$, 
    where $j = \lfloor i/2 \rfloor$. 
    Now, observe that for natural $i$,  
    \[ 
        \left\lfloor \frac{i}{2} \right\rfloor = \left\lceil \frac{i-1}{2} \right\rceil. 
    \] 
    Therefore,  
    \begin{align*} 
        \matel{\bar{\bv{b}}_i^{(n+1)}}{1} = \matel{\bar{\bv{b}}_{\lfloor i/2 \rfloor+1}^{(n)}}{1} &\eqann{a} 1+ \left\lceil \frac{\lfloor i/2 \rfloor}{2^n} \right\rceil  
        =  1+ \left\lceil \frac{\lceil (i-1)/2 \rceil}{2^n} \right\rceil \\  &\eqann{b} 1+ \left\lceil \frac{ (i-1)/2 }{2^n} \right\rceil  =
 1+ \left\lceil \frac{i-1}{2^{n+1}} \right\rceil, 
 \end{align*} 
 where~\eqannref{a} is by the induction assumption and~\eqannref{b} is by~\cite[equation 3.11]{concreteMathematics}. 

    \emph{Proof of the left-hand side of~\eqref{eq_maxmin dj}:}
    For $n=0$ and any $i \in \medset{0}$, trivially $\matel{\bar{\bv{b}}_i^{(0)}}{2^0} = \lfloor i \cdot 2^{-0}\rfloor = i$. 
    Assume that the left-hand side of~\eqref{eq_maxmin dj} holds for some $n \geq 0$; we will show it holds for $n+1$ as well.  
    Let $i \in \medset{n+1}$; by~\eqref{eq_base-vector recursion med}, $\matel{\bar{\bv{b}}_i^{(n+1)}}{2^{n+1}} = \matel{\bar{\bv{b}}_{j}^{(n)}}{2^n}$, 
    where $j = \lfloor i/2 \rfloor$. 
    Therefore, 
    \[ 
        \matel{\bar{\bv{b}}_i^{(n+1)}}{2^{n+1}} = \matel{\bar{\bv{b}}_{\lfloor i/2 \rfloor}^{(n)}}{2^n} \eqann{a} \left\lfloor \frac{\lfloor i/2
        \rfloor}{2^n} \right\rfloor \eqann{b}   \left\lfloor \frac{ i/2}{2^n} \right\rfloor =\left\lfloor \frac{ i}{2^{n+1}} \right\rfloor, 
    \] 
 where~\eqannref{a} is by the induction assumption and~\eqannref{b} is by~\cite[equation 3.11]{concreteMathematics}. 
\end{IEEEproof}

\begin{corollary} \label{cor_num of elements in detvec}
    If $n \geq 1$ then for any $i \in \medset{n}$, 
\[ 
    1 \leq \max_{\ell} \matel{\bar{\bv{b}}^{(n)}_i}{\ell} - \min_{\ell} \matel{\bar{\bv{b}}^{(n)}_i}{\ell} \leq 2. 
\] 
\end{corollary}

\begin{IEEEproof}
    This is an immediate consequence of \Cref{lem_BST base-vector}. 
Specifically, if $\lfloor i/2^n \rfloor = r$ then                                 
\[
     r \leq \frac{i}{2^n} < r+1 \, \Rightarrow \, r - \frac{1}{2^n} \leq \frac{i-1}{2^n}
    < r+1-\frac{1}{2^n}.
\]                      
The ceiling operation $\lceil \cdot \rceil$ is monotonically increasing.
Thus, we apply it to the three terms on the right-hand side to yield
$r \leq \lceil (i-1)/2^n \rceil \leq r+1$. 
\end{IEEEproof}

\subsection{The Observation-Truncated \BST} \label{sec_otbst}
The Observation-Truncated \BST (\OTBST in short) is a variation on the \BST that will be useful for analysis. 
It is defined recursively, just like the \BST, but with a different initialization.

The \BST may be looked at as a recursively-defined sequence of functions. 
Let $F_1^{N_n} \sarrow G_1^{N_n}$ be the output of a level-$n$ \BST with parameters $L_0$ and $M_0$ of \soblock $X_1^{N_n} \sarrow Y_1^{N_n}$.
Recall that $X_i \in\mathcal{X} =\{0,1\}$ and $Y_i \in \mathcal{Y}$ for any $i$, where $\mathcal{Y}$ is some finite alphabet. 
For any $i \in \medset{n}$ there exist functions 
\begin{align*}
    f_{n,i}:\mathcal{X}^{N_n} &\to \mathcal{X}, \\  
    g_{n,i}:\mathcal{X}^{N_n}\times\mathcal{Y}^{N_n} &\to \mathcal{X}^{i-1}\times\mathcal{Y}^{N_n},
\end{align*}
 such that $f_{n,i}(X_1^{N_n}) = F_i$ and $g_{n,i}(X_1^{N_n}, Y_1^{N_n}) = G_i$. 
From~\eqref{eq_level0},~\eqref{eq_Fj for medial indices}, and~\eqref{eq_Gj for medial indices}, they are recursively defined as follows. 
Initialization for any $i \in \medset{0}$: 
    \begin{IEEEeqnarray}{rCl} \IEEEyesnumber \label{eq_fjgj init}
        f_{0,i}(X_1^{N_0}) &=& X_i, \IEEEyessubnumber \label{eq_fjgj init fj}  \\
        g_{0,i}(X_1^{N_0},Y_1^{N_0}) &=& (X_1^{i-1}, Y_1^{N_0}).  \IEEEyessubnumber \label{eq_fjgj init gj}
\end{IEEEeqnarray}
Recursion for $f_{n+1,i}$ for any $i \in \medset{n+1}$: 
\begin{align}
    &f_{n+1,i}(X_1^{N_{n+1}}) \nonumber  \\ 
    & =\begin{cases} 
        f_{n,j+1}(X_1^{N_n}) + f_{n,j}(X_{N_n+1}^{2N_n}), & i = 2j, \\ 
        f_{n,j}(X_{N_n+1}^{2N_n}), & i = 2j+1,\ j \in \medsetMinus{n}, \\ 
        f_{n,j+1}(X_1^{N_n}),      & i = 2j+1,\ j \in \medsetPlus{n}. 
        \end{cases} \label{eq_fj iteration}
    \end{align}
Recursion for $g_{n+1,i}$ for any $i \in \medset{n+1}$: 
\begin{align}
    &g_{n+1,i}(X_1^{N_{n+1}}, Y_1^{N_{n+1}}) \nonumber \\ 
    &\mathord=\begin{cases} 
        \Big(g_{n,j}(X_{N_n+1}^{2N_n}, Y_{N_n+1}^{2N_n}), g_{n,j+1}(X_1^{N_n}, Y_1^{N_n}) \Big), & 
        \begin{aligned} &i = 2j,  \\  & j \in \medsetMinus{n},  \end{aligned} \\[0.3cm]
        \Big(g_{n,j+1}(X_1^{N_n}, Y_1^{N_n}), g_{n,j}(X_{N_n+1}^{2N_n}, Y_{N_n+1}^{2N_n}) \Big), & \begin{aligned} &i = 2j,  \\  & j \in
        \medsetPlus{n},  \end{aligned} \\[0.3cm]
        \Big(f_{n+1,i-1}(X_1^{N_{n+1}}), g_{n+1,i-1}(X_1^{N_{n+1}},Y_1^{N_{n+1}}) \Big), & i = 2j+1. 
        \end{cases}\label{eq_gj iteration}
\end{align}
In the recursion for $g_{n+1,i}(X_1^{N_{n+1}}, Y_1^{N_{n+1}})$ where $i=2j$ we differentiate between the cases $j \in \medsetMinus{n}$ and $j\in \medsetPlus{n}$ to
ensure that, for even $i$, the first part of
the observation is an observation from $\medsetMinus{n}$ and the second part is an observation from $\medsetPlus{n}$. 
This is an artifact of the medial indices alternating between blocks, see \Cref{fig_level n+1 of universal construction}. 
This subtlety will be important for a technicality in the proof of \Cref{lem_all medial have the same entropy} below. 
For all other purposes, the reader is encouraged to disregard this rather technical distinction. 

    We concentrate here only on medial indices, because our analysis will focus on medial indices. 
    The recursion~\eqref{eq_fj iteration},~\eqref{eq_gj iteration} is well-defined, as
    medial indices are only ever generated from medial indices (see \Cref{rem_medial only}), so nowhere in the recursion will a non-medial index appear.

    The \emph{observation-truncated \BST} is also a recursively-defined sequence of functions $\tilde{f}_{n,i}$ and $\tilde{g}_{n,i}$. 
    The recursion for these functions is given by~\eqref{eq_fj iteration} and~\eqref{eq_gj iteration}, and is governed by
    the same two parameters, $L_0$ and $M_0$, as the \BST. 
However, the \OTBST has a different initialization than that of the \BST. 
The initialization for the \OTBST is, for any $i \in \medset{0}$,  
\begin{IEEEeqnarray}{rCl} \IEEEyesnumber  \label{eq_fjgj init trunc}
    \tilde{f}_{0,i}(X_1^{N_0}) &=& X_i, \IEEEyessubnumber  \label{eq_fjgj init trunc fj}\\
    \tilde{g}_{0,i}(X_1^{N_0},Y_1^{N_0}) &=& (X_{i-L_0}^{i-1}, Y_{i-L_0}^{i+L_0}).  \IEEEyessubnumber  \label{eq_fjgj init trunc gj}
\end{IEEEeqnarray}

By comparing~\eqref{eq_fjgj init} and~\eqref{eq_fjgj init trunc}, two observations are made. 
First, $f_{n,i} = \tilde{f}_{n,i}$ for any $i \in \medset{n}$. 
Second, there exists a mapping $\gamma_{n,i}$ from $g_{n,i}$ to $\tilde{g}_{n,i}$. 
That is, given $G_i = g_{n,i}(X_1^{N_n}, Y_1^{N_n})$, one may compute 
\[ 
    \tilde{g}_{n,i}(X_1^{N_n}, Y_1^{N_n})= \gamma_{n,i}(g_{n,i} (X_1^{N_n}, Y_1^{N_n}))= \gamma_{n,i}(G_i). 
\] 
This is clear from the initialization step, and for the remaining steps it follows from the recursive definition~\eqref{eq_gj iteration} and since
$f_{n,i} = \tilde{f}_{n,i}$.

The domains for $f_{n,i}, \tilde{f}_{n,i}, g_{n,i}, \tilde{g}_{n,i}$ are over specified. 
Not all inputs of these functions are relevant. 
The relevant domain of these functions may be expressed using the base-vector of $i$. 
To this end, we recall the following notation. 
For any vector of indices $\bv{i} = \begin{bmatrix} i_1 & i_2 & \cdots & i_k \end{bmatrix}$,
natural numbers $L,M$, and a sequence of random variables $X_j$, we denote
\begin{IEEEeqnarray}{rCl} \IEEEyesnumber \label{eq_Svec notation}
    X_{\bv{i}} &=& (X_{i_1}, X_{i_2}, \ldots, X_{i_k}), \IEEEyessubnumber \\ 
    X_{\bv{i}-L} &=& (X_{i_1 - L}, X_{i_2 - L}, \ldots, X_{i_k - L}), \IEEEyessubnumber \\ 
    X_{\bv{i}-L}^{\bv{i}+M} &=& (X_{i_1 - L}^{i_1+M}, X_{i_2 - L}^{i_2+M}, \ldots, X_{i_k - L}^{i_2+M}). \IEEEyessubnumber 
\end{IEEEeqnarray}

Now, let $\bv{b}$ be the base-vector of level-$n$ index $i$. 
Then,   $f_{n,i}$ and $\tilde{f}_{n,i}$ are actually functions of $X_{\bv{b}}$. 
This follows from the recursive definitions of the functions and the base-vector. 
With some abuse of notation we henceforth write
\[ 
    f_{n,i}(X_1^{N_n}) = f_{n,i}(X_{\bv{b}}). 
\] 
Similarly, by~\eqref{eq_fjgj init gj},~\eqref{eq_gj iteration}, and~\eqref{eq_fjgj init trunc gj}, 
\begin{align*}
    g_{n,i}(X_1^{N_n}, Y_1^{N_n}) &= g_{n,i}(X_{\bv{a}}^{\bv{b}}, Y_{\bv{a}}^{\bv{z}}), \\ 
    \tilde{g}_{n,i}(X_1^{N_n}, Y_1^{N_n}) &= \tilde{g}_{n,i}(X_{\bv{b}-L_0}^{\bv{b}}, Y_{\bv{b}-L_0}^{\bv{b}+L_0}),  
\end{align*}
where we denoted
\begin{align*}
    \bv{a} &= \begin{bmatrix} 1 & N_0+1 & 2N_0 +1 & \cdots & (2^{n}-1)N_0 + 1\end{bmatrix}, \\ 
    \bv{z} &= \begin{bmatrix} N_0  & 2N_0 & 3N_0 & \cdots & 2^{n}N_0\end{bmatrix}.
\end{align*}
Note that $Y_{\bv{a}}^{\bv{z}} = Y_1^{N_n}$. 

\begin{example}[continues=ex_level 3 BST]
    For a level-$3$ \BST initialized with $L_0 = 3, M_0 =6$, consider $f_{3,34}$ and $f_{3,35}$. 
The base-vector for either index $34$ or $35$ is 
 \[
     \bv{b} = \begin{bmatrix} 6 & 17 & 29 & 40 & 53 & 64 & 77 & 88 \end{bmatrix}.
 \] 
 We have (see \Cref{fig_level 3 BST}): 
 \begin{align*}
     F_{34} &= f_{3,34}(X_{\bv{b}}) =  X_6 + X_{17} + X_{40} + X_{77} + X_{88}, \\ 
     F_{35} &= f_{3,35}(X_{\bv{b}}) =  X_6 + X_{17} + X_{40}. 
 \end{align*}
 \end{example}

Recall that $\bv{b}$ is the base-vector of level-$n$ index $i$. 
From the recursive definition~\eqref{eq_gj iteration}, we observe that we can compute
$X_{\bv{b}-L_0}^{\bv{b}-1}$ from $\tilde{g}_{n,i}(X_{\bv{b}-L_0}^{\bv{b}}, Y_{\bv{b}-L_0}^{\bv{b}+L_0})$. 
This is easily shown by induction. 
It is trivially true for $n=0$. 
Assume that this holds for $n \geq 0$ for any medial index; we will show it holds for $n+1$ as well. 
Indeed, write $\bv{b} = \begin{bmatrix} \bv{b}_1 & \bv{b}_2 \end{bmatrix}$, where $\bv{b}_1$ and $\bv{b}_2$ are of length $2^{n-1}$.
By the recursive definition of $\bv{b}$,~\eqref{eq_base-vector recursion med}, the recursion~\eqref{eq_gj iteration} becomes
\begin{align*}
    &\tilde{g}_{n+1,i}(X_{\bv{b}-L_0}^{\bv{b}}, Y_{\bv{b}-L_0}^{\bv{b}+L_0}) \\ 
    &\mathord=\begin{cases} 
        \Big(\tilde{g}_{n,j}(X_{\bv{b}_2-L_0}^{\bv{b}_2}, Y_{\bv{b}_2-L_0}^{\bv{b}_2+L_0}), 
            \tilde{g}_{n,j+1}(X_{\bv{b}_1-L_0}^{\bv{b}_1}, Y_{\bv{b}_1-L_0}^{\bv{b}_1+L_0}) \Big), & 
            \begin{subarray}{r}  i = 2j,  \\  j \in \medsetMinus{n}, \end{subarray} \\[0.3cm]
        \Big(\tilde{g}_{n,j+1}(X_{\bv{b}_1-L_0}^{\bv{b}_1}, Y_{\bv{b}_1-L_0}^{\bv{b}_1+L_0}), 
            \tilde{g}_{n,j}(X_{\bv{b}_2-L_0}^{\bv{b}_2}, Y_{\bv{b}_2-L_0}^{\bv{b}_2+L_0})\Big), &
            \begin{subarray}{r}  i = 2j,  \\  j \in \medsetPlus{n}, \end{subarray} \\[0.3cm]
        \Big(\tilde{f}_{n+1,i-1}(X_{\bv{b}-L_0}^{\bv{b}}), \tilde{g}_{n+1,i-1}(X_{\bv{b}-L_0}^{\bv{b}},Y_{\bv{b}-L_0}^{\bv{b}+L_0}) \Big), & i = 2j+1. 
        \end{cases}
\end{align*}
By the induction hypothesis, we can compute $X_{\bv{b}_1-L_0}^{\bv{b}_1-1}$ from $\tilde{g}_{n,j+1}(X_{\bv{b}_1-L_0}^{\bv{b}_1})$, and
$X_{\bv{b}_2-L_0}^{\bv{b}_2-1}$ from $\tilde{g}_{n,j}(X_{\bv{b}_2-L_0}^{\bv{b}_2}, Y_{\bv{b}_2-L_0}^{\bv{b}_2+L_0})$.
In other words, we can compute $X_{\bv{b}-L_0}^{\bv{b}-1}$  from $\tilde{g}_{n+1,i}(X_{\bv{b}-L_0}^{\bv{b}}, Y_{\bv{b}-L_0}^{\bv{b}+L_0})$.
Of course, one can also compute $Y_{\bv{b}-L_0}^{\bv{b}+L_0}$ from $\tilde{g}_{n,i}(X_{\bv{b}-L_0}^{\bv{b}}, Y_{\bv{b}-L_0}^{\bv{b}+L_0})$. 
Therefore, 
recalling that `$\equiv$' between two vectors means that there is a one-to-one mapping between either one and the other that is independent of either
vector,
\begin{equation} \label{eq_gjequiv}
    \tilde{g}_{n,i}(X_{\bv{b}-L_0}^{\bv{b}}, Y_{\bv{b}-L_0}^{\bv{b}+L_0}) \equiv \Big(\tilde{g}_{n,i}(X_{\bv{b}-L_0}^{\bv{b}},
    Y_{\bv{b}-L_0}^{\bv{b}+L_0}), X_{\bv{b}-L_0}^{\bv{b}-1}, Y_{\bv{b}-L_0}^{\bv{b}+L_0} \Big). 
\end{equation}

We saw above that given $G_i=g_{n,i}(X_{\bv{a}}^{\bv{b}}, Y_{\bv{a}}^{\bv{z}} )$ one can compute $\at{G}_i=\tilde{g}_{n,i}(X_{\bv{b}-L_0}^{\bv{b}},
Y_{\bv{b}-L_0}^{\bv{b}+L_0})$. 
In fact, more is true. 
We can compute from $G_i$ two quantities: $\at{G}_i$, which is a function of $(X_{\bv{b}-L_0}^{\bv{b}}, Y_{\bv{b}-L_0}^{\bv{b}+L_0})$, and $\ut{G}_i$,
which consists of $(X_{\bv{a}}^{\bv{b}-L_0-1}, Y_{\bv{a}}^{\bv{b}-L_0-1}, Y_{\bv{b}+L_0+1}^{\bv{z}})$.
Thus, we may write
\begin{equation} \label{eq_Gj splitting}
    G_i = g_{n,i}(X_{\bv{a}}^{\bv{b}}, Y_{\bv{a}}^{\bv{z}}) \equiv (\at{G}_i, \ut{G}_i), 
\end{equation} 
where 
\begin{align*}
    \at{G}_i &= \tilde{g}_{n,i}(X_{\bv{b}-L_0}^{\bv{b}}, Y_{\bv{b}-L_0}^{\bv{b}+L_0}), \\ 
    \ut{G}_i &= (X_{\bv{a}}^{\bv{b}-L_0-1}, Y_{\bv{a}}^{\bv{b}-L_0-1}, Y_{\bv{b}+L_0+1}^{\bv{z}}).  
\end{align*}
This follows by induction similar to the one above. 
Indeed, this is obvious for the initialization step by comparing~\eqref{eq_fjgj init gj} and~\eqref{eq_fjgj init trunc gj}, and the induction step
follows, as above, from the recursive definition of the base-vector~\eqref{eq_base-vector recursion med} and from~\eqref{eq_gj iteration}.

\begin{remark}
    At this point, the reader may be wondering why we used the notation $\at{G}_i, \ut{G}_i$ rather than $\tilde{G}_i, \util{G}_i$. 
    The reason is that we reserve the latter notation to the result of the \OTBST when applied for a different process, the block-independent process, that we
    introduce in \Cref{sec_biprocess}.
    The notation for the block-independent process will use tildes. 
    Our main use of the \OTBST will be for the block-independent process. 
\end{remark}

We conclude this section with a note on terminology. 
The \OTBST is \emph{not} an \htransform. 
That said, we borrow some terminology from \htransforms and apply it to the \OTBST. 
Specifically, for level-$n$ index $i$ with base-vector $\bv{b}$ we call $\tilde{f}_{n,i}(X_{\bv{b}})$ an \ottransformed index. 
The conditional entropy of \ottransformed level-$n$ index $i$ is 
$H( \tilde{f}_{n,i}(X_{\bv{b}}) | \tilde{g}_{n,i}(X_{\bv{b}-L_0}^{\bv{b}}, Y_{\bv{b}-L_0}^{\bv{b}+L_0}))$. 
Finally, for $\eta>0$ and index sets $\mathcal{L}, \mathcal{H} \subseteq \{1,2,\ldots, N_n\}$, the \OTBST is $(\eta, \mathcal{L}, \mathcal{H})$-monopolarizing if either
$H( \tilde{f}_{n,i}(X_{\bv{b}}) | \tilde{g}_{n,i}(X_{\bv{b}-L_0}^{\bv{b}}, Y_{\bv{b}-L_0}^{\bv{b}+L_0})) < \eta$ for all $i \in \mathcal{L}$, or 
$H( \tilde{f}_{n,i}(X_{\bv{b}}) | \tilde{g}_{n,i}(X_{\bv{b}-L_0}^{\bv{b}}, Y_{\bv{b}-L_0}^{\bv{b}+L_0})) > 1-\eta$ for all $i \in \mathcal{H}$.  

\section{The \BST is Monopolarizing} \label{sec_polarizable soprocesses}
For a suitable family of \soprocesses, the \BST is monopolarizing. 
We now describe this family and establish that the \BST is monopolarizing for it. 

\subsection{A Probabilistic Model with Memory} \label{sec_probabilistic model with memory}
The \soprocesses for which we prove that the \BST is monopolarizing share a certain probabilistic structure. 
That is, the distribution of the \soprocess $X_{\star} \sarrow Y_{\star}$ has a specific form: it depends on an underlying Markov sequence, $S_j,
j \in \mathbb{Z}$.  
We assume throughout that, for any $j$, $X_j$ is binary, $Y_j \in \mathcal{Y}$, and $S_j \in \mathcal{S}$, where $\mathcal{Y}, \mathcal{S}$ are finite alphabets.

\begin{definition}[FAIM process] \label{def_FAIM process}
    A strictly stationary process $(S_j,X_j,Y_j)$, $j \in \mathbb{Z}$ is called a  \emph{Finite-State, Aperiodic, Irreducible, Markov} (FAIM) process if, for any any $j$,
\begin{equation} \label{eq_markov property of FAIM} P_{S_j, X_j, Y_j| S_{-\infty}^{j-1},
X_{-\infty}^{j-1}, Y_{-\infty}^{j-1}} = P_{S_j, X_j, Y_j | S_{j-1}} = P_{S_j|S_{j-1}}\cdot P_{X_j,Y_j|S_j},\end{equation} 
        and $S_j, j \in \mathbb{Z}$ is a finite-state, homogeneous, irreducible, and aperiodic stationary Markov chain.

An \soprocess $X_{\star} \sarrow Y_{\star}$ whose joint distribution is derived from a FAIM process $(S_j, X_j, Y_j)$ is called a \emph{FAIM-derived
\soprocess}. 
\end{definition}
Equation~\eqref{eq_markov property of FAIM} implies that conditioned on $S_{j-1}$, the random variables $S_k, X_k, Y_k$ are independent of
$S_{l-1}, X_l, Y_l$, for any $l < j \leq k$.  
Furthermore, $X_j, Y_j$ are a function (possibly probabilistic) of $S_j$. 
FAIM processes are described in detail in~\cite{Shuval_Tal_Memory_2017}.

\begin{remark}
    The definition of FAIM processes in~\cite{Shuval_Tal_Memory_2017} did not include the rightmost equality of~\eqref{eq_markov property of FAIM}.
    However, by suitably redefining the state of the process (for example, take $(S_j, S_{j-1})$ as the state at time $j$),\footnote{Indeed, the
        redefined Markov chain remains finite-state, aperiodic, and irreducible. The redefined state $\bar{S}_j$ takes values in alphabet $\bar{\mathcal{S}}=\{ (s_{j},s_{j-1})
        \ | \ s_{j},s_{j-1}\in \mathcal{S},  \ P_{S_j|S_{j-1}}(s_j|s_{j-1}) > 0\}$. It assumes the value $\bar{S_j} = (s_j,s_{j-1})$ whenever $S_j = s_j, S_{j-1}
        = s_{j-1}$. Since $|\mathcal{S}|< \infty$, so is $|\bar{\mathcal{S}}| < \infty$. The original Markov chain is aperiodic and irreducible if and
        only if there exists $k>0$ such that $P_{S_k|S_0}(s_k|s_0) > 0$ for any $s_0, s_k \in \mathcal{S}$. For this $k$ and any $\bar{s}_0 = (s_0,
        s_{-1}) \in \bar{\mathcal{S}}$ and
        $\bar{s}_{k+1} = (s_{k+1}, s_k) \in \bar{\mathcal{S}}$, we have $P_{\bar{S}_{k+1} | \bar{S}_0}(\bar{s}_{k+1} | \bar{s}_{0}) > 0$. Thus, the redefined Markov process
    remains finite-state, aperiodic, and irreducible.} we may obtain the rightmost equality of~\eqref{eq_markov property of FAIM} from its
    leftmost equality. 
    Therefore, there is no loss of generality in the definition of a FAIM process given here as compared to the one in~\cite{Shuval_Tal_Memory_2017}. 
\end{remark}

In the following lemma we prove an important property of FAIM processes. 
Informally, it implies that two \soblocks that are sufficiently far apart --- that is, the last index of the first \soblock is sufficiently less than
the first index of the second \soblock --- are approximately independent. 

\begin{lemma} \label{lem_FAIM is psi mixing}
    If $X_{\star} \sarrow Y_{\star}$ is a FAIM-derived \soprocess, there exist sequences $\psi_k, \phi_k$, $k \geq 0$,  such that 
    for any $L \leq M \in \mathbb{Z}$,  
    \begin{IEEEeqnarray}{rCl} \IEEEyesnumber \label{eq_psiphi mixing}
        P_{X_{-\infty}^{L}, Y_{-\infty}^{L}, X_{M+1}^{\infty}, Y_{M+1}^{\infty}} &\leq& \psi_{M-L}\cdot
        P_{X_{-\infty}^{L}, Y_{-\infty}^{L}}\cdot P_{X_{M+1}^{\infty}, Y_{M+1}^{\infty}}, \IEEEeqnarraynumspace \IEEEyessubnumber \label{eq_psi mixing} \\
        P_{X_{-\infty}^{L}, Y_{-\infty}^{L}, X_{M+1}^{\infty}, Y_{M+1}^{\infty}} &\geq& \phi_{M-L}\cdot
        P_{X_{-\infty}^{L}, Y_{-\infty}^{L}}\cdot P_{X_{M+1}^{\infty}, Y_{M+1}^{\infty}}. \IEEEyessubnumber \label{eq_phi mixing}
\end{IEEEeqnarray} 
The sequence $\psi_k$ is nonincreasing and the sequence $\phi_k$ is nondecreasing. 
Both $\psi_k$ and $\phi_k$ tend to $1$ exponentially fast as $k \to \infty$. 
\end{lemma}

The sequences $\psi_k$ and $\phi_k$ are called \emph{mixing sequences}. 
Part of the lemma, namely~\eqref{eq_psi mixing}, was established in~\cite[Lemma 5]{Shuval_Tal_Memory_2017}, and the proof
for~\eqref{eq_phi mixing} is similar. 
For completeness, we provide a proof in Appendix~\ref{ap_auxiliary proofs for sec3A}. 
We note at this point that for $k \geq 1$ we may take 
\begin{align*}
    \psi_k &= \max_{s,\sigma}  \frac{\Prob{S_0 = s, S_k = \sigma}}{\Prob{S_0 = s}\Prob{S_k = \sigma}}, \\ 
    \phi_k &= \min_{s,\sigma}  \frac{\Prob{S_0 = s, S_k = \sigma}}{\Prob{S_0 = s}\Prob{S_k = \sigma}}
\end{align*}
in~\eqref{eq_psiphi mixing}. 
These are well-defined because the Markov chain $S_j$, $j \in \mathbb{Z}$ is finite-state, irreducible, and aperiodic. 
As a result, its stationary distribution is positive: $\Prob{S_k = s} > 0$ for any $s \in \mathcal{S}$ and $k \in \mathbb{Z}$, \cite[Theorem 4.2]{Iosifescu}. 

It is immediately evident that for any $k\geq 1$, $1 \leq \psi_k < \infty$ and $0 \leq \phi_k \leq 1$. 
It is possible, however, that for small values of $k$, we will have $\phi_k = 0$. 
Nevertheless, \Cref{lem_FAIM is psi mixing} ensures that if $k$ is large enough, $\phi_k$ will be positive; in fact, by increasing $k$ it can be as close to $1$ as
desired. 

\Cref{lem_FAIM is psi mixing} ensures that \soblocks of a FAIM-derived process become almost independent when sufficiently far apart. 
We will need a separate property that explores what happens when a single \soblock of a FAIM process is large enough.
Specifically, we will be interested in FAIM processes that, in a sense, ``forget'' their past. 
In a forgetful FAIM process, the initial and final states of a sufficiently large \soblock  are almost independent both when given 
just the observations or when given the symbols and observations jointly. 
A precise definition of a forgetful FAIM process follows.

\begin{definition}[Forgetful FAIM process] \label{def_memlength}
    A FAIM process $(S_j, X_j, Y_j)$, $j \in \mathbb{Z}$ is said to be \emph{forgetful} if
    for any $\epsilon > 0$ there exists a natural number $\memlength$ such that if $k \geq \memlength$ then
    \begin{IEEEeqnarray}{rCl} \IEEEyesnumber \label{eq_memory length inequality}
        I(S_1; S_k | X_1^k, Y_1^k) &\leq& \epsilon, \IEEEyessubnumber \label{eq_memory length inequality XY} \\
        I(S_1; S_k | Y_{1}^{k}) &\leq& \epsilon. \IEEEyessubnumber \label{eq_memory length inequality Y}
    \end{IEEEeqnarray}
    For a given $\memlength$, the infimal $\epsilon$ satisfying the above  is called the $\memlength$-\emph{forgetfulness} of the \soprocess, and is denoted $\epsilon_\memlength$. Clearly, $\epsilon_{\lambda}$ is nonincreasing with $\lambda$ and converges to $0$. Conversely, for a given $\epsilon$, the minimal $\memlength$ is called the $\epsilon$-\emph{recollection} of the process. 

    We say that FAIM-derived \soprocess $X_{\star} \sarrow Y_{\star}$ is forgetful if it is derived from a forgetful FAIM process. 
\end{definition}

Several remarks are in order. 
\begin{enumerate}
    \item A sufficient condition  for a FAIM process to be forgetful (\Cref{cond_kaijser}), as well as how to compute the recollection for a given $\epsilon$, 
        are detailed in \Cref{sec_HMM} (see also \Cref{ex_memory length for SXY} in that section). 
	In particular, forgetful FAIM processes do exist. 
    \item Somewhat unintuitively, a FAIM process need not  to be forgetful.
        See \Cref{ex_kaijser counter example} below for an example of a FAIM process that is \emph{not} forgetful. 
    \item Both conditions~\eqref{eq_memory length inequality XY} and~\eqref{eq_memory length inequality Y} are required: neither condition implies the
        other. 
        We demonstrate this unintuitive fact in \Cref{ex_function counter example} below. 
    \item \label{rem_information for other klm}
        In \Cref{lem_information for other klm} below, we show that \eqref{eq_memory length inequality} together with the Markov property \eqref{eq_markov property of FAIM} imply that for any $k \geq \memlength$, $\ell \leq 1$, and $m \geq k$,  
        \begin{IEEEeqnarray}{rCl} \IEEEyesnumber \label{eq_memory length inequality 0}
            I(S_{\ell}; S_m | X_1^k, Y_1^k) &\leq& \epsilon, \IEEEyessubnumber \label{eq_memory length inequality XY 0} \\
            I(S_{\ell}; S_m | Y_{1}^{k}) &\leq& \epsilon. \IEEEyessubnumber \label{eq_memory length inequality Y 0}
        \end{IEEEeqnarray}
\end{enumerate}

The following lemma is proved in Appendix~\ref{ap_auxiliary proofs for sec3A}. 
\begin{lemma} \label{lem_information for other klm}
    Let  $(S_j, X_j, Y_j)$, $j \in \mathbb{Z}$ be a FAIM process. Then, for any $\ell \leq 1$ and $m \geq k \geq \lambda \geq 1$,  we have
    \begin{IEEEeqnarray} {rCl} \IEEEyesnumber \label{eq_memory length inequality general}
        I(S_1; S_{\lambda} | X_1^{\lambda}, Y_1^{\lambda}) & \geq & I(S_{\ell}; S_m | X_1^k, Y_1^k) ;  \IEEEyessubnumber \label{eq_memory length inequality general a} \\
        I(S_1; S_{\lambda} | Y_1^{\lambda}) & \geq & I(S_{\ell}; S_m | Y_1^k). \IEEEeqnarraynumspace \IEEEyessubnumber \label{eq_memory length inequality general b} 
    \end{IEEEeqnarray}
\end{lemma}

\begin{example}\label{ex_kaijser counter example}
    This example is due to~\cite[Section 10]{kaijser1975}. 
 In \Cref{fig_HMM example} we illustrate the process $(S_j, Y_j)$.
Specifically, the Markov chain $S_j$ has transition matrix 
\[ 
    \mat{M}  = \begin{bmatrix} 
                  1/2 & 0   & 1/2 & 0 \\
                   0  & 1/2 & 0   & 1/2 \\ 
                  1/2 & 0   & 0   & 1/2 \\ 
                   0  & 1/2 & 1/2 & 0 
               \end{bmatrix},
\]
and the observation $Y_j$ is given by 
\begin{equation} \label{eq_kaijser observation}
    Y_j = \begin{cases} 
        a, & \text{if } S_j \in\{1,2\}, \\ 
        b, & \text{if } S_j \in \{3,4\}. 
          \end{cases} 
\end{equation} 
In this example we will not be interested in $X_j$. 
This is a FAIM process: the Markov chain $S_j$ is finite-state, aperiodic, and irreducible; indeed, $\mat{M}^3 > 0$.  
\begin{figure}[t] 
    \begin{center}
        \begin{tikzpicture}[> = latex]

            \node[draw, circle, inner sep = 0pt, minimum size = 1cm] (A1) at (0,0) {$1$}; 
            \node[draw, circle, inner sep = 0pt, minimum size = 1cm, right = 2 of A1] (A2) {$2$}; 
            \node[draw, circle, inner sep = 0pt, minimum size = 1cm, below = 2 of A1] (A3) {$3$}; 
            \node[draw, circle, inner sep = 0pt, minimum size = 1cm, right = 2 of A3] (A4) {$4$}; 
            
            \draw[dashed] ($(A1)!0.5!(A3)-(1.5,0)$) -- ($(A2)!0.5!(A4)+(1.5,0)$) node[above]{$a$} node[below]{$b$}; 
            \draw[->] (A1.250) to[bend right] (A3.110); 
            \draw[->] (A3.70)  to[bend right] (A1.290); 
            \draw[->] (A2.250) to[bend right] (A4.110); 
            \draw[->] (A4.70)  to[bend right] (A2.290); 
            \draw[->] (A3.340) to[bend right] (A4.200); 
            \draw[->] (A4.160) to[bend right] (A3.20); 

            \draw[->] (A1) edge[in = 170, out = 100,min distance = 5mm, looseness=5] (A1); 
            \draw[->] (A2) edge[in = 10, out = 80,min distance = 5mm, looseness=5] (A2); 

        \end{tikzpicture}
    \end{center}
    \caption{The Markov chain $S_j$ has four states. 
             The possible transitions are depicted using arrows; the probability of choosing any transition is $1/2$. 
             The observation $Y_j$ is `$a$' if $S_j\in \{1,2\}$ or `$b$' if $S_j \in \{3,4\}$.}
             \label{fig_HMM example}
\end{figure}

From the observation $Y_j$ we can infer whether state $S_j$ is in the top half or the bottom half of \Cref{fig_HMM example}.
For two consecutive observations to differ, the process must transition from a state in one half of \Cref{fig_HMM example} to the other. 
Given a sequence of observations, our best guess for the next state is equi-probable among two states. 
For example, given the observation sequence $Y_1 = a, Y_2 = b, \ldots, Y_{k} = b$, we know that $S_k \in \{3,4\}$, but $S_{k}$ could be either $3$ or
$4$ with equal probability. 

Assume now that, in addition to the observation sequence, we are told the state at time $1$.
Say, $S_1 = 1$ (accordingly, $Y_1 = a$). 
The observations are tied to transitions from one half of \Cref{fig_HMM example} to the other half, 
so that one can trace the state: $Y_2 = a$ implies that $S_2 = 1$. 
Then, $Y_3 = b$ implies that $S_3 = 3$, and so on. 
In this manner, we are able to find $S_k$ precisely. 

We have demonstrated that in this example, $I(S_1; S_{k} | Y_1^k)$ cannot vanish with $k$, so this process is not forgetful. 
\end{example}

\begin{example} \label{ex_function counter example}
    Let $S_j$ be as in \Cref{ex_kaijser counter example}.
    We now construct two FAIM processes. 
    For the first process, $I(S_1; S_{k}|X_1^k, Y_1^k)$ will vanish with $k$ but $I(S_1; S_{k} | Y_1^k)$ will not.
    For the second process, $I(S_1; S_{k}|X_1^k, Y_1^k)$ will not vanish with $k$ but $I(S_1; S_{k} | Y_1^k)$ will.
    \begin{itemize}
        \item Let $X_j = S_j$ and $Y_j$ as in~\eqref{eq_kaijser observation}.
            Then, $I(S_1; S_{k} | X_1^k, Y_1^k) = I(S_1; S_{k}|S_1^k) = 0$ trivially. 
            On the other hand, as shown in \Cref{ex_kaijser counter example}, $I(S_1; S_{k} | Y_1^k)$ does not vanish for any $k$.  
        \item Let $X_j$ be given by~\eqref{eq_kaijser observation} (that is, $X_j = a$ if $S_j \in \{1,2\}$ and $X_j = b$ otherwise) and $Y_j = 0$.
            Then, $I(S_1; S_{k} | X_1^k, Y_1^k)$ cannot vanish with $k$, as shown in \Cref{ex_kaijser counter example}.
            On the other hand, $I(S_1; S_{k} | Y_1^k) = I(S_1; S_{k}) \to 0$, since the Markov chain $S_j$ is finite-state, aperiodic, and irreducible
            (see, e.g.,~\cite[Theorem 4.3]{Iosifescu}). 
\end{itemize}
\end{example}

Assume we have a forgetful FAIM process, and we apply to it a level-$0$ \BST (i.e.\ \eqref{eq_level0}), initialized with $L_0$
that is greater than its $\epsilon$-recollection. 
We expect that in this case, all medial \sopairs will have approximately the same conditional
entropy. 
This is indeed the case, as we will soon show in \Cref{lem_conditional entropy for finite
memlength}. 
Moreover, we will see in \Cref{cor_entropy rate with finite memory length} that this conditional
entropy cannot veer much from the conditional entropy rate of the \soprocess. 
First, however, we require an additional lemma. 

\begin{lemma} \label{lem_finite memory for HMM}
    Let $(S_j, X_j, Y_j)$ be a forgetful FAIM process.
    Then, for every $\epsilon >0$ there exists a natural number $\memlength$ such that for any integers $m,\ell,k$ such that $\min\{m,\ell\} \geq k \geq \memlength$ we have
    \begin{equation} \label{eq_I S0 Sn S-n inequality}
        I(S_0; S_{-k},S_k | X_{-\ell}^{-1}, Y_{-\ell}^{m}) \leq 2\epsilon. 
    \end{equation}
\end{lemma}
This is a consequence of~\eqref{eq_memory length inequality}.
To prove it, we take $\memlength$ as the $\epsilon$-recollection of the process, and make multiple uses of~\eqref{eq_DPI}, which are possible due to the
Markov property~\eqref{eq_markov property of FAIM}. 
A detailed proof can be found in Appendix~\ref{ap_auxiliary proofs for sec3A}. 

\Cref{lem_finite memory for HMM} shows that the mutual information between a state and two surrounding states vanishes when given a sequence
of observations between the surrounding states. 
The following corollary shows that this is also the case when considering the mutual information between a sequence of states and a sequence of surrounding states. 
This will be useful in the sequel. 

\begin{corollary} \label{cor_forgetting multi blocks}
    Let $(S_j, X_j, Y_j)$ be a forgetful FAIM process.
    Then, for every $\epsilon > 0$ there exists a natural number $\memlength$ such that for any positive natural numbers $k$, $i_1, i_2, \ldots, i_k$,
    and $L_0$ that
    satisfy $L_0 \geq \memlength$ and
    \[ 
         i_1 -L_0 \leq i_1 \leq i_1 + L_0 \leq i_2 -L_0 \leq i_2 \leq \cdots \leq i_k \leq i_k +L_0
    \]                                                                                                           
    we have
    \[ 
        I(S_{\bv{i}} ; S_{\bv{i}-L_0}, S_{\bv{i}+L_0} | X_{\bv{i}-L_0}^{\bv{i}-1},
        Y_{\bv{i}-L_0}^{\bv{i}+L_0}) \leq k \cdot 2\epsilon, 
    \] 
    where
    \[
        \bv{i} = \begin{bmatrix} i_1 & i_2 & \cdots & i_k \end{bmatrix}.
    \]
\end{corollary}
In the statement of the corollary, we used the notation of~\eqref{eq_Svec notation}.
The proof of the corollary is relegated to Appendix \ref{ap_auxiliary proofs for sec3A}.

In the next lemma, we show that, for a forgetful FAIM-derived \soprocess, all medial \sopairs in a level-$0$ \BST have approximately the same conditional entropy,
\begin{equation} \label{eq_def of H0}
\Hwind \triangleq H(X_i| X_{i-L_0}^{i-1}, Y_{i-L_0}^{i+L_0}). 
\end{equation}
By stationarity, $\Hwind$ is indeed independent of $i$.

\begin{lemma} \label{lem_conditional entropy for finite memlength}
    Let $X_{\star} \sarrow Y_{\star}$ be a forgetful FAIM-derived \soprocess with $\epsilon$-recollection $\memlength$.  
    Let $L_0 \geq \memlength$ and $M_0 \geq 1$, and denote $N_0 = 2L_0 + M_0$. 
    Then, for any $L_0 + 1 \leq i \leq L_0 + M_0$ we have
    \begin{equation} \label{eq_conditional entropy with and without future}
         0 \leq \Hwind - H(X_i | X_{1}^{i-1}, Y_{1}^{N_0})  \leq 2\epsilon.
    \end{equation}  
\end{lemma}

\begin{IEEEproof}
    Observe that 
    \begin{align*}
        &\Hwind - H(X_i | X_{1}^{i-1}, Y_{1}^{N_0}) \\
        &\quad=  H(X_i | X_{i-L_0}^{i-1}, Y_{i-L_0}^{i+L_0}) - H(X_i | X_{1}^{i-1}, Y_{1}^{N_0})   \\ 
        &\quad= I\left(X_i; \big(X_{1}^{i-L_0-1}, Y_1^{i-L_0-1}, Y_{i+L_0+1}^{N_0} \big)\big| X_{i-L_0}^{i-1},
            Y_{i-L_0}^{i+L_0}\right).
    \end{align*}
    This right-hand side is nonnegative. 
    It remains to upper-bound it by $2\epsilon$ to establish~\eqref{eq_conditional entropy with and without future}. 

    Let $(S_j, X_j, Y_j)$ be the FAIM process from which $X_{\star} \sarrow Y_{\star}$ is derived. 
    By stationarity and \Cref{lem_finite memory for HMM},
    for any $m,\ell, k$ such that $\min\{m,\ell\} \geq k \geq \memlength$, 
    \begin{equation} \label{eq_memory length inequality i}
        I(S_i ; S_{i-k}, S_{i+k} | X_{i-\ell}^{i-1}, Y_{i-\ell}^{i+m}) \leq 2\epsilon.
    \end{equation}
    Setting $\ell = m =  k = L_0$ in~\eqref{eq_memory length inequality i} yields 
    \[ 
        I(S_i; S_{i-L_0}, S_{i+L_0} | X_{i-L_0}^{i-1}, Y_{i-L_0}^{i+L_0}) \leq 2\epsilon.
    \]

    By~\eqref{eq_markov property of FAIM} and the data processing inequality~\eqref{eq_DPI} used
    twice,
    we obtain
    \begin{align*}
        2\epsilon & \geq I\left(S_i; S_{i-L_0}, S_{i+L_0} | X_{i-L_0}^{i-1}, Y_{i-L_0}^{i+L_0}\right) \\ 
                  & \eqann[\geq]{a}  I\left(X_i; S_{i-L_0}, S_{i+L_0} | X_{i-L_0}^{i-1}, Y_{i-L_0}^{i+L_0}\right) \\ 
                  & \eqann[\geq]{b}  I\left(X_i; X_1^{i-L_0-1}, Y_1^{i-L_0-1}, Y_{i+L_0+1}^{N_0} | X_{i-L_0}^{i-1},
    Y_{i-L_0}^{i+L_0}\right).
    \end{align*}
    We now detail the Markov chains used for the inequalities, both using~\eqref{eq_markov property of FAIM}. 
    Inequality~\eqannref{a} is due to 
    \[ 
        (S_{i-L_0}, S_{i+L_0}) \markov (S_i, X_{i-L_0}^{i-1}, Y_{i-L_0}^{i-1}) \markov X_i, 
    \] 
    and inequality~\eqannref{b} is due to 
    \[ 
            X_i \markov (\ns S_{i-L_0}, S_{i+L_0}, X_{i-L_0}^{i-1}, Y_{i-L_0}^{i+L_0}) \markov (\ns X_1^{i-L_0-1}, Y_1^{i-L_0-1},
            Y_{i+L_0+1}^{N_0}\ns). 
        \]
    This completes the proof. 
\end{IEEEproof}

The following corollary shows that, for a forgetful FAIM-derived \soprocess, $\Hwind$ is approximately equal to the conditional entropy rate of 
the \soprocess. 

\begin{corollary} \label{cor_entropy rate with finite memory length}
    Under the same setting as \Cref{lem_conditional entropy for finite memlength},
    \begin{equation} \label{eq_H0 and ENTI difference}
        \left|\ENT{X_{\star} | Y_{\star}} - \Hwind \right| \leq 2\epsilon,
    \end{equation}
\end{corollary}

\begin{IEEEproof}
    For any $\xi > 0$, let $N = N(\xi) > 2L_0 $ be large enough so that $|\ENT{X_{\star} | Y_{\star} } - H(X_1^{N} | Y_1^{N})/N | \leq \xi/2$ and
    $2L_0/N \leq \xi/2$. 
    Then, 
    \begin{align*}
    & | \ENT{X_{\star} | Y_{\star}} - \Hwind| \\  
    & \quad \eqann[\leq]{a} \left|\ENT{X_{\star} | Y_{\star}} - \frac{1}{N}H(X_1^{N} | Y_1^{N})  \right| +  \left| \frac{1}{N}H(X_1^{N} | Y_1^{N})  - \Hwind \right| \\ 
    & \quad \eqann[\leq]{b} \frac{\xi}{2} +  \frac{1}{N} \sum_{i=1}^{N} \left| H(X_i | X_1^{i-1}, Y_1^{N}) - \Hwind \right|\\
    & \quad \eqann[\leq]{c} \frac{\xi}{2} + \frac{2 L_0}{N} +  \frac{1}{N}\sum_{i=L_0+1}^{N - L_0}
        \left| H(X_i|X_1^{i-1},
    Y_1^{N}) - \Hwind \right| \\ 
    &\quad \eqann[\leq]{d} \xi + \frac{N - 2L_0}{N} 2\epsilon \\ 
    &\quad \leq 2\epsilon+ \xi,
    \end{align*} 
where~\eqannref{a} and~\eqannref{b} are by the triangle inequality;~\eqannref{c} is because
$|H(X_i|X_1^{i-1}, Y_1^{N})-\Hwind| \leq \max\{\Hwind, H(X_i|X_1^{i-1} , Y_1^N)\} \leq 1$, where the latter inequality holds since $X_i$ is binary; finally,~\eqannref{d} is by
    \Cref{lem_conditional entropy for finite memlength}, with $N_0$ replaced with $N$. 
    The above holds for any $\xi >0$, so it holds for $\xi = 0$ as well. 
\end{IEEEproof}

\subsection{The Block-Independent Process} \label{sec_biprocess}
We will prove in \Cref{sec_monopolarization for FAIM derived processes} that the \BST is monopolarizing with the help of another process, the
block-independent process, that we now introduce. 
We will show that an \OTBST is monopolarizing when applied to the block-independent process. 
It turns out that the result of an \OTBST applied to the block-independent process is approximately the same as the result of a \BST applied to
a forgetful FAIM-derived process, provided that the transform parameters are carefully chosen. 
Therefore, monopolarization of the \OTBST of the block-independent process will be of vital importance in proving that the \BST is monopolarizing. 

Let $N_n = 2^n N_0$, where $N_0 = 2L_0+M_0$. 
Denote by $P_{X_1^{N_n},Y_1^{N_n}}$ the joint distribution of $(X_1^{N_n}, Y_1^{N_n})$. 
 By marginalizing $P_{X_1^{N_n},Y_1^{N_n}}$, we obtain the distribution of a single \bblock, $P_{X_{(\ell-1)N_0+1}^{\ell N_0}, Y_{(\ell-1)N_0+1}^{\ell N_0}}$, which, by
 stationarity, is independent of $\ell$. 

 \begin{definition}[Block-Independent Process] \label{def_biprocess}
The \emph{block-independent process} (\biprocess) $\tilde{X}_{\star}  \sarrow \tilde{Y}_{\star}$ with parameter $N_0$, is distributed according to 
\[ 
    (\tilde{X}_1^{N_n}, \tilde{Y}_1^{N_n}) \sim \prod_{\ell=1}^{2^n} P_{X_{(\ell-1)N_0+1}^{\ell N_0}, Y_{(\ell-1)N_0+1}^{\ell N_0}}. 
\] 
That is, \bblocks of length $N_0$ are independent in this distribution. 
\end{definition}

If $\bv{b} = \begin{bmatrix} b_1 & b_2 & \cdots & b_{2^n} \end{bmatrix}$ 
is the base-vector of a level-$n$ medial index, we have
\begin{equation} \label{eq_tilde process}
    (\tilde{X}_{\bv{b}-L_0}^{\bv{b}}, \tilde{Y}_{\bv{b}-L_0}^{\bv{b}+L_0}) \sim \prod_{\ell=1}^{2^n} 
            P_{X_{b_{\ell}- L_0}^{b_{\ell}}, Y_{b_{\ell}-L_0}^{b_{\ell}+L_0} }, 
        \end{equation} 
where $P_{X_{b_{\ell}- L_0}^{b_{\ell}}, Y_{b_{\ell}-L_0}^{b_{\ell}+L_0} }$ is obtained from  $P_{X_{(\ell-1)N_0+1}^{\ell N_0}, Y_{(\ell-1)N_0+1}^{\ell
N_0}}$ by marginalization. 
Note that since each $b_{\ell}$ is medial, $(X_{b_{\ell-L_0}}^{b_{\ell}}, Y_{b_{\ell}-L_0}^{b_{\ell}+L_0})$ is wholly contained in a \bblock with
    \bblock number $\ell$. 

    \begin{remark} \label{rem_windowed BIprocess is iid}
	    Observe that the RHS of \eqref{eq_tilde process} consists of a product of distributions of ``windows'' of the same size. Each window, by stationarity, has the same distribution. Moreover, each window is in a different \bblock. By block-independence, therefore, the RHS of~\eqref{eq_tilde process} is independent of $\bv{b}$. Put another way, these windows are i.i.d.  This is the crux of the results that follow: the transforms operate on an i.i.d.\ process.\footnote{In~\cite{sasoglu_2016_universal}, the underlying process was i.i.d.\ to begin with; here, where there is memory, we need more intricate mechanics: the \OTBST and the \biprocess.} In fact, the results of this section hold also for a BST operating on i.i.d. \sopairs. This observation will be useful in \Cref{sec_monopolarization for FAIM derived processes cascade}, where we consider a cascade of BSTs, in which a step of the cascade operates on such \sopairs. 
    \end{remark}

Throughout this section index $i \in \medset{n}$ has base-vector $\bv{b} = \begin{bmatrix} b_1 & b_2 & \cdots & b_{2^n} \end{bmatrix}$,
and index $j \in \medset{n}$ has base-vector $\bv{d} = \begin{bmatrix} d_1 & d_2 & \cdots & d_{2^n} \end{bmatrix}$. 
We also denote
\begin{align*}
    \bv{a} &= \begin{bmatrix} 1 & N_0+1 & 2N_0 +1 & \cdots & (2^{n}-1)N_0 + 1\end{bmatrix}, \\ 
    \bv{z} &= \begin{bmatrix} N_0  & 2N_0 & 3N_0 & \cdots & 2^{n}N_0\end{bmatrix}. 
\end{align*}
Recalling the definitions of $\tilde{f}_{n,i}$ and $\tilde{g}_{n,i}$ at the beginning of \Cref{sec_otbst}, 
we define 
\begin{IEEEeqnarray}{rCl?rCl} \IEEEyesnumber \label{eq_FiGiFjGjtilde} 
    \tilde{F}_i &=& \tilde{f}_{n,i}(\tilde{X}_{\bv{b}}),  &\tilde{G}_i &=& \tilde{g}_{n,i}(\tilde{X}_{\bv{b}-L_0}^{\bv{b}}.
    \tilde{Y}_{\bv{\bv{b}}-L_0}^{\bv{b}+L_0}).  \IEEEyessubnumber \label{eq_FiGitilde} \\ 
    \tilde{F}_j &=& \tilde{f}_{n,j}(\tilde{X}_{\bv{d}}), &\tilde{G}_{j} &=& \tilde{g}_{n,j}(\tilde{X}_{\bv{d}-L_0}^{\bv{d}},
    \tilde{Y}_{\bv{d}-L_0}^{\bv{d}+L_0}).  \IEEEyessubnumber \label{eq_FjGjtilde}
\end{IEEEeqnarray}
The joint distribution of $(\tilde{X}_{\bv{b}-L_0}^{\bv{b}}, \tilde{Y}_{\bv{b}-L_0}^{\bv{b}+L_0})$ is given by~\eqref{eq_tilde process} with $\bv{b}$ as the base-vector of $i$. 
The joint distribution of $(\tilde{X}_{\bv{d}-L_0}^{\bv{d}}, \tilde{Y}_{\bv{d}-L_0}^{\bv{d}+L_0})$ is given by~\eqref{eq_tilde process} with $\bv{b}$
set to $\bv{d}$, the base-vector of $j$.

Recall from~\eqref{eq_def of H0} that we denoted $\Hwind = H(X_i|X_{i-L_0}^{i-1}, Y_{i-L_0}^{i+L_0})$,
which, by stationarity, is independent of $i$. 
We wish to show that there exists $\delta_n \geq 0$, independent of $i$, such that if $i \in \medsetMinus{n}$ then $H(\tilde{F}_i | \tilde{G}_i) = \Hwind
+ \delta_n$ and if $i \in \medsetPlus{n}$ then $H(\tilde{F}_i | \tilde{G}_i) = \Hwind -\delta_n$. 
This will follow as a corollary to the following lemma. 

\begin{lemma} \label{lem_all medial have the same entropy}
Suppose that either $i, j \in \medsetMinus{n}$ or $i,j \in \medsetPlus{n}$. 
Then, the joint distribution of $(\tilde{F}_i, \tilde{G}_i)$ is the same as the joint distribution of $(\tilde{F}_j, \tilde{G}_j)$.  
\end{lemma}
\begin{IEEEproof}
    We use induction. 
    For $ n = 0$, the claim is true by stationarity and the initialization of the \OTBST,~\eqref{eq_fjgj init trunc}.
    Indeed, in this case, $\tilde{F}_i = \tilde{X}_i$, $\tilde{F}_j = \tilde{X}_j$, $\tilde{G}_i = (\tilde{X}_{i-L_0}^{i-1}, \tilde{Y}_{i-L_0}^{i+L_0})$, and $\tilde{G}_j
    = (\tilde{X}_{j-L_0}^{j-1}, \tilde{Y}_{j-L_0}^{j+L_0})$. 
    Stationarity implies that the joint distribution of $(\tilde{F}_i,\tilde{G}_i)$ is the same as the joint distribution of $(\tilde{F}_j, \tilde{G}_j)$. 

    Assume the claim is true for some $n-1\geq 0$. 
    We now show it holds for $n$. 

    Denote $i' = \lfloor i/2 \rfloor$ and $j' =\lfloor j/2 \rfloor$. 
    We write $\bv{b} = \begin{bmatrix} \bv{b}_1 & \bv{b}_2 \end{bmatrix}$ and $\bv{d} = \begin{bmatrix} \bv{d}_1 & \bv{d}_2 \end{bmatrix}$, where
    $\bv{b}_1, \bv{b}_2, \bv{d}_1, \bv{d}_2$ are vectors of length $2^{n-1}$. 
    Then, $\bv{b}_1$ is the base-vector of $i'+1$, and $\bv{b}_2$ is the base-vector of $i'$, see~\eqref{eq_base-vector recursion med}. 
    Similarly, $\bv{d}_1$ is the base-vector of $j'+1$, and $\bv{d}_2$ is the base-vector of $j'$. 
    Denote 
    \begin{IEEEeqnarray}{rCl/rCl} \IEEEyesnumber \label{eq_UVQRtilde} 
            \tilde{U}_{i'+1} &=& \tilde{f}_{n-1,i'+1}(\tilde{X}_{\bv{b}_1}), &\tilde{Q}_{i'+1} &=& 
        \tilde{g}_{n-1,i'+1}(\tilde{X}_{\bv{b}_1-L_0}^{\bv{b}_1}, \tilde{Y}_{\bv{b}_1 -L_0}^{\bv{b}_1 +L_0}),  \IEEEyessubnumber \IEEEeqnarraynumspace \\ 
        \tilde{V}_{i'} &=& \tilde{f}_{n-1,i'}(\tilde{X}_{\bv{b}_2}), &\tilde{R}_{i'} &=& 
        \tilde{g}_{n-1,i'}(\tilde{X}_{\bv{b}_2-L_0}^{\bv{b}_2}, \tilde{Y}_{\bv{b}_2 -L_0}^{\bv{b}_2 +L_0}). \IEEEyessubnumber  
\end{IEEEeqnarray}
Of the two \sopairs $\tilde{U}_{i'+1} \sarrow \tilde{Q}_{i'+1}$ and $\tilde{V}_{i'} \sarrow \tilde{R}_{i'}$, one is in $\medsetMinus{n-1}$ and the
other in $\medsetPlus{n-1}$. 
We denote by $\tilde{T}_i^-$ the pair that is in $\medsetMinus{n-1}$ and by $\tilde{T}_i^+$ the pair that is in $\medsetPlus{n-1}$.   
That is, 
\[
    \tilde{T}_i^- = \begin{cases} 
    (\tilde{V}_{i'}, \tilde{R}_{i'}), & i' \in \medsetMinus{n-1}, \\  
    (\tilde{U}_{i'+1}, \tilde{Q}_{i'+1}), & i' \in \medsetPlus{n-1} 
\end{cases}  
\]
and
\[
\tilde{T}_i^+ = \begin{cases} 
    (\tilde{U}_{i'+1}, \tilde{Q}_{i'+1}), & i' \in \medsetMinus{n-1},  \\ 
    (\tilde{V}_{i'}, \tilde{R}_{i'}), & i' \in \medsetPlus{n-1}.   
\end{cases} 
\] 
We similarly define $\tilde{U}_{j'+1}$, $\tilde{V}_{j'}$, $\tilde{Q}_{j'+1}$, $\tilde{R}_{j'}$, $\tilde{T}_j^-$, and $\tilde{T}_j^+$ (with $\bv{b}$
replaced with $\bv{d}$ and $i'$ replaced with $j'$).

    For the \biprocess, \bblocks are independent. 
    In particular, by~\eqref{eq_tilde process}, $(\tilde{X}_{\bv{b}_1-L_0}^{\bv{b}_1}, \tilde{Y}_{\bv{b}_1 -L_0}^{\bv{b}_1+L_0})$ is independent of
    $(\tilde{X}_{\bv{b}_2-L_0}^{\bv{b}_2}, \tilde{Y}_{\bv{b}_2 - L_0}^{\bv{b}_2+L_0})$. 
    Hence, $\tilde{T}_i^-$ and $\tilde{T}_i^+$ are independent. 
    Similarly, $\tilde{T}_j^-$ and $\tilde{T}_j^+$ are independent.
    By the induction hypothesis, $\tilde{T}_i^-$ and $\tilde{T}_j^-$ have the same distribution; 
    $\tilde{T}_i^+$ and $\tilde{T}_j^+$ are also equi-distributed. 
    By block-independence, the joint distribution of $(\tilde{T}_i^-, \tilde{T}_i^+)$ is the same as the joint distribution of 
 $(\tilde{T}_j^-, \tilde{T}_j^+)$.

    Assume first that $i,j \in \medsetMinus{n}$.
    We then have, by~\eqref{eq_fj iteration} and~\eqref{eq_gj iteration},  
    \begin{equation} \label{eq_FiGi for i even}
        \tilde{F}_i = \tilde{U}_{i'+1}+ \tilde{V}_{i'}, \quad \tilde{G}_i =\begin{cases} 
            (\tilde{R}_{i'}, \tilde{Q}_{i'+1}), & i' \in \medsetMinus{n-1},  \\ 
            (\tilde{Q}_{i'+1}, \tilde{R}_{i'}), & i' \in \medsetPlus{n-1}, 
        \end{cases}
    \end{equation}
    and 
    \begin{equation} \label{eq_FjGj for j even}
        \tilde{F}_j = \tilde{U}_{j'+1}+ \tilde{V}_{j'}, \quad \tilde{G}_j =\begin{cases} 
            (\tilde{R}_{j'}, \tilde{Q}_{j'+1}), & j' \in \medsetMinus{n-1},  \\ 
            (\tilde{Q}_{j'+1}, \tilde{R}_{j'}), & j' \in \medsetPlus{n-1}.
        \end{cases}
    \end{equation}

 Comparing~\eqref{eq_FiGi for i even} and~\eqref{eq_FjGj for j even}, the mapping from $(\tilde{T}_i^-, \tilde{T}_i^+)$ to $(\tilde{F}_i,
 \tilde{G}_i)$ is the same as the mapping from   $(\tilde{T}_j^-, \tilde{T}_j^+)$ to $(\tilde{F}_j, \tilde{G}_j)$. 
 We conclude that the joint distribution of $(\tilde{F}_i, \tilde{G}_i)$ is the same
    as the joint distribution of $(\tilde{F}_j, \tilde{G}_j)$. 

    For the case where $i,j \in \medsetPlus{n}$, we have
    by~\eqref{eq_fj iteration}, 
    \[ 
        \tilde{F}_i = \begin{cases} 
            \tilde{V}_{i'}, & i' \in \medsetMinus{n-1},  \\ 
            \tilde{U}_{i'+1}, & i' \in \medsetPlus{n-1}. 
        \end{cases}
    \]
    Observe that $\tilde{F}_i$ is always a symbol in $\medsetMinus{n-1}$. 
    Further recall from~\eqref{eq_fj iteration} that, since $i-1 \in \medsetMinus{n}$, we have
    $\tilde{F}_{i-1} = \tilde{U}_{i'+1} + \tilde{V}_{i'}$, so that $\tilde{F}_i + \tilde{F}_{i-1}$
    is a symbol from $\medsetPlus{n-1}$. 

    By~\eqref{eq_fj iteration} and~\eqref{eq_gj iteration}, 
    \begin{equation} \label{eq_FiGitilde equiv}
        (\tilde{F}_i, \tilde{G}_{i})=  (\tilde{F}_i, \tilde{F}_{i-1},\tilde{G}_{i-1}) \equiv 
        (\tilde{F}_i, \tilde{F}_i + \tilde{F}_{i-1}, \tilde{G}_{i-1}), 
    \end{equation}
    Similarly, 
    \begin{equation} \label{eq_FjGjtilde equiv}
        (\tilde{F}_j, \tilde{G}_{j}) =(\tilde{F}_j, \tilde{F}_{j-1},\tilde{G}_{j-1}) \equiv (\tilde{F}_j, \tilde{F}_j + \tilde{F}_{j-1},
        \tilde{G}_{j-1}). 
   \end{equation}
   The mappings on the right-hand sides of~\eqref{eq_FiGitilde equiv} and~\eqref{eq_FjGjtilde equiv} are the same. 
   Moreover, by~\eqref{eq_gj iteration}, the mapping between $(\tilde{F}_i, \tilde{F}_i + \tilde{F}_{i-1}, \tilde{G}_{i-1})$ and 
   $(\tilde{T}_i^-, \tilde{T}_i^+)$
   is the same as the mapping between  $(\tilde{F}_j, \tilde{F}_j + \tilde{F}_{j-1}, \tilde{G}_{j-1})$ and  
   $(\tilde{T}_j^-, \tilde{T}_j^+)$.
   Since $(\tilde{T}_i^-, \tilde{T}_i^+)$ and $(\tilde{T}_j^-, \tilde{T}_j^+)$ are equi-distributed, so are $(\tilde{F}_i, \tilde{G}_i)$ and
   $(\tilde{F}_j, \tilde{G}_j)$. 
\end{IEEEproof}

\begin{corollary} \label{cor_medial plus minus delta}
    There exists a nondecreasing sequence $\delta_n\geq 0$, independent of $i$, such that if $i \in \medsetMinus{n}$ then
    $H(\tilde{F}_i | \tilde{G}_i) = \Hwind + \delta_n$  
    and  $H(\tilde{F}_{i+1} | \tilde{G}_{i+1}) = \Hwind - \delta_n$. 
\end{corollary}

Observe from~\eqref{eq_def of medA} and~\eqref{eq_def of medB} that \Cref{cor_medial plus minus delta} implies that there exists a nondecreasing
sequence $\delta_n \geq 0$ such that 
\begin{equation}  \label{eq_HFi given Gi deltan}
    H(\tilde{F}_i | \tilde{G}_i) = \begin{cases} \Hwind + \delta_n, & i \in \medsetMinus{n}, \\ 
                                                 \Hwind - \delta_n, & i \in \medsetPlus{n}.
                                             \end{cases}
\end{equation}
Further observe that~\Cref{cor_medial plus minus delta} implies that for any $i \in \medsetMinus{n}$ and $j \in \medsetPlus{n}$ we have
\begin{equation} \label{eq_sum of HFiGi minus plus}
    H(\tilde{F}_i | \tilde{G}_i)  + H(\tilde{F}_j | \tilde{G}_j) = 2\Hwind. 
\end{equation}

\begin{IEEEproof}
    We show this using induction. 
    The claim is true for $n=0$ with $\delta_0 = 0$. 
    For $n > 0$, we assume the claim is true for $n-1$ and show it also holds for $n$. 

    Let $i \in \medsetMinus{n}$ with base-vector $\bv{b}$.
    Since $n\geq 1$, $i$ is even (see \Cref{rem_sets alt}), and we denote $i' = i/2$. 
    Let $\tilde{F}_i, \tilde{G}_i$, as well as $\tilde{F}_{i+1}, \tilde{G}_{i+1}$, be defined as in~\eqref{eq_FiGitilde} and let $\tilde{U}_{i'+1},
    \tilde{V}_{i'}, \tilde{Q}_{i'+1}, \tilde{R}_{i'}$
    be defined as in~\eqref{eq_UVQRtilde}. 
    We have, by~\eqref{eq_fj iteration} and \eqref{eq_gj iteration},
    \begin{IEEEeqnarray}{rCl}
        H(\tilde{F}_i | \tilde{G}_i) + H(\tilde{F}_{i+1} | \tilde{G}_{i+1}) &=& H(\tilde{F}_i, \tilde{F}_{i+1} | \tilde{Q}_{i'+1}, \tilde{R}_{i'})
        \IEEEnonumber \\
        &=& H(\tilde{U}_{i'+1}, \tilde{V}_{i'} | \tilde{Q}_{i'+1}, \tilde{R}_{i'})  \IEEEnonumber\\ 
        &=& H(\tilde{U}_{i'+1} | \tilde{Q}_{i'+1}) + H(\tilde{V}_{i'} | \tilde{R}_{i'}), \IEEEeqnarraynumspace \IEEEyesnumber \label{eq_HFGUVQR tilde}
    \end{IEEEeqnarray}
    where the last equality is by block independence. 
    By the induction assumption and stationarity there exists $\delta_{n-1} \geq 0$ such that
    \[
        H(\tilde{U}_{i'}|\tilde{Q}_{i'}) = H(\tilde{V}_{i'}|\tilde{R}_{i'}) = \begin{cases}
            \Hwind + \delta_{n-1}, & i'\in \medsetMinus{n-1}, \\ 
            \Hwind - \delta_{n-1}, & i'\in \medsetPlus{n-1}. 
        \end{cases}
    \]
    Thus, 
    \begin{equation}\label{eq_sum of HFiGi and HFiGi1}
        H(\tilde{F}_i | \tilde{G}_i) + H(\tilde{F}_{i+1} | \tilde{G}_{i+1}) = 2\Hwind.
    \end{equation} 

    By~\eqref{eq_def of medA} and~\eqref{eq_def of medB} and since $i \in \medsetMinus{n}$, we have $i+1 \in \medsetPlus{n}$. 
    Recall from \Cref{rem_sets alt} that since $n \geq 1$ then $i$ is even and $i+1$ is odd. 
    By~\eqref{eq_fj iteration},~\eqref{eq_gj iteration}, and since conditioning reduces entropy, we have 
    \begin{IEEEeqnarray}{rCl} 
        H(\tilde{F}_{i+1} | \tilde{G}_{i+1}) &\leq& \min\{ H(\tilde{U}_{i'+1} | \tilde{Q}_{i'+1}), H(\tilde{V}_{i'+1} | \tilde{R}_{i'+1})\}
        \IEEEnonumber\\ 
        &=& \Hwind -\delta_{n-1}. \IEEEyesnumber \label{eq_HFiGi1 less than H0}
    \end{IEEEeqnarray}
    From~\eqref{eq_sum of HFiGi and HFiGi1} and~\eqref{eq_HFiGi1 less than H0}, we conclude that there must exist $\delta_n \geq \delta_{n-1}
    \geq 0$ such that $H(\tilde{F}_i | \tilde{G}_i) = \Hwind + \delta_{n}$ and $H(\tilde{F}_{i+1}
    | \tilde{G}_{i+1}) = \Hwind - \delta_n$.
    Finally, by \Cref{lem_all medial have the same entropy}, $\delta_n$ must be independent of $i$. 
\end{IEEEproof}

Recall that we wish to prove that the \OTBST is monopolarizing  for the \biprocess. 
From the proof of \Cref{cor_medial plus minus delta} it follows that $\delta_n \geq \delta_{n-1}$ for any $n$. 
This is not sufficient for monopolarization; to show monopolarization we must show that, unless we have already monopolarized, $\delta_n
> \delta_{n-1} + \Delta$ for some $\Delta > 0$ independent of $n$. 
This is the role of \Cref{lem_otbst monopolarizes helper} that follows. 
To this end, we will need an auxiliary lemma.  

The binary entropy function $h_2$, defined in~\eqref{eq_def of h2}, is monotone increasing over $[0,1/2]$. 
Denote the (cyclic) convolution of two numbers $0 \leq \alpha, \beta \leq 1/2$ by 
\begin{equation*} \label{eq_def of convolution}
    \alpha * \beta = \alpha(1-\beta) + \beta(1-\alpha). 
\end{equation*} 
Since 
\begin{equation} \label{eq_convolution is monotone} 
        \alpha*\beta = \alpha + \beta(1-2\alpha) = \beta+\alpha(1-2\beta),
\end{equation}
 we have $h_2(\alpha*\beta) \geq  h_2(\beta)$ for any $\alpha,\beta\in [0,1/2]$. 
More precisely, we have the following lemma;  its proof can be found in Appendix~\ref{ap_auxiliary proofs for otbst}. 

\begin{lemma} \label{lem_binary rvs}
    Let $0 \leq \alpha_a , \beta_b \leq 1/2$, $a,b=1,2,\ldots, k$ and let $p_a, q_b \geq 0$ such that $\sum_{a=1}^k p_a = \sum_{b=1}^k q_b = 1$. 
    If, for some $\xi_1, \xi_2 > 0$, 
    \begin{equation} \label{eq_max min entropy}
        \sum_{a=1}^k p_a h_2(\alpha_a) \geq \xi_1, \quad \sum_{b=1}^k q_b h_2(\beta_b) \leq \xi_2,
    \end{equation}
    then there exists $\Delta(\xi_1,\xi_2)> 0$ such that
    \[ 
        \sum_{a=1}^k \sum_{b=1}^k p_a q_b \left( h_2(\alpha_a*\beta_b) - h_2(\beta_b)\right) \geq \Delta(\xi_1,\xi_2). 
    \]
\end{lemma}

Recall that $i \in \medset{n}$, with base-vector $\bv{b} = \begin{bmatrix} \bv{b}_1 & \bv{b}_2 \end{bmatrix}$, 
where $\bv{b}_1$ and $\bv{b}_2$ are of length $2^{n-1}$.  
Assume further that $i \in \medsetMinus{n}$, so that $i$ is even, and $i' = i/2$. 
We define $\tilde{F}_i, \tilde{G}_i$ as in~\eqref{eq_FiGitilde}, and $\tilde{U}_{i'+1}, \tilde{V}_{i'}, \tilde{Q}_{i'+1}, \tilde{R}_{i'}$ as
in~\eqref{eq_UVQRtilde}.
\begin{lemma} \label{lem_otbst monopolarizes helper}
    For all $\xi>0$, if $i \in \medsetMinus{n}$ and 
    \begin{equation} \label{eq_HUQVR xi}
        H(\tilde{U}_{i'+1} | \tilde{Q}_{i'+1}), H(\tilde{V}_{i'}|\tilde{R}_{i'}) \in (\xi, 1-\xi)
   \end{equation} 
    then 
    \begin{equation*} \label{eq_entropy difference inequality theta}
        H(\tilde{F}_i|\tilde{G}_i) - \max\{H(\tilde{U}_{i'+1} | \tilde{Q}_{i'+1}), H(\tilde{V}_{i'} | \tilde{R}_{i'})\} \geq \Delta(\xi,1-\xi).
    \end{equation*} 
\end{lemma}

\begin{IEEEproof}
There is nothing to prove if $\xi \geq 1/2$. 
Therefore, we assume that $\xi < 1/2$. 
We show the proof for the case where $H(\tilde{V}_{i'}|\tilde{R}_{i'}) \geq H(\tilde{U}_{i'+1} | \tilde{Q}_{i'+1})$. 
The proof of the other case is similar and omitted.

We will use the simplified notation 
\[ 
    \tilde{p}(u,v,q,r) = \Prob{\tilde{U}_{i'+1}=u, \tilde{V}_{i'}=v, \tilde{Q}_{i'+1}=q, \tilde{R}_{i'}=r}. 
\] 
Since $(\tilde{U}_{i'+1}, \tilde{Q}_{i'+1})$ and $(\tilde{V}_{i'}, \tilde{R}_{i'})$ are independent,  
we have 
\[ 
    \tilde{p}(u,v,q,r) = \tilde{p}(u,q) \tilde{p}(v,r). 
\] 
We also introduce the shorthand 
\begin{align*}
    \alpha_{q}     &= \min_u \Prob{\tilde{U}_{i'+1} = u | \tilde{Q}_{i'+1} = q} = \min_u \tilde{p}(u|q), \\ 
    \beta_{r} &= \min_v \Prob{\tilde{V}_{i'} = v | \tilde{R}_{i'} = r} = \min_v \tilde{p}(v|r). 
\end{align*}
Recall that $\tilde{U}_{i'+1}, \tilde{V}_{i'}$ are binary, so the minimizations are between two terms. 
As a result, $0 \leq \alpha_{q}, \beta_{r} \leq 1/2$. 
With this notation and by~\eqref{eq_HUQVR xi} we have 
\begin{align*}
    H(\tilde{U}_{i'+1} |\tilde{Q}_{i'+1}) &= \sum_q \tilde{p}(q)h_2(\alpha_q) \geq \xi, \\ 
    H(\tilde{V}_{i'} |\tilde{R}_{i'}) &= \sum_r \tilde{p}(r)h_2(\beta_r) \leq 1-\xi.
\end{align*}
Thus, by~\eqref{eq_FiGi for i even} and the independence of $(\tilde{U}_{i'+1}, \tilde{Q}_{i'+1})$ and $(\tilde{V}_{i'}, \tilde{R}_{i'})$, we obtain 
\begin{align*}
    H(\tilde{F}_i | \tilde{G}_i) - H(\tilde{V}_{i'} | \tilde{R}_{i'})  
    & = H(\tilde{U}_{i'+1} + \tilde{V}_{i'} | \tilde{Q}_{i'+1}, \tilde{R}_{i'}) - H(\tilde{V}_{i'} | \tilde{R}_{i'}) \\ 
    &=\sol{\sum_{q,r}} \tilde{p}(q)\tilde{p}(r) \left( h_2(\alpha_{q} * \beta_{r}) - h_2(\beta_{r}) \right) \\ 
    &\geq \Delta(\xi,1-\xi),
\end{align*}
where the inequality is by \Cref{lem_binary rvs}. 
\end{IEEEproof}

We are now ready to show that the \OTBST is monopolarizing for the \biprocess. 
To this end, recall that $\Hwind$ was defined in~\eqref{eq_def of H0}. 
\begin{proposition} \label{prop_OTBST is monopolarizing}
    For every $\xi > 0$, there exists a threshold value $\nth \geq 0$ such that if $n \geq \nth$ then a level-$n$ \OTBST with any parameters $L_0, M_0$ is $(\xi,
    \medsetPlus{n}, \medsetMinus{n})$-monopolarizing for any \biprocess $\tilde{X}_{\star}\sarrow \tilde{Y}_{\star}$ with parameter $N_0 = 2L_0 + M_0$. 

    \looseness=-1
    Specifically, let $\tilde{F}_{1}^{N_{n}} \sarrow \tilde{G}_{1}^{N_{n}}$ be an \ottransformed \soblock of a level-$n$
    \OTBST initialized with $L_0$ and $M_0$ as above, where $n \geq \nth$. 
    Then:
    \begin{itemize}
        \item if $\Hwind \leq 1/2$ then
\begin{itemize}
	\item	$H(\tilde{F}_{i} | \tilde{G}_{i}) < \xi$,  \tabto{4.3cm} $\forall i \in \medsetPlus{n}$; 
	\item	$H(\tilde{F}_{i} | \tilde{G}_{i}) > 2\Hwind - \xi$,  \tabto{4.3cm} $\forall i \in \medsetMinus{n}$; 
\end{itemize}
        \item if $\Hwind \geq 1/2$ then 
\begin{itemize}
	\item	$H(\tilde{F}_{i} | \tilde{G}_{i}) > 1-\xi$,\tabto{4.3cm} $\forall i \in \medsetMinus{n}$; 
	\item	$H(\tilde{F}_{i} | \tilde{G}_{i}) <  2\Hwind - (1-\xi)$,\tabto{4.3cm} $\forall i \in \medsetPlus{n}$. 
\end{itemize}
    \end{itemize}
\end{proposition}

\begin{IEEEproof}
    We first consider a specific selection of the transform parameters $L_0$, $M_0$, and a specific \biprocess $\tilde{X}_{\star}\sarrow \tilde{Y}_{\star}$ with parameter $N_0 = 2L_0 + M_0$. For these selections we will find a threshold $\nth'$. We will then find a global upper bound $\nth$ on $\nth'$, which is independent of these selections.

    Denote the indicator functions 
    \begin{align*}
        \mathcal{M}_n^{-} &= \begin{cases} 1, &  H(\tilde{F}_i|\tilde{G}_i) > 1- \xi, \;\; \forall i \in \medsetMinus{n}, \\
                                            0, & \text{otherwise}, 
                            \end{cases} \\ 
        \mathcal{M}_n^{+} &= \begin{cases} 1, & H(\tilde{F}_i|\tilde{G}_i) <  \xi, \;\; \forall i \in \medsetPlus{n}, \\
                                           0, & \text{otherwise}.
                             \end{cases}
    \end{align*}
    Observe by \Cref{cor_medial plus minus delta} that if $\mathcal{M}_n^{-} = 1$ then also $H(\tilde{F}_{i} | \tilde{G}_{i}) < 2\Hwind - (1-\xi)$, and if $\mathcal{M}_n^{+}=1$ then also $H(\tilde{F}_{i} | \tilde{G}_{i}) > 2\Hwind - \xi$.

    Further denote
    \[
    \mathcal{M}_n = \begin{cases} 1, & \mathcal{M}_n^{-} = 1 \text{ or } \mathcal{M}_n^+ = 1, \\
                                  0, & \text{otherwise}. 
                     \end{cases}
    \]
    Observe that $\mathcal{M}_n = 1$ if and only if the \OTBST has $(\xi, \medsetPlus{n}, \medsetMinus{n})$-monopolarized for the \biprocess. Moreover, by \Cref{cor_medial plus minus delta}, $\mathcal{M}_n = 1$ if and only if the bulleted items in the claim hold. 

    We define
    \[ 
        \nth' = \min\left\{ n \in \mathbb{N} \ | \ \mathcal{M}_n = 1 \right\},
    \] 
    the first index $n$ for which $\mathcal{M}_n=1$. 
    We will show that $\nth'$ is finite by upper-bounding it. 

    By \Cref{cor_medial plus minus delta}, there exists a nondecreasing sequence $\delta_n\geq0$ such that~\eqref{eq_HFi given Gi deltan} holds. 
    Since $\delta_n$ is a nondecreasing sequence, $\mathcal{M}_n = 1$ for every $n \geq \nth'$. 
The entropy of a binary random variable is bounded between $0$ and $1$; thus for any $n$, $0 \leq \Hwind -\delta_n \leq \Hwind + \delta_n
\leq 1$.
Hence,  $\delta_n \leq \min \{\Hwind, 1-\Hwind\}$. 
We conclude that if $\Hwind \leq 1/2$ and $n \geq \nth'$ then $\mathcal{M}_n^{+}=1$, and if $\Hwind \leq 1/2$ and $n\geq \nth'$ then
$\mathcal{M}_n^{-}=1$.
It now remains to upper-bound $\nth'$. 

If $\mathcal{M}_0 = 1$, then we may take $\nth'=0$ and we are done. 
Otherwise, we assume that $\mathcal{M}_0 = 0$. 

    If, for some $n \geq 0$, $\mathcal{M}_n =0 $, then by~\eqref{eq_HFi given Gi deltan} and by definition of $\mathcal{M}_n^-$, $\mathcal{M}_n^+$, we
    obtain 
    \[ 
        \xi \leq \Hwind - \delta_{n} \leq \Hwind + \delta_{n} \leq 1-\xi.
    \] 
    Rearranging, this yields 
    \begin{equation} \label{eq_bound on delta n01}
        \mathcal{M}_n = 0 \Rightarrow \delta_{n} \leq  \min\{\Hwind, 1-\Hwind\} -\xi. 
    \end{equation} 

    On the other hand, by~\eqref{eq_HFi given Gi deltan} and \Cref{lem_otbst monopolarizes helper}, if $\mathcal{M}_{n-1}=0$ for some $n\geq 1$, we have 
    \[ 
        \Hwind + \delta_{n} - (\Hwind + \delta_{n-1}) \geq \Delta(\xi,1-\xi) \Rightarrow \delta_{n} \geq \delta_{n-1} + \Delta(\xi,1-\xi). 
    \] 
    Continuing in this manner and recalling that $\delta_0 =0$, we obtain
    \begin{equation}  \label{eq_bound on delta n02}
        \mathcal{M}_{n-1} = 0 \Rightarrow \delta_{n} \geq n \Delta(\xi,1-\xi). 
    \end{equation} 

    Now, let 
    \begin{equation}\label{eq_n1}
        n_1 =  1+ \left \lfloor \frac{\min\{\Hwind, 1-\Hwind\}-\xi}{\Delta(\xi,1-\xi)} \right \rfloor, 
    \end{equation} 
    and assume to the contrary that $\nth' > n_1$. 
    In particular, $\mathcal{M}_{n_1} = \mathcal{M}_{n_1-1} =0$. 
    Thus, by~\eqref{eq_bound on delta n01} and~\eqref{eq_bound on delta n02} we obtain 
    \[ 
        n_1 \Delta(\xi,1-\xi) \leq \delta_{n_1} \leq \min\{\Hwind,1-\Hwind\} - \xi. 
    \] 
    Since $\Delta(\xi,1-\xi) > 0$, we rearrange and obtain 
    \[ 
        n_1 \leq \frac{\min\{\Hwind, 1-\Hwind\}-\xi}{\Delta(\xi,1-\xi)},
    \] 
    which contradicts~\eqref{eq_n1} (see, e.g.,~\cite[Equation 3.3]{concreteMathematics}). 
    We conclude that we must have $\nth' \leq n_1$.

    We have found an upper bound for $\nth'$, which is given by the RHS of \eqref{eq_n1}. Note that this bound is indeed positive, thus it holds for both cases of $\mathcal{M}_0$ discussed above. Next, observe that
    \[
       1+ \left \lfloor \frac{\min\{\Hwind, 1-\Hwind\}-\xi}{\Delta(\xi,1-\xi)} \right \rfloor \leq  
       1+ \left \lfloor \frac{1/2-\xi}{\Delta(\xi,1-\xi)} \right \rfloor .
    \]
    Thus, defining 
    \begin{equation}
        \label{eq_nth}
     \nth = 1+ \left \lfloor \frac{1/2-\xi}{\Delta(\xi,1-\xi)} \right \rfloor 
 \end{equation}
    suffices.
\end{IEEEproof}

\begin{remark} \label{rem_n1 instead of nth}
    Note that if $\Hwind$ is given to us, then $\nth$ can be taken as the RHS of \eqref{eq_n1}.
\end{remark}

The following corollary is a restatement of \Cref{prop_OTBST is monopolarizing} that will be useful in the sequel for proving~\Cref{prop_cascade of OTBST is monopolarizing}. 
\begin{corollary}
	\label{cor_alpha beta prime prime}
	Let $\Hwind$, $H(\tilde{F}_{i} | \tilde{G}_{i})$, $\xi$ and $n \geq \nth$ be as in \Cref{prop_OTBST is monopolarizing}. Define 
	\begin{align*}
		\alpha(\Hwind) &= \min\{2\Hwind,1\}, \\
        \alpha'(\Hwind) &= \alpha(\Hwind)-\xi,  \\
		\beta(\Hwind) &= \max\{2\Hwind-1,0\}, \\
        \beta'(\Hwind) &= \beta(\Hwind)+\xi.
	\end{align*}
	Then,
    \begin{subnumcases}
        {H(\tilde{F}_{i} | \tilde{G}_{i}) \in}
	\left(\alpha'(\Hwind),\alpha(\Hwind) \right] & $i \in \medsetMinus{n}$ \label{eq_HFiGi in alpha beta minus} \\[0.2cm]
	\left[\beta(\Hwind),\beta'(\Hwind) \right) & $i \in \medsetPlus{n}.$ \label{eq_HFiGi in alpha beta plus}
		\end{subnumcases}
\end{corollary}
\begin{IEEEproof}
    Observe that $\alpha(\Hwind)+\beta(\Hwind) = 2\Hwind$, by considering separately the cases $\Hwind \leq 1/2$ and $\Hwind \geq 1/2$. Clearly, $\alpha'(\Hwind) + \beta'(\Hwind) = 2\Hwind$ as well. This implies, by~\eqref{eq_sum of HFiGi minus plus}, that if~\eqref{eq_HFiGi in alpha beta minus} holds for some $\Hwind$ then~\eqref{eq_HFiGi in alpha beta plus} also holds, and vice versa.

    Consider first the case $\Hwind \leq 1/2$ and let $i \in \medsetPlus{n}$. In this case
    \[ 
		\beta(\Hwind) = 0, \; \beta'(\Hwind) = \xi.
    \]
    By~\Cref{prop_OTBST is monopolarizing}, $H(\tilde{F}_i | \tilde{G}_i) \in [\beta(\Hwind),\beta'(\Hwind) )$. Hence, in this case~\eqref{eq_HFiGi in alpha beta plus} holds, and by the discussion above~\eqref{eq_HFiGi in alpha beta minus} must also hold for this case. 

    Next, consider the case $\Hwind \geq 1/2$ and let $i \in \medsetMinus{n}$. In this case
    \[ 
		\alpha(\Hwind)  = 1 , \; \alpha'(\Hwind) = 1 - \xi. 
    \]
    By~\Cref{prop_OTBST is monopolarizing}, $H(\tilde{F}_i | \tilde{G}_i) \in (\alpha'(\Hwind),\alpha(\Hwind) ]$. Hence, in this case~\eqref{eq_HFiGi in alpha beta minus} holds, and by the discussion above~\eqref{eq_HFiGi in alpha beta plus} must also hold for this case. 
\end{IEEEproof}

The following corollarly will be used in the proof of~\Cref{thm_BST is monopolarizing}.
\begin{corollary}\label{cor_OTBST monopolarizing}
    For a given $\xi>0$, let $L_0, M_0$, and $\nth$ be as in \Cref{prop_OTBST is monopolarizing}.
    Then, under the same setting as \Cref{prop_OTBST is monopolarizing}, for any $0 \leq \zeta \leq 1$ and $n \geq \nth$ we have
    \begin{itemize}
        \item if $\Hwind \leq \frac{1+\zeta}{2}$ then $H(\tilde{F}_{i} | \tilde{G}_{i}) < \xi+\zeta$,  \tabto{5.8cm} $\forall i \in
            \medsetPlus{n}$,\vphantom{$\displaystyle \sum_a$}
        \item if $\Hwind \geq \frac{1-\zeta}{2}$ then $H(\tilde{F}_{i} | \tilde{G}_{i}) > 1-\xi-\zeta$,\tabto{5.8cm} $\forall i \in \medsetMinus{n}$. 
    \end{itemize}
\end{corollary}
\begin{IEEEproof}
    This corollary follows from \Cref{prop_OTBST is monopolarizing} and \Cref{cor_medial plus minus delta}. 
    Recall that by \Cref{cor_medial plus minus delta}, there exists $\delta_n\geq0$ such that~\eqref{eq_HFi given Gi deltan} holds. 

    We only prove the corollary for the case where $\Hwind \leq (1+\zeta)/2$. 
    The case $\Hwind \geq (1-\zeta)/2$ is similar and omitted. 

    If $\Hwind \leq 1/2$, we are done by \Cref{prop_OTBST is monopolarizing}. 
    Otherwise, $\Hwind \geq 1/2$, so by \Cref{prop_OTBST is monopolarizing} and~\eqref{eq_HFi given Gi deltan},  
    \[ 
        i\in \medsetMinus{n} \Rightarrow H(\tilde{F}_i|\tilde{G}_i) =  \Hwind + \delta_n > 1-\xi.
    \] 
    Rearranging, we obtain $\delta_n > 1- \Hwind - \xi$. 
    Now, by~\eqref{eq_HFi given Gi deltan}, 
    \begin{align*}
        i\in \medsetPlus{n} \Rightarrow H(\tilde{F}_i|\tilde{G}_i) =  \Hwind - \delta_n &< \Hwind - (1- \Hwind -\xi) \\ 
        &= \xi + 2\Hwind - 1 \\ 
        &\leq \xi + (1+\zeta) - 1 \\ 
        &= \xi+\zeta,
    \end{align*}
    where the final inequality is due to our assumption that $\Hwind \leq (1+\zeta)/2$. 
\end{IEEEproof}

The upper bound for $\nth$ given in \Cref{prop_OTBST is monopolarizing} is pessimistic.
It is based on the \emph{minimal} change that must occur at every step of the \OTBST. 
The change at every \OTBST step is typically larger, and thus the actual required value of $\nth$ is expected to be much smaller.
We adapt \cite[Proposition 2]{sasoglu_2016_universal} to give better bounds on the required number of \OTBST steps to ensure monopolarization. 
To this end, we define, for $y \in [0,1]$ and $x \in[0,\min\{y,1-y\}]$, the functions 
\begin{align*}
    c(x,y) &= h_2(h_2^{-1}(y+x) * h_2^{-1}(y-x)) - y,  \\  
    d(x,y) &=  y - (y+x)(y-x),
\end{align*}
where $h_2^{-1}:[0,1] \to [0,1/2]$ is the inverse of $h_2$. 
Since $h_2$ is concave-$\cap$ and increasing over $[0,1/2]$, $h_2^{-1}$ is convex-$\cup$ and increasing over $[0,1]$. 
We also define the sequence of functions
\begin{align*}
    C_0(y) &= D_0(y) = 0, \\
    C_n(y) &= c(C_{n-1}(y),y), &n=1,2,\ldots,\\
    D_n(y) &= d(D_{n-1}(y),y), &n=1,2,\ldots.
\end{align*}     

\begin{lemma} \label{lem_WS iterations}
    Let $n \geq 0$. 
    If $i \in \medsetMinus{n}$ then 
    \[ 
        C_n(\Hwind) \leq H(\tilde{F}_i | \tilde{G}_i) - \Hwind \leq D_n(\Hwind).
    \]
    If $i \in \medsetPlus{n}$ then 
    \[ 
        C_n(\Hwind) \leq \Hwind - H(\tilde{F}_i | \tilde{G}_i) \leq D_n(\Hwind).
    \]
\end{lemma}
\begin{IEEEproof}               
    In light of \Cref{cor_medial plus minus delta}, denote, for any $n \geq 0$ and arbitrary $i \in \medsetMinus{n}$
    \[ 
        \delta_{n} = H(\tilde{F}_i | \tilde{G}_i) - \Hwind. 
    \] 
    Observe that for arbitrary $i \in \medsetPlus{n}$, by \Cref{cor_medial plus minus delta} we have $\delta_n =  \Hwind  - H(\tilde{F}_i | \tilde{G}_i)$. 
    Our goal is thus to show that for any $n\geq 0$, 
    \begin{equation}  \label{eq_Hn induction}
        C_n(\Hwind) \leq \delta_n \leq D_n(\Hwind).
    \end{equation} 

    The remainder of the proof mirrors the proof of \cite[Proposition 2]{sasoglu_2016_universal}. 
    We prove the claim by induction. 
    If $n = 0$, the claim is trivially true. 
    Assume that the claim holds for some $n \geq 0$, and we will show it also holds for $n+1$. 

    By block-independence of the \biprocess we may use \cite[Lemma 2.1]{sasoglu_thesis}, by which
    \begin{align*}
        \Hwind + \delta_{n+1} &\geq h_2(h_2^{-1}(\Hwind +\delta_n) * h_2^{-1}(\Hwind-\delta_n)),   \\
        \Hwind + \delta_{n+1} &\leq (\Hwind + \delta_n) + (\Hwind - \delta_n) - (\Hwind+\delta_n)(\Hwind-\delta_n). 
    \end{align*}
    Rearranging, we obtain 
    \begin{equation} \label{eq_cn deltan dn}
        c(\delta_n, \Hwind) \leq \delta_{n+1} \leq d(\delta_n, \Hwind). 
    \end{equation} 
    Now, $d(x,y) = x^2 -y^2 + y$ is increasing in $x$ whenever $x \geq 0$. 
    The function $c(x,y)$ is also increasing for $x \in [0, \min\{y,1-y\}]$. 
    To see this, it suffices to show that $c_y(x) = h_2^{-1}(y+x)*h_2^{-1}(y-x)$ is increasing, as $h_2$ is increasing. 
    Denoting $r(x) = h_2^{-1}(x)$ we obtain that 
    \begin{align*} 
        \frac{\dd c_y(x)}{\dd x} &= r'(y+x)(1-2r(y-x)) - r'(y-x)(1-2r(y+x)) \\ 
        &\eqann[\geq]{a} (r'(y+x) - r'(y-x)) (1-2r(y+x)) \\
        &\eqann[\geq]{b} 0,
    \end{align*} 
    where~\eqannref{a} is because $r(\cdot)$ is increasing, and~\eqannref{b} is because $r(\cdot)$ is convex so its derivative $r'(\cdot)$ is increasing and
    since $r(\cdot) \leq 1/2$ by definition. 
    Thus, by~\eqref{eq_cn deltan dn} and the induction hypothesis~\eqref{eq_Hn induction}, 
    \begin{IEEEeqnarray*}{rCcCl}
        \delta_{n+1} \geq c(\delta_n, \Hwind) &\geq& c(C_n(\Hwind), \Hwind) &=& C_{n+1}(\Hwind), \\ 
        \delta_{n+1} \leq d(\delta_n, \Hwind) &\leq& d(D_n(\Hwind), \Hwind) &=& D_{n+1}(\Hwind), 
    \end{IEEEeqnarray*}
    which completes the proof. 
\end{IEEEproof}

\begin{example}
    \label{ex_our bounds are weak}
    Consider a \biprocess with $\Hwind  = 0.2$. 
    We wish to find $\nth$ that will ensure that the \OTBST is $(0.004,\medsetPlus{n}, \medsetMinus{n})$-monopolarizing for
    the \biprocess whenever $n \geq \nth$. 

    \Cref{prop_OTBST is monopolarizing}, even when using the tighter bound in \Cref{rem_n1 instead of nth},  gives the upper bound 
    \[ 
        \nth \leq 1+ \left \lfloor \frac{\Hwind - \xi}{\Delta(\xi,1-\xi)} \right \rfloor = 40162. 
    \] 
    This is a prohibitive value. 
    Thankfully, it is also unnecessarily pessimistic. 
    To obtain a practical value for $\nth$, we turn to \Cref{lem_WS iterations}, by which 
    \begin{align*} 
        2.22\cdot 10^{-5}  &\leq H(\tilde{F}_i | \tilde{G}_i) \leq 0.0041, & i &\in \medsetPlus{9}, \\ 
        8.89\cdot 10^{-6} &\leq H(\tilde{F}_i | \tilde{G}_i) \leq 0.0031, & i &\in \medsetPlus{10}.
    \end{align*}
    Therefore, when $\Hwind = 0.2$, $\nth = 10$ suffices to ensure  $(0.004,\medsetPlus{n}, \medsetMinus{n})$-monopolarization for $n \geq \nth$. 
\end{example}

\subsection{Monopolarization for FAIM-derived Processes, for a Single BST} \label{sec_monopolarization for FAIM derived processes}
This subsection is devoted to proving \Cref{thm_BST is monopolarizing} below. That is, we show that the \BST is monopolarizing for suitably chosen $\eta$, $\mathcal{L}$, $\mathcal{H}$ when applied to a set of forgetful\footnote{An interesting open problem is whether the forgetfulness condition is necessary.}  FAIM-derived \soprocesses. The theorem holds for a set of \soprocesses that satisfies a set of rather lax conditions. The conditions are defined in terms of the $L$-forgetfulness of the \soprocess, $\epsilon_L$, see \Cref{def_memlength}; and the mixing parameters $\psi_M$ and $\phi_M$ from \Cref{lem_FAIM is psi mixing}.

Recall that if a \emph{specific} FAIM-derived \soprocess satisfies \Cref{cond_kaijser}, it is forgetful (see \Cref{ex_memory length for SXY}). Thus, by \Cref{def_memlength}, we have that $\epsilon_L \xrightarrow[L \to \infty]{} 0\vspace{0.1cm}$. Furthermore, for this FAIM-derived \soprocess, by \Cref{lem_FAIM is psi mixing}, $\psi_M \xrightarrow[M \to \infty]{} 1$, and $\phi_M \xrightarrow[M \to \infty]{} 1\vspace{0.1cm}$, where  $\psi_M$ is nonincreasing and $\phi_M$ is nondecreasing. Our conditions for the set of forgetful FAIM-derived \soprocesses are that there exist sequences $\bar{\epsilon}_L$, $\bar{\psi}_M$, and $\bar{\phi}_M$ such that for \emph{any} \soprocess in the set we have
\begin{IEEEeqnarray}{rCcCl} \IEEEyesnumber \label{eq_conditions for universal set of processes}
	\epsilon_L & \leq &\bar{\epsilon}_L & \xrightarrow[L \to \infty]{} & 0, \IEEEyessubnumber \label{eq_conditions for universal set of processes epsilon} \\[0.1cm]
	\psi_M & \leq &\bar{\psi}_M & \xrightarrow[M \to \infty]{} & 1, \IEEEyessubnumber \label{eq_conditions for universal set of processes psi} \\[0.1cm]
	\phi_M & \geq &\bar{\phi}_M & \xrightarrow[M \to \infty]{} & 1, \IEEEyessubnumber \label{eq_conditions for universal set of processes phi} \\[0.1cm]
	\IEEEeqnarraymulticol{5}{c}{\bar{\psi}_{M} \ \mathrm{ nonincreasing}, \quad \bar{\phi}_M \ \mathrm{ nondecreasing}.} \IEEEyessubnumber \label{eq_conditions for universal set of processes monotone}
\end{IEEEeqnarray}
Clearly, for a given finite set $\mathcal{C}$ of forgetful FAIM-derived \soprocesses, conditions~\eqref{eq_conditions for universal set of processes} are easy to satisfy. Namely, for any $L$ and $M$, take $\bar{\epsilon}_L = \max_{c \in \mathcal{C}} \epsilon_L(c)$, $\bar{\psi}_M = \max_{c \in \mathcal{C}} \psi_M(c)$, and $\bar{\phi}_M = \min_{c \in \mathcal{C}} \phi_M(c)$. For an infinite set $\mathcal{C}$, we may replace $\max$ and $\min$ above by $\sup$ and $\inf$, respectively.

\begin{theorem} \label{thm_BST is monopolarizing}
Fix sequences $\bar{\epsilon}_L$, $\bar{\psi}_M$, and $\bar{\phi}_M$ that satisfy the limits in \eqref{eq_conditions for universal set of processes epsilon}--\eqref{eq_conditions for universal set of processes phi}, as well as the conditions in~\eqref{eq_conditions for universal set of processes monotone}.
    Let $X_{\star} \sarrow Y_{\star}$ be a forgetful FAIM-derived \soprocess that satisfies the inequalities in~\eqref{eq_conditions for universal set of processes epsilon}--\eqref{eq_conditions for universal set of processes phi}.
    For every $\eta >0$ there exist $\Lth$, $\Mth$, and $\nth$, \emph{independent of the process},  such that if $L_0\geq \Lth$, $M_0\geq \Mth$, and $n \geq \nth$ then
    a level-$n$ \BST initialized with parameters $L_0$ and $M_0$ is $(\eta, \medsetPlus{n},
    \medsetMinus{n})$-monopolarizing. 

    Specifically, let $F_{1}^{N_{n}} \sarrow G_{1}^{N_{n}}$ be a transformed \soblock of a level-$n$
    \BST initialized with $L_0$ and $M_0$ as above.
    Then: 
    \begin{itemize}
        \item if $\ENT{X_{\star} | Y_{\star}}\leq 1/2$ then $H(F_{i} | G_{i})<\eta$, \hphantom{$1;$}
            $\forall i \in
    \medsetPlus{n}$; 
        \item if $\ENT{X_{\star} | Y_{\star}}\geq 1/2$ then $H(F_{i} | G_{i})>1-\eta$, $\forall i \in \medsetMinus{n}$. 
    \end{itemize}
\end{theorem}

This theorem will follow as a corollary to \Cref{prop_tilde non tilde} below. 
We will show in \Cref{prop_tilde non tilde} that, when $\Lth$ and $\Mth$ are suitably chosen, there is a close relationship between the \BST of a
forgetful FAIM-derived \soprocess and the \OTBST of a \biprocess. 
Since, by \Cref{prop_OTBST is monopolarizing}, the \OTBST of a \biprocess is monopolarizing, this will imply that the \BST is also monopolarizing.


The parameters $\Lth$, $\Mth$, and $\nth$ for given sequences $\bar{\epsilon}_L$, $\bar{\psi}_M$, and $\bar{\phi}_M$ are determined in the proof of \Cref{thm_BST is monopolarizing}. 
For future reference, they are detailed in the following remark.

\begin{remark}
    \label{rem_bst parameters for monopolarization}
    We are given $\eta > 0$, and our goal is to set the parameters $\nth$, $\Lth$, and $\Mth$. We do this via the following steps.
    \begin{enumerate}
        \item Fix $\varepsilon_1 < \eta/12$ and $\varepsilon_2 < \eta^2/32$, in line with \eqref{eq_varepsilon12} below.
        \item Let $\xi$ be as in \eqref{eq_choice of xi} below.
        \item Set $\nth$ using this $\xi$ in \eqref{eq_nth}.
        \item Take $\epsilon = \varepsilon_1 2^{-\nth}$, in line with \eqref{eq_epsilon as a function of varepsilon_1}.
        \item Find $\Lth$ such that \eqref{eq_memory length inequality} holds with this $\epsilon$, and $\Lth$ in place of $k$.
        \item Find $\Mth$ such that \eqref{eq_varepsilon2 and psiphi} holds with $\Mth$ in place of $M$, and $\nth$ in place of $n$.
    \end{enumerate}
These steps will follow from the proof of \Cref{thm_BST is monopolarizing}.
\end{remark}

Recall our notation from \Cref{sec_otbst} for the \BST and \OTBST. 
We will only consider medial indices. 
The \BST is expressed using the sequence of functions $f_{n,i}$, $g_{n,i}$, where $i\in \medset{n}$. 
The \OTBST is expressed using the sequence of functions $\tilde{f}_{n,i}$, $\tilde{g}_{n,i}$. 

Let $i \in \medset{n}$; its base-vector $\bv{b}$ is given by
\[ 
    \bv{b} = \begin{bmatrix} b_1 & b_2 & \cdots & b_{2^n} \end{bmatrix},
\] 
We also denote
\begin{align*}
    \bv{a} &= \begin{bmatrix} 1 & N_0+1 & 2N_0 +1 & \cdots & (2^{n}-1)N_0 + 1\end{bmatrix}, \\ 
    \bv{z} &= \begin{bmatrix} N_0  & 2N_0 & 3N_0 & \cdots & 2^{n}N_0\end{bmatrix}. 
\end{align*}
We further define for index $i \in \medset{n}$: 
\begin{IEEEeqnarray}{rCl?rCl} \IEEEyesnumber \label{eq_FiGiFiGiFiGi}
    F_i &=& f_{n,i}(X_{\bv{b}}), &G_i &=& g_{n,i}(X_{\bv{a}}^{\bv{b}}, Y_{\bv{a}}^{\bv{z}}),  \IEEEyessubnumber \label{eq_FiGiFiGiFiGi1} \\ 
    \at{F}_i &=& \tilde{f}_{n,i}(X_{\bv{b}}),  &\at{G}_i &=& \tilde{g}_{n,i}(X_{\bv{b}-L_0}^{\bv{b}}, Y_{\bv{b}-L_0}^{\bv{b}+L_0}),
    \IEEEyessubnumber \label{eq_FiGiFiGiFiGi2}  \\ 
    \tilde{F}_i &=& \tilde{f}_{n,i}(\tilde{X}_{\bv{b}}),  &\tilde{G}_i &=& \tilde{g}_{n,i}(\tilde{X}_{\bv{b}-L_0}^{\bv{b}},
    \tilde{Y}_{\bv{\bv{b}}-L_0}^{\bv{b}+L_0}).  \IEEEyessubnumber \label{eq_FiGiFiGiFiGi3}  
    \end{IEEEeqnarray}
    In words:
    \begin{itemize} 
        \item $F_i \sarrow G_i$ is a transformed \sopair obtained after applying a level-$n$ \BST to the FAIM-derived process; 
        \item $\at{F}_i \sarrow \at{G}_i$ is an \ottransformed \sopair obtained after applying a level-$n$ \OTBST to the FAIM-derived process; 
        \item  $\tilde{F}_i \sarrow \tilde{G}_i$ is an \ottransformed \sopair obtained after applying a level-$n$ \OTBST to the \biprocess.  
    \end{itemize}

    The following proposition, as well as \Cref{lem_lean bblock,lem_independent bblock}, are stated for a forgetful FAIM-derived \soprocess, $X_{\star} \sarrow Y_{\star}$, that satisfies
    \eqref{eq_conditions for universal set of processes} for some sequences $\bar{\epsilon}_L$, $\bar{\psi}_M$, and $\bar{\phi}_M$.
\begin{proposition}\label{prop_tilde non tilde}
    Fix $n \geq 0$, $\varepsilon_1>0$, and $0 < \varepsilon_2 < \frac{1}{6}$. 
    There exist $\Lth$ and $\Mth$ such that 
    a level-$n$ \BST initialized with parameters $L_0 \geq \Lth, M_0 \geq \Mth$ satisfies: 
    \begin{IEEEeqnarray}{rCl} \IEEEyesnumber \label{eq_HFiGi tilde inequality}
    |H(F_{i}|G_{i}) - H(\tilde{F}_{i} | \tilde{G}_{i})| &\leq& 2 \varepsilon_1 + \frac{\varepsilon_2}{2} - 3 \varepsilon_2 \log{\frac{3\varepsilon_2}{2}} \IEEEeqnarraynumspace  \IEEEyessubnumber \label{eq_HFiGi tilde inequality a}\\  &<& 2\varepsilon_1 + \sqrt{8\varepsilon_2}.\IEEEyessubnumber  \label{eq_HFiGi tilde inequality b}                                                                                                               \end{IEEEeqnarray}

    Furthermore, we have
    \begin{equation} \label{eq_HFiGiS0Sn tilde inequality}
	    |H(F_{i}|G_{i}, S_0, S_{N_n}) - H(\tilde{F}_{i} | \tilde{G}_{i})| \leq 2 \varepsilon_1 + \frac{\varepsilon_2}{2} - 3 \varepsilon_2 \log{\frac{3\varepsilon_2}{2}}. 
    \end{equation}

\end{proposition}

\begin{IEEEproof}
    Denote
    \begin{align*}
        \at{P} &= P_{X_{\bv{b}-L_0}^{\bv{b}}, Y_{\bv{b}-L_0}^{\bv{b}+L_0}}\,\, ,  \\ 
        \tilde{P} &= \prod_{\ell=1}^{2^n} P_{X_{b_\ell - L_0}^{b_{\ell}}, Y_{b_{\ell}-L_0}^{b_{\ell}+L_0} }\, . 
    \end{align*}
    Then, $(X_{\bv{b}-L_0}^{\bv{b}}, Y_{\bv{b}-L_0}^{\bv{b}+L_0})$ is distributed according to $\at{P}$ and $(\tilde{X}_{\bv{b}-L_0}^{\bv{b}},
    \tilde{Y}_{\bv{b}-L_0}^{\bv{b}+L_0})$ is distributed according to $\tilde{P}$. 

In \Cref{lem_lean bblock} that follows we show that there exists $\Lth$ such that if $L_0 \geq \Lth$ then
\begin{equation}  \label{eq_HFiGi FiGiat inequality}
    |H(F_{i}|G_{i}) - H(\at{F}_{i} | \at{G}_{i})| \leq 2 \varepsilon_1. 
\end{equation} 
Next, in \Cref{lem_independent bblock} that follows we show that there exists $\Mth$ such that if $M_0 \geq \Mth$ then  
\[ 
    (1-\varepsilon_2) \tilde{P} \leq \at{P} \leq (1+ \varepsilon_2) \tilde{P}.
\] 
This will enable us to use \Cref{lem_entropy difference if distributions close} below with $f = \tilde{f}_{n,i}$ and $g = \tilde{g}_{n,i}$ to obtain
\[ 
    |H(\at{F}_{i} | \at{G}_{i}) - H(\tilde{F}_{i} | \tilde{G}_{i}) | \leq \frac{\varepsilon_2}{2} - 3 \varepsilon_2 \log{\frac{3\varepsilon_2}{2}}  < \sqrt{8\varepsilon_2}. 
\] 
Hence, we conclude that 
\begin{align*}
    &|H(F_{i} | G_{i}) - H(\tilde{F}_{i} | \tilde{G}_{i})|\\ 
    &\quad \leq |H(F_i|G_i) - H(\at{F}_i | \at{G}_i)| + |H(\at{F}_i|\at{G}_i) - H(\tilde{F}_i
    | \tilde{G}_i) | \\ 
    &\quad \leq 2\varepsilon_1 +  \frac{\varepsilon_2}{2} - 3 \varepsilon_2 \log{\frac{3\varepsilon_2}{2}} \\ 
    &\quad < 2\varepsilon_1 +  \sqrt{8\varepsilon_2}.
\end{align*}
This proves \eqref{eq_HFiGi tilde inequality}.

Finally, to show~\eqref{eq_HFiGiS0Sn tilde inequality}, by \eqref{eq_entropy UQ with and without tildes S0SN} in \Cref{lem_lean bblock} we have
\[
	|H(F_{i}|G_{i},S_0,S_{N_n}) - H(\at{F}_{i} | \at{G}_{i})| \leq 2 \varepsilon_1. 
\]
We insert the above in place of \eqref{eq_HFiGi FiGiat inequality}, and by the same arguments as before,
\begin{align*}
    &|H(F_{i} | G_{i}, S_0, S_{N_n}) - H(\tilde{F}_{i} | \tilde{G}_{i})|\\ 
    &\quad \leq |H(F_i|G_i, S_0, S_{N_n}) - H(\at{F}_i | \at{G}_i)| + |H(\at{F}_i|\at{G}_i) - H(\tilde{F}_i
    | \tilde{G}_i) | \\ 
    &\quad \leq 2\varepsilon_1 +  \frac{\varepsilon_2}{2} - 3 \varepsilon_2 \log{\frac{3\varepsilon_2}{2}}.
\end{align*}
    This completes the proof.
\end{IEEEproof}
In the sequel, we will refer to the right-hand side of~\eqref{eq_HFiGi tilde inequality a}. To this end, for $\varepsilon_1>0$ and $0<\varepsilon_2 < \frac{1}{6}$, we denote
\begin{equation}\label{eq_definition of varepsilon_3}
    \varepsilon_3 = \varepsilon_3(\varepsilon_1,\varepsilon_2) \triangleq 2 \varepsilon_1 + \frac{\varepsilon_2}{2} - 3 \varepsilon_2 \log{\frac{3\varepsilon_2}{2}}.   
\end{equation}

We now state and prove \Cref{lem_lean bblock,lem_independent bblock,lem_entropy difference if distributions close}. 

For the lemma below, recall that a BST is initialized with parameters $L_0$ and $M_0$. This lemma is concerned with $L_0$ and applies for any $M_0$. 
\begin{lemma}  \label{lem_lean bblock}
    Fix $n \geq 0$ and $\varepsilon_1 > 0$. 
    There exists $\Lth$ such that if $L_0 \geq \Lth$ then for all $M_0$,
    \begin{equation}  \label{eq_entropy UQ with and without tildes}
        0 \leq  H(\at{F}_i | \at{G}_i) - H(F_i | G_i) \leq  2\varepsilon_1
   \end{equation} 
   and
    \begin{equation}  \label{eq_entropy UQ with and without tildes S0SN}
	    0 \leq  H(\at{F}_i | \at{G}_i) - H(F_i | G_i, S_0, S_{N_n}) \leq  2\varepsilon_1.
   \end{equation} 
\end{lemma}


\begin{IEEEproof}
    By~\eqref{eq_Gj splitting}, $G_i \equiv (\at{G}_i, \ut{G}_i)$, 
    where 
    \begin{align} 
        \at{G}_i &= \tilde{g}_{n,i}(X_{\bv{b}-L_0}^{\bv{b}},Y_{\bv{b}-L_0}^{\bv{b}+L_0}),   \notag \\
        \ut{G}_i &= (X_{\bv{a}}^{\bv{b}-L_0-1},Y_{\bv{a}}^{\bv{b}-L_0-1}, Y_{\bv{b}+L_0+1}^{\bv{z}}). \label{eq_utGidef}
    \end{align} 
    Since $f_{n,i} = \tilde{f}_{n,i}$, we have $F_i = \at{F}_i$. 
    Therefore, 
    \begin{equation} \label{eq_HFiGi to HFiatGi} 
        H(F_i | G_i) = H(\at{F}_i | \at{G}_i, \ut{G}_i) \leq H(\at{F}_i | \at{G}_i), 
    \end{equation} 
    where the inequality is because conditioning reduces entropy. 
    This proves the left-hand side of~\eqref{eq_entropy UQ with and without tildes}. 

    We now turn to proving the right-hand side of~\eqref{eq_entropy UQ with and without tildes}.
    Utilizing  the left-hand side of \eqref{eq_conditions for universal set of processes epsilon} and \Cref{cor_forgetting multi blocks} with $\bv{i} = \bv{b}$, $\memlength = \Lth$,  and $L_0 \geq \Lth$, we obtain
    \begin{equation} 
	    \label{eq_mutual information between states in a window}
        I(S_{\bv{b}} ; S_{\bv{b}-L_0}, S_{\bv{b}+L_0} \  | \ X_{\bv{b}-L_0}^{\bv{b}-1}, Y_{\bv{b}-L_0}^{\bv{b}+L_0}) \leq 2^n\cdot 2\bar{\epsilon}_L.
\end{equation} 
    Using the right-hand side of \eqref{eq_conditions for universal set of processes epsilon}, we take $\Lth$ large enough so that
    \begin{equation}
        \label{eq_epsilon as a function of varepsilon_1}
        \bar{\epsilon}_{\Lth} \leq \varepsilon_1 \cdot 2^{-n} . 
    \end{equation}
    Hence, 
    \begin{IEEEeqnarray}{rCl} \IEEEnonumber
	    2\varepsilon_1 &\geq& 
        I(S_{\bv{b}} ; S_{\bv{b}-L_0}, S_{\bv{b}+L_0} \ | \ X_{\bv{b}-L_0}^{\bv{b}-1}, Y_{\bv{b}-L_0}^{\bv{b}+L_0}) \IEEEnonumber
\\ 
			   &\eqann[\geq]{a}& I(\at{F}_i, \at{G}_i ; S_{\bv{b}-L_0}, S_{\bv{b}+L_0} \ |\  X_{\bv{b}-L_0}^{\bv{b}-1}, Y_{\bv{b}-L_0}^{\bv{b}+L_0}) \IEEEnonumber
\\ 
			   &\eqann[\geq]{b}& I(\at{F}_i; S_{\bv{b}-L_0}, S_{\bv{b}+L_0}\ | \  \at{G}_i, X_{\bv{b}-L_0}^{\bv{b}-1}, Y_{\bv{b}-L_0}^{\bv{b}+L_0}) \IEEEnonumber
\\ 
			   &\eqann{c}& I(\at{F}_i; S_{\bv{b}-L_0}, S_{\bv{b}+L_0}\ | \  \at{G}_i) \IEEEnonumber
\\ 
			   &=& H(\at{F}_i | \at{G}_i) - H(\at{F}_i | \at{G}_i, S_{\bv{b}-L_0}, S_{\bv{b}+L_0}) \IEEEnonumber
\\ 
			   &\eqann{d}& H(\at{F}_i | \at{G}_i) - H(\at{F}_i | \at{G}_i, \ut{G}_i, S_{\bv{b}-L_0}, S_{\bv{b}+L_0}) \IEEEnonumber
\\ 
			   &\eqann{e}& H(\at{F}_i | \at{G}_i) - H(F_i | G_i, S_{\bv{b}-L_0}, S_{\bv{b}+L_0} )  \IEEEyesnumber \label{eq_2varepsilon inequality}\\ 
			   &\eqann[\geq]{f}& H(\at{F}_i | \at{G}_i) - H(F_i | G_i), \IEEEnonumber
\end{IEEEeqnarray}	
    where:
    \begin{itemize}
        \item \eqannref{a} is due to~\eqref{eq_DPI}. 
            By~\eqref{eq_markov property of FAIM}, $X_{\bv{b}}$ is a probabilistic function of $S_{\bv{b}}$; by \eqref{eq_FiGiFiGiFiGi2},
            $\at{F}_i$ is a function of $X_{\bv{b}}$, and $\at{G}_i$ is a function of $(X_{\bv{b}-L_0}^{\bv{b}}, Y_{\bv{b}-L_0}^{\bv{b}+L_0})$.
            Thus, we have the Markov chain 
            \begin{align*} 
                (S_{\bv{b}-L_0}, S_{\bv{b}+L_0}) &\markov (S_{\bv{b}}, X_{\bv{b}-L_0}^{\bv{b}-1}, Y_{\bv{b}-L_0}^{\bv{b}+L_0})\\ 
                & \markov (X_{\bv{b}}, X_{\bv{b}-L_0}^{\bv{b}-1}, Y_{\bv{b}-L_0}^{\bv{b}+L_0}) \markov (\at{F}_i, \at{G}_{i}).
            \end{align*}
            Specifically, we have the Markov chain 
            \[ 
                (S_{\bv{b}-L_0}, S_{\bv{b}+L_0}) \markov (S_{\bv{b}}, X_{\bv{b}-L_0}^{\bv{b}-1}, Y_{\bv{b}-L_0}^{\bv{b}+L_0}) 
                \markov (\at{F}_i, \at{G}_{i}),
            \]
            for which we use~\eqref{eq_DPI}. 
        \item \eqannref{b} is by the chain rule.
        \item \eqannref{c} is since $\at{G}_i \equiv (\at{G}_i, X_{\bv{b}-L_0}^{\bv{b}-1}, Y_{\bv{b}-L_0}^{\bv{b}+L_0})$, which holds due
            to~\eqref{eq_gjequiv} and~\eqref{eq_FiGiFiGiFiGi2}. 
        \item \eqannref{d} is by the Markov property~\eqref{eq_markov property of FAIM},~\eqref{eq_FiGiFiGiFiGi2}, and~\eqref{eq_utGidef}: $\at{F}_i$
            and $\at{G}_i$ are probabilistic functions of states $S_{\bv{b}-L_0}^{\bv{b}+L_0}$, whereas 
            $\ut{G}_i$ is a probabilistic function of states $S_{\bv{a}}^{\bv{b}-L_0-1}$ and $S_{\bv{b}+L_0+1}^{\bv{z}}$. 
        \item \eqannref{e} is because $\at{F}_i = F_i$ and because $G_i \equiv (\at{G}_i, \ut{G}_i)$ by~\eqref{eq_Gj splitting}. 
        \item \eqannref{f} is because conditioning reduces entropy. 
    \end{itemize}
    This shows~\eqref{eq_entropy UQ with and without tildes}.

    Finally, to show~\eqref{eq_entropy UQ with and without tildes S0SN} we use the same steps. Namely, we obtain the left-hand side of~\eqref{eq_entropy UQ with and without tildes S0SN} by replacing~\eqref{eq_HFiGi to HFiatGi} with
    \[ 
	    H(F_i | G_i, S_0, S_{N_n}) = H(\at{F}_i | \at{G}_i, \ut{G}_i, S_0, S_{N_n}) \leq H(\at{F}_i | \at{G}_i).
\] 
For the right-hand side of~\eqref{eq_entropy UQ with and without tildes S0SN}, observe that by the penultimate step of~\eqref{eq_2varepsilon inequality} we have 
\begin{align*}
	2\varepsilon_1 &\geq H(\at{F}_i | \at{G}_i) - H(F_i | G_i, S_{\bv{b}-L_0}, S_{\bv{b}+L_0} ) \\ 
		       &= H(\at{F}_i | \at{G}_i) - H(F_i | G_i, S_{\bv{b}-L_0}, S_{\bv{b}+L_0}, S_0, S_{N_n} ) \\ 
		       &\geq H(\at{F}_i | \at{G}_i) - H(F_i | G_i, S_0, S_{N_n} ).
\end{align*}
The first equality is by Markov property \eqref{eq_markov property of FAIM}. 
The final inequality is because conditioning reduces entropy. 

Lastly, note that there are no restrictions on $M_0$ throughout the proof --- its only role is setting the parameters of the BST --- and thus the claim holds for any $M_0$.
\end{IEEEproof}

For the lemma below, again recall that a BST is initialized with parameters $L_0$ and $M_0$. This lemma is concerned with $M_0$ and applies for any $L_0$. 
\begin{lemma} \label{lem_independent bblock}
    Fix $n \geq 0$, $\varepsilon_2 > 0$, and $L_0$.
    There exists $\Mth$ such that if 
    $M_0 \geq \Mth$ then 
    \begin{IEEEeqnarray}{rCl} \IEEEyesnumber \label{eq_probability inequality}
        P_{X_{\bv{b}-L_0}^{\bv{b}}, Y_{\bv{b}-L_0}^{\bv{b}+L_0}} &\leq& (1+\varepsilon_2)
        \prod_{\ell=1}^{2^n} P_{X_{b_{\ell}-L_0}^{b_{\ell}}, Y_{b_{\ell}-L_0}^{b_{\ell+L_0}}}, \IEEEyessubnumber \label{eq_probability inequality A}\\
        P_{X_{\bv{b}-L_0}^{\bv{b}}, Y_{\bv{b}-L_0}^{\bv{b}+L_0}} &\geq& (1-\varepsilon_2)
        \prod_{\ell=1}^{2^n} P_{X_{b_{\ell}-L_0}^{b_{\ell}}, Y_{b_{\ell}-L_0}^{b_{\ell+L_0}}}. \IEEEyessubnumber \label{eq_probability inequality B}
    \end{IEEEeqnarray}
\end{lemma}

\begin{IEEEproof}
	By the right-hand sides of~\eqref{eq_conditions for universal set of processes psi} and~\eqref{eq_conditions for universal set of processes phi},
    we may choose $\Mth$ such that 
    \begin{IEEEeqnarray}{rCl}         \IEEEyesnumber \label{eq_varepsilon2 and psiphi}
        (\bar{\psi}_{\Mth-2})^{(2^n)} &\leq& 1 + \varepsilon_2, \IEEEyessubnumber\\ 
        (\bar{\phi}_{\Mth-2})^{(2^n)} &\geq& 1 - \varepsilon_2. \IEEEyessubnumber 
    \end{IEEEeqnarray} 
    For any $M_0 \geq \Mth$, by~\eqref{eq_conditions for universal set of processes monotone},  we thus have
    \begin{IEEEeqnarray}{rCcCl} \IEEEyesnumber \label{eq_psiphiM02n bound}
        (\bar{\psi}_{M_0-2})^{(2^n)} &\leq& (\bar{\psi}_{\Mth-2})^{(2^n)} &\leq& 1+\varepsilon_2,\IEEEyessubnumber  \label{eq_psiM02n bound} \\ 
        (\bar{\phi}_{M_0-2})^{(2^n)} &\geq& (\bar{\phi}_{\Mth-2})^{(2^n)} &\geq& 1-\varepsilon_2.\IEEEyessubnumber  \label{eq_phiM02n bound}
    \end{IEEEeqnarray}

    Denote by $\bar{\bv{b}} = \begin{bmatrix} \bar{b}_1 & \bar{b}_2 & \cdots & \bar{b}_{2^n} \end{bmatrix}$ the modulo-base-vector of $i$. 
By \Cref{cor_num of elements in detvec}, for any $1 \leq \ell < 2^n$ we have $1 \leq |\bar{b}_{\ell+1} - \bar{b}_{\ell}| \leq 2$. 
    Hence, by~\eqref{eq_modulo determining}, and recalling that $N_0 =  2L_0 + M_0$, 
    \begin{align} 
        (b_{\ell+1} - L_0) - (b_{\ell}+L_0) &= \ell N_0 - (\ell-1)N_0 -2L_0 +\bar{b}_{\ell+1} - \bar{b}_{\ell} \nonumber \\ 
        &= M_0 + (\bar{b}_{\ell+1} - \bar{b}_{\ell})  \nonumber\\
        &\geq M_0- |\bar{b}_{\ell+1} - \bar{b}_{\ell}|  \nonumber\\
        &\geq M_0 - 2. \label{eq_dlp1 dl diff} 
    \end{align} 
    
    The vector $X_{\bv{b}-L_0}^{\bv{b}}, Y_{\bv{b}-L_0}^{\bv{b}+L_0}$ contains symbols with indices in $\mathcal{B} = \cup_{\ell} \mathcal{B}_{\ell}$,
    where $\mathcal{B}_{\ell} = \{b_{\ell}-L_0, b_{\ell}-L_0 + 1, \ldots, b_{\ell}+L_0 \}$, $1 \leq \ell \leq 2^n$.
    Each set $\mathcal{B}_{\ell}$ is a contiguous subsequence of $\mathcal{B}$. 
    The greatest index in $\mathcal{B}_{\ell}$ is $b_{\ell}+L_0$ and the smallest index in $\mathcal{B}_{\ell+1}$ is $b_{\ell+1}-L_0$; 
    see \Cref{fig_new process} for an illustration. 
    By~\eqref{eq_dlp1 dl diff}, any two consecutive sets $\mathcal{B}_{\ell}$ and $\mathcal{B}_{\ell+1}$ are separated by at least $M_0
    -2$ indices.

\begin{figure}[t]
    \begin{center}
            \begin{tikzpicture}[>=latex]
                \node[rectangle, minimum height = 0.25cm, minimum width = 1.2cm, fill = red!15!white] (R1) at (1.1,0) {};  
                \node[rectangle, minimum height = 0.25cm, minimum width = 1.2cm, fill = red!15!white] (R2) at (3.2,0) {};  
                \node[rectangle, minimum height = 0.25cm, minimum width = 1.2cm, fill = red!15!white] (R3) at (5.3,0) {};  
                \node[rectangle, minimum height = 0.25cm, minimum width = 1.2cm, fill = red!15!white] (R4) at (7.4,0) {};  

                \draw[|-|] (0,0) node[above=2pt]{\tiny $1$}  -- (8.4,0) node[above=2pt] {\tiny $N_n$}; 

                \foreach \x in {2.15,4.25,6.35}
                {
                    \draw[gray] (\x,0.5) -- (\x,-0.9); 
                }

                \foreach \x in {1,2,3,4}
                {
                    \fill[pattern=north east lines, pattern color = green!50!black] (R\x.north) rectangle (R\x.south west); 
                    \draw[blue, thick] (R\x.90) node[below=0.7cm,black](D\x){\footnotesize$\ell=\x$} node[above] {\tiny $b_{\x}$} -- (R\x.270); 
                    \draw[red] (R\x.south west) node[below](B1\x){\tiny $b_{\x}\!\!-\!\!L_0$} -- (R\x.north west) node(B1t\x){} ; 
                    \draw[red] (R\x.south east) node[below](B2\x){\tiny $b_{\x}\!\!+\!\!L_0$} -- (R\x.north east) node(B2t\x){}; 
                }

                \draw[dotted] ($(B1t2.center)+(0,0.5)$) -- (B1t2.center); 
                \draw[dotted] ($(B2t2.center)+(0,0.5)$) -- (B2t2.center); 
                \draw[decorate, decoration={brace}, black] ($(B1t2.center)+(0,0.5)$) -- node[above, font=\fontsize{5pt}{6pt}]{
                        ${\color{green!50!black}X_{b_2-L_0}^{b_2}}, {\color{red}
                    Y_{b_2-L_0}^{b_2+L_0}}$} ($(B2t2.center)+(0,0.5)$); 

                \draw[dotted] ($(B2t3.center)+(0,0.5)$) -- (B2t3.center); 
                \draw[dotted] ($(B1t4.center)+(0,0.5)$) -- (B1t4.center); 
                \draw[<->]     ($(B2t3.center)+(0,0.45)$) -- node[above]{\tiny $b_4-b_3-2L_0$} ($(B1t4.center)+(0,0.45)$);

        \end{tikzpicture}
        \caption{Illustration of a level-$2$ \BST. 
            There are four \bblocks, with \bblock numbers $\ell =1,2,3,4$. 
            The base-index (in blue) in \bblock $\ell$ is $b_{\ell}$. 
            The red boxes in the illustration correspond to $X_{\bv{b}-L_0}^{\bv{b}}, Y_{\bv{b}-L_0}^{\bv{b}+L_0}$, 
            where $X$ is only available to the left of the blue lines (shown in green).
            Each red box represents a contiguous set of indices, and there are $2^n$ such sets; they are separated in time. 
        }
        \label{fig_new process}
    \end{center}
\end{figure}

    Using the left-hand sides of~\eqref{eq_conditions for universal set of processes psi} and~\eqref{eq_conditions for universal set of processes phi}, \Cref{lem_FAIM is psi mixing}, and a straightforward induction argument, we conclude that 
    \begin{align*}
        P_{X_{\bv{b}-L_0}^{\bv{b}}, Y_{\bv{b}-L_0}^{\bv{b}+L_0}} &\leq (\bar{\psi}_{M_0-2})^{(2^n)} 
        \prod_{\ell=1}^{2^n} P_{X_{b_{\ell}-L_0}^{b_{\ell}}, Y_{b_{\ell}-L_0}^{b_{\ell+L_0}}}, \\ 
        P_{X_{\bv{b}-L_0}^{\bv{b}}, Y_{\bv{b}-L_0}^{\bv{b}+L_0}} &\geq (\bar{\phi}_{M_0-2})^{(2^n)} 
        \prod_{\ell=1}^{2^n} P_{X_{b_{\ell}-L_0}^{b_{\ell}}, Y_{b_{\ell}-L_0}^{b_{\ell+L_0}}}. 
    \end{align*}
    Thus, by~\eqref{eq_psiphiM02n bound}, 
    \begin{align*}
        P_{X_{\bv{b}-L_0}^{\bv{b}}, Y_{\bv{b}-L_0}^{\bv{b}+L_0}} &\leq (1+\varepsilon_2)
        \prod_{\ell=1}^{2^n} P_{X_{b_{\ell}-L_0}^{b_{\ell}}, Y_{b_{\ell}-L_0}^{b_{\ell+L_0}}}, \\ 
        P_{X_{\bv{b}-L_0}^{\bv{b}}, Y_{\bv{b}-L_0}^{\bv{b}+L_0}} &\geq (1-\varepsilon_2)
        \prod_{\ell=1}^{2^n} P_{X_{b_{\ell}-L_0}^{b_{\ell}}, Y_{b_{\ell}-L_0}^{b_{\ell+L_0}}}, 
    \end{align*}
    which is~\eqref{eq_probability inequality}. 
\end{IEEEproof}


In \Cref{lem_independent bblock} we saw that $\at{P}$ and $\tilde{P}$ are close in the sense of~\eqref{eq_probability inequality}. 
The following lemma, whose proof can be found in Appendix~\ref{ap_auxiliary proofs for sec3B}, 
translates this proximity to conditional entropies. 
\begin{lemma}  \label{lem_entropy difference if distributions close}
    Let $A$ and $\tilde{A}$ be two discrete random variables over the same finite alphabet $\mathcal{A}$. 
    Denote $\Prob{A=a} = p(a)$ and $\Prob{\tilde{A}=a} = q(a)$ for all $a \in \mathcal{A}$. 
    Assume that for some $0 \leq \varepsilon < \frac{1}{6}$,
    \begin{equation}\label{eq_pq inequality}
        (1-\varepsilon)q(a) \leq p(a) \leq (1+\varepsilon)q(a), \quad \forall a \in \mathcal{A}. 
    \end{equation}
    Then,  for any $f:\mathcal{A} \to \{0,1\}$ and $g:\mathcal{A} \to \mathcal{G}$, where $\mathcal{G}$ is some finite alphabet, we have
    \begin{align*}
        \big| H(f(A) | g(A)) - H(f(\tilde{A})|g(\tilde{A})) \big| & \leq \frac{\varepsilon}{2} - 3 \varepsilon \log \frac{3\varepsilon}{2} \\ 
                                                                  &< \sqrt{8\varepsilon}. 
    \end{align*}
    %
\end{lemma}

We are now ready to prove \Cref{thm_BST is monopolarizing}. 
\begin{IEEEproof}[Proof of \Cref{thm_BST is monopolarizing}]
    Choose $\varepsilon_1>0$ and $0 < \varepsilon_2 < \frac{1}{6}$ small enough such that 
    \begin{equation} \label{eq_choice of xi}
        \xi \triangleq \eta - 4\varepsilon_1 - (2\varepsilon_1 + \sqrt{8\varepsilon_2}) > 0. 
    \end{equation} 
    For example, one may take
    \begin{IEEEeqnarray}{rCl}
        \IEEEyesnumber \label{eq_varepsilon12}
        \varepsilon_1 &<& \frac{\eta}{12}, \IEEEyessubnumber \\[0.1cm]
        \varepsilon_2 &<& \frac{\eta^2}{32} .\IEEEyessubnumber 
    \end{IEEEeqnarray}

    Take $\nth$ large enough so that \Cref{prop_OTBST is monopolarizing} holds with $\xi$ as above. 
    Recall that \Cref{prop_OTBST is monopolarizing} holds for any $L_0$ and $M_0$, so we are free to set them as desired. 

    By \Cref{prop_tilde non tilde}, for $\nth$, $\varepsilon_1$, and $\varepsilon_2$ above, there exist $\Lth$ and $\Mth$ such that~\eqref{eq_HFiGi tilde
    inequality} holds for $L_0 \geq \Lth$ and $M_0 \geq \Mth$.
    That is, 
    \begin{IEEEeqnarray}{r}
        -(2\varepsilon_1 + \sqrt{8\varepsilon_2}) \leq H(F_i|G_i) - H(\tilde{F}_i|\tilde{G}_i) \leq (2\varepsilon_1 + \sqrt{8\varepsilon_2}).
        \IEEEeqnarraynumspace \label{eq_HFGtilde eps}
    \end{IEEEeqnarray}
    In fact, we choose $L_0 \geq \Lth$ as in the proof of \Cref{lem_lean bblock}.
    This ensures that the $L_0$-forgetfulness of the \soprocess is upper-bounded by $\varepsilon_1$. 
    Thus, by \Cref{cor_entropy rate with finite memory length},~\eqref{eq_H0 and ENTI difference} holds with $\epsilon \leq \varepsilon_1$, so that
    \[ 
        -2 \varepsilon_1 \leq \ENT{X_{\star} | Y_{\star}} - \Hwind \leq 2\varepsilon_1. 
    \] 
    Hence, if $\ENT{X_{\star} | Y_{\star}} \leq 1/2$ then $\Hwind \leq (1+4\varepsilon_1)/2$ and if $\ENT{X_{\star} | Y_{\star}} \geq 1/2$ then
    $\Hwind \geq (1-4\varepsilon_1)/2$. 
    Consequently, by \Cref{cor_OTBST monopolarizing} with $\zeta = 4\varepsilon_1$, if $n \geq \nth$ then 
    \begin{IEEEeqnarray*}{rCl/l}
        \ENT{X_{\star}|Y_{\star}} \leq 1/2 &\Rightarrow& H(\tilde{F}_i | \tilde{G}_i) < \xi + 4\varepsilon_1,  &\forall i \in \medsetPlus{n}, \\
        \ENT{X_{\star}|Y_{\star}} \geq 1/2 &\Rightarrow& H(\tilde{F}_i | \tilde{G}_i) > 1- \xi-4\varepsilon_1, &\forall i \in \medsetMinus{n}.
    \end{IEEEeqnarray*}
    Combining the above with~\eqref{eq_choice of xi} and~\eqref{eq_HFGtilde eps} we obtain that for $n \geq \nth$, 
    \begin{IEEEeqnarray*}{rCl/l}
        \ENT{X_{\star}|Y_{\star}} \leq 1/2 &\Rightarrow& H(F_i | G_i) < \eta, &\forall i \in \medsetPlus{n}, \\
        \ENT{X_{\star}|Y_{\star}} \geq 1/2 &\Rightarrow& H(F_i | G_i) > 1- \eta, &\forall i \in \medsetMinus{n}.
    \end{IEEEeqnarray*}
    This completes the proof. 
\end{IEEEproof}
\begin{remark}
We have proved~\Cref{thm_BST is monopolarizing} using \eqref{eq_HFiGi tilde inequality b}, which is looser than \eqref{eq_HFiGi tilde inequality a}. Hence, the proof would also hold if we were to define $\xi$ in \eqref{eq_choice of xi} as $\eta - 4\varepsilon_1 - \varepsilon_3$. 
The looser definition of $\xi$ circumvents a cumbersome upper bound on $\varepsilon_3$, and by proxy on $\varepsilon_2$, that involves the Lambert $W$ function~\cite[p. 332]{LambertW_1996}. 
\end{remark}

\subsection{Monopolarization for FAIM-derived Processes, for a Cascade of  BSTs} \label{sec_monopolarization for FAIM derived processes cascade}
The previous subsection considered a single BST. However, in practice, one may cascade several BSTs to obtain a universal polar code of rate different from $1/2$. In this subsection, we extend the previous results to a cascade of BSTs.

A cascade of BSTs is defined by the following:
\begin{itemize}
	\item the number of BSTs in the cascade, $t$ (that is, the overall transform is comprised of $t$ stages); 
	\item parameter $L_0$, which defines the number of lateral indices in the  level-$0$ block of the first BST in the cascade;
	\item parameters $\Mcascade{1}, \Mcascade{2}, \ldots, \Mcascade{t}$, where $\Mcascade{i}$ defines the number of medial indices in the  level-$0$ block of the $i$th BST in the cascade;
	\item recursion depths $n_1,n_2,\ldots,n_t$, of the $t$ BSTs in the cascade;
	\item a binary vector $\bv{c} = \begin{bmatrix} c_1&c_2&\ldots&c_{t-1}\end{bmatrix}$ of length $t-1$. Each stage of the cascade applies a BST operation on a subset of indices from the previous stage. This is determined by $\bv{c}$; informally, $c_i=0$ ($c_i=1$) implies that stage $i+1$ is the result of applying a BST on the newly formed medial-minus (plus) indices of stage $i$.
\end{itemize}

The cascade is constructed recursively. For $t=1$, we are in the single BST case. This BST is defined through $L_0$, parameter $M_0=\Mcascade{1}$, and recursion depth $n_1$. That is, we transform $X_1^{\Ncascade{1}}$ to $U_1^{\Ncascade{1}}$, where $\Ncascade{1} = (2L_0 + \Mcascade{1}) \cdot 2^{n_1}$. Recall that medial indices at level $n_1$ of the BST are split into $\medsetMinus{n_1}$ and $\medsetPlus{n_1}$, see~\eqref{eq_defs of lat-med sets}. We denote the medial sets of stage $1$ of the cascade as
\[
\medsetMinusBrace{1} = \medsetMinus{n_1}, \quad  \medsetPlusBrace{1} = \medsetPlus{n_1}.
\]
The remaining indices are lateral. We further define the two sets $\numedsetMinus{1}$ and $\numedsetPlus{1}$, which we call ``the new medial-minus and medial-plus sets of stage $1$ of the cascade.'' For this base case, they coincide with $\medsetMinusBrace{1}$ and $\medsetPlusBrace{1}$, respectively. That is,
\[
\numedsetMinus{1} = \medsetMinusBrace{1}, \quad  \numedsetPlus{1} = \medsetPlusBrace{1}.
	\]

When moving from stage $i$ to stage $i+1$ of the cascade, we first make $\Mcascade{i+1} \cdot 2^{n_{i+1}}$ copies of the stage-$i$ cascade. That is, the length of the stage-($i+1$) cascade is  
\begin{equation} \label{eq_Ncascade recursion}
	\Ncascade{i+1}=\Ncascade{i} \cdot \Mcascade{i+1} \cdot 2^{n_{i+1}}.
\end{equation}
For $1 \leq \ell \leq \Mcascade{i+1} \cdot 2^{n_{i+1}}$ and $1 \leq j \leq \Ncascade{i}$, Denote
\begin{align*}
    X_j^{(\ell)} &= X_{j+(\ell-1)\Ncascade{i}}, \\ 
    U_j^{(\ell)} &= U_{j+(\ell-1)\Ncascade{i}},  
\end{align*}
and their vector versions
\begin{align*}
    \bs{X}_{\ell} &= \begin{bmatrix} X_1^{(\ell)} & X_2^{(\ell)} & \cdots & X_{\Ncascade{i}}^{(\ell)} \end{bmatrix}, \\ 
    \bs{U}_{\ell} &= \begin{bmatrix} U_1^{(\ell)} & U_2^{(\ell)} & \cdots & U_{\Ncascade{i}}^{(\ell)} \end{bmatrix}.
\end{align*}
Copy $\ell$ of the stage-$i$ cascade transforms $\bs{X}_{\ell}$ to $\bs{U}_{\ell}$. 
Next, for each $1 \leq j \leq N^{(i)}$, we take the $j$th index of each copy and apply, as described below, an operation: either a BST or a pass-through. That is, for   $1 \leq j \leq \Ncascade{i}$ and $1 \leq \ell \leq \Mcascade{i+1} \cdot 2^{n_{i+1}}$, further denote
\begin{equation} \label{eq_def of V ell j}
V_{\ell}^{(j)} = V_{\ell + (j-1) \cdot \Mcascade{i+1} \cdot 2^{n_{i+1}}},
\end{equation}
and the vectors
\begin{IEEEeqnarray}{rCl} \IEEEyesnumber \label{eq_def of Uj and Vj underline}
	\underline{\bs{U}}_{j} &=& \begin{bmatrix} U_j^{(1)} & U_j^{(2)} & \cdots & U_{j}^{(\Mcascade{i+1} \cdot 2^{n_{i+1}})} \end{bmatrix}, \IEEEyessubnumber \label{eq_def of Uj underline} \\ 
    \underline{\bs{V}}_{j} &=& \begin{bmatrix} V_1^{(j)} & V_2^{(j)} & \cdots & V_{\Mcascade{i+1} \cdot 2^{n_{i+1}}}^{(j)} \end{bmatrix}\IEEEyessubnumber \label{eq_def of Vj underline}.
\end{IEEEeqnarray}
The operation transforms $\underline{\bs{U}}_{j}$ to $\underline{\bs{V}}_{j}$. The output of the stage-$(i+1)$ cascade is $\underline{\bs{V}}_{1}$, $\underline{\bs{V}}_{2}$, $\ldots$, $\underline{\bs{V}}_{\Ncascade{i}}$. In other words, we operate on a single symbol from each copy, and the result is a single contiguous block on the output side. 
This ordering is amenable to successive-cancellation decoding. 

What remains to define is which operation, BST or pass-through, to apply to which index, and to determine the various medial sets. At stage $i$ of the cascade, each index is either medial or lateral. The medial indices are split into $\medsetMinusBrace{i}$ and $\medsetPlusBrace{i}$. These sets will be defined as part of the recursion. Important subsets of these sets are $\numedsetMinus{i}$ and $\numedsetPlus{i}$, respectively. These two subsets will also be defined as part of the recursion. BST operations will be applied to exactly one of these subsets, called the ``active set.'' On all other indices, the pass-through operation will be applied. 

When moving from stage $i<t$ to stage $i+1$, define the ``active set'' $\sigma_i$ as
\begin{equation} \label{eq_def of active set}
\sigma_i =
\begin{cases}
	\numedsetMinus{i}, & \textrm{if} \;  c_i = 0, \\
	\numedsetPlus{i}, & \textrm{if} \;  c_i = 1.
\end{cases}
\end{equation}
For $i=t$ we technically define the active set as the descendants of the active set of the previous stage, that is 
\begin{equation} \label{eq_def of active set t}
	\sigma_t = \numedsetMinus{t} \cup \numedsetPlus{t}.
\end{equation}

For each index $1 \leq j \leq \Ncascade{i}$, we do the following:
\begin{itemize}
	\item If $j \in \sigma_i$, apply a BST operation to $\underline{\bs{U}}_{j}$. The BST is defined by parameters $L_0=0$, $M_0=\Mcascade{i+1}$, and has recursion depth $n_{i+1}$.
		\begin{itemize}
			\item If $V_{\ell}^{(j)}$ is a medial-minus symbol, then (see~\eqref{eq_def of V ell j}) \[\ell + (j-1) \cdot \Mcascade{i+1}\cdot 2^{n_{i+1}} \in \numedsetMinus{i+1}.\]
			\item If $V_{\ell}^{(j)}$ is a medial-plus symbol, then \[\ell + (j-1) \cdot \Mcascade{i+1}\cdot 2^{n_{i+1}} \in \numedsetPlus{i+1}.\]
			\item All indices in $\numedsetMinus{i+1}$ are also in $\medsetMinusBrace{i+1}$. 
			\item All indices in $\numedsetPlus{i+1}$ are also in $\medsetPlusBrace{i+1}$. 

		\end{itemize}
	\item Otherwise, apply a pass-through operation to $\underline{\bs{U}}_{j}$, that is $\underline{\bs{V}}_{j} = \underline{\bs{U}}_{j}$, i.e., by~\eqref{eq_def of Uj and Vj underline}, $V_{\ell}^{(j)} = U_j^{(\ell)}$ for each $1 \leq \ell \leq \Mcascade{i+1} \cdot 2^{n_{i+1}}$. 
		\begin{itemize}
			\item If $j \in \medsetMinusBrace{i}$, then all the indices in $\underline{\bs{V}}_{j}$ (see~\eqref{eq_def of V ell j} and~\eqref{eq_def of Vj underline}), are in $\medsetMinusBrace{i+1}$.
			\item If $j \in \medsetPlusBrace{i}$, then all the indices in $\underline{\bs{V}}_{j}$ are in $\medsetPlusBrace{i+1}$.
		\end{itemize}
	\item Any index that is not in  $\medsetMinusBrace{i}$ or $\medsetPlusBrace{i}$ is lateral.
\end{itemize}

We introduce the following definition --- a specialization of the above --- to simplify the statements of the claims in this subsection.
\begin{definition}
	\label{def_tcLMn cascade}
	A $(t,\bv{c};L_0,M_0,n)$-cascade is a cascade of BSTs as above, with $\Mcascade{i} = M_0$ and $n_i = n$, for $1 \leq i \leq t$.
\end{definition}
The length of a $(t,\bv{c};L_0,M_0,n)$-cascade, by the recursion~\eqref{eq_Ncascade recursion} and recalling that $\Ncascade{1} = (2L_0+M_0)2^n$ is given by
\begin{equation} \label{eq_length of a (t,c,L0,M0,n) cascade}
	\Ncascade{t} = (2L_0+M_0)M_0^{t-1}2^{nt}.
\end{equation}

Our plan for the rest of this subsection is as follows. Denote the cascade threshold entropy as
\begin{equation} \label{eq_threshold entropy of a cascade}
h(\bv{c}) = \frac{1 + \sum_{i=1}^{t-1} c_{i} 2^{t-i} }{2^t}.
\end{equation}
In \Cref{lem_fraction of medial minus indices out of medial indices in a cascade}, we show that the fraction of medial-minus indices out of all medial indices of the cascade approaches $h(\bv{c})$. We  further show that the fraction of medial indices out of all indices approaches $1$. Then, in \Cref{thm_cascade of BSTs}, we show that $h(\bv{c})$ is indeed a threshold entropy of the cascade. That is,  for an \soprocess with conditional entropy rate less than $h(\bv{c})$, the medial-plus indices monopolarize; and for an \soprocess with conditional entropy rate greater than $h(\bv{c})$, the medial-minus indices monopolarize.

\begin{lemma}
	\label{lem_fraction of medial minus indices out of medial indices in a cascade}
Consider a $(t,\bv{c};L_0,M_0,n)$-cascade.  Then, 
\begin{multline} \label{eq_full expression for med ratio}
	\frac{|\medsetMinusBrace{t}|}{|\medsetPlusBrace{t} \cup \medsetMinusBrace{t}| }  \\ 
=\frac{\displaystyle{
	1+ \sum_{i=1}^{t-1}c_i2^{t-i}\left( \frac{1}{1-2(1-2^{-n})M_0^{-1}} \right)^{t-i}}}{\displaystyle{
2+\sum_{i=1}^{t-1}2^{t-i}\left( \frac{1}{1-2(1-2^{-n})M_0^{-1}} \right)^{t-i}}}.
\end{multline}
Moreover, 
\begin{multline} \label{eq_full expression for med minus cup med plus}
	\frac{|\medsetPlusBrace{t} \cup \medsetMinusBrace{t}| }{\Ncascade{t}}  \\ 
=\frac{\displaystyle{
		\left(1 - \frac{2-2^{1-n}}{M_0}\right)^t\left(	
2+\sum_{i=1}^{t-1}2^{t-i}\left( \frac{1}{1-2(1-2^{-n})M_0^{-1}} \right)^{t-i}\right)}}{2^t(2L_0+M_0)M_0^{-1}}.
\end{multline}

\end{lemma}
Observe that when $M_0$ is large, the right-hand side of~\eqref{eq_full expression for med ratio} approaches $h(\bv{c})$, as the terms in parentheses approach $1$ and  $\sum_{i=1}^{t-1}2^{t-i}=2^t-2$. Furthermore, when $M_0$ is also large with respect to $L_0$, the right-hand side of~\eqref{eq_full expression for med minus cup med plus} approaches $1$.  

\begin{IEEEproof}
	Define $\nu_0 = 1$, and denote by $\nu_i$ the number of new medial plus indices (which is the same as the number of new medial minus indices) at the output of stage $i$ of the cascade. By the cascade construction, the first stage is merely a BST of depth $n$ with \bblocks consisting of $M_0$ medial indices and $2 L_0$ lateral indices each. Recalling that the number of medial indices in a BST of depth $n$ is given in~\eqref{eq_number of lateral and medial indices of a BST medial}, we have 
	\[ 
		\nu_1 = \frac{1}{2} \nu_0\cdot\left(2^{n} M_{0} - 2(2^n-1)\right).
	\] 
	Note that the factor $1/2$ stems from counting the number of new medial plus indices, which is half the number of new medial indices.

	When moving from step $i-1$ of the cascade to step $i$, we perform $\nu_{i-1}$ BSTs of depth $n$, each with \bblocks consisting of $M_0$ medial indices and $0$ lateral indices. Hence, the number of new medial plus indices at the end of stage $i$ is
	\begin{equation} 
		\label{eq_nu i recursive}
		\nu_i = \frac{1}{2} \nu_{i-1}\cdot\left(2^{n} M_{0} - 2(2^n-1)\right).
\end{equation}
Thus,
\begin{equation}
\label{eq_nu i non-recursive} 
		\nu_i = \frac{1}{2^i} \left(2^{n} M_{0} - 2(2^n-1)\right)^i.
\end{equation}

	We now claim that for $1 \leq i \leq t$,
	\begin{IEEEeqnarray}{rCl} \IEEEyesnumber \label{eq_size of medsetBrace}
		|\medsetMinusBrace{i}| &=& \nu_i + \sum_{j=1}^{i-1}c_j \nu_j (2^n M_0)^{i-j} \IEEEyessubnumber \label{eq_size of medsetMinusBrace} \\ 
		|\medsetPlusBrace{i}| &=& \nu_i + \sum_{j=1}^{i-1}(1-c_j) \nu_j (2^n M_0)^{i-j}. \IEEEyessubnumber \label{eq_size of medsetPlusBrace}
	\end{IEEEeqnarray}
	Indeed, by the construction above, if $c_{i}=0$,
	\begin{align*}
		|\medsetMinusBrace{i+1}| &= \nu_{i+1} + 2^{n} M_0 |\medsetMinusBrace{i} \setminus \numedsetMinus{i}|,\\ 
		|\medsetPlusBrace{i+1}| &= \nu_{i+1} + 2^{n} M_0 |\medsetPlusBrace{i}|,
	\end{align*}
	and if $c_{i} = 1$ then
	\begin{align*}
		|\medsetMinusBrace{i+1}| &= \nu_{i+1} + 2^{n} M_0 |\medsetMinusBrace{i} |,\\ 
		|\medsetPlusBrace{i+1}| &= \nu_{i+1} + 2^{n} M_0 |\medsetPlusBrace{i}\setminus \numedsetPlus{i}|.
	\end{align*}
	Recalling that $|\numedsetMinus{i}| = |\numedsetPlus{i}| = \nu_{i}$, we can use the above to prove \eqref{eq_size of medsetBrace} by induction on $i$. 
	
	Next, observe that 
	\begin{IEEEeqnarray}{rCl}
		|\medsetMinusBrace{i} \cup \medsetPlusBrace{i}| &=& |\medsetMinusBrace{i}| + |\medsetPlusBrace{i}| \IEEEnonumber \\
								&=& 2 \nu_i  + \sum_{j=1}^{i-1} \nu_j (2^n M_0)^{i-j}. \IEEEyesnumber \label{eq_union of med minus plus brace}
	\end{IEEEeqnarray}
	
	Denote $\alpha = 2(2^n-1)$. By~\eqref{eq_nu i non-recursive} and \eqref{eq_size of medsetMinusBrace},
	\begin{IEEEeqnarray*}{rCl}
		\IEEEeqnarraymulticol{3}{l}{|\medsetMinusBrace{t}| = \nu_t + \sum_{i=1}^{t-1}c_i \nu_i (2^n M_0)^{t-i}} \\
				       &=& \frac{(2^nM_0-\alpha)^t}{2^t}+ \sum_{i=1}^{t-1}c_i2^{-i}(2^nM_0-\alpha)^{i}(2^nM_0)^{t-i} \\ 
				       &=& \frac{(2^nM_0-\alpha)^t}{2^t}+ \frac{(2^nM_0-\alpha)^t}{2^t}\sum_{i=1}^{t-1}c_i2^{t-i}\left( \frac{2^nM_0}{2^nM_0-\alpha} \right)^{t-i} \\
				       &=& \frac{(2^nM_0-\alpha)^t}{2^t}\left(1+ \sum_{i=1}^{t-1}c_i2^{t-i}\left( \frac{1}{1-\alpha2^{-n}M_0^{-1}} \right)^{t-i}\right).
	\end{IEEEeqnarray*}
	Similarly, by~\eqref{eq_nu i non-recursive} and \eqref{eq_union of med minus plus brace},  
	\begin{IEEEeqnarray}{rCl} 
		\IEEEeqnarraymulticol{3}{l}{|\medsetMinusBrace{t} \cup \medsetPlusBrace{t}| = 2 \nu_t + \sum_{i=1}^{t-1} \nu_i (2^nM_0)^{t-i}} \IEEEnonumber \\
				       &=& 2\cdot\frac{(2^nM_0-\alpha)^t}{2^t}+\sum_{i=1}^{t-1}2^{-i}(2^nM_0-\alpha)^{i}(2^nM_0)^{t-i} \IEEEnonumber \\ 
				       &=& 2\cdot\frac{(2^nM_0-\alpha)^t}{2^t}+ \frac{(2^nM_0-\alpha)^t}{2^t}\left(\sum_{i=1}^{t-1}2^{t-i}\left( \frac{2^nM_0}{2^nM_0-\alpha} \right)^{t-i}\right)\IEEEnonumber  \\
				       &=& \frac{(2^nM_0-\alpha)^t}{2^t}\left(2+\sum_{i=1}^{t-1}2^{t-i}\left( \frac{1}{1-\alpha2^{-n}M_0^{-1}} \right)^{t-i}\right).  \IEEEyesnumber
				       \label{eq_expression for med minus brace cup med plus brace}
	\end{IEEEeqnarray}

	Combining the above two expressions, we obtain
\begin{IEEEeqnarray*}{rCl}
	\IEEEeqnarraymulticol{3}{l}{\frac{|\medsetMinusBrace{t}|}{|\medsetPlusBrace{t} \cup \medsetMinusBrace{t}| }}  \\ 
	\qquad &=& \frac{\displaystyle{
		 1+ \sum_{i=1}^{t-1}c_i2^{t-i}\left( \frac{1}{1-\alpha2^{-n}M_0^{-1}} \right)^{t-i}}}{\displaystyle{
 2+\sum_{i=1}^{t-1}2^{t-i}\left( \frac{1}{1-\alpha2^{-n}M_0^{-1}} \right)^{t-i}}},\\[0.2cm]
	\qquad &=& \frac{\displaystyle{
	1+ \sum_{i=1}^{t-1}c_i2^{t-i}\left( \frac{1}{1-2(1-2^{-n})M_0^{-1}} \right)^{t-i}}}{\displaystyle{
2+\sum_{i=1}^{t-1}2^{t-i}\left( \frac{1}{1-2(1-2^{-n})M_0^{-1}} \right)^{t-i}}},
 \end{IEEEeqnarray*}
 where in the last equality we recalled that $\alpha = 2(2^n-1)$. This proves~\eqref{eq_full expression for med ratio}. 

 To prove~\eqref{eq_full expression for med minus cup med plus}, we divide the expression in~\eqref{eq_expression for med minus brace cup med plus brace} by the expression 
 in~\eqref{eq_length of a (t,c,L0,M0,n) cascade}. Since $\alpha = 2(2^n-1)$, the ratio between the term preceding the large parentheses in~\eqref{eq_expression for med minus brace cup med plus brace} and $\Ncascade{t}$ is
 \begin{IEEEeqnarray*}{rCl}
	 \frac{(2^nM_0-\alpha)^t}{2^t \Ncascade{t}} &=& \frac{(2^n(M_0-2)+2)^t}{2^t (2L_0+M_0)M_0^{t-1}2^{nt}}  \\ 
						    &=& \frac{(M_0-2+2^{1-n})^t}{2^t (2L_0+M_0)M_0^{t-1}}\\
						    &=& \frac{(1-(2-2^{1-n})M_0^{-1})^t}{2^t (2L_0+M_0) M_0^{-1}}.
 \end{IEEEeqnarray*}
 This completes the proof.
\end{IEEEproof}


The following simple observations on $h(\bv{c})$ will be useful in the proof of \Cref{prop_cascade of OTBST is monopolarizing} below.

\begin{lemma} \label{lem_simple observation on hc}
	Let $\bv{c}$ be a binary vector of length $t-1$ and let its length-$(t-2)$ suffix obtained by removing $c_1$ be $\bv{c}^{1}$ Then: 
    \begin{align}
        h(\bv{c}^{1}) &= 2h(\bv{c}) - c_1, \label{eq_hctag vs hc} \\ 
        \frac{1}{2^{t-1}} \leq h(\bv{c}^{1}) &\leq 1- \frac{1}{2^{t-1}}. \label{eq_hctag bounds}
\end{align}
\end{lemma}

Observe that by~\eqref{eq_hctag vs hc} and~\eqref{eq_hctag bounds} we have $h(\bv{c}) \leq 1/2 - 2^{-t}$ if $c_{1} = 0$ and $h(\bv{c}) \geq 1/2 + 2^{-t}$ if $c_{1} = 1$. 

\begin{IEEEproof}
    The proof follows from simple algebra. Indeed,  
    \begin{align*}
	    h(\bv{c}^{1}) &= \frac{1 + \sum_{i=1}^{t-2} c_{i+1} 2^{t-1-i} }{2^{t-1}} 
                   = 2\cdot \left( \frac{1 + \sum_{i=1}^{t-1} c_{i} 2^{t-i} }{2^{t}} - \frac{c_1}{2} \right)  
		 \\    & = 2h(\bv{c})-c_1, 
    \end{align*}
    which is~\eqref{eq_hctag vs hc}. 
    Next, by setting $c_{i+1} = 0$ ($c_{i+1}=1$) for all $1\leq i\leq t-2$ in  the expression for $h(\bv{c}^{1})$, i.e.,  the first equality above, we obtain the lower (upper) bound in~\eqref{eq_hctag bounds}. 
\end{IEEEproof}
In the sequel we will denote by $\bv{c}^{\ell}$ the suffix of $\bv{c}$ after removing its first $\ell$ elements. That is, 
\begin{equation}\label{eq_def of cell}
	\bv{c}^{\ell} = \begin{bmatrix} c_{\ell+1} & c_{\ell+2} & \cdots & c_{t-1} \end{bmatrix}.
\end{equation}
Note that $\bv{c}^{t-1}$ is an empty vector, and by \eqref{eq_threshold entropy of a cascade} we have $h(\bv{c}^{t-1}) = 1/2$.

The following is the cascade equivalent of \Cref{thm_BST is monopolarizing}.
\begin{theorem} \label{thm_cascade of BSTs}
    Let the number of stages of the cascade $t$, and the binary vector $\bv{c}$ of length $t-1$ be given.
Fix sequences $\bar{\epsilon}_L$, $\bar{\psi}_M$, and $\bar{\phi}_M$ that satisfy the limits in \eqref{eq_conditions for universal set of processes epsilon}--\eqref{eq_conditions for universal set of processes phi}, as well as the conditions in~\eqref{eq_conditions for universal set of processes monotone}.
    Let $X_{\star} \sarrow Y_{\star}$ be a forgetful FAIM-derived \soprocess that satisfies the inequalities in~\eqref{eq_conditions for universal set of processes epsilon}--\eqref{eq_conditions for universal set of processes phi}.
    For every $\eta >0$ there exist $\Lth$, $\Mth$, and $\nth$, \emph{independent of the process},  such that if $L_0 \geq \Lth$, $M_0 \geq \Mth$, and $n \geq \nth$ then
    a $(t,\bv{c}; L_0, M_0, n)$-cascade is $(\eta, \medsetPlusBrace{t}, \medsetMinusBrace{t})$-monopolarizing. 

    Specifically, let $F_{1}^{\Ncascade{t}} \sarrow G_{1}^{\Ncascade{t}}$ be a transformed \soblock of a $(t,\bv{c}; L_0, M_0, n)$-cascade as in \Cref{def_tcLMn cascade}.
    Then: 
    \begin{itemize}
        \item if $\ENT{X_{\star} | Y_{\star}}\leq h(\bv{c})$ then $H(F_{i} | G_{i})<\eta$, \hphantom{$1;$}
            $\forall i \in
    \medsetPlusBrace{t}$; 
        \item if $\ENT{X_{\star} | Y_{\star}}\geq h(\bv{c})$ then $H(F_{i} | G_{i})>1-\eta$, $\forall i \in \medsetMinusBrace{t}$. 
    \end{itemize}
\end{theorem}

	The proof will follow along the same general lines of \Cref{thm_BST is monopolarizing}. We first consider the simple case of an observation-truncated transform applied to a \biprocess. For this simple case, we generalize \Cref{prop_OTBST is monopolarizing}.

	Recall that in a \biprocess with parameter $N_0 = 2L_0 + M_0$, contiguous symbol and observation blocks of length $N_0$ are independent. In our setting, a \biprocess is defined as in \Cref{def_biprocess},  with the transform length $N_n$ replaced by $\Ncascade{t}$ and the number of copies of the level-0 block of length $N_0$ ($2^n$ in the definition) is  $\Ncascade{t}/N_0$.  Further recall that an observation-truncated transform is defined in \Cref{sec_otbst}. The key point to note is that the ``observation-truncated'' property is defined based on a truncation at a level-$0$ block of a BST, see \eqref{eq_fjgj init trunc}. In other words, this property is determined at the ``input-output'' level of the process. Consequently, in a cascade of more than one BST, the overall transform is observation truncated if the first BST is. We call such a cascade an observation-truncated cascade.

	The following is a generalization of \Cref{prop_OTBST is monopolarizing}. Recall that $\Hwind$ was defined in \eqref{eq_def of H0}, and that $\sigma_{t}$ was defined 
	in~\eqref{eq_def of active set t}.

\begin{proposition} \label{prop_cascade of OTBST is monopolarizing}
    Fix cascade parameters $t$ and $\bv{c}$.
    For every $\zeta > 0$, there exists a threshold value $\nth \geq 0$ such that if $n \geq \nth$ then an observation-truncated $(t,\bv{c}; L_0, M_0, n)$-cascade with any parameters $L_0, M_0$ is $(\zeta, \medsetPlusBrace{t}, \medsetMinusBrace{t})$-monopolarizing for any \biprocess $\tilde{X}_{\star}\sarrow \tilde{Y}_{\star}$ with parameter $N_0 = 2L_0 + M_0$. 

    \looseness=-1
    Specifically, let $\tilde{F}_{1}^{\Ncascade{t}} \sarrow \tilde{G}_{1}^{\Ncascade{t}}$ be an \ottransformed \soblock of the observation-truncated $(t,\bv{c}; L_0, M_0, n)$-cascade, where $n \geq \nth$. 
    Then:
    \begin{IEEEeqnarray}{rCl'l} \IEEEyesnumber \label{eq_Hwind leq geq h(c)}
	    \Hwind \leq h(\bv{c}) &\Rightarrow& H(\tilde{F}_{i} | \tilde{G}_{i}) < \zeta,  & \forall i \in \medsetPlusBrace{t}, \IEEEyessubnumber\label{eq_Hwind leq h(c) zeta}\\
	    \Hwind \geq h(\bv{c}) &\Rightarrow& H(\tilde{F}_{i} | \tilde{G}_{i}) > 1-\zeta, &  \forall i \in \medsetMinusBrace{t}. \IEEEyessubnumber\IEEEeqnarraynumspace   \label{eq_Hwind geq h(c) 1-zeta}
\end{IEEEeqnarray}
    %
    In fact, we can strengthen the above:
    \begin{IEEEeqnarray}{rCl/l} \IEEEyesnumber \label{eq_Hwind leq geq h(c) strong}
	    \Hwind \leq h(\bv{c}) &\Rightarrow& H(\tilde{F}_{i} | \tilde{G}_{i}) < \frac{\zeta}{2^t}, &  \forall i \in \medsetPlusBrace{t}\!\setminus\! \sigma_{t}, \IEEEyessubnumber\label{eq_Hwind leq h(c) zeta strong}\\
	    \Hwind \geq h(\bv{c}) &\Rightarrow& H(\tilde{F}_{i} | \tilde{G}_{i}) > 1-\frac{\zeta}{2^t}, &  \forall i \in \medsetMinusBrace{t}\!\setminus\! \sigma_{t}. \IEEEyessubnumber\IEEEeqnarraynumspace   \label{eq_Hwind geq h(c) 1-zeta strong}
\end{IEEEeqnarray}
    %
\end{proposition}

\begin{IEEEproof}
First observe that if the proposition holds for some $\zeta < 1$, then it clearly holds for all $\zeta \geq 1$ with the same $\nth$. Thus, assume that $\zeta < 1$.

We start by choosing $\nth$ such that \Cref{prop_OTBST is monopolarizing} holds with $\xi = \zeta/2^t$. First consider the case $\Hwind = h(\bv{c})$; the more general case will follow by monotonicity. We now track the evolution of the cascade stages, by using \Cref{cor_alpha beta prime prime}.

	If $c_1 = 0$, then, by~\Cref{lem_simple observation on hc}, $h(\bv{c})\leq 1/2$. Hence, by \Cref{cor_alpha beta prime prime} and~\eqref{eq_hctag vs hc},
	\[ 
		H(\tilde{F}_i | \tilde{G}_i) \in \begin{cases} 
			(h(\bv{c}^{1})-\xi,h(\bv{c}^{1})], & i \in \medsetMinusBrace{1}, \\
			[0,\xi), & i \in \medsetPlusBrace{1}. 
		\end{cases}
		\] 
	Similarly, if $c_1 = 1$, then, by~\Cref{lem_simple observation on hc}, $h(\bv{c})\geq 1/2$. Hence, by \Cref{cor_alpha beta prime prime} and~\eqref{eq_hctag vs hc},
	\[ 
		H(\tilde{F}_i | \tilde{G}_i) \in \begin{cases} 
			(1-\xi,1], & i \in \medsetMinusBrace{1}, \\  
			[h(\bv{c}^{1}),h(\bv{c}^{1})+\xi), & i \in \medsetPlusBrace{1}. 
		\end{cases}
		\] 
	Recall that by the cascade construction, $\numedsetMinus{1} = \medsetMinusBrace{1}$ and $\numedsetPlus{1}= \medsetPlusBrace{1}$. Thus, the active set $\sigma_1$ (see~\eqref{eq_def of active set}) is not polarized, whereas the remaining new medial set is polarized in the sense of~\eqref{eq_Hwind leq geq h(c) strong}. 
	In particular, regardless of whether $c_1=0$ or $c_1=1$, we have
	\[
		i \in \sigma_1 \Longrightarrow H(\tilde{F}_i|\tilde{G}_i) \in (h(\bv{c}^{1})-\xi, h(\bv{c}^{1})+\xi).
	\]

	This will form the basis of the following claim, which we prove by induction: after $\ell < t$ stages of the cascade,
	\begin{multline*}
		H(\tilde{F}_i | \tilde{G}_i) \in \\  \begin{cases}
			(h(\bv{c}^{\ell})-(2^{\ell}-1)\xi, h(\bv{c}^{\ell})+(2^\ell-1)\xi), & i \in \sigma_{\ell}, \\ 
			(1-\xi,1], & i \in \medsetMinusBrace{\ell}  \!\setminus \! \sigma_{\ell}, \\ 
			[0,\xi), & i \in \medsetPlusBrace{\ell}  \!\setminus \! \sigma_{\ell}.
			\end{cases}
	\end{multline*}
	The claim implies that after $\ell < t$ stages of the cascade, all the medial indices are polarized in the sense of~\eqref{eq_Hwind leq geq h(c) strong}, except for the active set.

	As shown above, the claim is indeed true for the basis case, $\ell=1$. For the induction step, assume the claim is true after $\ell<t-1$ stages. To prove that it is also true after $\ell+1$ stages, we call upon \Cref{cor_alpha beta prime prime} and \Cref{rem_windowed BIprocess is iid}.  That is, recall that to form stage $\ell+1$, we make $\Mcascade{\ell+1} \cdot 2^{n_{\ell+1}}$ copies of a symbol in the active set, and apply a BST to the copies. Further recall that we are in a \biprocess setting. That is, these copies are i.i.d. By \Cref{rem_windowed BIprocess is iid} and the induction hypothesis, \Cref{cor_alpha beta prime prime} holds with $\Hwind$ replaced by some value
	\begin{equation}
		\label{eq_eta range}
		\eta \in (h(\bv{c}^{\ell})-(2^{\ell}-1)\xi, h(\bv{c}^{\ell})+(2^\ell-1)\xi). 
		\end{equation}

	For $\eta$ as in~\eqref{eq_eta range}, if $c_{\ell+1} = 0$, then, by~\Cref{lem_simple observation on hc}, $h(\bv{c}^{\ell})\leq 1/2 - 2^{-(t-\ell)}$. By our assumption that $\zeta < 1$, we have $\xi = \zeta/2^t < 2^{-t}$. Hence, by \eqref{eq_eta range},
	\[
		\eta < h(\bv{c}^{\ell})+(2^\ell-1)\xi 
		     \leq \frac{1}{2} + 2^{\ell}\cdot(\xi - 2^{-t}) - \xi 
		     < \frac{1}{2}.
	\]
		Therefore, by \Cref{cor_alpha beta prime prime},
	\begin{equation}
		\label{eq_HFiGi for ell plus 1 and c zero}
		H(\tilde{F}_i | \tilde{G}_i) \in \begin{cases} 
			(2\eta-\xi,2\eta], & i \in \numedsetMinus{\ell+1}, \\
			[0,\xi), & i \in \numedsetPlus{\ell+1}. 
		\end{cases}
		\end{equation}
	Using~\eqref{eq_hctag vs hc} and~\eqref{eq_eta range}, equation \eqref{eq_HFiGi for ell plus 1 and c zero} implies
	\begin{multline*}
		i \in \numedsetMinus{\ell+1} \Longrightarrow \\ 
		H(\tilde{F}_i | \tilde{G}_i) \in  
			(h(\bv{c}^{\ell+1})-(2^{\ell+1}-1)\xi,h(\bv{c}^{\ell+1})+(2^{\ell+1}-2)\xi).
		\end{multline*}

	Similarly, for $\eta$ as in~\eqref{eq_eta range}, if $c_{\ell+1} = 1$, then $\eta > 1/2$, so by \Cref{cor_alpha beta prime prime},
	\[ 
		H(\tilde{F}_i | \tilde{G}_i) \in \begin{cases} 
			(1-\xi,1], & i \in \numedsetMinus{\ell+1}, \\  
			[2\eta-1,2\eta-1+\xi), & i \in \numedsetPlus{\ell+1}. 
		\end{cases}
		\] 
		Again, Using~\eqref{eq_hctag vs hc} and~\eqref{eq_eta range}, we obtain that in this case
	\begin{multline*}
		i \in \numedsetPlus{\ell+1} \Longrightarrow \\ 
		H(\tilde{F}_i | \tilde{G}_i) \in  
			(h(\bv{c}^{\ell+1})-(2^{\ell+1}-2)\xi,h(\bv{c}^{\ell+1})+(2^{\ell+1}-1)\xi).
		\end{multline*}

	In other words, combining both of the above cases and recalling the definition of the active set~\eqref{eq_def of active set}, we have
	\begin{multline*}
		i \in \sigma_{\ell+1} \Longrightarrow \\ 
		H(\tilde{F}_i | \tilde{G}_i) \in (h(\bv{c}^{\ell+1})-(2^{\ell+1}-1)\xi,h(\bv{c}^{\ell+1})+(2^{\ell+1}-1)\xi).
	\end{multline*}
	Recalling the cascade construction, and specifically how $\medsetMinusBrace{\ell+1}$ and $\medsetPlusBrace{\ell+1}$ are obtained from $\medsetMinusBrace{\ell}$, $\medsetPlusBrace{\ell}$, $\numedsetMinus{\ell+1}$, and $\numedsetPlus{\ell+1}$, we obtain the remainder of the inductive claim. Using~\eqref{eq_def of active set t}, this proves~\eqref{eq_Hwind leq geq h(c) strong}.

	For the last stage of the cascade, $t$, we call upon \Cref{cor_alpha beta prime prime}. By the cascade construction, we need only consider the active set $\sigma_{t-1}$, as all other indices are polarized (in the sense of~\eqref{eq_Hwind leq geq h(c) strong} and thus also in the sense of~\eqref{eq_Hwind leq geq h(c)}). Recall that $h(\bv{c}^{t-1})=1/2$, and thus, by the inductive claim, and since $\zeta = \xi\cdot2^t$,
	\[
		i \in \sigma_{t-1} \Longrightarrow  
		H(\tilde{F}_i | \tilde{G}_i) \in \left(\frac{1}{2}-\frac{\zeta}{2}+\xi,\frac{1}{2}+\frac{\zeta}{2}-\xi\right).
	\]
	By~\eqref{eq_HFiGi in alpha beta minus} and the monotonicity of $\alpha(\cdot)$ and $\alpha'(\cdot)$, 
	\begin{align*} 
		i \in \numedsetMinus{t} \Longrightarrow H(\tilde{F}_i |\tilde{G}_i) & > \alpha'(1/2-\zeta/2+\xi) \\ 
										    &= 1-\zeta+2\xi - \xi \\ &> 1-\zeta.
	\end{align*}
	Similarly, by~\eqref{eq_HFiGi in alpha beta plus} and the monotonicity of $\beta(\cdot)$ and $\beta'(\cdot)$, 
	\begin{align*} 
		i \in \numedsetPlus{t} \Longrightarrow H(\tilde{F}_i |\tilde{G}_i) &< \beta'(1/2+\zeta/2-\xi) \\ 
										    &= 1+\zeta-2\xi - 1  + \xi \\ &< \zeta.
	\end{align*}

	We have proved the claim for $\Hwind = h(\bv{c})$. The general case follows by monotonicity. That is, recall that the derivation above relies on repeated applications of the functions $\alpha,\alpha',\beta, \beta'$ in \Cref{cor_alpha beta prime prime}. These functions are monotone. Thus, since we have proved \eqref{eq_Hwind leq h(c) zeta} and~\eqref{eq_Hwind leq h(c) zeta strong} for  $i \in \medsetPlusBrace{t}$ when $\Hwind = h(\bv{c})$, this must also be the case for $\Hwind \leq h(\bv{c})$. The case $\Hwind \geq h(\bv{c})$ follows similarly.
\end{IEEEproof}

The following corollary of \Cref{prop_cascade of OTBST is monopolarizing} is the analog of \Cref{cor_OTBST monopolarizing} for the cascade case.
\begin{corollary}\label{cor_OTBST monopolarizing cascade}
    For a given $\zeta>0$, let $t$, $\bv{c}$, $L_0, M_0$, and $\nth$ be as in \Cref{prop_cascade of OTBST is monopolarizing}.
    Then, under the same setting as \Cref{prop_cascade of OTBST is monopolarizing}, for any $0 \leq \tau \leq 1$ and $n \geq \nth$ we have
    \begin{IEEEeqnarray}{rCl.l} \IEEEyesnumber \label{eq_Hwind leq geq h(c) + tau/2^t}
	    \Hwind \!\leq\! h(\bv{c})+ \frac{\tau}{2^t}\! &\Rightarrow& H(\tilde{F}_{i} | \tilde{G}_{i}) \!<\! 2\zeta +\tau,  & \forall i \!\in\! \medsetPlusBrace{t}, \IEEEeqnarraynumspace \IEEEyessubnumber\label{eq_Hwind leq geq h(c) + tau/2^t plus}\\
	    \Hwind \!\geq\! h(\bv{c})- \frac{\tau}{2^t}\! &\Rightarrow& H(\tilde{F}_{i} | \tilde{G}_{i}) \!>\! 1-2\zeta-\tau, &  \forall i \!\in\! \medsetMinusBrace{t}. \IEEEyessubnumber\IEEEeqnarraynumspace   \label{eq_Hwind geq h(c) + tau/2^t minus}
\end{IEEEeqnarray}
\end{corollary}
\begin{IEEEproof}
	Recall that in the cascade construction, when moving from stage $i < t$ to stage $i+1$, we operate only on the new medial indices in the active set $\sigma_i$, defined in \eqref{eq_def of active set}. The remaining new medial indices belong to the set $\bar{\sigma}_i$, defined as:
	\[
\bar{\sigma}_i =
\begin{cases}
	\numedsetMinus{i}, & \textrm{if} \;  c_i = 1, \\
	\numedsetPlus{i}, & \textrm{if} \;  c_i = 0.
\end{cases}
	\]
	By assumption, we are in a \biprocess setting. By \Cref{lem_all medial have the same entropy}, the conditional entropy corresponding to any index in $\sigma_i$ is the same, and hence we denote it by $a_i$. Similarly, we denote by $b_i$ the conditional entropy corresponding to an arbitrary index in $\bar{\sigma}_i$. For stage $t$, we denote by $a_t^-$ and $a_t^+$ the conditional entropies corresponding to indices in $\numedsetMinus{t}$ and $\numedsetPlus{t}$, respectively. Furthermore, conservation of conditional entropy holds, by \Cref{cor_medial plus minus delta}. That is,
	\begin{IEEEeqnarray*}{rCl'l}
		a_1 + b_1 &=& 2 \Hwind, \\
		a_i + b_i &=& 2a_{i-1},  & 2 \leq i < t, \\
		a_t^- + a_t^+ &=& 2a_{t-1}.
	\end{IEEEeqnarray*}
	From the above, we easily get by induction that
	\[
		2^t \Hwind = \left(a_t^-+ a_t^+\right)+ \sum_{i=1}^{t-1} 2^{t-i}b_i .  
	\]
	From the above and \eqref{eq_threshold entropy of a cascade}, we have
	\[
		\Hwind - h(\bv{c}) = \frac{a_t^- + a_t^+ - 1 + \sum_{i=1}^{t-1} 2^{t-i} (b_i - c_i) }{2^t} . 
	\]

	We now prove \eqref{eq_Hwind leq geq h(c) + tau/2^t plus}. If $\Hwind \leq h(\bv{c})$, then the result follows trivially from \eqref{eq_Hwind leq h(c) zeta} in \Cref{prop_cascade of OTBST is monopolarizing}. Hence, assume that
	\begin{equation}
		\label{eq_assumption on Hwind}
		h(\bv{c}) < \Hwind \leq h(\bv{c}) + \tau/2^t .
	\end{equation}
	By~\eqref{eq_assumption on Hwind},~\eqref{eq_Hwind geq h(c) 1-zeta strong}, and the definition of $\bar{\sigma}_i$, observe that if $c_i = 1$ then $b_i > 1-\zeta/2^t$. Moreover, 
	by~\eqref{eq_Hwind geq h(c) 1-zeta} and~\eqref{eq_assumption on Hwind}, $a_t^- > 1-\zeta$. Thus, 
	\begin{align*}
		\tau & \geq 2^t(\Hwind - h(\bv{c}) ) \\ 
		     & = a_t^- + a_t^+ - 1 + \sum_{i=1}^{t-1} 2^{t-i} (b_i - c_i) \\
		     & = \left( a_t^+ + \sum_{i, c_i = 0} 2^{t-i} b_i \right) + \left(a_t^- - 1 + \sum_{i, c_i=1} 2^{t-i}(b_i-1) \right) \\ 
		     & > \left( a_t^+ + \sum_{i, c_i = 0} 2^{t-i} b_i \right) -\left( 1+ \sum_{i, c_i=1} 2^{-i} \right)\zeta \\ 
		     & > \left( a_t^+ + \sum_{i, c_i = 0} 2^{t-i} b_i \right) - 2\zeta.
	\end{align*}
	Rearranging the above yields
	\[
		a_t^+ + \sum_{i, c_i = 0} 2^{t-i} b_i < 2\zeta + \tau .
	\]
	By the non-negativity of conditional entropy, this implies that $a_t^+ < 2\zeta + \tau$ and for $i$ such that $c_i=0$, we have $b_i < 2\zeta + \tau$. Finally, recalling the cascade construction, the definition of $\medsetPlusBrace{t}$, and the definition of $\bar{\sigma}_i$, the conditional entropy of any index in $\medsetPlusBrace{t}$ is either $a_t^{+}$ or some $b_i$ where $i$ is such that $c_i=0$. This 
	yields~\eqref{eq_Hwind leq geq h(c) + tau/2^t plus}. 
	
	The proof for~\eqref{eq_Hwind geq h(c) + tau/2^t minus} is similar. If $\Hwind \geq h(\bv{c})$, then the result follows trivially from \eqref{eq_Hwind geq h(c) 1-zeta} in \Cref{prop_cascade of OTBST is monopolarizing}. Hence, assume that
	\begin{equation}
		\label{eq_assumption on Hwind second direction}
		h(\bv{c}) - \tau/2^t  \leq \Hwind < h(\bv{c}).
	\end{equation}
	By~\eqref{eq_assumption on Hwind second direction},~\eqref{eq_Hwind leq h(c) zeta strong}, and the definition of $\bar{\sigma}_i$, observe that if $c_i = 0$ then $b_i < \zeta/2^t$. Moreover, 
	by~\eqref{eq_Hwind leq h(c) zeta} and~\eqref{eq_assumption on Hwind second direction}, $a_t^+ < \zeta$. Thus, 
	\begin{align*}
		-\tau & \leq 2^t(\Hwind - h(\bv{c}) ) \\ 
		     & = a_t^- + a_t^+ - 1 + \sum_{i=1}^{t-1} 2^{t-i} (b_i - c_i) \\
		     & = \left( a_t^+ + \sum_{i, c_i = 0} 2^{t-i} b_i \right) + \left(a_t^- - 1 + \sum_{i, c_i=1} 2^{t-i}(b_i-1) \right) \\ 
		     & < \left( \zeta + \sum_{i, c_i = 0} 2^{-i} \zeta \right) + \left(a_t^- - 1 + \sum_{i, c_i=1} 2^{t-i}(b_i-1) \right) \\ 
		     & < 2\zeta + \left( (a_t^- - 1) + \sum_{i, c_i=1} 2^{t-i}(b_i-1) \right).
	\end{align*}
	Rearranging the above yields
	\[
		- 2\zeta -\tau  <  (a_t^- - 1) + \sum_{i, c_i=1} 2^{t-i}(b_i-1).
	\]
Since conditional entropy is upper-bounded by $1$, we have $(a_t^- - 1) \leq 0$ and $(b_i - 1) \leq 0$. The above inequality thus implies that $(a_t^- -1) > -2\zeta - \tau$ and for $i$ such that $c_i=1$, we have $(b_i -1) > -2\zeta - \tau$. Finally, recalling the cascade construction, the definition of $\medsetMinusBrace{t}$, and the definition of $\bar{\sigma}_i$, the conditional entropy of any index in $\medsetMinusBrace{t}$ is either $a_t^{-}$ or some $b_i$ where $i$ is such that $c_i=1$. Simple rearranging yields~\eqref{eq_Hwind geq h(c) + tau/2^t minus}.
\end{IEEEproof}
	
The following is the analog of \Cref{prop_tilde non tilde} for the cascade case. Similar to \Cref{prop_tilde non tilde}, we state the following for a forgetful FAIM-derived \soprocess, $X_{\star} \sarrow Y_{\star}$, that satisfies \eqref{eq_conditions for universal set of processes} for some sequences $\bar{\epsilon}_L$, $\bar{\psi}_M$, and $\bar{\phi}_M$.

\begin{proposition}\label{prop_tilde non tilde cascade}
    Let the number of stages of the cascade $t$, and the binary vector $\bv{c}$ of length $t-1$ be given.
    Fix $n \geq 0$, $\varepsilon_1>0$, and $0 < \varepsilon_2 < \frac{1}{6}$. 
    There exist $\Lth$ and $\Mth$ such that for any $L_0 \geq \Lth, M_0 \geq \Mth$, 
    a $(t,\bv{c}; L_0, M_0, n)$-cascade satisfies:
    \begin{IEEEeqnarray*}{rCl} 
    |H(F_{i}|G_{i}) - H(\tilde{F}_{i} | \tilde{G}_{i})| &\leq& 2 \varepsilon_1 + \frac{\varepsilon_2}{2} - 3 \varepsilon_2 \log{\frac{3\varepsilon_2}{2}} \\  &<& 2\varepsilon_1 + \sqrt{8\varepsilon_2}.
\end{IEEEeqnarray*}
    Furthermore, we have
    \begin{equation*} 
	    |H(F_{i}|G_{i}, S_0, S_{N_n}) - H(\tilde{F}_{i} | \tilde{G}_{i})| \leq 2 \varepsilon_1 + \frac{\varepsilon_2}{2} - 3 \varepsilon_2 \log{\frac{3\varepsilon_2}{2}}. 
    \end{equation*}
\end{proposition}

    \begin{IEEEproof}
	    The proof follows along the same lines as \Cref{prop_tilde non tilde}. In the cascade case, the base-vector $\bv{b}$ of index $i$ is of length $2^{n \cdot t}$ as opposed to $2^{n}$ in the non-cascade case. This vector holds a single medial index from each of the \bblocks on the RHS of the transform, just as for the non-cascade case. All that remains are minor adaptations to the proof, to account for this change. That is, throughout the proof, including in the underlying lemmas, we make the following changes. 
	    \begin{itemize}
		    \item Replace $2^n$ and $2^{-n}$ with $2^{n \cdot t}$ and $2^{-n \cdot t}$, respectively.
		    \item Replace $N_n$ with $\Ncascade{t}$.
		    \item Extend $\tilde{f}_{n,i}$ and $\tilde{g}_{n,i}$ to the cascade case according to the construction in \Cref{sec_monopolarization for FAIM derived processes cascade}.
	    \end{itemize} 
	    This completes the proof.
    \end{IEEEproof}

\begin{IEEEproof}[Proof of \Cref{thm_cascade of BSTs}]
    Choose $\varepsilon_1>0$ and $0 < \varepsilon_2 < \frac{1}{6}$ small enough such that 
    \begin{equation} \label{eq_choice of zeta}
	    \zeta \triangleq \eta - 2^{t+1}\varepsilon_1 - (2\varepsilon_1 + \sqrt{8\varepsilon_2}) > 0. 
    \end{equation} 
    For example, one may take
    \begin{IEEEeqnarray*}{rCl}
        \varepsilon_1 &<& \frac{\eta}{4(1+2^t)}, \\[0.1cm]
        \varepsilon_2 &<& \frac{\eta^2}{32} . 
    \end{IEEEeqnarray*}

    Take $\nth$ large enough so that \Cref{prop_cascade of OTBST is monopolarizing} holds with $\zeta$ as above. 
    Recall that \Cref{prop_cascade of OTBST is monopolarizing} holds for any $L_0$ and $M_0$, so we are free to set them as desired. 

    By \Cref{prop_tilde non tilde cascade}, for $\nth$, $\varepsilon_1$, and $\varepsilon_2$ above, there exist $\Lth$ and $\Mth$ such that~\eqref{eq_HFiGi tilde
    inequality} holds for $L_0 \geq \Lth$ and $M_0 \geq \Mth$.
    That is, 
    \begin{IEEEeqnarray}{r}
        -(2\varepsilon_1 + \sqrt{8\varepsilon_2}) \leq H(F_i|G_i) - H(\tilde{F}_i|\tilde{G}_i) \leq (2\varepsilon_1 + \sqrt{8\varepsilon_2}).
        \IEEEeqnarraynumspace \label{eq_HFGtilde eps cascade}
    \end{IEEEeqnarray}
    In fact, we choose $L_0 \geq \Lth$ as in the proof of \Cref{lem_lean bblock} (adapted to the cascade case as detailed in the proof of~\Cref{prop_tilde non tilde cascade}).
    This ensures that the $L_0$-forgetfulness of the \soprocess is upper-bounded by $\varepsilon_1$. 
    Thus, by \Cref{cor_entropy rate with finite memory length},~\eqref{eq_H0 and ENTI difference} holds with $\epsilon \leq \varepsilon_1$, so that
    \[ 
        -2 \varepsilon_1 \leq \ENT{X_{\star} | Y_{\star}} - \Hwind \leq 2\varepsilon_1. 
    \] 
    Hence, if $\ENT{X_{\star} | Y_{\star}} \leq h(\bv{c})$ then $\Hwind \leq h(\bv{c})+2\varepsilon_1$ and if $\ENT{X_{\star} | Y_{\star}} \geq h(\bv{c})$ then
    $\Hwind \geq h(\bv{c})-2\varepsilon_1$. 
    Consequently, by \Cref{cor_OTBST monopolarizing cascade} with $\tau = 2^{t+1}\varepsilon_1$, if $n \geq \nth$ then 
    \begin{IEEEeqnarray*}{rCl/l}
        \ENT{X_{\star}|Y_{\star}} \leq h(\bv{c}) &\Rightarrow& H(\tilde{F}_i | \tilde{G}_i) < 2\zeta + \tau,  &\forall i \in \medsetPlus{n}, \\
        \ENT{X_{\star}|Y_{\star}} \geq h(\bv{c}) &\Rightarrow& H(\tilde{F}_i | \tilde{G}_i) > 1- 2\zeta - \tau, &\forall i \in \medsetMinus{n}.
    \end{IEEEeqnarray*}
    Combining the above with~\eqref{eq_choice of zeta} and~\eqref{eq_HFGtilde eps cascade} we obtain that for $n \geq \nth$, 
    \begin{IEEEeqnarray*}{rCl/l}
	\ENT{X_{\star}|Y_{\star}} \leq h(\bv{c}) &\Rightarrow& H(F_i | G_i) < \eta, &\forall i \in \medsetPlus{n}, \\
        \ENT{X_{\star}|Y_{\star}} \geq h(\bv{c}) &\Rightarrow& H(F_i | G_i) > 1- \eta, &\forall i \in \medsetMinus{n}.
    \end{IEEEeqnarray*}
    This completes the proof. 
\end{IEEEproof}

\section{Decoding the Universal Polar Code} \label{sec_decoding}
The universal polar code consists of a concatenation of a \BST cascade and \arikan's seminal codes, that is fast transforms. 
Ultimately, the code consists of recursive applications of \arikan transforms, which can be decoded efficiently using successive-cancellation decoding. 
The difference between the slow and fast stages lies in the order in which the \arikan transforms are connected.
Therefore, both the slow and fast polarization stages are decoded using successive-cancellation decoding, performed in lockstep. 

Specifically, the decoder estimates the transformed bits (the $\hat{U}$ in \Cref{fig_notation for how to construct}) in succession, assuming previous decoding decisions are correct. 
To decode a symbol, the decoder computes its likelihood ratio; this is performed recursively. 
If the symbol is ``frozen,'' the decoder returns its frozen value.
In a non-symmetric case, this might employ some common randomness shared between the encoder and decoder, see~\cite{Honda_Yamamoto_2013} for details.

Due to the memory in the \soprocess, the recursive computation of likelihoods is done via the successive-cancellation trellis decoding
of~\cite{wang2014joint} and~\cite{Wang_2015}. 
In this variation of successive-cancellation decoding, the decoder is cognizant of the existence of an underlying state connecting two blocks, and averages over it when
computing likelihoods. 
This results in a slight increase in complexity; in a seminal polar code, when there are $|\mathcal{S}|$ states and the code length is $\hat{N}$, the decoding
complexity is $O(|\mathcal{S}|^3 \hat{N}\cdot \log \hat{N})$, see~\cite[Theorem 2]{Wang_2015}

The overall codelength of the universal polar code is $ \Lambda = N\cdot \hat{N}$ (see \Cref{sec_fast stage}), so its decoding complexity using successive-cancellation
trellis decoding is $O(|\mathcal{S}|^3 \Lambda \cdot \log(\Lambda ))$. 
In the proof of~\Cref{thm_main} below, we show that the overall decoding error of this scheme is upper-bounded by $2^{-\Lambda^{\beta}}$ for any $\beta < 1/2$ and $\hat{N}$ large enough. 

As we will see in \Cref{sec_numerical results}, the decoding performance may be enhanced by using a successive cancellation list decoder. Such a decoder, tailored to the universal polar coding scheme of this paper is described in \cite{ShuvalTal:20c}. For a list of size $L$, the decoding complexity increases by a factor of $L$, with respect to (plain) successive cancellation decoding.

To prove~\Cref{thm_main} we will need some notation for the inputs and outputs of the slow and fast stages of the overall transform. 
The notation is illustrated in \Cref{fig_notation for how to construct}. The construction consists of a layer of $\hat{N}$ copies of a \BST cascade, each of length $N$, which is concatenated to a layer of $N$ fast transforms, each of length $\hat{N}$. A vector comprising the $j$th output of each \BST cascade is the input to fast transform $j$. Let $1 \leq \ell \leq \hat{N}$, $1 \leq i \leq \hat{N}$, and $ 1 \leq j \leq N$. Denote
\begin{align*}
    X_j^{(\ell)} &= X_{j+(\ell-1)N}, \\ 
    Y_j^{(\ell)} &= Y_{j+(\ell-1)N},  
\end{align*}
and their vector versions
\begin{align*}
    \bs{X}_{\ell} &= \begin{bmatrix} X_1^{(\ell)} & X_2^{(\ell)} & \cdots & X_N^{(\ell)} \end{bmatrix}, \\ 
    \bs{Y}_{\ell} &= \begin{bmatrix} Y_1^{(\ell)} & Y_2^{(\ell)} & \cdots & Y_N^{(\ell)} \end{bmatrix}.
\end{align*}
That is, the \soprocess relevant for BST cascade $\ell$ is $\bs{X}_{\ell} \sarrow \bs{Y}_{\ell}$. 
Output $j$ of BST cascade $\ell$ is $F_j^{(\ell)}$. 
    The input to fast transform $j$ is the vector
    \[ \bs{F}_j = \begin{bmatrix} F_j^{(1)} & F_j^{(2)} & \cdots & F_j^{(\hat{N})} \end{bmatrix}.\] 
Output $i$ of fast transform $j$ is $\hat{U}_i^{(j)}= \hat{U}_{i+(j-1)\hat{N}}$.  
The overall output of fast transform $j$ is the vector 
\[
    \hat{\bs{U}}_{j} = \begin{bmatrix}  \hat{U}_1^{(j)}  & \hat{U}_2^{(j)} & \cdots & \hat{U}_{\hat{N}}^{(j)}\end{bmatrix}.
\]
    Our notation for a vector of ordered elements (see \Cref{sec_notation conventions and reminders}) carriers over to an ordered vector of vectors, e.g., $\hat{\bs{U}}_a^b = \begin{bmatrix} \hat{\bs{U}}_a & \hat{\bs{U}}_{a+1} & \cdots & \hat{\bs{U}}_b \end{bmatrix}$. Moreover, $\hat{U}_1^{N\hat{N}} = \hat{\bs{U}}_1^{N}$.  

\begin{figure} 
    \begin{center}
\begin{tikzpicture}[>=latex]
    \node[draw, rectangle, minimum width = 1.6cm, minimum height = 1cm, text width = 0.9cm, align = center, label = above:{$\hat{N}$ copies}, font = \scriptsize, fill=green!30!white] (F0) at (0,0) {BST cascade $1$}; 
    \foreach \x/\y/\z in {2.5cm/1/$1$, 2.5cm/2/$\ell$, 5.0cm/3/$\hat{N}$}
    {
        \begin{scope}[yshift = -\x]
            \node[draw, rectangle, minimum width = 1.6cm, minimum height = 1cm, text width = 0.9cm, align = center, font = \scriptsize, fill=green!30!white] (F\y) at (0,0) {BST cascade \z}; 
        \end{scope}
    }

    \foreach \y/\z in {0/1,1/\ell,3/\hat{N}}
    {
        \node[right = 0.2 of  F\y] {$\bs{X}_{\z} \sarrow \bs{Y}_{\z}$}; 
    }
    \node[circle,draw,fill, minimum size = 2pt, inner sep = 0] (Cx) at ($(F0)!0.5!(F1)$){}; 
    \node[circle,draw,fill, minimum size = 2pt, inner sep = 0] at ($(F0)!0.7!(F1)$){}; 
    \node[circle,draw,fill, minimum size = 2pt, inner sep = 0] at ($(F0)!0.3!(F1)$){}; 
    
    \node[circle,draw,fill, minimum size = 2pt, inner sep = 0] (C) at ($(F1)!0.5!(F3)$){}; 
    \node[circle,draw,fill, minimum size = 2pt, inner sep = 0] at ($(F1)!0.7!(F3)$){}; 
    \node[circle,draw,fill, minimum size = 2pt, inner sep = 0] at ($(F1)!0.3!(F3)$){}; 

    \draw[<->] ($(F0.north east) + (0.1,0)$) -- node[right = -0.1, pos = 0.75]{\tiny $N$}($(F0.south east) + (0.1,0)$) ; 

    \node[draw, rectangle, minimum width = 1.6cm,minimum height = 1.1cm, text width = 1.0cm, left = 2 of F0, label=above:{$N$ copies}, align=center, font = \scriptsize, fill=orange!30!white ] (fa) {Fast transform $1$}; 
    \foreach \x/\y in {0/20,1/0,3/-20}
    {
        \draw[->] ($(F\x.north west)!0.1!(F\x.south west)$) --  (fa.\y); 
        \draw[->] (fa.180) -- +(-0.4,0); 
        \draw[->] (fa.210) -- +(-0.4,0); 
        \draw[->] (fa.150) -- +(-0.4,0); 
    }
    \draw[<->] ($(fa.north west) - (0.6,0)$) -- node[left=-0.1, pos = 0.75]{\tiny$\hat{N}$}($(fa.south west) - (0.6,0)$) ; 
    %
    \node[draw, rectangle, minimum width = 1.6cm,minimum height = 1.1cm, text width = 1.0cm, below = 1.2 of fa, align=center, font = \scriptsize, fill=orange!30!white] (fb) {Fast transform $j$}; 
    \foreach \x/\y/\z/\w in {0/20/1/0.05,1/0/\ell/0.05,3/-20/\hat{N}/-0.08}
    {
        \draw[->, red] ($(F\x.north west)!0.5!(F\x.south west)$) node[right=-0.1] {\tiny $j$}  --node[above=-\w, pos = 0.25]{\tiny $F_j^{(\z)}$} (fb.\y); 
        \draw[->, blue] (fb.180) node[right = -0.05 ] {\tiny $i$} -- +(-0.4,0) node[left = -0.1] {\tiny $\hat{U}_i^{(j)}$}; 
        \draw[->] (fb.210) -- +(-0.4,0); 
        \draw[->] (fb.150) -- +(-0.4,0); 
    }

    \node[draw, rectangle, minimum width = 1.6cm,minimum height = 1.1cm, text width = 1.0cm, below = 3.4 of fa, align=center, font = \scriptsize, fill=orange!30!white] (fc) {Fast transform $N$}; 
    \foreach \x/\y in {0/20,1/0,3/-20}
    {
        \draw[->] ($(F\x.north west)!0.9!(F\x.south west)$) -- (fc.\y); 
        \draw[->] (fc.180) -- +(-0.4,0); 
        \draw[->] (fc.210) -- +(-0.4,0); 
        \draw[->] (fc.150) -- +(-0.4,0); 
    }
    \foreach \x/\y in {fa/fb,fb/fc}
    {
    \node[circle,draw,fill, minimum size = 2pt, inner sep = 0] (C) at ($(\x)!0.5!(\y)$){}; 
    \node[circle,draw,fill, minimum size = 2pt, inner sep = 0] at ($(\x)!0.65!(\y)$){}; 
    \node[circle,draw,fill, minimum size = 2pt, inner sep = 0] at ($(\x)!0.35!(\y)$){}; 
    }
\end{tikzpicture}
\end{center}
\caption{Notation for inputs and outputs of the slow and fast stages in our universal construction.}
\label{fig_notation for how to construct}
\end{figure}

Note that under this notation, the decoder decodes the bit vectors $\hat{\bs{U}}_j$, $j = 1,\ldots,N$ in order using successive-cancellation trellis decoding. Specifically, it first decodes $\hat{\bs{U}}_1$, then $\hat{\bs{U}}_2$, and so on. The decoding order in a vector $\hat{\bs{U}}_j$ is as expected: $\hat{U}_1^{(j)}$, then $\hat{U}_2^{(j)}$, and so forth up to  $\hat{U}_{\hat{N}}^{(j)}$. Namely, when decoding $\hat{U}_i^{(j)}$, we do this after having decoded $\bs{\hat{U}}_1^{j-1}$ as well as $\hat{U}_{1}^{(j)},\hat{U}_{2}^{(j)},\ldots,\hat{U}_{i-1}^{(j)}$.

\begin{IEEEproof}[Proof of \Cref{thm_main}]
	Our proof is divided into four parts:
	\begin{enumerate}[label=\Roman*]
		\item \label{thm_main:part:defining coding scheme} Defining the coding scheme, i.e., the indices on which information bits are transmitted.
		\item \label{thm_main:part:complexity} Proving item \ref{thm_main:complexity} in the theorem statement: encoding and decoding complexity.
		\item \label{thm_main:part:error prob} Proving item \ref{thm_main:error probability} in the theorem statement: vanishing error probability and the error exponent.
		\item \label{thm_main:part:rate} Proving item \ref{thm_main:rate} in the theorem statement: the code rate $R$ is achievable.
	\end{enumerate}

	\emph{Part \ref{thm_main:part:defining coding scheme} -- Coding Scheme Definition}

	Our sequence of codes is parametrized by $R < I^*$ and $\beta < 1/2$, both given in the theorem statement. It is based on the universal construction of this paper. That is, it consists of a concatenation of a cascade of BSTs and a fast transform. To fully specify a member of this sequence, we must define the following:
	\begin{itemize}
		\item The slow transform parameters: $t,\bv{c}, L_0, M_0$, and $n$.
		\item The number of stages in the fast transform: $\hat{n}$.
		\item The set of indices $\candidateA$ over which information bits are transmitted.
	\end{itemize}
	The sequence is formed by increasing $\hat{n}$ and keeping the slow transform parameters fixed. The parameter $\hat{n}$ must be large enough; conditions on its size are given in this part, as well as in Parts~\ref{thm_main:part:error prob} and \ref{thm_main:part:rate}. All these conditions must be met.

	Let $\mathcal{C}$ be the family of \soprocesses. If $I^* = 0$, the theorem is trivially correct; hence, we assume that $I^* > 0$. Let $\bar{\epsilon}_L$, $\bar{\psi}_M$, and $\bar{\phi}_M$ be sequences such that \eqref{eq_conditions for universal set of processes} holds for any \soprocess in $\mathcal{C}$. Every \soprocess $X_{\star}\sarrow Y_{\star}$ in $\mathcal{C}$ has the same input distribution; we denote the entropy rate of the input process by $\ENT{X_{\star}}$. Denote
	\begin{equation}
		\label{eq_h ast definition}
h^* = \ENT{X_{\star}} - I^* .
	\end{equation}
	By definition of $I^*$, any \soprocess $X_{\star}\sarrow Y_{\star}$ in $\mathcal{C}$ satisfies $I^* \leq \ENT{X_{\star}} - \ENT{X_{\star}|Y_{\star}}$. In other words, we always have
	\begin{equation}
		\label{eq_h ast inequality}
		\ENT{X_{\star}|Y_{\star}} \leq h^* .
	\end{equation}
	Denote
	\begin{equation}
		\label{eq_def of rcdiff}
		\rcdiff = I^* - R.
	\end{equation}

	Looking forward, in part~\ref{thm_main:part:error prob} we will use the union bound to upper-bound the probability of error. Thus, we will require fast polarization, for which we turn to \Cref{prop_universal fast polarization}. All processes in $\mathcal{C}$ satisfy the Bhattacharyya recursion \eqref{eq_bhattacharyya recursion} with 
	\begin{equation} \label{eq_kappa thm 1} 
		\kappa = 2 \bar{\psi}_0, 
	\end{equation}
	see \eqref{eq_kappa} and \eqref{eq_conditions for universal set of processes psi}. Next, fix $\beta'$ such that 
	\begin{equation}
		\label{eq:betaBetaPrime}
		\beta < \beta' < 1/2
	\end{equation}
	and let
	\begin{equation}
		\label{eq:rcdiffprime}
		\rcdiffprime = \rcdiff/5 .
	\end{equation}
	By \Cref{prop_universal fast polarization}, there exist $\eta$ and $n_0$ such that if $Z_0 \leq \eta$, then \eqref{eq_fast polarization z eta} below holds with $\beta'$ and $\delta'$ in place of $\beta$ and $\delta$. This will be utilized in Part~\ref{thm_main:part:rate}. Parameters $\eta$ and $n_0$ are used as follows. 
	\begin{itemize}
		\item Parameter $\eta$ will be the monopolarization goal for the BST cascade in \Cref{thm_cascade of BSTs}. 
		\item Parameter $n_0$ will be one of the lower bounds on the number of fast polarization steps, $\hat{n}$. 
\end{itemize}

	To apply \Cref{thm_cascade of BSTs}, we now set the parameters $t$ and $\bv{c}$.
	We set $t$ and $\bv{c}$ such that 
	\begin{equation}
		\label{eq_hc bounds}
		h^* \leq h(\bv{c}) \leq h^* + \rcdiffprime.
	\end{equation}
	Indeed, this can be done by \eqref{eq_threshold entropy of a cascade}, recalling that $I^* > 0$ implies by \eqref{eq_h ast definition} that $h^* < 1$.

	Recall from \Cref{def_tcLMn cascade} that a cascade of BSTs is defined by five parameters: $t,\bv{c}, L_0, M_0$, and $n$. We have already set $t$ and $\bv{c}$. For the remaining parameters, we will utilize \Cref{thm_cascade of BSTs}. Namely, given $t, \bv{c}$, and $\eta$, there exist $\Lth, \Mth$, and $\nth$ (see \Cref{rem_bst parameters for monopolarization}) such that if $L_0 \geq \Lth$, $M_0 \geq \Mth$, and $n \geq \nth$ then
	a $(t,\bv{c}; L_0, M_0, n)$-cascade is $(\eta, \medsetPlusBrace{t}, \medsetMinusBrace{t})$-monopolarizing, for any \soprocess in $\mathcal{C}$. Specifically, since $\ENT{X_{\star}|Y_{\star}} \leq h(\bv{c})$ by \eqref{eq_h ast inequality} and \eqref{eq_hc bounds}, the medial-plus indices are the ones that monopolarize. Thus, due to \Cref{prop_universal fast polarization}, after the fast transform almost all of them will have a very low Bhattacharyya parameter. 

	We set $n=\nth$. To set $L_0$ and $M_0$, we further call upon \Cref{lem_fraction of medial minus indices out of medial indices in a cascade}. Namely,   we set $L_0 = \Lth$ and take $M_0 \geq \Mth$ large enough such that the fraction of medial-plus indices out of all indices in our BST cascade is at least 
	\begin{equation}\label{eq_lower bound on med plus in thm 1}
		\frac{|\medsetPlusBrace{t}|}{N}\geq 1 - h(\bv{c}) - \rcdiffprime.
	\end{equation}
	We can do this due to the combination of \eqref{eq_full expression for med ratio} and \eqref{eq_full expression for med minus cup med plus}, and by recalling the observation following \eqref{eq_full expression for med minus cup med plus}.

	Up to this point, we have defined all the parameters of the BST cascade: $t, \bv{c}, L_0, M_0, n$. Recall that the sequence is formed by increasing $\hat{n}$, and that $\hat{n}$ must satisfy a set of conditions that we have yet to fully specify. For now, consider some $\hat{n}$ large enough.

	Using \eqref{eq_length of a (t,c,L0,M0,n) cascade}, let $N = (2L_0+M_0)M_0^{t-1}2^{nt}$ and denote
	\begin{equation} \label{eq_Nhat definition} 
		\hat{N} = 2^{\hat{n}}. 
	\end{equation}
		The total codelength is thus
	\begin{equation} \label{eq_length of code in sequence} 
		\Lambda = N \cdot \hat{N} = (2L_0+M_0)M_0^{t-1}2^{nt} \cdot 2^{\hat{n}}.
	\end{equation}
	
	 We now define the set $\candidate$ as all indices $k = i + (j-1) \hat{N}$, $1 \leq i \leq \hat{N}$, $1 \leq j \leq N$ such that the following conditions hold:
	 \begin{enumerate}
		 \item $j \in \medsetPlusBrace{t}$, where we recall that $\medsetPlusBrace{t}$ is the medial-plus set for our $(t,\bv{c}; L_0, M_0, n)$-cascade of \BSTs;
		 \item $\bar{Z}_{\hat{n}} \leq 2^{-\hat{N}^{\beta'}}$, where $\bar{Z}_0 = \eta$, and the process $\bar{Z}$ satisfies
			 \eqref{eq_bhattacharyya recursion} with equality instead of inequality, $\kappa=2\bar{\psi}_0$ (see 
			 \eqref{eq_kappa thm 1}), and we take $B_{m+1}$ as the $m$th bit in the binary representation of $i$.
	 \end{enumerate}

	 \begin{remark} \label{rem:D is a good set in terms of Z}
	Observe that if $k \in \candidate$, then for any \soprocess in $\mathcal{C}$, we have $Z(\hat{U}_k | \hat{U}_1^{k-1}, Y_1^\Lambda) \leq 2^{-\hat{N}^{\beta'}}$. This follows from the second condition in the definition of $\candidate$, which upper-bounds the Bhattacharyya process for any \soprocess in $\mathcal{C}$.
\end{remark}

	We now partition $\candidate$ into three disjoint sets, $\candidateA$, $\candidateB$, and $\candidateC$.
	\begin{IEEEeqnarray*}{rCl}
		\candidateA &=& \left\{ k \in \candidate : K(\hat{U}_k | \hat{U}_1^{k-1}) \leq 2^{-{\hat{N}}^{\beta'}}  \right\} , \\
		\candidateB &=& \left\{ k \in \candidate : K(\hat{U}_k | \hat{U}_1^{k-1}) \geq 1- 2^{-{\hat{N}}^{\beta'}}  \right\} , \\
		\candidateC &=& \left\{ k \in \candidate : 2^{-{\hat{N}}^{\beta'}} < K(\hat{U}_k | \hat{U}_1^{k-1}) < 1- 2^{-{\hat{N}}^{\beta'}}  \right\} ,
	\end{IEEEeqnarray*}
	where $K$ is the total variation distance, see \cite[Definition 3]{ShuvalTal:19.2p}.
	Thus,
	\begin{equation} \label{eq_partitioning of D by size}
|\candidate| = |\candidateA| + |\candidateB| + |\candidateC| .
	\end{equation}
	Following the Honda-Yamamoto scheme \cite{Honda_Yamamoto_2013}, we take $\candidateA$ as the set of information indices.

	\emph{Part \ref{thm_main:part:complexity} -- Complexity}

	We have already discussed the decoding complexity of our scheme in the beginning of this section, and showed that it is $O(|\mathcal{S}|^3 \Lambda \cdot \log(\Lambda ))$. This is also the encoding complexity, since encoding uses the successive cancellation trellis algorithm \cite{Wang_2015} as well. This proves item \ref{thm_main:complexity} in the theorem statement.

	\emph{Part \ref{thm_main:part:error prob} -- Error Probability}

	To upper-bound the probability of error, we will use the union bound. For this, we will need a second condition on $\hat{n}$: it is large enough such that
	\begin{equation}
		2^{-\hat{N}^{\beta'}} \leq \frac{1}{2\Lambda} 2^{-\Lambda^\beta}. \label{eq_towardsAUnionBound Z}  
	\end{equation}
It is possible to satisfy \eqref{eq_towardsAUnionBound Z} for all $\hat{n}$ large enough since $\beta' > \beta$ by \eqref{eq:betaBetaPrime}, $N$ is fixed, and $\Lambda = N\cdot\hat{N}$ by~\eqref{eq_length of code in sequence}. 

Recall that the input distribution is fixed over the set $\mathcal{C}$. Denote this input distribution by $P_{X_1^\Lambda}$. Since $\hat{U}_1^\Lambda$ is the result of a transform we have specified over $X_1^\Lambda$, we denote the corresponding distribution of $\hat{U}_1^\Lambda$ as $P_{\hat{U}_1^\Lambda}$.

The claim on the probability of error follows from~\cite{Honda_Yamamoto_2013}; for completeness we show this directly here. 
\begin{itemize}
	\item We encode $\hat{u}_k$ sequentially.
		\begin{itemize}
	\item If $k \in \candidateA$, encode the next information bit into $\hat{u}_k$. We assume that the information bits are i.i.d.\ Bernoulli$(1/2)$.
	\item If $k \notin \candidateA$, we draw $\hat{u}_k$ according to a binary random variable with distribution $P_{\hat{U}_k|\hat{U}_1^{k-1}}( u | \hat{u}_1^{k-1} )$.
\end{itemize}
\item This results in a $\hat{u}_1^\Lambda$ that is sampled from a distribution $Q_{\hat{U}_1^\Lambda}$ that is different from the distribution $P_{\hat{U}_1^\Lambda}$. 
\item By \Cref{rem:D is a good set in terms of Z} and \eqref{eq_towardsAUnionBound Z}, had $\hat{u}_1^\Lambda$ been sampled from $P_{\hat{U}_1^\Lambda}$, the union bound would yield a total probability of error of at most $\frac{1}{2} 2^{-\Lambda^\beta}$.
\item We abuse notation and denote by $P(\hat{u}_1^\Lambda,y_1^{\Lambda})$ the joint probability of $\hat{u}_1^\Lambda$ sampled from $P_{\hat{U}_1^\Lambda}$ and $y_1^\Lambda$ the resulting channel output. Similarly, $Q(\hat{u}_k,y_1^{\Lambda})$ denotes this joint probability if $\hat{u}_1^\Lambda$ were sampled from $Q_{\hat{U}_1^\Lambda}$. Under this notation, from the previous bullet, we have
	\begin{equation}
		\label{eq:p error from distribution P}
		\sum_{\hat{u}_1^\Lambda, y_1^\Lambda} P(\hat{u}_1^\Lambda, y_1^\Lambda) \cdot \left[\mathrm{Dec}(y_1^\Lambda) \neq \hat{u}_1^\Lambda \right] \leq \frac{1}{2} 2^{-\Lambda^\beta} ,
	\end{equation}
	where $[ \cdot ]$ is the Iverson notation \cite[p. 11]{Iverson:62b}: equal to $1$ if the condition holds and to $0$ otherwise, and `$\mathrm{Dec}$' is the successive cancellation decoder.

\item The true probability of error of this scheme, i.e., under $Q$, is given by
	\begin{IEEEeqnarray*}{rCl}
		P_\mathrm{e} &=& \sum_{\hat{u}_1^\Lambda, y_1^\Lambda} Q(\hat{u}_1^\Lambda, y_1^\Lambda) \cdot \left[\mathrm{Dec}(y_1^\Lambda) \neq \hat{u}_1^\Lambda \right] \\
			     & = & \sum_{\hat{u}_1^\Lambda, y_1^\Lambda} P(\hat{u}_1^\Lambda, y_1^\Lambda) \cdot \left[\mathrm{Dec}(y_1^\Lambda) \neq \hat{u}_1^\Lambda \right] \\
			     & & {} + \sum_{\hat{u}_1^\Lambda, y_1^\Lambda} \left(Q(\hat{u}_1^\Lambda, y_1^\Lambda) - P(\hat{u}_1^\Lambda, y_1^\Lambda)\right) \cdot \left[\mathrm{Dec}(y_1^\Lambda) \neq \hat{u}_1^\Lambda \right] \\
			     & \leq &  \frac{1}{2} 2^{-\Lambda^\beta}  + \sum_{\hat{u}_1^\Lambda, y_1^\Lambda} \left|Q(\hat{u}_1^\Lambda, y_1^\Lambda) - P(\hat{u}_1^\Lambda, y_1^\Lambda)\right| ,
	\end{IEEEeqnarray*}
	where the inequality is by \eqref{eq:p error from distribution P}, and since $[ \cdot ] \leq 1$.
\item Following \cite[eq. 57]{Honda_Yamamoto_2013} (see also \cite[Lemma 3.5]{korada2009}), we can bound the second term above as
	\begin{IEEEeqnarray*}{rCl}
		\IEEEeqnarraymulticol{3}{l}{\sum_{\hat{u}_1^\Lambda, y_1^\Lambda} \left|Q(\hat{u}_1^\Lambda, y_1^\Lambda) - P(\hat{u}_1^\Lambda, y_1^\Lambda)\right| } \\
		\quad & \eqann[\leq]{a} & \sum_{k \in \candidateA} \sum_{\hat{u}_1^{k}} P(\hat{u}_1^{k-1}) \left| Q(\hat{u}_k|\hat{u}_1^{k-1}) - P(\hat{u}_k|\hat{u}_1^{k-1}) \right| \\
		      & \eqann{b} & \sum_{k \in \candidateA} \sum_{\hat{u}_1^{k-1}} P(\hat{u}_1^{k-1}) \sum_{\hat{u}_k}  \left| \frac{1}{2} - P(\hat{u}_k|\hat{u}_1^{k-1}) \right| \\ 
		      & \eqann{c} &  \sum_{k \in \candidateA} \sum_{\hat{u}_1^{k-1}} P(\hat{u}_1^{k-1})  \left| P(0|\hat{u}_1^{k-1})- P(1|\hat{u}_1^{k-1}) \right| \\ 
		      & \eqann{d} &  \sum_{k \in \candidateA} K(\hat{U}_k | \hat{U}_1^{k-1})  \\ 
		      & \eqann[\leq]{e} &  |\candidateA| 2^{-\hat{N}^{\beta'}} \\ 
		      & \leq & \Lambda \cdot 2^{-\hat{N}^{\beta'}} \\ 
		      & \eqann[\leq]{f} & \frac{1}{2}2^{-\Lambda^\beta},
	\end{IEEEeqnarray*} 
	where in \eqannref{a} we abused notation and used $P$ and $Q$ to denote $P_{\hat{U}_1^{k-1}}$, $Q_{\hat{U}_k | \hat{U}_1^{k-1}}$, and $P_{\hat{U}_k | \hat{U}_1^{k-1}}$ ; in \eqannref{b} we recalled from the topmost bullet that for $k \in \candidateA$ we draw $\hat{u}_k$ from a Bernoulli$(1/2)$ random variable; in \eqannref{c} we used the equality $1/2=(P(0|\hat{u}_1^{k-1}) + P(1|\hat{u}_1^{k-1}))/2$; in \eqannref{d} we used \cite[Definition 3]{Shuval_Tal_Memory_2017}; \eqannref{e} follows from the definition of $\candidateA$; finally, \eqannref{f} is by~\eqref{eq_towardsAUnionBound Z}.
\item Combining the above two bullets, we obtain
	\[ P_{\mathrm{e}} \leq 2^{-\Lambda^\beta}. \] 
\end{itemize}
This proves item \ref{thm_main:error probability} in the theorem statement.

	\emph{Part \ref{thm_main:part:rate} -- Rate}

	Our goal is to show that 
	\begin{equation} \label{eq: lower bound on size of candidate A}
		|\candidateA| \geq \Lambda R. 
	\end{equation}

	 The size of $\candidate$ is lower-bounded by 
	 \begin{IEEEeqnarray}{rCl} \label{eq_lower bound on D}
		|\candidate| & \eqann[\geq]{a} & N \cdot (1 - h(\bv{c}) - \rcdiffprime) \cdot \hat{N} \cdot (1-\rcdiffprime) \IEEEnonumber \\
			     & \eqann{b} & \Lambda (1 - h(\bv{c}) - \rcdiffprime) \cdot (1-\rcdiffprime)  \IEEEnonumber\\
			     & \eqann[>]{c} & \Lambda (1 - h(\bv{c}) - 2 \rcdiffprime)  \IEEEnonumber\\
			     & \eqann[\geq]{d} & \Lambda (1 - h^* - 3 \rcdiffprime)  \IEEEnonumber\\
			     & \eqann{e} & \Lambda (I^*  + 1 - \mathcal{H}(X_\star)  - 3 \rcdiffprime) ,  \IEEEyesnumber
	\end{IEEEeqnarray}
	where
	\begin{itemize}
		\item \eqannref{a} is by \eqref{eq_lower bound on med plus in thm 1} and by~\eqref{eq_fast polarization z eta} with $\beta'$ and $\rcdiffprime$ in place of $\beta$ and $\rcdiff$, recalling~\eqref{eq_kappa thm 1} and the discussion following~\eqref{eq:rcdiffprime}.
		\item \eqannref{b} is by \eqref{eq_length of code in sequence}.
		\item \eqannref{c} is since $\delta' > 0$. See \eqref{eq:rcdiffprime}, \eqref{eq_def of rcdiff}, and the text preceding \eqref{eq_def of rcdiff}.
		\item \eqannref{d} is by \eqref{eq_hc bounds}.
		\item \eqannref{e} is by \eqref{eq_h ast definition}.
	\end{itemize}

	 To lower-bound the size of $\candidateA$, we define the sets $\candidateA'$, $\candidateB'$, $\candidateC'$, which are defined similarly to $\candidateA$, $\candidateB$, and $\candidateC$, but we do not limit the indices $k$ to be from $\candidate$. That is,
	\begin{IEEEeqnarray*}{rCl}
		\candidateA' &=& \left\{1 \leq k \leq \Lambda : K(\hat{U}_k | \hat{U}_1^{k-1}) \leq 2^{-{\hat{N}}^{\beta'}}  \right\} , \\
		\candidateB' &=& \left\{1 \leq k \leq \Lambda : K(\hat{U}_k | \hat{U}_1^{k-1}) \geq 1- 2^{-{\hat{N}}^{\beta'}}  \right\} , \\
		\candidateC' &=& \left\{1 \leq k \leq \Lambda : 2^{-{\hat{N}}^{\beta'}} < K(\hat{U}_k | \hat{U}_1^{k-1}) < 1- 2^{-{\hat{N}}^{\beta'}}  \right\} .
	\end{IEEEeqnarray*}
	Clearly, $|\candidateB| \leq |\candidateB'|$ and $|\candidateC| \leq |\candidateC'|$, thus 
	\begin{equation}
		\label{eq:B and C prime bound B and C}
	|\candidateB| + |\candidateC| \leq |\candidateB'| + |\candidateC'|. 
\end{equation}
	We now prove that
	\begin{equation}
		\label{eq:bound on size of B prime plus C prime}
|\candidateB'| + |\candidateC'| \leq \Lambda ( 1 - \mathcal{H}(X_\star)  + 2 \rcdiffprime) .
\end{equation}

First, let us show that for $\hat{n}$ large enough,
\begin{equation} \label{eq:upper bound on size of C prime}
|\candidateC'| \leq \Lambda \rcdiffprime .
\end{equation}
Recall from \Cref{fig_notation for how to construct} that the transform consists of a concatenation of copies of BST cascades with fast transforms. Specifically, for fixed $1 \leq j \leq N$, all indices of the form $k = i + (j-1) \hat{N}$, $1 \leq i \leq \hat{N}$ belong to the same fast transform. We claim that for a fixed $j$ there exists an $\hat{n}$ large enough such that the fraction of such indices $k$ belonging to $\candidateC'$ is at most $\rcdiffprime$. Indeed, this follows from fast polarization results in \cite{ShuvalTal:19.2p}. To see this, first denote $Z_k = Z(\hat{U}_k | \hat{U}_1^{k-1})$ and $K_k = K(\hat{U}_k | \hat{U}_1^{k-1})$. From \cite[(4a) and (4b)]{ShuvalTal:19.2p}, if $Z_k \leq 2^{-\hat{N}^\beta}$ then
\[
	K_k \geq 1- Z_k \geq 1- 2^{-\hat{N}^\beta} .
\]
Combining the above with (28) in \cite[Theorem 7]{ShuvalTal:19.2p} and the first two displayed equations in the proof of \cite[Theorem 13]{ShuvalTal:19.2p}, we see the claim.

Now, to see \eqref{eq:upper bound on size of C prime}, note that since we have fixed $N$, the number of possible indices $j$ is finite. Hence, we take $\hat{n}$ as being at least the maximum $\hat{n}$ over all such $j$. This yields \eqref{eq:upper bound on size of C prime}.

We now show that for $\hat{n}$ large enough,
\begin{equation} \label{eq:lower bound of size of A prime}
|\candidateA'| \geq \Lambda \left( \mathcal{H}(X_\star) - 2 \rcdiffprime \right) .
\end{equation}
Indeed, this follows by 
\begin{IEEEeqnarray*}{rCl}
	\IEEEeqnarraymulticol{3}{l}{\Lambda \mathcal{H}(X_\star)} \\
	 \quad &\eqann[\leq]{a} & \sum_{k=1}^\Lambda H(\hat{X}_k| \hat{X}_1^{k-1}) \\
	 &\eqann{b} & \sum_{k=1}^\Lambda H(\hat{U}_k| \hat{U}_1^{k-1}) \\
			     &=& \sum_{k \in \candidateA'} H(\hat{U}_k| \hat{U}_1^{k-1}) + \sum_{k \in \candidateB'} H(\hat{U}_k| \hat{U}_1^{k-1}) + \sum_{k \in \candidateC'} H(\hat{U}_k| \hat{U}_1^{k-1}) \\
			     & \eqann[\leq]{c} & |\candidateA'| + \sum_{k \in \candidateB'} H(\hat{U}_k| \hat{U}_1^{k-1}) + |\candidateC'| \\
			     & \eqann[\leq]{d} & |\candidateA'| + \sum_{k \in \candidateB'} H(\hat{U}_k| \hat{U}_1^{k-1}) + \Lambda \rcdiffprime \\
			     & \eqann[\leq]{e} & |\candidateA'| + |\candidateB'| \rcdiffprime + \Lambda \rcdiffprime \\
			     & \leq & |\candidateA'| + \Lambda \rcdiffprime + \Lambda \rcdiffprime \\
			     & = & |\candidateA'| + 2 \Lambda \rcdiffprime ,
\end{IEEEeqnarray*}
where:
\begin{itemize}
	\item \eqannref{a} follows from \cite[Theorem 4.2.2]{cover_thomas}, which states that the summands in the series are a non-increasing sequence with limit $\mathcal{H}(X_\star)$. 
	\item \eqannref{b} follows since our overall transform is invertible.
	\item \eqannref{c} follows since the summed entropies are upper bounded by $1$.
	\item \eqannref{d} follows from \eqref{eq:upper bound on size of C prime}.
	\item To see \eqannref{e}, we utilize \cite[(4c)]{ShuvalTal:19.2p}. That is, the inequality
		\[
			H(\hat{U}_k | \hat{U}_1^{k-1}) \leq \sqrt{1 - (K(\hat{U}_k | \hat{U}_1^{k-1}))^2} .
		\]
		Since for every $k \in \candidateB'$ we have $K(\hat{U}_k | \hat{U}_1^{k-1}) \geq 1- 2^{-{\hat{N}}^{\beta'}}$, we can take $\hat{N}$ large enough such that $H(\hat{U}_k | \hat{U}_1^{k-1}) \leq \delta'$.
\end{itemize}
Rearranging yields \eqref{eq:lower bound of size of A prime}.

Finally, to obtain~\eqref{eq:bound on size of B prime plus C prime}, observe that 
\[ 
\Lambda = |\candidateA'| + |\candidateB'| + |\candidateC'|.
\]
Thus, using~\eqref{eq:lower bound of size of A prime} we obtain
\[
	|\candidateB'|+|\candidateC'| \leq \Lambda( 1 - \mathcal{H}(X_{\star}) + 2 \rcdiffprime), 
\] 
which is~\eqref{eq:bound on size of B prime plus C prime}.

Having proved \eqref{eq:bound on size of B prime plus C prime}, we utilize \eqref{eq_partitioning of D by size} to conclude that 
	\begin{IEEEeqnarray*}{rCl}
	|\candidateA| & = & |\candidate| - |\candidateB| - |\candidateC| \\ 
		      &\eqann[\geq]{a} & |\candidate| - |\candidateB'| - |\candidateC'| \\ 
		      &\eqann[\geq]{b} & |\candidate| - \Lambda(1-\mathcal{H}(X_\star) + 2\rcdiffprime) \\ 
		      &\eqann[\geq]{c} & \Lambda(I^* + 1 - \mathcal{H}(X_\star) - 3\rcdiffprime) - \Lambda(1-\mathcal{H}(X_\star) + 2\rcdiffprime) \\ 
		      &=& \Lambda(I^* - 5\rcdiffprime) \\ 
		      &\eqann{d}& \Lambda (I^* - \rcdiff) \\
		      &\eqann{e}& \Lambda R ,
	\end{IEEEeqnarray*}
	where:
	\begin{itemize}
		\item \eqannref{a} is by~\eqref{eq:B and C prime bound B and C}.
		\item \eqannref{b} is by~\eqref{eq:bound on size of B prime plus C prime}. 
		\item \eqannref{c} is by~\eqref{eq_lower bound on D}.
		\item \eqannref{d} is by~\eqref{eq:rcdiffprime}.
		\item \eqannref{e} is by~\eqref{eq_def of rcdiff}.
	\end{itemize}

	Thus, the size of $\candidateA$ is lower-bounded by $R$, proving item \ref{thm_main:rate} in the theorem statement, and completing the proof. 
\end{IEEEproof}

\section{How to Construct Universal Polar Codes} \label{sec_construction}

In this section, we explain how one would construct universal polar codes in practice. Until this point, we have shown that a set of processes with memory that satisfy certain conditions can be decoded using a universal polar code. The code's parameters depend on these conditions and must be sufficiently large to attain a specified coding rate and error probability. A naive approach would be to set these parameters using \Cref{rem_bst parameters for monopolarization} and \Cref{sec_fast stage}. However, this might needlessly result in an impracticably large codelength --- recall \Cref{ex_our bounds are weak} that suggests a BST comprising of $40162$ layers, whereas in fact just $10$ layers suffice.

In what follows, for simplicity, we consider a universal transform whose slow stage is a single BST layer (i.e., a cascade with $t=1$ layers). Generalizing to cascades with larger $t$ is straightforward.

In a practical application, the code is to be used over some set of \soprocesses $\mathcal{C}$. All processes in $\mathcal{C}$ share the same fixed input distribution, which is known. We are further given a size constraint on the codelength.

To continue, we first decide the code parameters:
\begin{itemize}
    \item The parameters $L_0$ and $M_0$ that respectively define the number of lateral and medial indices in the first level of the BST, see~\Cref{sec_slow polarization stage}.
    \item The number of levels of the BST, $n$. Thus, using $L_0$ and $M_0$ above, the length of each BST in the slow stage is $N = (2 L_0 + M_0)\cdot 2^n$.
    \item The number of levels of the fast transform, $\hat{n}$. Recall that this also determines the number of copies of the BST in the slow stage, $\hat{N}= 2^{\hat{n}}$. 
\end{itemize}
With these parameters set, the codelength is given by $\Lambda = N \cdot \hat{N}$, see~\eqref{eq_length of code in sequence}. 
We can now construct the transform, but we must still determine the frozen and non-frozen indices, and thus set the code rate. To this end, we must supply some minimal information on $\mathcal{C}$. 
This information consists of a set of \emph{bounds} on the following parameters, which must hold for \emph{all processes} in $\mathcal{C}$.
\begin{itemize}
    \item Upper bound on $H(X_0|Y_{-L_0}^{L_0}, X_{-L_0}^{-1})$.
    \item Recalling \eqref{eq_memory length inequality}, an upper bound $\bar{\epsilon}_{L_0}$ on both $I(S_1; S_{L_0} | X_1^{L_0}, Y_1^{L_0})$ and $I(S_1; S_{L_0} | Y_{1}^{L_0})$. 
    \item Upper bound $\bar{\psi}_{M_0 - 2}$ and lower bound $\bar{\phi}_{M_0 - 2}$, on $\psi_{M_0-2}$ and $\phi_{M_0-2}$, respectively, see \Cref{lem_FAIM is psi mixing}.  
    \item Upper bound $\bar{\psi}_0$ on $\psi_0$, which by~\eqref{eq_kappa thm 1} leads to an upper bound on $\kappa$.
\end{itemize}

In step \ref{slow second substep} of the construction below, it will become apparent that we would like $\bar{\epsilon}_{L_0}$ to be close to $0$ and both $\bar{\psi}_{M_0-2}$ and $\bar{\phi}_{M_0-2}$ to be close to $1$. This can be achieved by taking $L_0$ and $M_0$ large enough. Next, we also must ensure that the fraction of medial indices, $\alpha_n$ in \eqref{eq_def of alphan}, is close to $1$. Recalling \Cref{lem_proportion of lateral sopairs}, this can be achieved by ensuring that $M_0$ is sufficiently larger than $L_0$.

The above set of  bounds is the key to computing upper bounds on the Bhattacharyya parameters of the \sopairs at the last level of the overall universal transform. These are part of what is required for determining the frozen set.

We must also take into consideration the fixed input process. We employ the Honda-Yamamoto \cite{Honda_Yamamoto_2013} scheme to this end. Namely, we need to compute upper bounds on the total variation of the \sopairs at the last level of the overall universal transform for the input process. Practically, since the input process is fixed and independent of the channel, Monte-Carlo simulation may be used to obtain these bounds.
These bounds, together with the above bounds on Bhattacharyya parameters, are used to determine the frozen set. 

    We now proceed via the following steps, which use the notation of~\Cref{fig_notation for how to construct}. Further recall that $G_j = (F_1^{j-1}, Y_1^N)$, see \Cref{def_H transform}. 

\begin{enumerate}
	\item \label{Bhattacharyya slow step} Compute upper bounds on the conditional entropies of the \sopairs at the last level of each BST of the slow stage. That is, for each $1 \leq j \leq N$, compute upper bounds on $H(F_j|G_j) = H(F_j|F_1^{j-1}, Y_1^N)$. Note that we have dropped the index $\ell$, since all BSTs are copies of one another, and we have assumed stationarity.  This step consists of two substeps.
        \begin{enumerate}
            \item \label{slow first substep} For the first substep, we momentarily assume a block-independent process regime (recall \Cref{sec_biprocess}). Using \Cref{lem_WS iterations} with the upper bound on $H(X_0|Y_{-L_0}^{L_0}, X_{-L_0}^{-1})$ in place of $\Hwind$, calculate upper bounds on $H(\tilde{F}_j|\tilde{G}_j)$ (see \eqref{eq_FiGiFiGiFiGi3}) 
                for each medial index $j$. Recall from \Cref{lem_all medial have the same entropy} that at each level $m$ of the BST, the entropy of all indices in $\medsetMinus{m}$ is identical and similarly the entropy of all indices $\medsetPlus{m}$ is also identical. The entropies of lateral indices are ``stuck'' and do not evolve further; at each level $m$ of the transform two new lateral indices are generated, one from $\medsetMinus{m}$ and one from $\medsetPlus{m}$, so the bounds on their entropies stop evolving and are simply taken from the previous level. 
	\item \label{slow second substep} To remove the assumption from the previous substep, we use \Cref{prop_tilde non tilde}. That is, to compute upper bounds on $H(F_j | G_j)$, we use 
                \eqref{eq_HFiGi tilde inequality a}. To this end, we compute an upper bound on $\varepsilon_3$, the right-hand side of \eqref{eq_HFiGi tilde inequality a}, using the bounds on the parameters above. Namely, in \eqref{eq_definition of varepsilon_3} we set
		$\varepsilon_1 \leftarrow \bar{\epsilon}_{L_0} \cdot 2^n$ and $\varepsilon_2 \leftarrow \max\{(\bar{\psi}_{M_0-2})^{2^n}-1,1-(\bar{\phi}_{M_0-2})^{2^n}\}$, in line with \eqref{eq_epsilon as a function of varepsilon_1} and \eqref{eq_psiphiM02n bound}, respectively. Observe that $\varepsilon_3$ bounds the difference between the results of this and the previous step; this difference may be made small by suitably increasing $L_0$ and $M_0$.  
        \end{enumerate}
    \item \label{Bhattacharyya fast step} Compute upper bounds on the conditional entropies of the \sopairs at the last level of each fast transform of the slow stage, $Z\left(\hat{U}_i^{(j)}\,\middle|\, (\hat{U}_{i'}^{(j)})_{i'=1}^{i-1},\hat{\bs{U}}_1^{j-1}, \bs{Y}_1^{\hat{N}}\right)$, for each $1 \leq j \leq N$ and $1 \leq i \leq \hat{N}$. We do this via the following substeps. 
    \begin{enumerate}
        \item \label{fast Bhattacharyya entropy bound} Using \cite[Lemma 1]{Shuval_Tal_Memory_2017} on the result of step \ref{slow second substep}, derive upper bounds on the Bhattacharyya parameters of the \sopairs entering the fast stage. That is, for each $j$, we compute the upper bound 
            \[ Z(F_j|F_1^{j-1},Y_1^N) \leq \sqrt{H(F_j|F_1^{j-1}, Y_1^N)}. \]  
\item \label{it_plan Z} From these, use~\eqref{eq_bhattacharyya recursion} from Appendix~\ref{ap_fast polarization} and the upper bound on $\kappa$ to derive upper bounds on the Bhattacharyya parameters of the \sopairs at the last level of the fast stage. That is, for each $1 \leq i \leq \hat{N}$ and $1 \leq j \leq N$, these are upper bounds on
	    \[
		    Z\left(\hat{U}_i^{(j)}\,\middle|\, (\hat{U}_{i'}^{(j)})_{i'=1}^{i-1},\hat{\bs{U}}_1^{j-1}, \bs{Y}_1^{\hat{N}}\right) .
	    \]
\end{enumerate}
\end{enumerate}

For the next step, we use the upper bounds on both the Bhattacharyya parameters and the total variations. That is, upper bounds on:
\begin{align*}
	& Z\left(\hat{U}_i^{(j)}\,\middle|\, (\hat{U}_{i'}^{(j)})_{i'=1}^{i-1},\hat{\bs{U}}_1^{j-1}, \bs{Y}_1^{\hat{N}}\right),\\
	&K\left(\hat{U}_i^{(j)}\, \middle|\, (\hat{U}_{i'}^{(j)})_{i'=1}^{i-1},\hat{\bs{U}}_1^{j-1}\right). 
\end{align*}

        \begin{enumerate}[resume]
	\item \label{it_loose upper bound} Take the non-frozen indices as those for which the sum of these upper bounds 
		is small enough. That is, the word error rate is upper-bounded by
	    \begin{multline*}
		    \sum_{i,j} \bigg( Z\left(\hat{U}_i^{(j)}\,\middle|\, (\hat{U}_{i'}^{(j)})_{i'=1}^{i-1},\hat{\bs{U}}_1^{j-1}, \bs{Y}_1^{\hat{N}}\right) \\ + 
		    K\left(\hat{U}_i^{(j)}\, \middle|\, (\hat{U}_{i'}^{(j)})_{i'=1}^{i-1},\hat{\bs{U}}_1^{j-1}\right)  \bigg), 
	    \end{multline*}
	    where the sum is over the indices $i,j$ for which $\hat{U}_i^{(j)}$ is non-frozen (contains an information bit). 
	    	
        Note that when the input distribution is uniform the total variation is always $0$, thus one may  consider only the sum of  upper bounds on the Bhattacharyya parameter to determine the non-frozen set. Accordingly, in this case, the word error rate is upper-bounded by 
	\[ 
		\sum_{i,j} Z\left(\hat{U}_i^{(j)}\,\middle|\, (\hat{U}_{i'}^{(j)})_{i'=1}^{i-1},\hat{\bs{U}}_1^{j-1}, \bs{Y}_1^{\hat{N}}\right), 
	\]
	where, again, the summation is over indices $i,j$ for which $\hat{U}_i^{(j)}$ is non-frozen. 
\end{enumerate}

\begin{remark} \label{rem_psi0 may be set to 1} In a process with memory, the parameter $\psi_0$ is typically greater than $1$ (see the discussion following \Cref{lem_FAIM is psi mixing}), and thus $\kappa$ from~\eqref{eq_kappa thm 1} is greater than $2$. Thus, the upper bounds on the Bhattacharyya parameter may be non-informative for many \sopairs, leading to very low rate codes. To see this, consider the Bhattacharyya recursion~\eqref{eq_bhattacharyya recursion} used in step \ref{it_plan Z}: in half of the steps, the recursion leads to the bound $Z_{n+1} \leq \kappa Z_n$. When $\kappa$ is large, the right-hand side may quickly exceed $1$, making this bound on the Bhattacharyya parameter non-informative. 
	In practice, we can do much better. In the fast stage of the universal transform, we combine transformed \sopairs from different BSTs. These BSTs are ``far apart:'' at the beginning and end of each BST are enough lateral symbols that are essentially a buffer that ensures forgetfulness ``takes place.'' Hence, practically we may assume that the two transformed \sopairs from different BSTs are independent and set $\psi_0 = 1$ (or $\kappa = 2$) in step \ref{it_plan Z}. This works well in practice, and we have followed this strategy for our numerical results below. The above heuristic can be made exact by adding an additional small buffer of symbols between BSTs. That is, due to the mixing property of FAIM processes, this buffer makes the BSTs as close to independent as desired. By \Cref{lem_FAIM is psi mixing}, the mixing coefficients tend to $1$ exponentially fast, so this buffer is negligible compared to the BST blocklength, incurring a vanishing rate loss.
\end{remark}

We end this section by contrasting the construction of universal polar codes described above to that of non-universal polar codes. Here, by definition, we must construct the code to work for a \emph{family} of \soprocesses. Thus, we cannot follow the process-specific technique described in \cite{Tal_2013}.\footnote{In fact, since we are dealing with states, the effective alphabet for the construction algorithm is non-binary, and we should have referred to \cite{Kartowsky_2017} and \cite{Ordentlich_2021} as well.} Hence, we resort to using the above bounding techniques. It is interesting to note that for the universal setting, asymptotically (for large enough blocklengths), the use of these bounds does not incur a rate penalty. This is in contrast to non-universal polar codes.

\section{Numerical Results} \label{sec_numerical results}
In this section, we provide simulation results for our universal polar code. These are given in \Cref{fig_bounds_0,fig_bounds_1}. The universal polar code was designed using the method in \Cref{sec_construction}. Namely, we selected the following code parameters:
\begin{itemize}
\item $L_0 = 6$ and $M_0 = 40$,
\item $n = 5$,
\item $\hat{n} = 7$.
\end{itemize}
Thus, the code length is
\[
	N \cdot \hat{N} = (2 L_0 + M_0) \cdot 2^{n + \hat{n}} = 212992 .
\]

We chose the memoryless uniform input distribution. Our code is to operate on two different Gilbert-Elliott channels (see \Cref{ex_M for Gilbert Elliott} in Appendix \ref{ap_markov model equivalence}). Indeed, the capacity-achieving input distribution for these symmetric channels is uniform. 
The Gilbert-Elliott channels have the following parameters: 
\begin{enumerate}
	\item Channel GE-one:
		\begin{itemize}
			\item crossover probability in good state: $0.01$;
			\item crossover probability in bad state: $0.175$;
			\item transition probability good state to bad state: $0.40$;
			\item transition probability bad state to good state: $0.40$.
		\end{itemize}
	\item Channel GE-two:
		\begin{itemize}
			\item crossover probability in good state: $0.01$;
			\item crossover probability in bad state: $0.170$;
			\item transition probability good state to bad state: $0.60$;
			\item transition probability bad state to good state: $0.55$.
		\end{itemize}
\end{enumerate}
For these channels, we have the following bounds:
\begin{itemize}
	\item $H(X_0|Y_{-L_0}^{L_0},X_{-L_0}^{-1}) \leq 0.45$,
	\item $\epsilon_{L_0} \leq 7.39 \cdot 10^{-8}$,  
	\item $\psi_{M_0-2} \approx \phi_{M_0-2} \approx 1$ (recall that these parameters approach $1$ exponentially fast with $M_0$, which we have set to $40$), 
	\item $\psi_0 \leq 2.1$.
\end{itemize}
From these bounds, in step \ref{slow second substep} of \Cref{sec_construction} one may compute: $\varepsilon_1 = 2.36\cdot 10^{-6}$, $\varepsilon_2 \approx 0$, and $\varepsilon_3 = 4.73 \cdot 10^{-6}$. 

We designed a set of universal polar codes using the procedure in \Cref{sec_construction}, with rates from $0.15$ to $0.30$. Specifically, in line with \Cref{rem_psi0 may be set to 1}, we have taken $\psi_0$ as $1$ in step \ref{it_plan Z}.
The simulation results of these of codes on both Gilbert-Elliott channels are shown in \Cref{fig_bounds_0}. The decoding of these codes was done using a list-decoder \cite{ShuvalTal:20c}. Indeed, utilizing a larger list size for decoding  improves the word error rate considerably.

Observe that the codes may operate on additional channels that satisfy the above bounds. Two examples are a BSC with crossover probability $0.09$, having capacity $0.44$, 
and BEC with erasure probability $0.45$, having capacity $0.45$. Both channels are memoryless so they trivially satisfy the other bounds. Simulation results of the \emph{same} codes for these channels, again using a list-decoder, are shown in \Cref{fig_bounds_1}. 

We wish to emphasize that the numerical results are far better than the upper bounds on decoding error (step \ref{it_loose upper bound} in the method of \Cref{sec_construction}). Indeed, the upper bound (sum of Bhattacharyya parameters of non-frozen indices) for rate $0.15$ is $1.9$ and the upper bound for rate $0.27$ is $3960.41$. Both are non-informative and extremely pessimistic. When constructing the codes, as non-frozen indices we have selected those with the lowest upper bounds on the Bhattacharyya parameter. 

\begin{figure}
\begin{tikzpicture}

\begin{axis}[%
width=7cm,
height=7cm,
scale only axis,
grid = major,
xlabel={Code rate},
ymode=log,
ylabel={Word error rate},
legend style={at={(0.99,0.02)},anchor=south east,legend cell align=left,align=left,fill=white, draw=none}
]

\addplot [color=blue, thick, solid,mark=x,mark options={solid}]
table[row sep=crcr]{%
.03   0 \\ 
.06   0 \\ 
.09   0 \\ 
.12   0 \\ 
.15   0 \\ 
.18   0.0205 \\ 
.21   0.204 \\ 
.24   0.976333333333333 \\ 
.27   1 \\ 
.30   1 \\ 
};

\addplot [color=blue, thick, solid,mark=o,mark options={solid}]
table[row sep=crcr]{%
.03   0 \\ 
.06   0 \\ 
.09   0 \\ 
.12   0 \\ 
.15   0 \\ 
.18   0.000333333333333333 \\ 
.21   0.00883333333333333 \\ 
.24   0.221833333333333 \\ 
.27   0.909666666666667 \\ 
.30   1 \\ 
};

\addplot [color=blue, thick, solid,mark=square,mark options={solid}]
table[row sep=crcr]{%
.03   0 \\ 
.06   0 \\ 
.09   0 \\ 
.12   0 \\ 
.15   0 \\ 
.18   0 \\ 
.21   0.000333333333333333 \\ 
.24   0.034 \\ 
.27   0.3465 \\ 
.30   1 \\ 
};

\addplot [color=blue, thick, solid,mark=triangle,mark options={solid}]
table[row sep=crcr]{%
.03   0 \\ 
.06   0 \\ 
.09   0 \\ 
.12   0 \\ 
.15   0 \\ 
.18   0 \\ 
.21   0.000166666666666667 \\ 
.24   0.005 \\ 
.27   0.101333333333333 \\ 
.30   1 \\ 
};

\addplot [color=blue, thick, solid,mark=star,mark options={solid}]
table[row sep=crcr]{%
.03   0 \\ 
.06   0 \\ 
.09   0 \\ 
.12   0 \\ 
.15   0 \\ 
.18   0 \\ 
.21   0.000166666666666667 \\ 
.24   0.00233333333333333 \\ 
.27   0.047 \\ 
.30   1 \\ 
};

\addplot [color=red, thick, dashed,mark=x,mark options={solid}]
table[row sep=crcr]{%
.03   0 \\ 
.06   0 \\ 
.09   0 \\ 
.12   0 \\ 
.15   0.000333333333333333 \\ 
.18   0.0235 \\ 
.21   0.23 \\ 
.24   0.9815 \\ 
.27   1 \\ 
.30   1 \\ 
};

\addplot [color=red, thick, dashed,mark=o,mark options={solid}]
table[row sep=crcr]{%
.03   0 \\ 
.06   0 \\ 
.09   0 \\ 
.12   0 \\ 
.15   0 \\ 
.18   0.000833333333333333 \\ 
.21   0.00816666666666667 \\ 
.24   0.257666666666667 \\ 
.27   0.936833333333333 \\ 
.30   1 \\ 
};

\addplot [color=red, thick, dashed,mark=square,mark options={solid}]
table[row sep=crcr]{%
.03   0 \\ 
.06   0 \\ 
.09   0 \\ 
.12   0 \\ 
.15   0 \\ 
.18   0 \\ 
.21   0.000333333333333333 \\ 
.24   0.0365 \\ 
.27   0.383333333333333 \\ 
.30   1 \\ 
};

\addplot [color=red, thick, dashed,mark=triangle,mark options={solid}]
table[row sep=crcr]{%
.03   0 \\ 
.06   0 \\ 
.09   0 \\ 
.12   0 \\ 
.15   0 \\ 
.18   0 \\ 
.21   0.000166666666666667 \\ 
.24   0.00766666666666667 \\ 
.27   0.121333333333333 \\ 
.30   1 \\ 
};

\addplot [color=red, thick, dashed,mark=star,mark options={solid}]
table[row sep=crcr]{%
.03   0 \\ 
.06   0 \\ 
.09   0 \\ 
.12   0 \\ 
.15   0 \\ 
.18   0 \\ 
.21   0.000166666666666667 \\ 
.24   0.003 \\ 
.27   0.0525 \\ 
.30   1 \\ 
};

\end{axis}
\end{tikzpicture}
\caption{Word error rate of a universal polar code of length $212992$, used on two different Gilbert-Elliott channels: 
    GE-one, in solid blue (\protect \tikz{\protect \draw[thick,blue, anchor=mid] (0,0) -- (0.3,0) ; \protect \node[inner sep = 0] at (0,-0.05) {}; }); and 
    GE-two, in dashed red (\protect \tikz{\protect \draw[thick,red,dashed, anchor=mid] (0,0) -- (0.3,0) ; \protect \node[inner sep = 0] at (0,-0.05) {}; }). 
We use a list-decoder for the decoding, employing list sizes $L=1$ ($\times$), $L=2$ ($\bigcirc$), $L=4$ ($\square$), $L=8$ ($\triangle$), $L=16$ ($\star$).}
\label{fig_bounds_0}
\end{figure}

\begin{figure}
\begin{tikzpicture}

\begin{axis}[%
width=7cm,
height=7cm,
scale only axis,
grid = major,
xlabel={Code rate},
ymode=log,
ylabel={Word error rate},
legend style={at={(0.99,0.02)},anchor=south east,legend cell align=left,align=left,fill=white, draw=none}
]

\addplot [color=red, thick, dashed,mark=x,mark options={solid}]
table[row sep=crcr]{%
.03   0 \\ 
.06   0 \\ 
.09   0 \\ 
.12   0 \\ 
.15   0.000166666666666667 \\ 
.18   0.00983333333333333 \\ 
.21   0.1655 \\ 
.24   0.946333333333333 \\ 
.27   1 \\ 
.30   1 \\ 
};

\addplot [color=red, thick, dashed,mark=o,mark options={solid}]
table[row sep=crcr]{%
.03   0 \\ 
.06   0 \\ 
.09   0 \\ 
.12   0 \\ 
.15   0 \\ 
.18   0.000166666666666667 \\ 
.21   0.00716666666666667 \\ 
.24   0.187333333333333 \\ 
.27   0.850833333333333 \\ 
.30   1 \\ 
};

\addplot [color=red, thick, dashed,mark=square,mark options={solid}]
table[row sep=crcr]{%
.03   0 \\ 
.06   0 \\ 
.09   0 \\ 
.12   0 \\ 
.15   0 \\ 
.18   0 \\ 
.21   0.000833333333333333 \\ 
.24   0.0253333333333333 \\ 
.27   0.293333333333333 \\ 
.30   1 \\ 
};

\addplot [color=red, thick, dashed,mark=triangle,mark options={solid}]
table[row sep=crcr]{%
.03   0 \\ 
.06   0 \\ 
.09   0 \\ 
.12   0 \\ 
.15   0 \\ 
.18   0 \\ 
.21   0.00116666666666667 \\ 
.24   0.00516666666666667 \\ 
.27   0.084 \\ 
.30   1 \\ 
};

\addplot [color=red, thick, dashed,mark=star,mark options={solid}]
table[row sep=crcr]{%
.03   0 \\ 
.06   0 \\ 
.09   0 \\ 
.12   0 \\ 
.15   0 \\ 
.18   0 \\ 
.21   0.000833333333333333 \\ 
.24   0.0025 \\ 
.27   0.0431666666666667 \\ 
.30   1 \\ 
};

\addplot [color=blue, thick, solid,mark=x,mark options={solid}]
table[row sep=crcr]{%
.03   0 \\ 
.06   0 \\ 
.09   0 \\ 
.12   0 \\ 
.15   0 \\ 
.18   0.0145 \\ 
.21   0.0641666666666667 \\ 
.24   0.972833333333333 \\ 
.27   1 \\ 
.30   1 \\ 
};

\addplot [color=blue, thick, solid,mark=o,mark options={solid}]
table[row sep=crcr]{%
.03   0 \\ 
.06   0 \\ 
.09   0 \\ 
.12   0 \\ 
.15   0 \\ 
.18   0 \\ 
.21   0.000166666666666667 \\ 
.24   0.0208333333333333 \\ 
.27   0.4045 \\ 
.30   1 \\ 
};

\addplot [color=blue, thick, solid,mark=square,mark options={solid}]
table[row sep=crcr]{%
.03   0 \\ 
.06   0 \\ 
.09   0 \\ 
.12   0 \\ 
.15   0 \\ 
.18   0 \\ 
.21   0 \\ 
.24   0.000333333333333333 \\ 
.27   0.0296666666666667 \\ 
.30   1 \\ 
};

\addplot [color=blue, thick, solid,mark=triangle,mark options={solid}]
table[row sep=crcr]{%
.03   0 \\ 
.06   0 \\ 
.09   0 \\ 
.12   0 \\ 
.15   0 \\ 
.18   0 \\ 
.21   0 \\ 
.24   0.000166666666666667 \\ 
.27   0.0175 \\ 
.30   1 \\ 
};

\addplot [color=blue, thick, solid,mark=star,mark options={solid}]
table[row sep=crcr]{%
.03   0 \\ 
.06   0 \\ 
.09   0 \\ 
.12   0 \\ 
.15   0 \\ 
.18   0 \\ 
.21   0 \\ 
.24   0 \\ 
.27   0.0156666666666667 \\ 
.30   1 \\ 
};

\end{axis}
\end{tikzpicture}
\caption{Word error rate of the same universal polar code as in \Cref{fig_bounds_0}, used on two additional channels: 
    BEC, in solid blue (\protect \tikz{\protect \draw[thick,blue, anchor=mid] (0,0) -- (0.3,0) ; \protect \node[inner sep = 0] at (0,-0.05) {}; }); and 
    BSC, in dashed red (\protect \tikz{\protect \draw[thick,red,dashed, anchor=mid] (0,0) -- (0.3,0) ; \protect \node[inner sep = 0] at (0,-0.05) {}; }). 
Again, we use a list-decoder for the decoding, employing list sizes $L=1$ ($\times$), $L=2$ ($\bigcirc$), $L=4$ ($\square$), $L=8$ ($\triangle$), $L=16$ ($\star$).}
\label{fig_bounds_1}
\end{figure}

\section{A Contraction Inequality} \label{sec_contraction}
In this section we introduce a contraction inequality that will be useful in proving a sufficient condition for forgetfulness in \Cref{sec_HMM}. 
To this end, we define a pseudo-metric $d$ between two nonnegative vectors that have the same support.  
We will show that if a matrix $\mat{M}$ satisfies a certain property called \emph{subrectangularity}, then it has a parameter $\Birkhoff{\mat{M}} <1$ such
that $d(\trp{\bv{x}}\mat{M}, \trp{\bv{y}}\mat{M}) \leq \Birkhoff{\mat{M}} d(\bv{x},\bv{y})$. 

This section invariably contains a large number of indices. 
For tractability, we adhere to the following notational convention in this section. Indices $i$ and $k$ denote indices of \emph{rows} of matrix
$\mat{M}$, and indices $j$, $l$ denote indices of \emph{columns} of matrix $\mat{M}$. 
Additionally, throughout this section, we implicitly assume that in any product of two matrices or a vector and a matrix, their dimensions match to
enable forming these products. 

Recall that the \emph{support} $\support{\bv{x}}$ of a vector $\bv{x}$ is the set of its nonzero indices. 
That is, $\support{\bv{x}} = \{ i \ | \ x_i \neq 0\}$. The following pseudo-metric~\cite[Chapter 3.1]{Seneta}, \cite[Section 2]{cohen_1979} is defined
for nonnegative vectors with the same support. 

\begin{definition}[Projective distance]  \label{def_projective distance}
    Let $\bv{x}$, $\bv{y}$ be two nonnegative nonzero vectors such that $\support{\bv{x}} = \support{\bv{y}}$. 
    The \emph{projective distance} $d$ between the two vectors is
    \begin{equation} \label{eq_hilbert projective distance} 
        d(\bv{x},\bv{y}) \triangleq \max_{j,l \in \support{\bv{x}}} \ln \frac{x_j/y_j}{x_l/y_l}
        = \ln \max_{j,l \in \support{\bv{x}}} \frac{x_j/y_j}{x_l/y_l}. 
    \end{equation}
    For row vectors we define $d(\trp{\bv{x}}, \trp{\bv{y}}) = d(\bv{x}, \bv{y})$. 
    If $\bv{x} = \bv{y} = \bv{0}$,  we define $d(\bv{x}, \bv{y}) = 0$. 
\end{definition}
The projective distance is usually defined for positive vectors. 
Our definition generalizes it slightly for nonnegative vectors, provided they have the same support. 
In other words, we may assume that the (joint) zero indices of $\bv{x}$ and $\bv{y}$ are deleted before computing this distance.
The projective distance is a pseudo-metric~\cite[Exercise 3.1]{Seneta}: it satisfies all of the properties of a metric over the nonnegative quadrant,
with the exception that $d(\bv{x}, \bv{y}) = 0$ if and only if $\bv{x} = c\bv{y}$ for some $c > 0 $.

The concept of a subrectangular matrix was introduced in~\cite{kaijser1975} for square nonnegative matrices. 
However, it is easily extended to arbitrary nonnegative matrices. 
In this work, therefore, a subrectangular matrix need not be square. 
Subrectangularity will play a key role in the contraction inequality we develop. 
\begin{definition}[Subrectangular matrix]  \label{def_subrectangular matrix}
    A nonnegative matrix $\mat{M}$ is called \emph{subrectangular} if $\matel{\mat{M}}{i,j} \neq 0$ and $\matel{\mat{M}}{k,l} \neq 0$ implies that
    $\matel{\mat{M}}{i,l} \neq 0$ and $\matel{\mat{M}}{k,j} \neq 0$. 
\end{definition}
We illustrate a subrectangular matrix in \Cref{fig_subrectangular}. 
To better understand the meaning of this concept, in the following lemma we introduce equivalent characterizations of a subrectangular matrix. 
To this end, we remind the reader that a nonzero row (column) of a matrix contains at least one nonzero element, and that for a matrix $\mat{M}$ we
denote its set of nonzero rows by $\nzrows{\mat{M}}$ and its set of nonzero columns by $\nzcols{\mat{M}}$. 

\begin{figure}[t]
    \begin{center} 
        \begin{tikzpicture}[>=latex]
            \draw (0,0) rectangle (5,5); 
            \filldraw[gray!50!white] (0.4,0.2) rectangle (1,2); 
            \filldraw[gray!50!white] (0.4,2.4) rectangle (1,3); 
            \filldraw[gray!50!white] (0.4,3.2) rectangle (1,4.6); 
                           
            \filldraw[gray!50!white] (1.6,0.2) rectangle (3,2); 
            \filldraw[gray!50!white] (1.6,2.4) rectangle (3,3); 
            \filldraw[gray!50!white] (1.6,3.2) rectangle (3,4.6); 
                           
            \filldraw[gray!50!white] (3.2,0.2) rectangle (3.6,2); 
            \filldraw[gray!50!white] (3.2,2.4) rectangle (3.6,3); 
            \filldraw[gray!50!white] (3.2,3.2) rectangle (3.6,4.6); 
                           
            \filldraw[gray!50!white] (4.4,0.2) rectangle (5.0,2); 
            \filldraw[gray!50!white] (4.4,2.4) rectangle (5.0,3); 
            \filldraw[gray!50!white] (4.4,3.2) rectangle (5.0,4.6); 

            \draw[step = 0.2cm] (0,0) grid (5,5);

            \draw[->] (-0.3,2.7) node[left]{$i$} -- (0,2.7); 
            \draw[->] (-0.3,0.7) node[left]{$k$} -- (0,0.7); 
            \draw[->] (2.3,5.3) node[above]{$j$} -- (2.3,5); 
            \draw[->] (4.5,5.3) node[above]{$l$} -- (4.5,5); 

            \draw[pattern=north east lines] (2.2,2.6) rectangle (2.4,2.8); 
            \draw[pattern=north west lines] (4.4,0.6) rectangle (4.6,0.8); 
            \draw[pattern=crosshatch] (2.2,0.6) rectangle (2.4,0.8); 
            \draw[pattern=crosshatch] (4.4,2.6) rectangle (4.6,2.8); 
        \end{tikzpicture}
    \end{center}
    \caption{An illustration of a subrectangular matrix. Each of the small squares is an element of the matrix. 
             The white squares contain zeros, whereas the filled squares contain positive values. 
             Elements $\matel{\mat{M}}{i,j}$ and $\matel{\mat{M}}{k,l}$, denoted with diagonal lines
             (\protect\tikz{\protect\draw[preaction={fill=gray!50!white},pattern=north east lines] (0,0) rectangle (0.2,0.2);} 
             and  
            \protect\tikz{\protect\draw[preaction={fill=gray!50!white},pattern=north west lines] (0,0) rectangle (0.2,0.2);} 
            respectively), are nonzero. 
            Therefore, elements $\matel{\mat{M}}{i,l}$ and $\matel{\mat{M}}{k,j}$, denoted with a crosshatch
            (\protect\tikz{\protect\draw[preaction={fill=gray!50!white},pattern=crosshatch] (0,0) rectangle (0.2,0.2);}),
            are also nonzero.
            In fact,  any matrix element in the support of a subrectangular matrix is nonzero. } 
    \label{fig_subrectangular}
\end{figure}

\begin{lemma}
    Let $\mat{M}$ be a nonnegative matrix. The following are equivalent:
    \begin{enumerate}
        \item The matrix $\mat{M}$ is subrectangular.
        \item If $\mat{M}$ contains a zero element, either the entire row containing it or the entire column containing it are all zeros: 
            \begin{equation} \label{eq_subrectangular zeros} 
                \matel{\mat{M}}{i,j} = 0 \iff i \notin \nzrows{\mat{M}} \text{ or } j \notin \nzcols{\mat{M}}.
            \end{equation} 
            \item The matrix $\mat{M}$ satisfies
                \begin{equation} \label{eq_subrectangular nonzero} 
                    \matel{\mat{M}}{i,j} \neq 0 \iff i \in \nzrows{\mat{M}} \text{ and } j \in \nzcols{\mat{M}}. 
                \end{equation} 
            \end{enumerate}
        \end{lemma}
\begin{IEEEproof}
    The second and third characterizations are clearly equivalent. 
    Hence, it suffices to show that 1 $\Rightarrow$ 2 and 3 $\Rightarrow$ 1.
    
    $\text{1} \Rightarrow \text{2}$: 
    Assume to the contrary that $\mat{M}$ is subrectangular but~\eqref{eq_subrectangular zeros} is not satisfied. 
    That is, there exist $i,j$ such that $\matel{\mat{M}}{i,j} = 0$ and $i \in \nzrows{\mat{M}}, j \in \nzcols{\mat{M}}$. 
    Since row $i$ and column $j$ of $\mat{M}$ are not all zeros, there exist $k, l$ such that $\matel{\mat{M}}{i,l} \neq 0$ and $\matel{\mat{M}}{k,j}
    \neq 0$.
    By subrectangularity of $\mat{M}$, $\matel{\mat{M}}{i,j}$ must also be nonzero, a contradiction. 

    $\text{3} \Rightarrow \text{1}$: 
    Assume that~\eqref{eq_subrectangular nonzero} holds. 
    If $\mat{M}$ is an all-zero matrix, or has just a single nonzero row (column), then $\mat{M}$ is obviously subrectangular.
    Assume, therefore, that $\mat{M}$ has at least two nonzero rows and at least two nonzero columns.
    That is, there exist $(i,j)$, $(k,l)$ such that $\matel{\mat{M}}{i,j} \neq 0$ and $\matel{\mat{M}}{k,l} \neq 0$.
    Thus, by~\eqref{eq_subrectangular nonzero}, $i,k \in \nzrows{\mat{M}}$ and $j,l \in \nzcols{\mat{M}}$. 
    Then, a second of use of~\eqref{eq_subrectangular nonzero} implies that $\matel{\mat{M}}{i,l} \neq 0$ and $\matel{\mat{M}}{k, j} \neq 0$.
    Therefore, $\mat{M}$ is subrectangular. 
\end{IEEEproof}

Observe from~\eqref{eq_subrectangular zeros} that if $\mat{M}$ is subrectangular and $\mat{M}'$ is obtained from $\mat{M}$ by multiplying some of its
rows or columns by $0$, then $\mat{M}'$ is also subrectangular. 
Similarly, if $\mat{M}''$ is obtained from $\mat{M}$ by deleting some of its rows or columns, then $\mat{M}''$ is also subrectangular. 
In particular, \eqref{eq_subrectangular nonzero} implies that the matrix formed by deleting all of the all-zero rows and columns of $\mat{M}$ is
positive --- it contains only positive elements.

\begin{lemma} \label{lem_vector times subrectangular} 
    If $\mat{M}$ is a nonzero subrectangular matrix and $\bv{x}, \bv{y}$ are nonnegative vectors such that $\norm{\trp{\bv{x}}\mat{M}}_1 >0$ and
    $\norm{\trp{\bv{y}}\mat{M}}_1 >0$, then $\support{\trp{\bv{x}}\mat{M}} = \support{\trp{\bv{y}}\mat{M}}$ and $\support{\mat{M} \bv{x}}
    = \support{\mat{M} \bv{y}}$. 
    \end{lemma}
We remark that this lemma holds even if $\support{\bv{x}} \neq \support{\bv{y}}$. 
In particular, it implies that if $\mat{M}$ is subrectangular and $\bv{x}, \bv{y}$ are arbitrary nonnegative vectors such that $\trp{\bv{x}}\mat{M}$
and $\trp{\bv{y}}\mat{M}$ are nonzero, then $d(\trp{\bv{x}}\mat{M}, \trp{\bv{y}}\mat{M})$ is well-defined. 
\begin{IEEEproof} 
    It suffices to prove the claim that $\support{\trp{\bv{x}}\mat{M}} = \support{\trp{\bv{y}}\mat{M}}$, for the second claim follows by noting that
    $\mat{M}$ is subrectangular if and only if $\trp{\mat{M}}$ is subrectangular. 
    Without loss of generality, we may assume that $\mat{M}$ does not have all-zero rows.
    For, if it had such rows, we could remove them and delete the corresponding indices from $\bv{x}$ and $\bv{y}$ without affecting any of the values
    involved. 
    This implies, by~\eqref{eq_subrectangular zeros}, that any column of $\mat{M}$ is either all positive or all zeros.
    Thus, for any nonnegative and nonzero vector $\bv{z}$, we have $\matel{\trp{\bv{z}}\mat{M}}{i} = 0$ if and only if column $i$ of $\mat{M}$ is an
    all-zero column.
    The claim follows since both $\bv{x}$ and $\bv{y}$ are nonnegative and nonzero.   
\end{IEEEproof}
The following corollary was stated as~\cite[Proposition 6.1]{kaijser1975} without proof. 
We provide a short proof. 
\begin{corollary}\label{cor_subrectangular}
    If $\mat{M}$ is a subrectangular matrix and $\mat{T}$, $\mat{T}'$ are some other nonnegative matrices (not necessarily subrectangular), then
    $\mat{T}\mat{M}$ and $\mat{M}\mat{T}'$ are subrectangular.
\end{corollary}
\begin{IEEEproof}
    \looseness=-1
    The case where either matrix is the zero matrix is trivial, so we assume they are both nonzero. 
    It suffices to consider the case $\mat{T}\mat{M}$, since that transpose of a subrectangular matrix remains subrectangular.
    By \Cref{lem_vector times subrectangular}, every row of $\mat{T}\mat{M}$ is either all-zeros, or has the same support as the other nonzero rows of
    $\mat{T}\mat{M}$.
    This implies, by~\eqref{eq_subrectangular nonzero}, that $\mat{T} \mat{M}$ is subrectangular. 
\end{IEEEproof}
We remark that a converse to \Cref{cor_subrectangular} does not hold. 
That is, if a product of two nonnegative matrices is subrectangular, this \emph{does not} imply that either of them is subrectangular. 
For example, if we denote by $*$ an arbitrary positive value in a matrix, then $\mat{T}_1, \mat{T}_2$ below are not subrectangular whereas their
product $\mat{T}_1 \mat{T}_2$ is: 
\[ 
    \mat{T}_1           = \begin{bmatrix} * & 0 \\ * & * \end{bmatrix}, \quad 
    \mat{T}_2           = \begin{bmatrix} * & * \\ 0 & * \end{bmatrix}, \quad 
    \mat{T}_1 \mat{T}_2 = \begin{bmatrix} * & * \\ * & * \end{bmatrix}. 
\] 
    
We now introduce a parameter that plays a key role in the contraction inequalities we develop. 
To this end, recall that the support $\support{\mat{M}}$ of a matrix $\mat{M}$ is the set of index pairs 
\[ 
    \support{\mat{M}} = \{ (i,j) \ | \ i  \in \nzrows{\mat{M}}, j \in \nzcols{\mat{M}} \}. 
\] 
By~\eqref{eq_subrectangular nonzero}, if $\mat{M}$ is subrectangular and $(i,j) \in \support{\mat{M}}$ then $\matel{\mat{M}}{i,j} > 0$. 
\begin{definition}[Birkhoff contraction coefficient]\label{def_birkhoff}
    Let $\mat{M}$ be a nonnegative matrix.  
    Its \emph{Birkhoff contraction coefficient} $\Birkhoff{\mat{M}}$ is defined as follows. 
    \begin{itemize}
        \item If $\mat{M}$ is subrectangular and nonzero, then
            \begin{equation} 
                \Birkhoff{\mat{M}} \triangleq \sup_{ \bv{x} >0, \, \bv{y} > 0} \frac{d(\trp{\bv{x}}\mat{M}, \trp{\bv{y}}\mat{M}) }{d(\bv{x},\bv{y})}.
                \label{eq_def of Birkhoff}
            \end{equation}
    \item If $\mat{M}$ is the zero matrix, then $\Birkhoff{\mat{M}} = 0$. 
    \item If $\mat{M}$ is not subrectangular, then $\Birkhoff{\mat{M}} = 1$. 
\end{itemize}
\end{definition}
By \Cref{lem_vector times subrectangular} and the positivity of $\bv{x}$ and $\bv{y}$, the numerator of~\eqref{eq_def of Birkhoff} is well-defined. 
That is, $\trp{\bv{x}}\mat{M}$ and $\trp{\bv{y}}\mat{M}$ have the same support. 
The denominator of~\eqref{eq_def of Birkhoff} is also well-defined, as $\bv{x}$ and $\bv{y}$ are positive and thus have the same support as well. 
Finally, to ensure that the ratio in~\eqref{eq_def of Birkhoff} is well-defined, we use the convention $0/0=0$.
Observe that the supremum in~\eqref{eq_def of Birkhoff} is obtained for $\bv{x} \neq c \bv{y}$ for $c > 0$.

The Birkhoff contraction coefficient~\cite[Chapter 3]{Seneta},~\cite{Artzrouni_Li_1995} is usually defined for matrices with no all-zero columns. 
We generalize here the definition slightly to apply also to matrices with columns that are all-zeros. 
In light of \Cref{def_projective distance,lem_vector times subrectangular}, the Birkhoff contraction coefficient of a matrix with some all-zero
columns is simply the Birkhoff contraction coefficient of the matrix obtained by deleting its all-zero columns. 
We note in passing that 
\begin{equation}\label{eq_Birkhoff transpose} 
    \Birkhoff{\mat{M}} = \Birkhoff{\trp{\mat{M}}},
\end{equation}
since $d(\trp{\bv{x}}\mat{M}, \trp{\bv{y}}\mat{M}) = d(\trp{\mat{M}}\bv{x}, \trp{\mat{M}}\bv{y})$. 

The following theorem is a restatement of~\cite[Section 3.4]{Seneta} (see~\cite[Theorem 1.1]{Artzrouni_Li_1995} for an alternative proof). 
\begin{theorem} \label{thm_birkhoff explicit form}
    If $\mat{M}$ is subrectangular and nonzero, then 
    \[ 
        \Birkhoff{\mat{M}} = \frac{1-\sqrt{\phi(\mat{M})}}{1+\sqrt{\phi(\mat{M})}} < 1,
    \] 
    where
    \begin{equation} \label{eq_def of phi} 
        \phi(\mat{M}) \triangleq \min_{\substack{i,k \in \nzrows{\mat{M}}, \\ j,l \in \nzcols{\mat{M}} }} 
                                 \frac{ \matel{\mat{M}}{i,j} \matel{\mat{M}}{k,l} }{ \matel{\mat{M}}{i,l} \matel{\mat{M}}{k,j} } > 0.
    \end{equation}
\end{theorem}
Since $\mat{M}$ is subrectangular and nonzero, all index pairs on the right-hand side of~\eqref{eq_def of phi} are in the support of $\mat{M}$, by
which $\phi(\mat{M}) > 0$.  
In other words, the Birkhoff contraction coefficient of a subrectangular matrix is the Birkhoff contraction coefficient of the positive matrix
obtained by deleting all of its all-zero rows and columns. 
The proofs of this theorem in~\cite[Section 3.4]{Seneta} and~\cite[Theorem 1.1]{Artzrouni_Li_1995} assume no all-zero columns in $\mat{M}$. 
However, as explained after \Cref{def_birkhoff}, they hold without change for our slightly generalized definition of the Birkhoff contraction coefficient. 

By \Cref{def_birkhoff,thm_birkhoff explicit form},  if $\bv{x}$ and $\bv{y}$ are \emph{positive} vectors and $\mat{M}$ is subrectangular, then 
\[ 
    d(\trp{\bv{x}}\mat{M}, \trp{\bv{y}}\mat{M}) \leq \Birkhoff{\mat{M}} d(\bv{x},\bv{y}). 
\] 
We now show that this holds in the more general case, where $\bv{x}$ and $\bv{y}$ are nonnegative vectors with the same support. 
\begin{corollary} \label{cor_projective distance inequality} 
    If $\bv{x}, \bv{y}$ are nonnegative vectors such that $\support{\bv{x}} = \support{\bv{y}}$ and $\mat{M}$ is subrectangular, then 
    \begin{equation} \label{eq_projective distance inequality}
        d(\trp{\bv{x}}\mat{M}, \trp{\bv{y}}\mat{M}) \leq \Birkhoff{ \mat{M} } d(\bv{x},\bv{y}).
    \end{equation} 
\end{corollary}
\begin{IEEEproof}
    The claim is trivial if $\bv{x} = \bv{y} = \bv{0}$. 
    If $\bv{x}, \bv{y}$ are positive, the claim follows from \Cref{def_birkhoff,thm_birkhoff explicit form}. 
    So, we assume that $\bv{x}$ and $\bv{y}$ are nonzero but have some zero elements.
    Denote by $\tilde{\bv{x}}, \tilde{\bv{y}}$ the vectors formed from $\bv{x}, \bv{y}$ by deleting their zero elements, and by $\tilde{\mat{M}}$ the
    matrix formed from $\mat{M}$ by deleting the rows corresponding to these indices.
    The resulting vectors are positive and the resulting matrix remains subrectangular.
    Therefore, 
    \begin{align*} 
        d(\trp{\bv{x}}\mat{M}, \trp{\bv{y}}\mat{M}) &= d(\trp{\tilde{\bv{x}}}\tilde{\mat{M}}, \trp{\tilde{\bv{y}}}\tilde{\mat{M}}) \\  
                                                    &\leq \Birkhoff{ \tilde{\mat{M}} } d(\tilde{\bv{x}},\tilde{\bv{y}}) 
                                                    = \Birkhoff{ \tilde{\mat{M}} } d(\bv{x},\bv{y}).
    \end{align*} 
    Finally, observe that $(1-\sqrt{x})/(1+\sqrt{x})$ is a decreasing function of $x$ when $x \geq 0$; this is easily seen by computing its
    derivative, $-(\sqrt{x}(1+\sqrt{x})^2)^{-1}$. 
    Since $\tilde{\mat{M}}$ is formed from $\mat{M}$ by deleting rows, $\phi(\tilde{\mat{M}}) \geq \phi(\mat{M})$.
    Thus, we must have $\Birkhoff{ \tilde{\mat{M}} } \leq \Birkhoff{ \mat{M} }$, which completes the proof. 
\end{IEEEproof}

In the following lemma we prove an inequality, adapted from the proof of \cite[Lemma 5]{cohen_1979}, that is useful in the sequel. 
\begin{lemma}  \label{lem_sum ratio inequality}
     Let $\alpha_i>0$, $\beta_i >0$, and $\gamma_i \geq 0$ for all $i$. 
     Assume that $\gamma_i > 0$ for some $i$.
     Then, 
     \begin{equation} \label{eq_sum ratio inequality} 
         \min_i \frac{\alpha_i}{\beta_i}\leq \frac{\sum_i \gamma_i \alpha_i}{\sum_i \gamma_i \beta_i} \leq \max_i \frac{\alpha_i}{\beta_i}. 
     \end{equation}
\end{lemma}
\begin{IEEEproof}
     Denoting $\rho_i = \alpha_i/\beta_i$, we have         
     \[ 
         \frac{\sum_i \gamma_i \alpha_i}{\sum_i \gamma_i \beta_i} 
             = \frac{\sum_i \gamma_i \beta_i \rho_i}{\sum_i \gamma_i \beta_i} 
             = \sum_i \frac{\gamma_i \beta_i}{\sum_{i'} \gamma_{i'} \beta_{i'}} \rho_i 
             = \sum_i \theta_i \rho_i, 
     \] 
     where $\theta_i \geq 0$ for all $i$ and $\sum_i \theta_i = 1$. 
     That is, the ratio on the left-hand side is a convex combination of the ratios $\rho_i$.
     Hence, it is lower-bounded by $\min_i \rho_i$ and upper-bounded by $\max_i \rho_i$, as required. 
\end{IEEEproof}

Armed with the above inequality, we can prove the following important property of the Birkhoff contraction coefficient. 
\begin{lemma}\label{lem_Birkhoff of matrix product}
    Let $\mat{M}$ be a subrectangular matrix and let $\mat{T}$ be a nonnegative matrix. 
    Then, 
    \[ 
        \Birkhoff{\mat{T}\mat{M}} \leq \Birkhoff{\mat{M}}.
    \] 
    If, in addition, $\mat{T}$ is subrectangular then 
    \begin{equation} \label{eq_Birkhoff of matrix product} 
        \Birkhoff{\mat{T}\mat{M}} \leq \Birkhoff{\mat{T}}\Birkhoff{\mat{M}}.
    \end{equation} 
\end{lemma}
\begin{remark}
    Two remarks are in order. 
    First, we note that~\eqref{eq_Birkhoff of matrix product} is adapted from~\cite[equation 3.7]{Seneta}.
    Second, there is nothing special about the ordering of the subrectangular and nonnegative matrix in the lemma.
    In particular, if the product $\mat{T}\mat{M}$ is replaced with the product $\mat{M}\mat{T}$ everywhere, the lemma holds unchanged. 
    Indeed, by~\eqref{eq_Birkhoff transpose}, $\Birkhoff{\mat{T}\mat{M}} = \Birkhoff{\trp{(\mat{T}\mat{M})}} = \Birkhoff{\trp{\mat{M}}\trp{\mat{T}}}$
    and $\mat{M}$ is subrectangular if and only if $\trp{\mat{M}}$ is subrectangular. 
\end{remark}
\begin{IEEEproof}
    There is nothing to prove if $\mat{T}\mat{M} = 0$, so we assume that $\mat{T}\mat{M}$ is nonzero.

    By \Cref{cor_subrectangular}, $\mat{T}\mat{M}$ is subrectangular. 
    For the first claim, let $i_0, k_0 \in \nzrows{\mat{T}\mat{M}}$ and $j_0, l_0 \in \nzcols{\mat{T}\mat{M}}$ achieve the minimum in~\eqref{eq_def of
    phi}; that is, be such that
    $\phi(\mat{T}\mat{M}) = (\matel{\mat{T}\mat{M}}{i_0,j_0}\matel{\mat{T}\mat{M}}{k_0,l_0})/(\matel{\mat{T}\mat{M}}{i_0,l_0}\matel{\mat{T}\mat{M}}{k_0,j_0})$.
    Thus, by~\eqref{eq_subrectangular nonzero}, $(i_0,j_0), (k_0,l_0) \in \support{\mat{T}\mat{M}}$.
    This implies that $j_0,l_0 \in \nzcols{\mat{M}}$   --- otherwise, for example, we would have
    $\matel{\mat{T}\mat{M}}{i_0,j_0} = \sum_r \matel{\mat{T}}{i_0,r}\matel{\mat{M}}{r,j_0} = 0$, 
    which contradicts $(i_0,j_0) \in \support{\mat{T}\mat{M}}$.   
    
    Hence, 
    \begin{align*}
        \phi(\mat{T}\mat{M}) 
        &= \frac{\matel{\mat{T}\mat{M}}{i_0,j_0}\matel{\mat{T}\mat{M}}{k_0,l_0} }{\matel{\mat{T}\mat{M}}{i_0,l_0}\matel{\mat{T}\mat{M}}{k_0,j_0}  } \\ 
        &= \frac{\sum\limits_i \matel{\mat{T}}{i_0,i} \matel{\mat{M}}{i,j_0}}{\sum\limits_i\matel{\mat{T}}{i_0,i}\matel{\mat{M}}{i,l_0}} \cdot
           \frac{\sum\limits\limits_k \matel{\mat{T}}{k_0,k} \matel{\mat{M}}{k,l_0}} { \sum\limits_k \matel{\mat{T}}{k_0,k}\matel{\mat{M}}{k,j_0}  } \\ 
        &= \frac{\sol[r]{\sum\limits_{i \in \nzrows{\mat{M}}}} \matel{\mat{T}}{i_0,i}
           \matel{\mat{M}}{i,j_0}}{\sol[r]{\sum\limits_{i \in \nzrows{\mat{M}}} } \matel{\mat{T}}{i_0,i}\matel{\mat{M}}{i,l_0}} \cdot
           \frac{\sol[r]{\sum\limits_{k \in \nzrows{\mat{M}}} } \matel{\mat{T}}{k_0,k} \matel{\mat{M}}{k,l_0}}{ \sol[r]{\sum\limits_{k \in \nzrows{\mat{M}}}}
           \matel{\mat{T}}{k_0,k}\matel{\mat{M}}{k,j_0}  } \\ 
        &\eqann[\geq]{a} \min_{i,k \in \nzrows{\mat{M}}} \frac{\matel{\mat{M}}{i,j_0} \matel{\mat{M}}{k,l_0}}{\matel{\mat{M}}{i,l_0}\matel{\mat{M}}{k,j_0}} \\ 
        &\eqann[\geq]{b} \min_{\substack{i,k \in \nzrows{\mat{M}} \\ j,l \in \nzcols{\mat{M} } }} \frac{\matel{\mat{M}}{i,j}
          \matel{\mat{M}}{k,l}}{\matel{\mat{M}}{i,l}\matel{\mat{M}}{k,j}} \\ 
        &= \phi(\mat{M}),
    \end{align*}
    where  \eqannref{a} is by the left-hand inequality of~\eqref{eq_sum ratio inequality}, used twice and since $j_0,l_0 \in \nzcols{\mat{M}}$ and the
    subrectangularity of $\mat{M}$;
    and in \eqannref{b} we minimize over a set of indices that contains $j_0, l_0$.
    Having established $\phi(\mat{T}\mat{M}) \geq \phi(\mat{M})$ and, since  $(1-\sqrt{x})/(1+\sqrt{x})$ is a decreasing function of $x$ for $x \geq
    0$ (see the proof of \Cref{cor_projective distance inequality}), we  conclude that $\Birkhoff{\mat{T}\mat{M}} \leq \Birkhoff{\mat{M}}$. 

    For the second claim, if $\mat{T}, \mat{M}$ are both subrectangular, then for any positive $\bv{x},\bv{y}$ we have
    $\support{\trp{\bv{x}}\mat{T}} = \support{\trp{\bv{y}}\mat{T}}$ and repeated applications of~\eqref{eq_projective distance inequality} yield  
    \begin{align*}
        d(\trp{\bv{x}}\mat{T} \mat{M}, \trp{\bv{y}}\mat{T} \mat{M})  
        &= d((\trp{\bv{x}}\mat{T}) \mat{M}, \trp{(\bv{y}}\mat{T}) \mat{M})  \\ 
        & \leq \Birkhoff{ \mat{M} } d(\trp{\bv{x}}\mat{T} , \trp{\bv{y}}\mat{T} ) \\ 
        &\leq \Birkhoff{ \mat{M} } \Birkhoff{ \mat{T} }  d(\bv{x},\bv{y}). 
    \end{align*}
    Thus, by~\eqref{eq_def of Birkhoff}, $\Birkhoff{\mat{T}\mat{M}} \leq \Birkhoff{\mat{T}}\Birkhoff{\mat{M}}$. 
\end{IEEEproof}

Applying \Cref{lem_Birkhoff of matrix product} to a product of $m$ subrectangular matrices $\mat{M}_1, \mat{M}_2, \ldots, \mat{M}_m$, we obtain
\begin{equation} \label{eq_birkhoff product inequality}
    \Birkhoff{ \mat{M}_1\mat{M}_2 \cdots\mat{M}_m } \leq \prod_{\ell=1}^m \Birkhoff { \mat{M}_{\ell} }.
\end{equation} 

\Cref{cor_projective distance inequality} required that $\bv{x},\bv{y}$ both have the same support.
For the cases where $\bv{x}$ and $\bv{y}$ have different supports, we have the following lemma. 
\begin{lemma} \label{lem_projective distance of ei and y times sr matrix}
    Let $\mat{M}$ be subrectangular and let $\mat{T}$ be an arbitrary nonnegative matrix.
    Then, for any two nonnegative vectors $\bv{x}$ and $\bv{y}$ such that $\norm{\trp{\bv{x}}\mat{T}\mat{M}}_1 > 0$ 
    and $\norm{\trp{\bv{y}}\mat{T}\mat{M}}_1 > 0$,
    \begin{equation} \label{eq_projective distance ei}
        d(\trp{\bv{x}}\mat{T}\mat{M}, \trp{\bv{y}}\mat{T}\mat{M}) \leq 
        4 \ln \left( \frac{1+ \Birkhoff{\mat{M}}}{1-\Birkhoff{\mat{M}}} \right). 
    \end{equation}
\end{lemma}
Since $\mat{M}$ is subrectangular, $\Birkhoff{\mat{M}}<1$, which implies that the right-hand side of~\eqref{eq_projective distance ei} is finite. 
\begin{IEEEproof}
    There is nothing to prove if $\mat{T}\mat{M} = 0$, so we assume that $\mat{T}\mat{M}$ is nonzero. 
    By \Cref{cor_subrectangular}, $\tilde{\mat{M}} = \mat{T}\mat{M}$ is subrectangular. 

    Fix any $i_0 \in \nzrows{\tilde{\mat{M}}}$. 
    Such an $i_0$ must exist because $\tilde{\mat{M}}$ is subrectangular and $\trp{\bv{x}}\tilde{\mat{M}}$ is nonzero by assumption.
    By \Cref{lem_vector times subrectangular}, and subrectangularity of $\tilde{\mat{M}}$,  
    \begin{equation} \label{eq_sigma mtilde}
        \support{\trp{\bv{e}_{i_0}}\tilde{\mat{M}}} = \support{\trp{\bv{x}} \tilde{\mat{M}}} = \nzcols{\tilde{\mat{M}}}. 
    \end{equation} 
    By the symmetry and triangle inequality properties of the projective distance~\cite[Exercise 3.1]{Seneta}, 
    \begin{equation*} 
        d(\trp{\bv{x}}\tilde{\mat{M}}, \trp{\bv{y}}\tilde{\mat{M}})  \leq
        d(\trp{\bv{e}_{i_0}}\tilde{\mat{M}}, \trp{\bv{x}}\tilde{\mat{M}}) + d(\trp{\bv{e}_{i_0}}\tilde{\mat{M}}, \trp{\bv{y}}\tilde{\mat{M}}).
    \end{equation*} 
    Thus, by
    \Cref{lem_Birkhoff of matrix product} and since $\ln((1+x)/(1-x))$ is monotone increasing for $0 \leq x < 1$, \eqref{eq_projective distance ei}
    will follow if we show that 
    \[ 
        d(\trp{\bv{e}_{i_0}}\tilde{\mat{M}}, \trp{\bv{x}}\tilde{\mat{M}}) 
        \leq \ln \left( \frac{1}{\phi(\tilde{\mat{M}})} \right) 
        = 2 \ln \left( \frac{1+ \Birkhoff{\tilde{\mat{M}}} }{ 1 - \Birkhoff{\tilde{\mat{M}}} }\right), 
    \]
    where $\phi$ is defined in~\eqref{eq_def of phi}.
    The right-hand equality follows directly from \Cref{thm_birkhoff explicit form}, so we concentrate on proving the inequality. 

    For any $j \in \nzcols{\tilde{\mat{M}}}$ denote  
            \[ 
                \rho_j 
                = \frac{ \matel{ \trp{\bv{e}_{i_0}} \tilde{\mat{M}} }{j} }{ \matel{ \trp{ \bv{x}} \tilde{\mat{M} }}{j} } 
                = \frac{ \matel{\tilde{\mat{M}}}{i_0,j} } { \sol[r]{\sum\limits_{k \in \nzrows{\tilde{\mat{M}}}} } x_k \matel{\tilde{\mat{M}}}{k,j} }.
            \] 
    The denominator is positive by~\eqref{eq_sigma mtilde}, so $\rho_j$ is well-defined. 
    Now, for $j, l \in \nzcols{\tilde{\mat{M}}}$,
    \begin{align} 
        \frac{\rho_j}{\rho_l} 
        &= \frac{  \sol[r]{\sum\limits_{k \in \nzrows{\tilde{\mat{M}}} }} x_k \matel{\tilde{\mat{M}}}{k,l}}
                {  \sol[r]{\sum\limits_{k \in \nzrows{\tilde{\mat{M}}} }} x_k \matel{\tilde{\mat{M}}}{k,j} }   \cdot 
                \frac{ \matel{\tilde{\mat{M}}}{i_0,j}  }{  \matel{\tilde{\mat{M}}}{i_0,l}   } \nonumber \\ 
        &\eqann[\leq]{a} \max_{k \in \nzrows{\tilde{\mat{M}}} }  \frac{ \matel{\tilde{\mat{M}}}{k,l}  }{  \matel{\tilde{\mat{M}}}{k,j}   } \cdot  
         \frac{ \matel{\tilde{\mat{M}}}{i_0,j}  }{  \matel{\tilde{\mat{M}}}{i_0,l}   } \nonumber\\
        &\eqann[\leq]{b} \max_{k \in \nzrows{\tilde{\mat{M}}} }  \frac{ \matel{\tilde{\mat{M}}}{k,l}  }{  \matel{\tilde{\mat{M}}}{k,j}   } \cdot 
         \max_{i \in \nzrows{\tilde{\mat{M}}}}  \frac{ \matel{\tilde{\mat{M}}}{i,j}  }{  \matel{\tilde{\mat{M}}}{i,l}   }, \label{eq_rhoj rhol ratio} 
    \end{align} 
    where \eqannref{a} is by \Cref{lem_sum ratio inequality} and in \eqannref{b} we maximize over a set that contains $i_0$. 

    Hence, recalling the definition of the projective distance,~\eqref{eq_hilbert projective distance},
    \begin{align*} 
        d(\trp{\bv{e}_{i_0}}\tilde{\mat{M}}, \trp{\bv{x}} \tilde{\mat{M}} ) 
        &= \ln \max_{j,l \in \nzcols{\tilde{\mat{M}}}}  \frac{\rho_j}{\rho_l}\\ 
        &\eqann[\leq]{a} \ln \max_{\substack{i,k \in \nzrows{\tilde{\mat{M}}}, \\ j,l \in \nzcols{\tilde{\mat{M}}} }} 
         \frac{ \matel{\tilde{\mat{M}}}{i,j} \matel{\tilde{\mat{M}}}{k,l} }{\matel{\tilde{\mat{M}}}{i,l} \matel{\tilde{\mat{M}}}{k,j} } \\  
        &\eqann{b} \ln \left( \frac{1}{\phi(\tilde{\mat{M}})} \right), 
    \end{align*} 
    where \eqannref{a} is by~\eqref{eq_rhoj rhol ratio} and \eqannref{b} follows from the definition of $\phi$ in~\eqref{eq_def of phi}.
    This completes the proof.
\end{IEEEproof}

The following proposition and the corollary that follows are a generalization of ideas from~\cite[Theorem 2]{Hochwald_Jelenkovic_Markov_1999}. 
\begin{proposition}\label{prop_hochwald generalization}
    Let $\mat{M}_1, \mat{M}_2, \ldots, \mat{M}_m, \mat{T}$ be a sequence of square nonzero nonnegative matrices, such that $\mat{M}_{\ell}$ are
    subrectangular for all $1 \leq \ell \leq m$, and let $\bv{x}, \bv{y}$ be two nonnegative nonzero vectors.
    Denote 
    \begin{align*} 
        \trp{\tilde{\bv{x}}} &= \trp{\bv{x}}\mat{M}_1, \\ 
        \trp{\tilde{\bv{y}}} &= \trp{\bv{y}}\mat{M}_1, \\ 
        \mat{M}_r^{s} &= \mat{M}_r \cdot \mat{M}_{r+1} \cdots \mat{M}_{s}, \quad r \leq s.
    \end{align*}
    If $\norm{\trp{\bv{x}}\mat{M}_1^m\mat{T}}_1 > 0$ and $\norm{\trp{\bv{y}}\mat{M}_1^m\mat{T}}_1 > 0$, then 
    \begin{equation} \label{eq_hochwald inequality}
        \ln\left(\frac{\norm{\trp{\bv{x}}\mat{M}_1^m\mat{T}}_1}{\norm{\trp{\bv{y}}\mat{M}_1^m\mat{T}}_1} \cdot
        \frac{\norm{\trp{\bv{y}}\mat{M}_1^{m}}_1}{\norm{\trp{\bv{x}}\mat{M}_1^{m}}_1}\right) 
        \leq d(\tilde{\bv{x}}, \tilde{\bv{y}}) \cdot \prod_{\ell=2}^{m}\Birkhoff{ \mat{M}_{\ell} }.
    \end{equation} 
\end{proposition}
\begin{IEEEproof}
    Since $\norm{\trp{\bv{x}}\mat{M}_1^m\mat{T}}_1 > 0$, we conclude that $\trp{\bv{x}}\mat{M}_1^{s}$ is nonzero for any $1 \leq s \leq m$, and the
    same holds if we replace $\bv{x}$ with $\bv{y}$.
    Thus, the left-hand side of~\eqref{eq_hochwald inequality} is well-defined. 
    We will show that
    \[ 
        \ln\left(\frac{\norm{\trp{\bv{x}}\mat{M}_1^m\mat{T}}_1}{\norm{\trp{\bv{y}}\mat{M}_1^m\mat{T}}_1} \cdot
        \frac{\norm{\trp{\bv{y}}\mat{M}_1^{m}}_1}{\norm{\trp{\bv{x}}\mat{M}_1^{m}}_1}\right) \leq
        d(\trp{\tilde{\bv{x}}}\mat{M}_2^{m},\trp{\tilde{\bv{y}}}\mat{M}_2^{m} ).
    \] 
    The right-hand side is well-defined since,  by \Cref{cor_subrectangular}, $\mat{M}_r^s$ is subrectangular for any $1 \leq r \leq s \leq m$ and by
    \Cref{lem_vector times subrectangular}.
    Then, as $\support{\tilde{\bv{x}}} = \support{\tilde{\bv{y}}}$ by \Cref{lem_vector times subrectangular},~\eqref{eq_hochwald inequality} will
    follow from \Cref{cor_projective distance inequality} and~\eqref{eq_birkhoff product inequality}. 

    Denote $J = \support{\trp{\tilde{\bv{x}}} \mat{M}_2^{m}} = \support{\trp{\tilde{\bv{y}}} \mat{M}_2^{m}} = \nzcols{\mat{M}_2^{m}}$, where the
    equalities are by \Cref{lem_vector times subrectangular} and subrectangularity.  
    By the right-hand inequality of~\eqref{eq_sum ratio inequality},
    \begin{align*} 
        \frac{\norm{\trp{\bv{y}}\mat{M}_1^{m}}_1}{\norm{\trp{\bv{x}}\mat{M}_1^{m}}_1} 
        &= \frac{\norm{\trp{\tilde{\bv{y}}}\mat{M}_2^{m}}_1}{\norm{\trp{\tilde{\bv{x}}}\mat{M}_2^{m}}_1}\\
        &= \frac{ \sum_{l\in J} 1 \cdot \matel{\trp{\tilde{\bv{y}}}\mat{M}_2^{m}}{l} }{ \sum_{l\in J} 1 \cdot 
           \matel{\trp{\tilde{\bv{x}}}\mat{M}_2^{m}}{l}} \leq 
           \max_{l \in J} \frac{\matel{\trp{\tilde{\bv{y}}}\mat{M}_2^{m}}{l}}{\matel{\trp{\tilde{\bv{x}}}\mat{M}_2^{m}}{l}}. 
    \end{align*} 

    Next, denote by $t_j = \norm{\matel{\mat{T}}{j,:}}_1$ the sum of the $j$th row of $\mat{T}$.
    Since $\mat{T}$ is nonzero, $t_j > 0$ for some $j$.
    Thus, a second application of the right-hand inequality of~\eqref{eq_sum ratio inequality} yields
    \[ 
        \frac{\norm{\trp{\bv{x}}\mat{M}_1^m\mat{T}}_1}{\norm{\trp{\bv{y}}\mat{M}_1^m\mat{T}}_1} 
        = \frac{ \sum_{j\in J} t_j \cdot \matel{\trp{\tilde{\bv{x}}}\mat{M}_2^{m}}{j} }{ \sum_{j\in J} t_j \cdot \matel{\trp{\tilde{\bv{y}}}\mat{M}_2^{m}}{j}} 
        \leq \max_{j \in J} \frac{\matel{\trp{\tilde{\bv{x}}}\mat{M}_2^{m}}{j}}{\matel{\trp{\tilde{\bv{y}}}\mat{M}_2^{m}}{j}}. 
    \]
    Combining, we obtain  
    \[ \frac{\norm{\trp{\bv{x}}\mat{M}_1^m\mat{T}}_1}{\norm{\trp{\bv{y}}\mat{M}_1^m\mat{T}}_1} \cdot
        \frac{\norm{\trp{\bv{y}}\mat{M}_1^{m}}_1}{\norm{\trp{\bv{x}}\mat{M}_1^{m}}_1} 
        \leq \max_{j,l \in J} \frac{\matel{\trp{\tilde{\bv{x}}}\mat{M}_2^{m}}{j} / \matel{\trp{\tilde{\bv{y}}}\mat{M}_2^{m}}{j}}
                                   {\matel{\trp{\tilde{\bv{x}}}\mat{M}_2^{m}}{l} / \matel{\trp{\tilde{\bv{y}}}\mat{M}_2^{m}}{l}}. 
    \] 
    Taking the logarithm of both sides, the right-hand side becomes $d(\trp{\tilde{\bv{x}}}\mat{M}_2^{m},\trp{\tilde{\bv{y}}}\mat{M}_2^{m} )$,
    which completes the proof.  
\end{IEEEproof}

Combining the above results we obtain the following corollary. 
\begin{corollary} \label{cor_contraction inequality}
    Let $\mat{M}_1, \mat{M}_2, \ldots, \mat{M}_m$ be a sequence of square nonzero subrectangular matrices, and let $\mat{T}$, as well as 
    $\mat{T}_1, \mat{T}_2, \ldots, \mat{T}_m$ be arbitrary square nonnegative and nonzero matrices.
    Denote
    \[ 
        \mat{R} = \mat{T}_1 \mat{M}_1 \mat{T}_2 \mat{M}_2 \cdots \mat{T}_m \mat{M}_m.  
    \]
    Then, for any two nonnegative nonzero vectors $\bv{x}, \bv{y}$ such that $\norm{\trp{\bv{x}}\mat{R}\mat{T}}_1 > 0$ and 
    $\norm{\trp{\bv{y}}\mat{R}\mat{T}}_1 > 0$
    we have 
    \begin{IEEEeqnarray}{l}
        \ln\left(\frac{\norm{\trp{\bv{x}}\mat{R}\mat{T}}_1}{\norm{\trp{\bv{y}}\mat{R}\mat{T}}_1} \cdot
        \frac{\norm{\trp{\bv{y}}\mat{R}}_1}{\norm{\trp{\bv{x}}\mat{R}}_1}\right)   
        \leq 4 \ln \left( \frac{1+ \Birkhoff{\mat{M}_1}}{1-\Birkhoff{\mat{M}_1}} \right) \cdot \prod_{\ell=2}^{m}\Birkhoff{ \mat{M}_{\ell} }.
        \IEEEeqnarraynumspace  \label{eq_contraction inequality}
    \end{IEEEeqnarray} 
\end{corollary}
\begin{IEEEproof}
The claim follows from \Cref{cor_subrectangular,lem_Birkhoff of matrix product,lem_projective distance of ei and y times sr matrix,prop_hochwald generalization}. 
\end{IEEEproof}
Observe that~\eqref{eq_contraction inequality} remains true if we replace `$\ln$' with `$\log$'. 

\begin{discussion}
    Our \Cref{prop_hochwald generalization,cor_contraction inequality} generalize \cite[Theorem 2]{Hochwald_Jelenkovic_Markov_1999} in several ways.
    In \cite[Theorem 2]{Hochwald_Jelenkovic_Markov_1999}, the matrices $\mat{M}_1, \ldots, \mat{M}_m, \mat{T}$ are all strictly positive.
    Each matrix corresponds to an observation of a hidden Markov model $(A_n, B_n)$, where the $(i,j)$ item of the matrix that corresponds to
    observation $b \in \mathcal{B}$ is the probability that $A_{n+1} = j$ and $B_{n+1} = b$ given that $A_n = i$.  
    In particular, \cite[Theorem 2]{Hochwald_Jelenkovic_Markov_1999} assumes that every observation $b \in \mathcal{B}$ can be emitted from the same
    number of states $a \in \mathcal{A}$,\footnote{We note that the authors of~\cite{Hochwald_Jelenkovic_Markov_1999} claim that this assumption can be
    relaxed with an appropriate extension, but they omit it and its derivation.} and that it is possible to transition between any two states of
    $\mathcal{A}$ in one step.   
    In this work, however, we are not confined to such assumptions.
    Our formulation allows for each observation to originate from a different number of states.
    Moreover, our formulation does not assume that one can move from every state of $\mathcal{A}$ to every other state of $\mathcal{A}$ in one step. 
\end{discussion}

\section{Hidden Markov Models that Forget their Initial State} \label{sec_HMM}
In this section we show that hidden Markov models that satisfy a mild requirement forget their initial state. 
Specifically, we will consider the mutual information between the state at time $n+1$ and the model's initial state given the observations in between.
The contraction inequality of \Cref{sec_contraction} will enable us to show that this mutual information vanishes with $n$. 
This enables us to obtain a sufficient condition --- \Cref{cond_kaijser} --- for forgetfulness. 
The development in this section is based on the techniques of~\cite{kaijser1975}.

\subsection{Hidden Markov Models}
A hidden Markov model (HMM) is a process $(A_n, B_n)$, where $A_n \in \mathcal{A}$ is a Markov chain and $B_n \in \mathcal{B}$ is an observation that
is a function of $A_n$. 
The alphabets $\mathcal{A}$ and $\mathcal{B}$ are assumed finite. 
Without loss of generality, $\mathcal{A} = \{1, 2, \ldots, |\mathcal{A}|\}$ and $\mathcal{B} = \{ 1,2,\ldots, |\mathcal{B}|\}$. 
A detailed description of the setting we consider follows.   

Let $A_n$, $n \in \mathbb{Z}$ be a homogeneous Markov process assuming values in some finite alphabet $\mathcal{A}$. 
Denote by $p(j|i)$ its transition probability function, which is independent of the time index $n$. 
That is, 
\[ 
    p(j | i) = \Prob{A_n = j | A_{n-1} =i}, \quad i,j \in \mathcal{A}.
\] 
The $|\mathcal{A}| \times |\mathcal{A}|$ transition probability matrix $\mat{M}$ of the Markov chain is defined by 
\[ 
    \matel{\mat{M}}{i,j} = p(j|i), \quad i,j \in \mathcal{A}.
\] 
This is a stochastic matrix: $\matel{\mat{M}}{i,j} \geq 0$ for all $i,j \in \mathcal{A}$ and for any $i$, $\sum_{j} \matel{\mat{M}}{i,j} = 1$. 
We assume that the process $A_n$ is aperiodic and irreducible (in some literature such Markov chains are called \emph{regular}). 
That is, we assume that the matrix $\mat{M}$ is aperiodic and irreducible (see, e.g., ~\cite[Proposition 4.1]{Iosifescu}).  
This implies~\cite[Theorems 1.9 and 4.2]{Iosifescu} that the process has a unique stationary distribution $\bs{\pi}$, which is positive.

Let $f: \mathcal{A} \to \mathcal{B}$ be a \emph{deterministic} function. 
For simplicity, we assume that $\mathcal{B}$ is finite. An observation of $A_n$  is $B_n = f(A_n)$. 
Denote, for any set $B \subseteq \mathcal{B}$, 
\[
    f^{-1}(B) = \{ i \in \mathcal{A} \, | \, f(i) = b, \, b \in B \}.
\]
Then, $\Prob{B_n = b} = \Prob{A_n \in f^{-1}(b)}$. 
We assume that $\mathcal{B}$ contains only observations that actually appear, that is, $\mathcal{B} = \{ b \,|\, f(i) = b, i \in \mathcal{A}\}$. 

The process $(A_n,B_n)$ described above is called a \emph{hidden Markov model}. 
We summarize this in the following definition. 
\begin{definition}[Hidden Markov model] \label{def_HMM}
    Let $A_n$ be a homogeneous Markov process taking values in $\mathcal{A}$ with transition probability matrix $\mat{M}$, which is aperiodic and
    irreducible. 
    Let $f:\mathcal{A} \to \mathcal{B}$ be a deterministic function, and let $B_n = f(A_n)$.
    The process $(A_n, B_n)$ is called a hidden Markov model. 
    Additionally, we use the following terminology: 
    \begin{itemize} 
        \item $A_n$ is the \emph{state} of the process, 
        \item $B_n$ is the \emph{observation} of the process. 
    \end{itemize}
 \end{definition}
 Typically, multiple states would have the same observation. That is, for $b \in \mathcal{B}$, the set $f^{-1}(b)$ typically contains multiple
 elements. 
 The actual state of the process is hidden, and the observation provides only partial information on the state.

The restriction to a deterministic function $f$, rather than a probabilistic one, seemingly presents a limitation. 
However, in appendix~\ref{ap_markov model equivalence} we show that there is no loss of generality in assuming that $f$ is deterministic. 
That is, we show that the deterministic and probabilistic settings are equivalent. 
We emphasize that taking the viewpoint of deterministic $f$ is done for convenience and to facilitate the derivation that follows. 
In particular, in our setting of a FAIM process, $(S_n, X_n, Y_n)$, without loss of generality one may assume that $(X_n, Y_n)$ is a deterministic
function of the state $S_n$.

The following notation, taken from~\cite{kaijser1975}, will be useful. 
Define the matrices $\mat{M}(b)$, $b \in \mathcal{B}$, by 
\begin{equation} \label{eq_definition of Mb}
    \matel{\mat{M}(b)}{i,j} = 
    \begin{cases} p(j|i), & \textrm{if } f(j) = b \\ 0, & \textrm{otherwise.} \end{cases} 
\end{equation} 
In words, $\matel{\mat{M}(b)}{i,j}$ is the probability of transitioning from state $i\in \mathcal{A}$ to state $j\in\mathcal{A}$ and observing
$b\in\mathcal{B}$ after having arrived at state $j$. 
That is, 
\[
    \matel{\mat{M}(b)}{i,j} = \Prob{A_n = j, B_n = b | A_{n-1} = i}.
\] 
For a sequence of observations $b_r^{s}$, $r \leq s$, we denote 
\[ 
    \mat{M}(b_r^{s}) \triangleq \mat{M}(b_r) \mat{M}(b_{r+1}) \cdots \mat{M}(b_{s}).
\] 
We call $\Birkhoff{\mat{M}(b_r^s)}$ the Birkhoff contraction coefficient \emph{induced} by the sequence $b_r^s$. 

The matrices $\mat{M}(b)$ are nonzero and  substochastic --- they are nonnegative with unequal row sums, all less than or equal to $1$. 
We can reconstruct $\mat{M}$ from them using
\[ 
    \mat{M} = \sum_b \mat{M}(b).
\] 
\Cref{ex_M for Gilbert Elliott} in appendix~\ref{ap_markov model equivalence} shows the matrix $\mat{M}$ and its decomposition to matrices
$\mat{M}(b)$ for a specific channel with memory. 

We also define for any $a \in \mathcal{A}$ the matrix $\Idmat_a$ by 
\[ 
    \matel{\Idmat_a}{i,j} = \begin{cases} 1, & \text{if } i=j = a\\ 0, & \text{otherwise}.  \end{cases} 
\] 
This matrix has a single nonzero element: `$1$' on the diagonal, at the $(a,a)$ position. 

The process $(A_n, B_n)$ is completely characterized by the matrices $\mat{M}(b)$, $b \in \mathcal{B}$, and its initial distribution. 
We assume that the process is stationary, so its initial distribution is $\bs{\pi}$, its unique stationary distribution. 
Thus, $\matel{\bs{\pi}}{i} = \Prob{A_0 = i}$ and
\begin{align*}
    \Prob{B_1 = b_1} &= \sum_{j \in \mathcal{A}} \Prob{A_1 = j, B_1 = b_1} \\ 
                     &= \sol{\sum_{i,j \in \mathcal{A}}} \Prob{A_1 = j, B_1 = b_1 | A_0 = i} \Prob{A_0 = i} \\
                     &= \norm{ \trp{\bs{\pi}}\mat{M}(b_1) }_1
\end{align*}
Moreover, the probability of observing the sequence $b_1^n$ is given by~\cite[Lemma 2.1]{kaijser1975}
\begin{align} 
        \Prob{B_1^n = b_1^n} &= \norm{\trp{\bs{\pi}}\mat{M}(b_1^n)}_1 \nonumber \\ 
                             &= \norm{\trp{\bs{\pi}}\mat{M}(b_1)\mat{M}(b_2)\cdots \mat{M}(b_n)}_1. \label{eq_prob of B sequence}
\end{align}
Similarly, for any $a\in \mathcal{A}$,  
\[
    \Prob{A_n = a, B_1^n = b_1^n} = \matel{\bs{\pi}^T\mat{M}(b_1^n)}{a} 
                                    = \norm{\trp{\bs{\pi}}\mat{M}(b_1^n) \Idmat_a}_1,
                                \] 
and
\begin{align}
    \Prob{A_{n+1} = a, B_1^n = b_1^n} 
    &= \matel{\bs{\pi}^T\mat{M}(b_1^n)\mat{M}}{a} \nonumber \\ 
    &= \norm{\trp{\bs{\pi}}\mat{M}(b_1^n) \mat{M}\Idmat_a}_1 \nonumber\\ 
    &= \norm{\trp{\bs{\pi}}\mat{M}(b_1^n) \mat{T}_{a} }_1, \label{eq_prob of A B sequence}
\end{align}
where we denoted for any $a \in \mathcal{A}$,  
\[ 
    \mat{T}_a \triangleq \mat{M} \Idmat_a.
\]

When $\Prob{B_1^n = b_1^n} > 0$ we further have by~\eqref{eq_prob of B sequence} and~\eqref{eq_prob of A B sequence}, 
\begin{align} 
    \Prob{A_{n+1} = a | B_1^{n} = b_1^{n}} 
    &= \frac{\Prob{ A_{n+1} = a, B_1^n = b_1^n}}{\Prob{B_1^{n} = b_1^{n}}} \nonumber \\ 
    &= \frac{ \norm{\trp{\bs{\pi}}\mat{M}(b_1^n)\mat{T}_{a}}_1 }{  \norm{\trp{\bs{\pi}}\mat{M}(b_1^{n})}_1 }.  \label{eq_prob of B conditioned}
\end{align}
This is well-defined because if $\Prob{B_1^n = b_1^n} > 0$ then the denominator on the right-hand side of~\eqref{eq_prob of B conditioned} must also be positive. 

Let us now consider the case where the initial state of the process is known. 
In this case, $\Prob{B_1 = b_1 | A_0 = a_0} = \norm{\trp{\bv{e}_{a_0}}\mat{M}(b_1)}_1$.  
Similar to the above, we obtain
\begin{IEEEeqnarray}{rCl}  
    \Prob{B_1^{n} = b_1^{n}| A_0 = a_0} &=&  \norm{\trp{\bv{e}_{a_0}}\mat{M}(b_1^n)}_1, \label{eq_prob of B sequence given A0} \\[0.1cm] 
    \Prob{A_{n+1} = a | B_1^{n} = b_1^{n}, A_0 = a_0} &=& 
    \frac{ \norm{\trp{\bv{e}_{a_0}}\mat{M}(b_1^n)\mat{T}_{a}}_1 }{  \norm{\trp{\bv{e}_{a_0}}\mat{M}(b_1^{n})}_1 },
    \IEEEeqnarraynumspace     \label{eq_prob of B sequence given A0 cond} 
 \end{IEEEeqnarray}
provided that the probability in~\eqref{eq_prob of B sequence given A0} is positive. 

In \eqref{eq_prob of B sequence}--\eqref{eq_prob of B sequence given A0 cond}, we have computed probabilities for particular realizations of
$A_0$, $B_1^n$ and $A_{n+1}$. 
Generally, however, these are random variables. 
They are jointly generated as follows. 
First, draw $A_0$ according to $\bs{\pi}$. 
Then, at time $n$, draw $A_n$ according to the $A_{n-1}$th row of $\mat{M}$ and compute $B_n = f(A_n)$. 

These random variables give rise to the random variables $\Prob{A_{n+1} |B_1^{n}}$ and $\Prob{A_{n+1} | B_1^{n}, A_0}$, obtained by plugging 
$A_{n+1}, B_1^n$, and $A_0$ for $a$, $b_1^n$, and $a_0$ respectively in the right-hand sides of~\eqref{eq_prob of B conditioned}
and~\eqref{eq_prob of B sequence given A0 cond}. 
They are well-defined with probability $1$. 
In other words, we can always compute their values via~\eqref{eq_prob of B conditioned} and~\eqref{eq_prob of B sequence given A0 cond}; with
probability $0$ will the denominators on the right-hand sides of these equations equal $0$. 
These random variables are of interest because
\begin{equation} \label{eq_mutual information expression}
    I(A_0 ; A_{n+1} | B_1^{n}) = \Exp{\log \frac{\Prob{A_{n+1} | B_1^{n}, A_0}}{\Prob{A_{n+1} | B_1^{n}}} }.
\end{equation} 
Using~\eqref{eq_prob of B conditioned} and~\eqref{eq_prob of B sequence given A0 cond} we write this as 
\begin{IEEEeqnarray}{l}
        I(A_0; A_{n+1} | B_1^{n} ) \IEEEnonumber\\  
     = \Exp{\log \left( 
            \frac{ \norm{\trp{\bv{e}_{A_0}}\mat{M}(B_1^n) \mat{T}_{A_{n+1}} }_1}{\norm{\trp{\bs{\pi}}\mat{M}(B_1^n) \mat{T}_{A_{n+1}} }_1}
            \cdot 
            \frac{\norm{\trp{\bs{\pi}}\mat{M}(B_1^{n})}_1}{\norm{\trp{\bv{e}_{A_0}}\mat{M}(B_1^{n})}_1 }
    \right) }. \IEEEeqnarraynumspace \label{eq_I Bn A0 given middle}
\end{IEEEeqnarray}
As above, the argument of the expectation is well-defined with probability $1$.

The Markov chain $A_n$ is finite-state, irreducible, and aperiodic. 
A classic result on such Markov chains~\cite[Theorem 4.3]{Iosifescu}, \cite[Theorem 8.9]{billingsley1995probability}, which harks back to the days of
A. A. Markov~\cite{seneta_markov}, is that the chain approaches its stationary distribution exponentially fast, regardless of its initial state. 
In particular, this implies that $I(A_0; A_{n+1}) \to 0$ as $n\to\infty$. 
By the Markov property we also have $I(A_0; A_{n+1} |A_1^n) = 0$. 
Our setting, however, is a hidden Markov setting, and we will be interested in whether $I(A_0; A_{n+1} | B_1^n) \to 0$. 
In general, the answer to this is negative --- even when $A_n$ is finite-state, aperiodic, and irreducible --- see \Cref{ex_kaijser counter example}
in \Cref{sec_probabilistic model with memory}, above.\footnote{Where the state is $A_n = S_n$ and the observation is $B_n = Y_n$.
It can be shown~\cite[Section 10]{kaijser1975} that this HMM does not satisfy \Cref{cond_kaijser}.} 

Our goal in the next subsection is to show that under a certain \Cref{cond_kaijser}, $I(A_0; A_{n+1}| B_1^{n}) \to 0$ as $n \to \infty$. 
This will employ~\eqref{eq_contraction inequality}, which bounds expressions of a form similar to the argument of the expectation 
in~\eqref{eq_I Bn A0 given middle}. 
\begin{remark}
    An expression similar to~\eqref{eq_mutual information expression} was pointed out in~\cite[Equation 3.7]{Hochwald_Jelenkovic_Markov_1999}, 
    in the proof of~\cite[Theorem 2]{Hochwald_Jelenkovic_Markov_1999}.  
    There, the goal  was to show that $I(A_0; B_n|B_1^{n-1}) \to 0$.
    This was done under a restrictive assumption that transitions between any two states in one step may happen with strictly positive probability.
    When put in our notation, this implies that the matrices $\mat{M}(b)$, $b \in \mathcal{B}$, contain only two types of columns: strictly positive
    columns and zero columns.\footnote{The assumption of \cite{Hochwald_Jelenkovic_Markov_1999} is that $\mat{M}$ is positive. Since $\mat{M}(b)$ is
        comprised of columns of $\mat{M}$ and zero columns, any nonzero column of $\mat{M}(b)$ must be positive.}  
    In this case, the matrices $\mat{M}(b)$ are all subrectangular, so their Birkhoff contraction coefficients are strictly less than $1$; 
    this allows one to use~\eqref{eq_contraction inequality} directly (with $\mat{T}_{\ell}
    = \Idmat$ for all $\ell$) and obtain that the mutual information indeed vanishes as $n$ grows. In this paper, we alleviate this restrictive
    assumption, and allow for a more general scenario where the individual matrices $\mat{M}(b)$ may also be \emph{not} subrectangular.
    We further remark that,  by the data processing inequality~\eqref{eq_DPI}, $I(A_0; A_{n+1} | B_1^{n}) \to 0$ implies that $I(A_0; B_{n+1} | B_1^n) \to 0$. 
\end{remark}

\subsection{Forgetting the Initial State}
We now show that under the following \emph{\Cref{cond_kaijser}} (so named in honor of Prof. Thomas Kaijser who had first suggested it
in~\cite{kaijser1975}), the mutual information $I( A_0;A_{n+1}  | B_1^{n})$ vanishes with $n$. 

\setcounter{condition}{10} 
\begin{condition}           \label{cond_kaijser}
    The HMM $(A_n, B_n)$ is characterized by matrices $\mat{M}(b)$, $b \in \mathcal{B}$ such that: 
    \begin{enumerate}
        \item The matrix $\mat{M}  = \sum_{b \in \mathcal{B}} \mat{M}(b)$ is aperiodic and irreducible. 
        \item There exists an ordered sequence $\beta_1, \beta_2, \ldots, \beta_l$ of elements of $\mathcal{B}$ such that the matrix 
              $\mat{M}(\beta_1^l) = \mat{M}(\beta_1) \mat{M}(\beta_2) \cdots \mat{M}(\beta_l)$ is nonzero and subrectangular. 
    \end{enumerate}
\end{condition}

The following are all examples where it is easy to check by inspection that \Cref{cond_kaijser} is satisfied: 
\begin{itemize}
    \item the transition matrix $\mat{M}$ is positive (or, more generally, subrectangular);
    \item there exists an observation $\beta$ for which $\mat{M}(\beta)$ has just a single column;
    \item there exists an observation $\beta$ for which $\mat{M}(\beta)$ is subrectangular. 
\end{itemize}
Generally, though, inspection may not suffice to declare that \Cref{cond_kaijser} is satisfied.
\begin{remark}
    The ability of a hidden Markov model to ``forget'' its initial state has also been studied under somewhat weaker assumptions than
    \Cref{cond_kaijser}. 
    The interested reader is invited to consult \cite{Kochman_Reeds,Chigansky_VanHandel}.
    It may be possible to generalize the results of this paper to processes that satisfy these weaker assumptions and do not satisfy \Cref{cond_kaijser}. 
    We leave such endeavors to future work. 
\end{remark}

\begin{theorem}\label{thm_upper bound on I epsilon} 
    Suppose the HMM $(A_n,B_n)$ satisfies \Cref{cond_kaijser}. 
    Then, for every $\epsilon > 0$ there exists an integer $\memlength$ such that if $n \geq \memlength$ then 
    \[ 
        I(A_0;A_{n+1}| B_1^{n}) \leq \epsilon.
    \] 
\end{theorem}
The proof is given in the next subsection, and will follow from \Cref{prop_upper bound on I accurate}, which provides a characterization of the rate at
which the mutual information vanishes. 
The idea is to use techniques similar to the ones used in the study of recurrence times of Markov chains. 
Namely, we bound the probability that in a long sequence of observations there will be sufficient non-overlapping occurrences of sequences that induce
a Birkhoff contraction coefficient below a certain threshold. 
Armed with this bound, we employ \Cref{cor_contraction inequality} in~\eqref{eq_I Bn A0 given middle} to obtain an upper bound on the mutual information. 

\begin{example}
 Let $A_n$ be a finite-state Markov chain with irreducible and aperiodic transition probability matrix $\mat{M}$. 
 Consider the case of no observations: $B_n = 0$ regardless of $A_n$.
 In this case, $\mat{M}(0) = \mat{M}$ and \Cref{cond_kaijser} is satisfied, as there exists $k_0$ such that $\mat{M}^{k_0} > 0$ \cite[Theorem 1.4]{Seneta}.
 Therefore, by~\Cref{thm_upper bound on I epsilon}, we have $I(A_0; A_{n+1}) \to 0$ as $n \to \infty$.
 As mentioned above, this is a well-known result for finite-state, irreducible, and aperiodic Markov chains.
 We note in passing that there exist other information-theoretic proofs that $I(A_0; A_{n+1}) \to 0$ as $n \to \infty$, see, e.g.,~\cite{kullback_markov}.  
\end{example}

\begin{corollary}\label{cor_HMM kaijser}
    Suppose the HMM $(A_n,B_n)$ satisfies \Cref{cond_kaijser}. 
    Then, for every $\epsilon > 0$ there exists an integer $\memlength$ such that if $n \geq \memlength$ then 
    \begin{equation} \label{eq_I An A1 inequality}
        I( A_1; A_{n} | B_1^{n}) \leq \epsilon. 
    \end{equation}
    and
    \begin{equation} \label{eq_I An A0 inequality}
        I( A_0; A_{n} | B_1^{n}) \leq \epsilon. 
    \end{equation}
\end{corollary}
\begin{IEEEproof}
    The conditions of \Cref{thm_upper bound on I epsilon} are satisfied.
    Let $\memlength$ be such that $I(A_1; A_n | B_2^{n-1}) \leq \epsilon$ for any $n \geq \memlength$. 

    We first show~\eqref{eq_I An A1 inequality}. 
    Recall that $B_j$ is a function of $A_j$ for any $j$. 
    Thus, for any $n \geq \memlength$, $I( A_1, B_1; A_n, B_n | B_2^{n-1}) = I(A_1; A_n |B_2^{n-1}) \leq \epsilon$.
    Therefore, 
    \begin{align*} 
        \epsilon &\geq I( A_1, B_1; A_n,B_n | B_2^{n-1}) \\
        &= I( B_1 ; A_n,B_n  | B_2^{n-1}) + I(A_1 ; B_n | B_1^{n-1}) + I(A_1; A_n|B_1^n).
    \end{align*} 
    Since mutual information is nonnegative, each of the summands on the right-hand side is upper-bounded by $\epsilon$. 
    This yields~\eqref{eq_I An A1 inequality}. 

    To see~\eqref{eq_I An A0 inequality}, since $A_0 \markov (A_1, B_1^n) \markov A_n$, we use~\eqref{eq_DPI} and obtain 
    \[ 
        I(A_0; A_n | B_1^n) \leq I(A_1; A_n | B_1^n) \leq \epsilon, 
    \] 
    as required. 
\end{IEEEproof}

We remark that under the same conditions as \Cref{cor_HMM kaijser} we also obtain $I(A_1; B_n | B_1^{n-1}) \leq \epsilon$ and $I(A_0; B_n | B_1^{n-1})
\leq \epsilon$.

Consider a Markov chain $A_n$ and \emph{two} HMMs it induces, $(A_n,B_n)$ and $(A_n, C_n)$, where $B_n = f(A_n)$ and $C_n
= g(B_n)$, for some deterministic functions $f,g$.
It is somewhat surprising, but even if one of the HMMs satisfies \Cref{cond_kaijser}, it does not imply that the other one does. 
See \Cref{ex_function counter example} in \Cref{sec_polarizable soprocesses}.\footnote{Taking $A_n = S_n$, $B_n = (X_n, Y_n)$, and $C_n
= Y_n$.} 
Suppose that both HMMs satisfy \Cref{cond_kaijser}. Then, by \Cref{cor_HMM kaijser}, for every $\epsilon>0$ there exists an integer $\memlength$ such that if
$n \geq \memlength$ then $I(A_1;A_n | B_1^n) \leq \epsilon$ and $I(A_1; A_n | C_1^n) \leq \epsilon$. 

\begin{example} \label{ex_memory length for SXY}
    Let $(S_n, X_n, Y_n)$ be a FAIM process. This is an HMM with state $A_n = (S_n, X_n, Y_n)$. 
    Clearly, there exist functions $f, g$ such that $(X_n, Y_n) = f(S_n)$ and $Y_n = g(X_n, Y_n)$. 
    Therefore, both $(A_n, (X_n, Y_n))$ and $(A_n, Y_n)$ are HMMs. 
    If each of the HMMs $(A_n, (X_n, Y_n))$ and $(A_n, Y_n)$ satisfies \Cref{cond_kaijser} then~\eqref{eq_I An A1 inequality} holds with 
    $B_n=(X_n,Y_n)$ or $B_n = Y_n$ for any $n$. 
    In particular, for any $\epsilon>0$ there exists an integer $\memlength$ such that for any $k \geq \memlength$ we have
    \begin{align*} 
        I(S_1; S_k | X_{1}^{k}, Y_{1}^k) &\leq \epsilon, \\ 
        I(S_1; S_k | Y_1^k) &\leq \epsilon. 
    \end{align*} 
    In other words, \Cref{cond_kaijser} is a sufficient condition for forgetfulness. 
\end{example}

\subsection{Proof of \Cref{thm_upper bound on I epsilon}}
The goal of this subsection is to prove \Cref{thm_upper bound on I epsilon}. 
To this end, we make the following definition. 
\begin{definition}[$(n_{\star}, \delta_{\star}, \tau_{\star})$-KHMM] \label{def_condK nlt}
    Let $n_{\star}$ be a positive integer, and $\delta_{\star}, \tau_{\star} \in [0,1)$. 
    The HMM $(A_n, B_n)$ is called an $(n_{\star}, \delta_{\star}, \tau_{\star})$-KHMM if it satisfies
    \begin{equation} \label{eq_condK implies nlt} 
        \Prob{\Birkhoff{\mat{M}(B_1^{n_{\star}})} \leq \tau_{\star} | A_0 = a_0 } \geq 1-\delta_{\star}, \quad \forall a_0 \in \mathcal{A}.
    \end{equation} 
    In words, the HMM has a probability at least $(1-\delta_{\star})$ of emitting by time $n_{\star}$ an observation sequence that induces a Birkhoff
    contraction coefficient at most $\tau_{\star}$, regardless of its initial state. 

    We say that an HMM is a KHMM if it is an $(n_{\star}, \delta_{\star}, \tau_{\star})$-KHMM for some $(n_{\star}, \delta_{\star}, \tau_{\star})$.
\end{definition}
Observe that if $(A_n,B_n)$ is an $(n_{\star}, \delta_{\star}, \tau_{\star})$-KHMM and $n_{\star} \leq n_{\star}'$, $\delta_{\star} \leq \delta_{\star}'$, 
and $\tau_{\star} \leq \tau_{\star}'$ then $(A_n, B_n)$ is also an $(n_{\star}', \delta_{\star}', \tau_{\star}')$-KHMM.   

In \Cref{lem_subrectangular matrices appear infinitely often}, adapted from~\cite[Lemma 8.2]{kaijser1975},  we show that if an HMM satisfies
\Cref{cond_kaijser}, then it is also a KHMM for some $(n_{\star}, \delta_{\star}, \tau_{\star})$.
This is because \Cref{cond_kaijser} ensures the existence of one sequence that induces a Birkhoff contraction coefficient less than $1$ (a ``good''
sequence). 
However, the HMM may very well have many ``good'' sequences, possibly shorter. 
Thus, a given HMM that satisfies \Cref{cond_kaijser} may be an $(n_{\star}, \delta_{\star}, \tau_{\star})$-KHMM for many different combinations of
$n_{\star}, \delta_{\star}, \tau_{\star}$. 
Since the bounds we develop are dependent on the values of $n_{\star}, \delta_{\star}, \tau_{\star}$, it is worthwhile to seek the combination that
yield the best bound.  

\begin{lemma} \label{lem_subrectangular matrices appear infinitely often} 
    If the HMM $(A_n, B_n)$ satisfies \Cref{cond_kaijser} then there exist a positive integer $n_{\star}$ and constants $\delta_{\star} < 1$ and 
    $0 \leq \tau_{\star} < 1$ such that~\eqref{eq_condK implies nlt} is satisfied. 
\end{lemma}
\begin{IEEEproof} 
    By \Cref{cond_kaijser} there exist positive integers $k_0, l_0$ and numbers $\gamma_0 > 0$ and $0 \leq \tau_{\star} < 1$ such that
    \begin{enumerate}
        \item For any $i,j \in \mathcal{A}$, $\matel{\mat{M}^{k_0}}{i,j} \geq \gamma_0$. 
            This follows from $\mat{M}$ being aperiodic and irreducible, so some power of it must be strictly positive~\cite[Theorem 1.4]{Seneta}. 
        \item For some sequence $\beta_1^{l_0}$ of elements of $\mathcal{B}$, the matrix $\mat{M}(\beta_1^{l_0})$ is nonzero and subrectangular.
            Existence of such sequences is guaranteed by \Cref{cond_kaijser}.
            We denote $\tau_{\star} = \Birkhoff{\mat{M}(\beta_1^{l_0})}$.
            Since $\mat{M}(\beta_1^{l_0})$ is subrectangular, $0 \leq \tau_{\star} < 1$.  
    \end{enumerate}
    Denote by $\mathcal{A}'$ the set of states that can lead to $f^{-1}(\beta_1)$ and then emit the observation sequence $\beta_1^{l_0}$, 
    \[ 
        \mathcal{A}' = \left \{ a \in \mathcal{A} \ \Big| \ \norm{\trp{\bv{e}_a} \mat{M}(\beta_1^{l_0})}_1 > 0 \right\}.
    \] 
    That is, there is positive probability that the next $l_0$ observations after any state in $\mathcal{A}'$ is the word
    $\beta_1^{l_0}$. 
    Since \Cref{cond_kaijser} is satisfied, $\mathcal{A}'$ is not empty, so that 
    \[ 
        \alpha_0 = \min_{a \in \mathcal{A}'}\norm{\trp{\bv{e}_a} \mat{M}(\beta_1^{l_0})}_1 >0.
    \] 

    We claim that~\eqref{eq_condK implies nlt} is satisfied with $n_{\star} = k_0 + l_0$ and $\delta_{\star} = 1-\alpha_0 \gamma_0 < 1$. 
    Indeed, for any $a_0 \in \mathcal{A}$,  
    \begin{align*}
        & \Prob{\Birkhoff{\mat{M}(B_1^{n_{\star}})} \leq \tau_{\star} \ \Big| A_0 = a_0} \\
        &\quad \eqann[ \geq ]{a} \Prob{\Birkhoff{\mat{M}(B_{k_0+1}^{k_0 + l_0})} \leq \tau_{\star} \ \Big| A_0 = a_0} \\ 
        &\quad \eqann[\geq]{b} \Prob{B_{k_0+1}^{k_0+l_0} = \beta_1^{l_0} \Big| A_0 = a_0}  \\ 
        &\quad = \sol{\sum_{a\in \mathcal{A}}} \Prob{B_{k_0+1}^{k_0+l_0} = \beta_1^{l_0}, A_{k_0} = a \Big| A_0 = a_0}  \\ 
        &\quad \eqann{c} \sol{\sum_{a\in \mathcal{A}'}} \Prob{B_{k_0+1}^{k_0+l_0} = \beta_1^{l_0}\Big| A_{k_0} = a} \cdot \Prob{A_{k_0} = a | A_0 = a_0}  \\ 
        &\quad \eqann{d} \sol{\sum_{a\in \mathcal{A}'}} \norm{\trp{\bv{e}_a} \mat{M}(\beta_1^{l_0})}_1 \cdot \matel{\mat{M}^{k_0}}{a_0, a} \\ 
        &\quad \geq \alpha_0 \gamma_0,
    \end{align*}
    where \eqannref{a} is by \Cref{cor_subrectangular,lem_Birkhoff of matrix product}, \eqannref{b} is by \Cref{cond_kaijser}, \eqannref{c} is by the
    Markov property, and \eqannref{d} is by~\eqref{eq_prob of B sequence given A0}.
\end{IEEEproof}

Let us now define the random variables $N_k(\tau)$, $k\geq 1$, by 
\begin{align*} 
    N_1( \tau) &= \min \{ n: \Birkhoff{\mat{M}(B_1^n)} \leq \tau \},\\
    N_{k+1}(\tau) &= \min\{ n: \Birkhoff{\mat{M}(B_{ N_k+1 }^{N_k + n})} \leq \tau\}, \quad k \geq 1.
\end{align*}
That is, the random variable $N_1( \tau)$ is the time of the first occurrence of a sequence that induces a Birkhoff contraction coefficient $\tau$ or
less. 
In other words, $N_1(\tau)$ is the smallest value of $n$ such that $\mat{M}(B_1^n)$ is a subrectangular matrix with Birkhoff contraction coefficient
$\tau$ or less.
Similarly, the random variable $N_k(\tau)$ is the gap between the $(k-1)$th and $k$th occurrences of such sequences. 

The following lemma and corollary are adapted from \cite[Lemma 8.3]{kaijser1975}, which was stated in~\cite{kaijser1975} without proof. 
\begin{lemma} \label{lem_upper bound on prob N1}
    Let $(A_n, B_n)$ be an $(n_{\star}, \delta_{\star}, \tau_{\star})$-KHMM. If $\delta_{\star} > 0$, there exist
    $\gamma > 0$ and $0\leq \rho<1$ such that for any positive integer $n_1$, 
    \begin{equation} \label{eq_upper bound on prob N1} 
        \Prob{N_1( \tau_{\star}) \geq n_1 \ | \ A_0 = a_0} \leq \gamma \rho^{n_1}, \quad \forall a_0 \in \mathcal{A}.
    \end{equation}                                   
\end{lemma}
\begin{IEEEproof}
    Let $T_0 = 1$ and denote, for any positive integer $k$, the random variable  $T_k = \Birkhoff{\mat{M}(B_1^{kn_{\star}})}$. 
    Observe that, by~\eqref{eq_condK implies nlt}, $\Prob{T_1 \leq \tau_{\star} | A_0 = a_0} \geq 1-\delta_{\star}$ for any $a_0 \in \mathcal{A}$. 

    We now show that for any positive integer $k$, and any $a_0 \in \mathcal{A}$, 
    \begin{equation}  \label{eq_Tk Tk1 inequality}
        \Prob{T_k > \tau_{\star}\  \big| \ T_{k-1} > \tau_{\star}, A_0 = a_0} \leq \delta_{\star}. 
    \end{equation} 
    We will demonstrate this for $k=2$, as the proof for all other values of $k$ is the same. 
    For any $a_0 \in \mathcal{A}$,
    \begin{align*}
        &\Prob{T_2 \leq \tau_{\star}\ \big| \ T_1 > \tau_{\star}, A_0 = a_0}  \\ 
        &\quad = \Prob{\Birkhoff{\mat{M}(B_{1}^{2n_{\star}})} \leq \tau_{\star} \ \big| \ T_1 > \tau_{\star}, A_0 = a_0}  \\ 
        &\quad\eqann[\geq]{a} \Prob{\Birkhoff{\mat{M}(B_{n_{\star}+1}^{2n_{\star}})} \leq \tau_{\star}\ \big|\  T_1 > \tau_{\star}, A_0 = a_0}  \\ 
        &\quad = \sum_a  \Prob{\Birkhoff{\mat{M}(B_{n_{\star}+1}^{2n_{\star}})} \leq \tau_{\star}, A_{n_{\star}} = a \ \big| \ T_1 > \tau_{\star}, A_0  = a_0}  \\ 
        &\quad\eqann{b}\sum_a \Prob{\Birkhoff{\mat{M}(B_{n_{\star}+1}^{2n_{\star}})} \leq \tau_{\star}\  \big| \ A_{n_{\star}} = a} p(a) \\ 
        &\quad\eqann{c} \sum_a \Prob{T_1 \leq \tau_{\star}\ | \ A_0 = a} p(a) \\ 
        &\quad \eqann[\geq]{d} 1-\delta_{\star}. 
    \end{align*}
    where \eqannref{a} is because, by \Cref{lem_Birkhoff of matrix product}, if $\Birkhoff{\mat{M}(B_m^n)} \leq \tau_{\star}$ then 
    $\Birkhoff{\mat{M}(B_1^n)} \leq \tau_{\star}$; 
    in \eqannref{b} we denoted $p(a) = \Prob{ A_{n_{\star}} = a | T_1 > \tau_{\star}, A_0 = a_0}$; 
    \eqannref{c} is by the Markov property; 
    and \eqannref{d} is by~\eqref{eq_condK implies nlt}.
    Rearranging yields~\eqref{eq_Tk Tk1 inequality}.  
    We remark that~\eqref{eq_Tk Tk1 inequality} is also true without conditioning on $\{T_{k-1} > \tau_{\star}\}$. 

    Thus, 
    \begin{align*}
        & \Prob{T_k > \tau_{\star} | A_0 = a_0} \\ 
        & = \Prob{T_k > \tau_{\star} | T_{k-1}> \tau_{\star}, A_0 = a_0} \cdot \Prob{T_{k-1} > \tau_{\star}| A_0 = a_0} \\ 
        &\quad + \Prob{T_k > \tau_{\star} | T_{k-1}\leq \tau_{\star}, A_0 = a_0} \cdot \Prob{T_{k-1} \leq \tau_{\star}| A_0 = a_0} \\ 
        &\eqann{a}  \Prob{T_k > \tau_{\star} | T_{k-1}> \tau_{\star}, A_0 = a_0} \cdot \Prob{T_{k-1} > \tau_{\star}| A_0 = a_0} \\ 
        &\eqann[\leq]{b} \delta_{\star} \Prob{T_{k-1} > \tau_{\star}| A_0 = a_0}, 
    \end{align*}
    where~\eqannref{a} is by \Cref{lem_Birkhoff of matrix product}, by which the second summand in the first equality must be $0$, and~\eqannref{b} is
    by~\eqref{eq_Tk Tk1 inequality}.  
    We conclude that for any integer $k$ and any $a_0 \in \mathcal{A}$, 
    \[ 
        \Prob{N_1(\tau_{\star}) > k n_{\star} | A_0 = a_0} = \Prob{T_k > \tau_{\star} | A_0 = a_0}  \leq \delta_{\star}^k.
    \]                                                                                                                   

    Hence, for any positive integer $n_1$ (not necessarily a multiple of $n_{\star}$) and any $a_0 \in \mathcal{A}$, 
    \[
        \Prob{N_1(\tau_{\star}) \geq n_1 | A_0 = a_0} \leq \delta_{\star}^{n_1/n_{\star} - 1}.
    \] 
    Rearranging, this yields
    \[ 
        \Prob{N_1(\tau_{\star}) \geq n_1 | A_0 = a_0} \leq \frac{1}{\delta_{\star}} \cdot \big(\delta_{\star}^{1/n_{\star}}\big)^{n_1}.
    \]  
    Thus, we obtain~\eqref{eq_upper bound on prob N1} with $\gamma = 1/\delta_{\star}$ and $\rho = \delta_{\star}^{1/n_{\star}}$.
    To complete the proof, observe that $0 \leq \rho < 1$ since $0 < \delta_{\star} < 1$. 
\end{IEEEproof}
We imposed $\delta_{\star} > 0$ in \Cref{lem_upper bound on prob N1} because this is the more interesting case. 
Clearly, \Cref{lem_upper bound on prob N1} also holds when $\delta_{\star} = 0$, albeit with different $\gamma, \rho$. 
However, we can do better in this case. 
Namely, if $\delta_{\star} = 0$ for some $n_{\star}$, this implies that at time $n_{\star}$ the sequence of observations is ensured to induce Birkhoff
contraction coefficient less than $\tau_{\star}$. 
In this case, we can obtain a much simpler bound on the mutual information. 
We will return to this point in the proof of \Cref{thm_upper bound on I epsilon}. 

The upper bound in~\eqref{eq_upper bound on prob N1} is independent of $a_0$. 
Therefore, whenever $(A_n,B_n)$ is an $(n_{\star}, \delta_{\star}, \tau_{\star})$-KHMM and $\delta_{\star}>0$, we conclude that
 \[ 
     \Prob{N_1(\tau_{\star}) \geq n_1} \leq \gamma \rho^{n_1}. 
 \] 
More generally, we have the following corollary. 
\begin{corollary}
    Let $(A_n, B_n)$ be an $(n_{\star}, \delta_{\star}, \tau_{\star})$-KHMM with $\delta_{\star} > 0$. 
    Then, there exist $\gamma > 0$ and $0\leq \rho<1$ such that for any positive integers  $n_1$, $n_2$, \ldots, $n_m$, 
    \begin{IEEEeqnarray}{l} 
        \Prob{N_{1}(\tau_{\star}) \geq n_1, N_{2}(\tau_{\star}) \geq n_2, \ldots ,N_{m}(\tau_{\star}) \geq n_m} \qquad\qquad \IEEEnonumber
    \\ \qquad\qquad\qquad\qquad\qquad\qquad \leq \gamma^m \rho^{n_1 + n_2 + \cdots + n_m}. \IEEEyesnumber \label{eq_upper bound on multiple N}
    \end{IEEEeqnarray} 
\end{corollary}
\begin{IEEEproof}
    For brevity, we denote $N_k = N_k(\tau_{\star})$. 
    Since
    \begin{IEEEeqnarray*}{l}
        \Prob{N_{1} \geq n_1, N_{2} \geq n_2, \ldots ,N_{m} \geq n_m}   \qquad\qquad 
        \\ \qquad\qquad \qquad\qquad\qquad = \prod_{k=1}^m \Prob{N_k \geq n_k | N_i \geq n_i, \ i < k},
    \end{IEEEeqnarray*}
    \eqref{eq_upper bound on multiple N} will follow if $ \Prob{N_{k} \geq n_{k} | N_{i} \geq n_i,\  i < k} \leq \gamma \rho^{n_k}$. 
    Indeed, for any $k$ we have 
    \begin{align*}
        &\Prob{N_{k} \geq n_{k} | N_{i} \geq n_i,\  i < k} \\ 
        &=\sol{\sum_{a}}\Prob{N_{k} \geq n_{k}, A_{N_{k-1}} = a | N_{i}
        \geq n_i, \ i < k} \\ 
        &=\sol{\sum_{a}}\Prob{N_{k}\geq n_{k}| A_{N_{k-1}} = a} \Prob{A_{N_{k-1}} = a| N_{i} \geq n_i, \ i < k} \\
        &\eqann{a}\sol{\sum_{a}}\Prob{N_{1}\geq n_{k}| A_{0} = a} \Prob{A_{N_{k-1}} = a| N_{i} \geq n_i, \ i < k} \\
        &\eqann[\leq]{b} \gamma \rho^{n_{k}} \sum_a \Prob{A_{N_{k-1}} = a| N_{i} \geq n_i, \ i < k} \\ 
        & = \gamma \rho^{n_{k}}, 
    \end{align*}
    where~\eqannref{a} is by definition of $N_{k}$ and~\eqannref{b}  is by~\eqref{eq_upper bound on prob N1}. 
\end{IEEEproof}

\begin{proposition} \label{prop_upper bound on I accurate} 
    Let $(A_n, B_n)$ be an $(n_{\star}, \delta_{\star}, \tau_{\star})$-KHMM with $\delta_{\star}>0$. 
    Denote 
    \[ 
        \gamma = \frac{1}{\delta_{\star}}, \quad \alpha = \gamma\cdot\log|\mathcal{A}|, \quad \rho = \delta_{\star}^{1/n_{\star}} < 1.
    \] 
    Then, for any $m \leq n$ we have 
    \begin{IEEEeqnarray}{r} \label{eq_upper bound on mutual information m n} 
        I(A_0;A_{n+1}| B_1^{n}) \leq 4 \log\left(\frac{1+\tau_{\star}}{1-\tau_{\star}}\right) \tau_{\star}^{m} + \alpha \frac{(\gamma n)^m}{ m!} \rho^{n+1}. 
    \IEEEeqnarraynumspace \end{IEEEeqnarray}
\end{proposition}
\begin{IEEEproof}
    Observe that the right-hand side of~\eqref{eq_upper bound on multiple N} depends only on the sum $n_1 + n_2 + \cdots + n_m$, and not the
    values of the individual values of $n_k$.
    Denote by $p(n,m)$ the number of positive integer $m$-tuples $(n_1,n_2,\ldots,n_m)$ such that $n = n_1 + n_2 + \cdots + n_m$, where each integer
    $n_k \geq 1$. 
    In \cite[p. 38]{Feller}, it is shown that $p(n,m) = {n-1 \choose m-1}$.
    Thus, by~\eqref{eq_upper bound on multiple N},
    \begin{align*} 
        \Prob{\sum_{k=1}^m N_k(\tau_{\star}) \geq n} &\leq p(n,m) \gamma^m \rho^n\\
        &= {n-1 \choose m-1} \gamma^m \rho^n \\ 
        &\leq \frac{(n-1)^{m-1}}{(m-1)!} \gamma^m \rho^n.
    \end{align*}

    Next, consider the matrix product $\mat{M}(B_1^n)$.
    We wish to count, in this product, the number of non-overlapping occurrences of contiguous sequences of matrices whose product has Birkhoff
    contraction coefficient at most $\tau_{\star}$.
    This is accomplished by the integer-valued random variable 
    \[ 
        \cntrv{n} = \cntrv{n}(\tau_{\star}) = \max\left\{ m : \sum_{k=1}^m N_k(\tau_{\star}) \leq n\right\}.
    \] 
    From the above discussion, 
    \begin{align}
        \Prob{\cntrv{n} \leq m} &= \Prob{\cntrv{n} < m+1}\nonumber \\
        &= \Prob{\sum_{k=1}^{m+1} N_k(\tau_{\star}) \geq n+1}  \nonumber \\
        &\leq\gamma \frac{(n\gamma)^m}{m!}\rho^{n+1}. \label{eq_upper bound on L}
    \end{align}

    Recall from~\eqref{eq_I Bn A0 given middle} that $I(A_0; A_{n+1}| B_1^n) = \Exp{J}$, where we have denoted, for brevity, 
    \[ 
        J  \triangleq \log \left(\frac{ \norm{\trp{\bv{e}_{A_0}}\mat{M}(B_1^n)\mat{T}_{A_{n+1}}}_1 }{ \norm{\trp{\bs{\pi}}\mat{M}(B_1^n)\mat{T}_{A_{n+1}} }_1} 
                      \cdot \frac{ \norm{\trp{\bs{\pi}}\mat{M}(B_1^{n})}_1}{ \norm{\trp{\bv{e}_{A_0}}\mat{M}(B_1^{n})}_1 }\right).
    \] 
    This is a conditional mutual information.
    In particular, for any fixed sequence $b_1^n$ we have 
    \begin{IEEEeqnarray}{c}  
        0 \leq I(A_0; A_{n+1} | B_1^n = b_1^n) = \Exp{J| B_1^n = b_1^n} \leq \log|\mathcal{A}|, \IEEEeqnarraynumspace
        \label{eq_mutual information bounds}
    \end{IEEEeqnarray} 
    where the inequalities are due to the properties of mutual information --- it is nonnegative and upper-bounded by the logarithm of the alphabet size. 
    The random variable $\cntrv{n}$ is a function of $B_1^n$ --- given any realization $b_1^n$ of $B_1^n$, we can compute the value of $\cntrv{n}$ precisely. 
    For any $m \leq n$, 
    \begin{align} 
            & \Exp{J|\cntrv{n}> m}\Prob{\cntrv{n} >m} \notag \\ 
            &\quad =  \sum_{b_1^n: \cntrv{n} > m} \Exp{J|B_1^n = b_1^n} \Prob{B_1^n = b_1^n} \notag\\ 
            &\quad \eqann{a}  \sum_{b_1^n: \cntrv{n} \geq m+1} \Exp{J|B_1^n = b_1^n} \Prob{B_1^n=b_1^n} \notag\\ 
            &\quad \eqann[\leq]{b}  4 \log\left(\frac{1+\tau_{\star}}{1-\tau_{\star}}\right) \cdot
            \tau_{\star}^{m}, \label{eq_I Ln gtr m}
    \end{align}
    where~\eqannref{a} is because $\cntrv{n}$ is integer valued and~\eqannref{b} is by  \Cref{lem_Birkhoff of matrix product} and 
    \Cref{cor_contraction inequality}.
    Moreover, 
    \begin{align} 
            & \Exp{J|\cntrv{n} \leq m}\Prob{\cntrv{n} \leq m} \notag \\ 
            & \quad =  \sum_{b_1^n: \cntrv{n} \leq m} \Exp{J|B_1^n = b_1^n} \Prob{B_1^n = b_1^n} \notag \\
            &\quad \eqann[\leq]{a} \log|\mathcal{A}|\cdot \Prob{\cntrv{n} \leq m} \notag \\ 
            &\quad \eqann[\leq]{b}  \log|\mathcal{A}| \cdot \gamma \frac{(n\gamma)^m}{m!}\rho^{n+1},  \label{eq_I Ln less m}
    \end{align}    
    where~\eqannref{a} is by the right-hand inequality of~\eqref{eq_mutual information bounds} and~\eqannref{b} is by~\eqref{eq_upper bound on L}. 

    Thus, for any $m \leq n$ we have by~\eqref{eq_I Ln gtr m} and~\eqref{eq_I Ln less m}, 
    \begin{align*}
        &I(A_0;A_{n+1}| B_1^{n})
        =  \Exp{J} \\ 
        & =  \Exp{J | \cntrv{n} > m}\Prob{\cntrv{n} > m} + \Exp{J| \cntrv{n} \leq m}\Prob{\cntrv{n}  \leq m}  \\ 
        & \leq \log\left(\frac{1+\tau_{\star}}{1-\tau_{\star}}\right) \cdot
        \tau_{\star}^{m} +(\gamma\cdot\log|\mathcal{A}|) \cdot \frac{(n\gamma)^m}{m!}\rho^{n+1}.
    \end{align*}
    Denoting $\alpha = \gamma \cdot \log|\mathcal{A}|$ completes the proof. 
\end{IEEEproof}
\begin{remark}
    We note in passing  that, if desired, one can set $m = \theta n$ in~\eqref{eq_upper bound on mutual information m n} and obtain an upper bound
    that vanishes with $n$, provided that $\theta$ is sufficiently small.
    To this end, we use the inequality $m! \geq (m/e)^m$, see~\cite[p.  52]{Feller}.
    We set $m = \theta n$, and upper-bound the second summand in the right-hand side of~\eqref{eq_upper bound on mutual information m n} to obtain
    \[ 
        \alpha \frac{(n\gamma)^m}{m!}\rho^{n+1} \leq \alpha\rho\cdot \left(\rho\left( \frac{\gamma e}{\theta} \right)^\theta\right)^n.
    \]                              
    The right-hand side of the above inequality vanishes with $n$ for small enough $\theta$.
    To see this, observe that $\lim_{\theta\to 0} (\gamma e/\theta)^{\theta} = 1$,\footnote{Indeed, since $(1/\theta)^{\theta} = e^{\theta
        \ln(1/\theta)}$ and by continuity of the exponential function at $0$, it suffices to show that $\lim_{\theta\to0} \theta \ln(1/\theta) = 0$. 
            This, in turn, holds by L'H\^o{}pital's rule: $\lim_{\theta \to 0} \theta \ln (1/\theta) = \lim_{\theta \to 0} \ln(1/\theta)/(1/\theta)
        = \lim_{\theta \to 0} (-1/\theta)/(-1/\theta^2) = \lim_{\theta \to 0} \theta = 0$.}
        so we are ensured that if $\theta$ is small enough,
    $\rho\cdot (\gamma e / \theta)^{\theta}  < 1$. 

    That said, taking $m = \theta n$ might not be the best strategy for minimizing $n$ in the right-hand side of~\eqref{eq_upper bound on mutual information m n}.
    A different strategy is outlined in the proof of \Cref{thm_upper bound on I epsilon}. 
\end{remark}

We are now ready to prove \Cref{thm_upper bound on I epsilon}. 
\begin{IEEEproof}[Proof of \Cref{thm_upper bound on I epsilon}]
    By \Cref{lem_subrectangular matrices appear infinitely often}, $(A_n,B_n)$ is an $(n_{\star}, \delta_{\star}, \tau_{\star})$-KHMM for some
    $n_{\star}$, $\delta_{\star}$, $\tau_{\star}$.
    Let 
    \[ 
        m = \left\lceil \log_{\tau_{\star}}\left( \frac{\epsilon}{2}\cdot\frac{1}{4\log\left(\frac{1+\tau_{\star}}{1-\tau_{\star}}\right)} \right) \right\rceil.
    \]

    \emph{Case 1:} 
    If $\delta_{\star} = 0$ then at time $\memlength = (m+1)n_{\star}$ the sequence $B_1^{\memlength}$ can be divided into $m+1$ contiguous sequences
    of length $n_{\star}$, each inducing a Birkhoff contraction coefficient less than $\tau_{\star}$.
    Therefore, using~\Cref{cor_contraction inequality} we obtain that in this case for any $n \geq \memlength$,  
    \[ 
        I(A_0;A_{n+1}| B_1^n) \leq 4 \log \left(\frac{1+\tau_{\star}}{1-\tau_{\star}}\right)\tau_{\star}^{m} \leq \frac{\epsilon}{2}.
    \] 
    
    \emph{Case 2:} 
    In the general case, $\delta_{\star} > 0$ and we turn to \Cref{prop_upper bound on I accurate}. 
    For $m$ fixed as above, we set $\memlength$ as the smallest integer greater than or equal to $m$ such that for any $n \geq \memlength$ we have 
    \[ 
        \frac{\alpha (\gamma n)^{m} \rho^{n+1}}{ m! } \leq \frac{\epsilon}{2},
    \]
    where $\gamma = 1/\delta_{\star}$, $\alpha = \gamma \cdot \log|\mathcal{A}|$, and $\rho = \delta_{\star}^{1/n_{\star}}$.
    Such $\memlength$ exists since $m$ is fixed and $\rho < 1$. 
    For this $m$ and any $n\geq \memlength$, the right-hand side of~\eqref{eq_upper bound on mutual information m n} is upper-bounded by $\epsilon$. 
\end{IEEEproof}
\begin{discussion}
    The upper bound in \Cref{prop_upper bound on I accurate} is generally quite loose. 
    We only count non-overlapping occurrences of ``good'' sequences, known to have Birkhoff contraction coefficient less than some $\tau_{\star}$,
    with lengths that are multiples of some $n_{\star}$.
    There may actually be many other subsequences --- possibly shorter --- that induce Birkhoff contraction coefficients less than $1$, and we ignore
    those.
    Moreover, most occurrences of ``good'' sequences appear as the suffix of longer sequences. By \Cref{lem_Birkhoff of matrix product}, the induced
    Birkhoff contraction coefficient of these longer sequences will be smaller than that of the ``good'' sequences. Moreover, the values of $\gamma$
    and $\rho$ are conservative. 

    A given KHMM may be associated with many combinations of $(n_{\star}, \delta_{\star}, \tau_{\star})$.
    Thus, one needs to carefully select the right combination of these parameters to minimize $\memlength$ in \Cref{thm_upper bound on I epsilon}. 
    A more refined analysis, that considers a KHMM for which multiple combinations $(n_{\star}, \delta_{\star}, \tau_{\star})$ are known may yield better bounds. 

    Nevertheless, even with this loose bound, we are able to ensure that the desired mutual information vanishes for sufficiently large $\memlength$. 
    In practice, for a given process, the mutual information will be below the desired threshold much earlier than promised in~\Cref{prop_upper bound on I accurate}. 
\end{discussion}

\appendices

\section{Proof of Fast Polarization} \label{ap_fast polarization}
In the fast stage of our construction, \arikan polar codes are designed based on recursive upper bounds on distribution parameters, such as the
Bhattacharyya parameter. 
In this appendix we show that this procedure leads to fast polarization universally. 
Fast polarization results are usually of the flavor: ``if the polar code length is large enough, then fast polarization is obtained.'' 
This ``large enough'' length is related to the process for which the polar code is designed. 
In a universal setting, however, we must design the fast stage before knowing which process the code is to be used for. 
We show that it is indeed possible to determine this length regardless of the process. 
This is afforded because the slow stage is $(\eta, \mathcal{L}, \mathcal{H})$-monopolarizing. 

Fast polarization is the phenomenon described in the following lemma. 
To keep the discussion focused, we present it for a special case of binary polar codes based on \arikan's kernel. 
\begin{lemma}[\cite{Arikan_Telatar_2009,sasoglu_thesis,Tal_2017_simple}]
    Let $B_1, B_2, \ldots$ be independent and identically distributed random variables with $\Probi{B_i = 0} = \Probi{B_i = 1} = 1/2$. 
    Let $Z_0, Z_1, \ldots$ be a $[0,1]$-valued random process such that 
    \begin{equation} \label{eq_bhattacharyya recursion}
        Z_{n+1} \leq \kappa \cdot\begin{cases} Z_n^2, & B_{n+1} = 0, \\ Z_{n}, & B_{n+1} = 1, \end{cases} \quad n \geq 0, 
    \end{equation}
    where $\kappa > 1$. 
    If $Z_n$ converges almost surely to a $\{0,1\}$-valued random variable $Z_{\infty}$ then for every $0<\beta < 1/2$, we have 
    \begin{equation} \label{eq_fast polarization z}
        \lim_{n \to \infty} \Prob{Z_n \leq 2^{-2^{n\beta}}} = \Prob{Z_{\infty} = 0}.
    \end{equation}
\end{lemma}
Fast polarization was first stated and proved in~\cite{Arikan_Telatar_2009}.
It was later generalized by \sasoglu (see, e.g.,~\cite[Lemma 4.2]{sasoglu_thesis}). 
A simpler proof of a stronger result\footnote{In which~\eqref{eq_fast polarization z} is replaced
    with $\lim_{n_0\to\infty} \Probi{\forall n \geq n_0, \, Z_n \leq 2^{-2^{n\beta}}}
= \Probi{Z_{\infty} =0}$.} for the general case can be found in~\cite{Tal_2017_simple}. 
Our fast polarization result is based on the proof of~\cite{Tal_2017_simple}. 

For example, $Z_n$ might be the Bhattacharyya parameter of a randomly-selected polarized \sopair
(tantamount to a synthetic channel, in a channel-coding setting), which is
an upper-bound on the probability of error of estimating the symbol from its observation. 
In  the memoryless case, the recursion~\eqref{eq_bhattacharyya recursion} for the Bhattacharyya parameter with $\kappa = 2$ was established in~\cite[Proposition 5]{Arikan_2009}. 
Under memory,~\eqref{eq_bhattacharyya recursion} was shown in~\cite[Theorem 2]{sasoglu_Tal_mem},
with
\begin{equation}
    \label{eq_kappa}
    \kappa = 2\psi_0,
\end{equation}
    where $\psi_0$ is a mixing parameter of the process; mixing parameters are
defined in \Cref{lem_FAIM is psi mixing}. 
Thus, the Bhattacharyya parameter polarizes fast to $0$ with or without memory. 

The proof in~\cite{Tal_2017_simple} establishes~\eqref{eq_fast polarization z} by showing that for
every $\delta>0$ there exists an $n_0$ such that 
\[ 
    \Prob{Z_{\infty} = 0} - \delta \leq \Prob{\forall n \geq n_0, \, Z_n \leq 2^{-2^{n\beta}}} \leq \Prob{Z_{\infty} = 0}. 
\] 
The magnitude of $n_0$ depends on two factors: the almost-sure convergence of $Z_n$ to $Z_{\infty}$ and the law of large numbers. 
The latter is independent of the process, but the former one is not. 
The proof utilizes the almost-sure convergence of $Z_n$ only for the following consequence. 
Recalling that $Z_n$ converges almost surely to a $\{0,1\}$-valued random variable, for any $\epsilon_a > 0$ and $\delta_a > 0$ there must be an $n_a$ such that 
\begin{equation} \label{eq_lb for Zn leq eps}
    \Prob{Z_n \leq \epsilon_a} \geq \Prob{Z_{\infty} = 0} - \delta_a, \quad \forall n \geq n_a. 
\end{equation}
We reiterate that $n_a$ is process-dependent. 

In our universal setting, the fast polarization stage occurs after the slow polarization stage. 
Specifically, it operates on \sopairs whose conditional entropy --- and thus also Bhattacharyya parameter\footnote{See~\cite[Lemma
    1]{Shuval_Tal_Memory_2017} for relationships between the Bhattacharyya parameter and the conditional entropy.} --- is universally smaller than
    $\eta$, which can be set as small as
    desired.\footnote{More generally, fast polarization of high-entropy indices may also be of interest, e.g., in source-coding applications.
        The universal stage also provides us with \sopairs whose conditional entropy  is as close
        to $1$ as desired. 
     Due to forgetfulness (see the proof of \Cref{lem_lean bblock}, stopping short of the last inequality,
    \eqannref{f}), this is true also when conditioning on the boundary states, by taking $L_0$ large enough. 
        Under memory, fast polarization of high-entropy \sopairs is obtained through boundary-state-informed parameters, namely the total variation distance
        (see~\cite{Shuval_Tal_Memory_2017}). 
        It was shown in~\cite[Proposition 12]{Shuval_Tal_Memory_2017} that the boundary-state-informed total variation distance undergoes a recursion similar
    to~\eqref{eq_bhattacharyya recursion}. 
The required connections between the boundary-state-informed conditional entropy and the boundary-state-informed total variation distance can be found
in~\cite[equation (4c)]{Shuval_Tal_Memory_2017}.}
        The ability to set $\eta$ as small as desired is the key to obtaining \emph{universal} fast polarization results. 
Namely, we prove the following proposition.

\begin{proposition} \label{prop_universal fast polarization}
    Let $B_1, B_2, \ldots$ be independent and identically distributed random variables with $\Probi{B_i = 0} = \Probi{B_i = 1} = 1/2$. 
    Let $Z_0, Z_1, \ldots$ be a $[0,1]$-valued random process that satisfies~\eqref{eq_bhattacharyya recursion} for some $\kappa > 1$. 
    Fix $0<\beta < 1/2$. Then, for every $\delta > 0$ there exist $\eta>0$ and $n_0$  such that if $Z_0 \leq \eta$ then
    \begin{equation} \label{eq_fast polarization z eta}
        \Prob{Z_n \leq 2^{-2^{n\beta}} \textup{ for all } n \geq n_0} \geq 1-\delta. 
    \end{equation}
\end{proposition}
Crucially, $\eta$ and $n_0$  depend on the process $Z_n$ only through $\kappa$. 
Inspection of the proof of~\cite{Tal_2017_simple} reveals that \Cref{prop_universal fast polarization} will be true once it is shown that  for any
$\epsilon_a > 0$ and $\delta'>0$ there exists $n_a$ such that 
\begin{equation} \label{eq_Zn below epsa}
    \Prob{Z_n \leq \epsilon_a \textup{ for all } n \geq n_a} \geq 1- \delta'. 
\end{equation}
The crux of our proof will be to show that we can set $\eta>0$ and $n_a$ such that the above holds.
We will need an auxiliary result, \Cref{cor_lh inequality}, which follows from \Cref{lem_lundberg hoeffding}, introduced and
proved below.

\begin{remark}
    Our statement of \Cref{prop_universal fast polarization} is for a fast polarization stage based
    on \arikan's kernel. 
    This is done for the sake of simplicity. 
    However, the lemma holds true for the more general case of other kernels. 
    The key technical tool in the proof, \Cref{lem_lundberg hoeffding}, is stated in a general
    manner, enabling its use for other kernels without change. 
\end{remark}

Let $T_1, T_2, \ldots$ be a sequence of independent and identically distributed (i.i.d.) random variables. 
Denote by $T$ a random variable distributed according to the same distribution as each of the random variables $T_i$, $i \in \mathbb{N}$. 
We assume that $T$ is bounded; in particular, there exist positive reals $a, b > 0$ such that 
\[ 
    - b \leq T \leq a, 
\]
and for every $\epsilon > 0$, $\Probi{T>a-\epsilon} >0$. 
We further assume that 
\begin{equation} \label{eq_expectation of T is negative}
    \mu \triangleq \Exp{T} < 0. 
\end{equation}
We define the random walk
\[ 
    J_n = \sum_{i=1}^n T_i, \quad n\in\mathbb{N}. 
\] 
For every $\alpha > 0$, define the events 
\[ 
    \mathcal{A}_{\alpha}(n) = \{ J_m \geq \alpha \text{ for some }  m \leq n \}
\] 
and
\[
    \mathcal{A}_{\alpha} = \{ J_m \geq \alpha \text{ for some } m \in \mathbb{N} \}.
\] 
Observe that $\mathcal{A}_{\alpha}(n) \subseteq \mathcal{A}_{\alpha}(n+1)$ and $\cup_{n=1}^{\infty}\mathcal{A}_{\alpha}(n)=\mathcal{A}_{\alpha}$, so
that by continuity of measure~\cite[Theorem 2.1]{billingsley1995probability}, 
\begin{equation} \label{eq_cont of measure}
    \Prob{\mathcal{A}_{\alpha}} = \lim_{n \to \infty} \Prob{\mathcal{A}_{\alpha}(n)}.
\end{equation}
We denote by $\mathcal{A}_{\alpha}^{\text{c}}$ the complementary event to $\mathcal{A}_{\alpha}$. 
That is, $\mathcal{A}_{\alpha}^{\text{c}} = \{ J_n < \alpha \text{ for all } n \in \mathbb{N}\}$. 
We then have the following lemma. 

\begin{lemma} \label{lem_lundberg hoeffding}
    There exists $r > 0$ such that for any $\alpha>0$,  
    \begin{equation} \label{eq_lundberg inequality} 
        \Prob{\mathcal{A}_{\alpha}} \leq e^{-r \alpha}. 
    \end{equation}
    Moreover, for any $0< \gamma < 1$ and $n \in \mathbb{N}$, 
    \begin{equation} \label{eq_hoeffding result}
        \Prob{J_n < n(1-\gamma)\mu} \geq 1 - e^{-2n
        \left(\frac{\gamma \mu}{a+b}\right)^2}. 
    \end{equation}
\end{lemma}

Since $\mu<0$ by~\eqref{eq_expectation of T is negative} and $0<\gamma<1$ by
assumption, then $n(1-\gamma)\mu < 0$ in~\eqref{eq_hoeffding result}. 
We will see in \Cref{cor_lh inequality} below that \Cref{lem_lundberg hoeffding} implies that for any negative threshold, there exists $n_a \in
\mathbb{N}$ and $\alpha>0$ such that with probability arbitrarily close  to $1$, $J_n$ drops below that threshold for every $n \geq n_a$ and never
(for any $n \in \mathbb{N}$) visits above $\alpha$. 
This will be key to obtaining~\eqref{eq_Zn below epsa}. 

\begin{IEEEproof}
    The proof combines two inequalities: ~\eqref{eq_lundberg inequality} is essentially the Lundberg
    inequality~\cite[equation 15]{Gerber1973} and for~\eqref{eq_hoeffding result} we call upon
    the Hoeffding inequality~\cite[Theorem 2]{Hoeffding1963}. 
    Since the proof of the Lundberg inequality in~\cite{Gerber1973} is for the continuous-time case, we provide a proof for the discrete-time case, adapted
    from the proof of~\cite{Gerber1973}. 

    Denote by $g(s)$ the moment-generating function of $T$. 
    That is, 
    \[ 
        g(s) = \Exp{e^{sT}}.
    \]
    The expectation is well-defined as $e^{sT}$ is a non-negative random variable~\cite[equation 15.3]{billingsley1995probability}. 
    Since $T$ is bounded by assumption, $g(s) < \infty$ for any $s \in \mathbb{R}$; hence, 
    $g(s)$ is continuous over $\mathbb{R}$, see~\cite[Theorem 9.3.3]{Rosenthal_2006}. 
    Observe that $g(0) = 1$ and, by~\cite[equation 21.23]{billingsley1995probability} and~\eqref{eq_expectation of T is negative}, $g'(0) = \Exp{T}
    < 0$.
    Thus, $g(s)$ is decreasing at $s=0$, so $g(s) < 1$ for $s$ small enough. 
    On the other hand, by assumption on $T$,  
    \[
        p \triangleq \Prob{T \geq a/2} = \Exp{\mathbf{1}\{T \geq a/2 \}} > 0,
    \] 
    where $\mathbf{1}\{\cdot\}$ is an indicator random variable. 
    Thus, 
    \[
        g(s) \geq \Exp{e^{sT}\cdot\mathbf{1}\{T\geq a/2\}} \geq e^{sa/2} p. 
    \]
    In particular, if $s > (2/a)\ln(1/p)$, then $g(s) > 1$. 
    Since $g(s)$ is continuous, there exists $s>0$ such that $g(s) = 1$.
    Thus, we define
    \begin{equation} \label{eq_def of theta lundberg}
        r \triangleq \max_{s>0} \left\{s : \Exp{e^{sT}} = 1 \right\}.
    \end{equation}

    For the $r$ found above, denote 
    \[ 
        \tilde{J}_n = e^{r J_n} = \prod_{i=1}^n e^{r T_i}. 
    \] 
    We claim that $\tilde{J}_n$, $n \in \mathbb{N}$, is a martingale. 
    Indeed, since the $T_i$ are independent, 
    \begin{align*}
        \Exp{\tilde{J}_n\,|\,\tilde{J}_{m},\ m <n } & = \Exp{e^{r T_n} \cdot \tilde{J}_{n-1} \,| \,\tilde{J}_{m},\ m <n }  \\ 
        &= \tilde{J}_{n-1} \Exp{e^{r T_n}} \\ 
        &= \tilde{J}_{n-1}, 
    \end{align*}
    where the last equality is by definition of $r$,~\eqref{eq_def of theta lundberg}. 
    Define the (possibly infinite) stopping time 
    \[ 
        \tau = \inf_{n} \{n:  J_n \geq \alpha \}.
    \] 
    Then, by~\cite[Section 10.9]{Williams_1991}, the stopped process 
    \[ 
        \tilde{J}_{n \wedge \tau} \triangleq \begin{cases} 
            \tilde{J}_n, & \tau > n, \\ 
            \tilde{J}_{\tau}, & \tau \leq n
        \end{cases}
    \]
    is also a martingale, and 
    \[ 
        \Exp{\tilde{J}_{n\wedge \tau}} = \Exp{\tilde{J}_1} = 1. 
    \]
    Observe that for any $n \in \mathbb{N}$, we have $\Probi{\mathcal{A}_{\alpha}(n)}
    = \Probi{\tau \leq n}$. 
    Thus, 
    \begin{align*}
        1 &= \Exp{\tilde{J}_{n \wedge \tau}} \\ 
            &= \Exp{\tilde{J}_{n \wedge \tau} | \tau \leq n} \cdot\Prob{\mathcal{A}_{\alpha}(n)} \\
            & \quad +\Exp{\tilde{J}_{n \wedge \tau} | \tau > n} \cdot \Big(1-\Prob{\mathcal{A}_{\alpha}(n)}\Big) \\
            &\eqann[\geq]{a} \Exp{\tilde{J}_{n \wedge \tau} | \tau \leq n} \Prob{\mathcal{A}_{\alpha}(n)} \\
            &\eqann{b} \Exp{\tilde{J}_{\tau} | J_{\tau} \geq \alpha, \tau \leq n} \Prob{\mathcal{A}_{\alpha}(n)} \\
            &= \Exp{e^{r J_{\tau}} | J_{\tau} \geq \alpha, \tau \leq n} \Prob{\mathcal{A}_{\alpha}(n)} \\
            &\eqann[\geq]{c} e^{r\alpha} \Prob{\mathcal{A}_{\alpha}(n)}. 
    \end{align*}
    where~\eqannref{a} is because $\tilde{J}_{n\wedge \tau} \geq 0$, \eqannref{b} is by definition of $\tau$ and of
    $\tilde{J}_{n\wedge\tau}$, and \eqannref{c} is because $r > 0$ by definition. 
    Rearranging, we obtain that for any $n \in \mathbb{N}$, 
    \[ 
        \Prob{\mathcal{A}_{\alpha}(n)} \leq e^{-r \alpha}. 
    \] 
    Thus, by~\eqref{eq_cont of measure},
    \[ 
        \Prob{\mathcal{A}_{\alpha}} = \lim_{n\to\infty} \Prob{\mathcal{A}_{\alpha}(n)} \leq e^{-r \alpha}. 
    \] 
    This completes the proof of~\eqref{eq_lundberg inequality}. 

    To prove~\eqref{eq_hoeffding result}, recall that by the Hoeffding inequality~\cite[Theorem 2]{Hoeffding1963}, for any $t > 0$ we have
    \[ 
        \Prob{J_n \geq n(\mu + t)} \leq e^{-2n\left(\frac{t}{a+b}\right)^2}. 
    \] 
    In particular, for any $0 < \gamma < 1$, we may choose $t = \gamma |\mu| = -\gamma \mu >0$ to obtain 
    \begin{align*} 
        \Prob{J_n < n(1-\gamma)\mu} &= 1- \Prob{J_n \geq n(\mu+\gamma|\mu|)} \\ 
        & \geq 1- e^{-2n\left(\frac{\gamma \mu}{a+b}\right)^2}. 
    \end{align*}
    This completes the proof. 
\end{IEEEproof}

\begin{corollary} \label{cor_lh inequality}
    Under the same setting as in \Cref{lem_lundberg hoeffding}, 
    for any $n_a \geq 0$, $\alpha>0$, and $0<\gamma<1$ we have 
    \begin{align} \label{eq_lundberg hoeffding inequality corr}
        &\Prob{ \{ \forall n \geq n_a, \, J_n < n_a(1-\gamma)\mu  \} \cap \mathcal{A}_{\alpha}^{\textup{c}}} \nonumber\\
        &\quad \geq 1- \left(1-e^{-2\left(\frac{\gamma\mu}{a+b}\right)^2}\right)^{-1}\cdot e^{-2n_a\left(\frac{\gamma\mu}{a+b}\right)^2} - e^{-r\alpha}. 
    \end{align}
\end{corollary}

\begin{IEEEproof}
    Note that 
    \begin{align}
        &\Prob{\forall n \geq n_a, \, J_n < n_a(1-\gamma)\mu } \nonumber\\
        &\quad = \Prob{ \bigcap_{n=n_a}^{\infty} \{J_n < n_a(1-\gamma)\mu\}} \nonumber\\ 
        &\quad\geq \Prob{ \bigcap_{n=n_a}^{\infty} \{J_n < n(1-\gamma)\mu\}} \nonumber\\ 
        &\quad= 1- \Prob{ \bigcup_{n=n_a}^{\infty} \{J_n \geq n(1-\gamma)\mu\}} \nonumber\\ 
        &\quad\eqann[\geq]{a} 1- \sum_{n=n_a}^{\infty} e^{-2n\left(\frac{\gamma\mu}{a+b}\right)^2} \nonumber\\ 
        &\quad= 1- \left( \frac{1}{1-e^{-2\left(\frac{\gamma\mu}{a+b}\right)^2}} \right)\cdot e^{-2n_a\left(\frac{\gamma\mu}{a+b}\right)^2}, 
        \label{eq_hoeffding cap}
    \end{align}
    where \eqannref{a} is by~\eqref{eq_hoeffding result} and the union bound. 
    Observing that 
    \begin{align*} 
        &\Prob{\forall n \geq n_a, \,J_n <  n(1-\gamma)\mu}\\
        &\quad= \Prob{\{\forall n \geq n_a, \,J_n <  n(1-\gamma)\mu\} \cap \mathcal{A}_{\alpha}}  \\
        &\qquad+  \Prob{\{\forall n \geq n_a, \,J_n < n(1-\gamma)\mu\} \cap \mathcal{A}_{\alpha}^{\textup{c}}}  \\ 
         &\quad\leq \Prob{\mathcal{A}_{\alpha}} 
         +  \Prob{\{\forall n \geq n_a, \,J_n <  n(1-\gamma)\mu\} \cap \mathcal{A}_{\alpha}^{\textup{c}}},  
    \end{align*}
    we obtain 
    \begin{align*}    
        &\Prob{\{\forall n \geq n_a, \,J_n <  n(1-\gamma)\mu\} \cap \mathcal{A}_{\alpha}^{\textup{c}}} \\ 
        &\quad\geq 
            \Prob{\forall n \geq n_a, \,J_n <  n(1-\gamma)\mu} - \Prob{\mathcal{A}_{\alpha}}.
    \end{align*}
    Combining this inequality with~\eqref{eq_lundberg inequality} and~\eqref{eq_hoeffding cap} yields~\eqref{eq_lundberg hoeffding inequality corr} and completes
    the proof. 
\end{IEEEproof}

\begin{IEEEproof}[Proof of \Cref{prop_universal fast polarization}] 
    By inspection of the proof of~\cite{Tal_2017_simple}, the lemma will be true once we show that for any $\epsilon_a > 0$ and $\delta' > 0$
    there exist $n_a$ and $\eta$ such that if $Z_0 \leq \eta$, then~\eqref{eq_Zn below epsa} holds. 
    Thus, we fix $\epsilon_a>0$ and $\delta'>0$, and work toward this goal. 

    Let the process $\bar{Z}_0, \bar{Z}_1, \ldots$ be defined as 
    \begin{align}
        \bar{Z}_0 &= \ln Z_0,  \nonumber\\ 
        \bar{Z}_{n+1} &= \begin{cases} 2\bar{Z}_{n} + \ln \kappa, & B_{n+1} = 0, \\ 
            \bar{Z}_n + \ln \kappa, & B_{n+1} =  1,
        \end{cases}         \quad n \geq 0. \label{eq_Znbar recursion}
    \end{align}
    Then, by~\eqref{eq_bhattacharyya recursion}, $\ln Z_n \leq \bar{Z}_n$ for any $n$. 
    Therefore,~\eqref{eq_Zn below epsa} will be true once we show that there exists $n_a$ and $\eta$ such that if $\bar{Z}_0 = \ln \eta$, then
    \[ 
        \Prob{\bar{Z}_n \leq \ln \epsilon_a \text{ for all } n \geq n_a} \geq 1-\delta'. 
\] 

Fix 
    \begin{equation} \label{eq_bounds on initial zeta} 
        0< \zeta < 1/\kappa^2
    \end{equation}
    such that $\bar{Z}_0 < \ln \zeta < 0$.  
    Since $\bar{Z}_0 = \ln \eta$  by assumption,  and since we may set $\eta$ as small as desired, we can ensure that this is possible. 
    We then have, by~\eqref{eq_bhattacharyya recursion}, 
    \[ 
        \bar{Z}_1 \leq \begin{cases}  
            \bar{Z}_0 + \ln \kappa + \ln \zeta, & B_n = 0, \\ 
            \bar{Z}_0 + \ln \kappa,             & B_n = 1.
        \end{cases}
    \]
    If, further, $\bar{Z}_1 < \ln \zeta$ then the above inequality holds when $\bar{Z}_1$ and $\bar{Z}_0$ are replaced with  $\bar{Z}_2$ and
    $\bar{Z}_1$, respectively. 
    More generally, we define the process $J_n$, $n \in \mathbb{N}$, by 
    \begin{align*}
        J_0 &= \bar{Z}_0 = \ln \eta, \\ 
        J_{n+1} &= J_n + T_{n+1}, \quad n \geq 0, 
    \end{align*}
    where 
    \[ 
        T_{n} = \begin{cases} \ln \kappa + \ln \zeta, & B_n = 0,\\
            \ln \kappa,  & B_n = 1,
        \end{cases}
        \quad n \geq 1.
    \]
    If $J_i < \ln \zeta$ for all $i \leq n$, then $\bar{Z}_n \leq J_n$. 

    Recall that $B_1, B_2, \ldots$ is a sequence of i.i.d.\ random variables with $\Probi{B_i = 0} = \Probi{B_i = 1} = 1/2$ for any $i$.
    Thus, $T_1, T_2, \ldots$ is a sequence of i.i.d.\ random variables. 
    Denoting by $T$ a random variable distributed according to their common distribution, we have $\Probi{T = \ln \kappa} = \Probi{T = \ln \kappa + \ln \zeta} = 
    1/2$.
    In particular, $T$ is bounded: 
    \[ 
        -\ln\left(\frac{1}{\kappa \zeta}\right) = -b \leq T \leq a = \ln \kappa.
    \]
    Both $a$ and $b$ are positive by~\eqref{eq_bounds on initial zeta} and since $\kappa > 1$ by assumption. 
    By definition, for any $\epsilon > 0$, $\Probi{T > a- \epsilon} \geq \Probi{T = a} = 1/2$. 
    Moreover, by~\eqref{eq_bounds on initial zeta}, 
    \[ 
        \mu = \Exp{T} = \frac{1}{2} \ln(\kappa^2 \zeta) < 0.
    \] 
    Consequently, \Cref{cor_lh inequality} holds for the random walk $J_n - J_0 = \sum_{i=1}^n T_i$, $n \in \mathbb{N}$. 

    Let $r > 0$ be the largest positive solution of the equation 
    \begin{equation} \label{eq_equation for r}
        \Exp{e^{rT}} = \frac{ (\kappa \zeta)^r + \kappa^r }{2} =1.
    \end{equation} 
    Such $r$ exists, as shown in the proof of \Cref{lem_lundberg hoeffding}. 
    Denote for brevity 
    \[ 
        \theta \triangleq \left|\frac{\mu}{a+b}\right|.
    \]
    By \Cref{cor_lh inequality}, for any $0<\gamma<1$ and $n_a \geq 0$ we have
    \begin{IEEEeqnarray*}{l}
        \Prob{\{\forall n \geq n_a, \,J_n - J_0  < n_a(1-\gamma)\mu\}\cap \mathcal{A}_{-J_0+\ln\zeta}^{\textup{c}} } \\
        \quad\eqann{a} \Prob{\{\forall n \geq n_a, \,J_n  < J_0 - n_a(1-\gamma)|\mu|\}\cap \mathcal{A}_{-J_0+\ln\zeta}^{\textup{c}} } \\
        \quad\geq 1 - (1-e^{-2\gamma^2 \theta^2})^{-1}e^{-2n_a \gamma^2 \theta^2} - e^{-r (-J_0+\ln\zeta)}, 
        \IEEEyesnumber \label{eq_JnJ0 lower bound}
    \end{IEEEeqnarray*}
    where~\eqannref{a} is because $\mu <0$. 

    Observe that since $J_n = J_0 + \sum_{i=1}^n T_i$ we have 
    \begin{align*} 
        \mathcal{A}_{-J_0+\ln\zeta}^{\textup{c}} &= \left\{ \sum_{i=1}^n T_i < -J_0+\ln \zeta \text{ for all } n \in \mathbb{N} \right\} \\
        & = \{ J_n < \ln \zeta \text{ for all } n \in \mathbb{N} \}. 
    \end{align*}
    Consequently, under the event $\mathcal{A}_{-J_0+\ln\zeta}^{\textup{c}}$, we have $\bar{Z}_n \leq J_n$ for any $n$. 
    Hence, 
    \[ 
        \Prob{\{\forall n \geq n_a, \,J_n < J_0 -n_a(1-\gamma)|\mu|\}\cap \mathcal{A}_{-J_0+\ln\zeta}^{\textup{c}} }
    \] 
    lower-bounds the probability that $\bar{Z}_n \leq J_0 +n_a(1-\gamma)\mu$ for all $n \geq n_a$. 

    Recall that $J_0 = \bar{Z}_0 = \ln \eta$. 
    It remains to set $\eta$ and $n_a$ such that $\ln \eta < \ln \zeta$, $J_0 - n_a(1-\gamma)|\mu| \leq \ln \epsilon_a$, and the right-hand side
    of~\eqref{eq_JnJ0 lower bound} exceeds $1-\delta'$.
    Below we show one selection of $\eta$ and $n_a$. 
    Observe that there is freedom in this selection, and generally it is desirable to find
    small $n_a$ and large $\eta$. We leave such optimization for future work. 

    We first set the parameters $\gamma$ and $\zeta$. 
    We take $\gamma = 1/2$ and $\zeta = 1/(2\kappa^2)$. 
    In this case, $|\mu| = (\ln2)/2$ and $\theta = \ln 2 / (2\ln(2\kappa^2))$. 
    Further, our plan is to split $\delta'$ equally among the two subtracted terms on the right-hand side of~\eqref{eq_JnJ0 lower bound}.
    We stress that these are arbitrary choices, and in practice should be optimized. 
    We plug $\zeta$ into~\eqref{eq_equation for r} and compute $r$, the largest positive solution of $\kappa^r + (2\kappa)^{-r} = 2$. 

    Next, we set $J_0$ so that $e^{-r(-J_0 + \ln \zeta)} \leq \delta'/2$;
    one choice is $J_0 = \ln \zeta + \frac{1}{r} \ln(\delta'/2)$. 
    Observe that indeed $J_0 = \ln \eta < \ln \zeta$ since $\delta' <1$ (there is nothing to prove if $\delta' \geq 1$). 
    We thus take 
    \[ 
        \eta = e^{J_0} = \frac{1}{2\kappa^2} \left(\frac{\delta'}{2}\right)^{1/r}.
    \]
    We set $n_a$ large enough such that both $J_0 - n_a|\mu|/2 \leq \ln\epsilon_a$ and $(1-e^{-2\gamma^2\theta^2})^{-1}e^{-2n_a\gamma^2\theta^2} \leq \delta'/2$.
    That is, $n_a = \lceil n_a' \rceil$, where
    \[
        n_a' = \max\left\{\frac{4}{\ln 2}(J_0-\ln \epsilon_a) , \frac{2}{\theta^2} \ln\left(\frac{2}{\delta'\cdot(1-e^{-\theta^2/2}) }\right) \right\}. 
\]
For the above $\eta$ and $n_a$, $\Probi{\bar{Z}_n \leq \ln\epsilon_a \text{ for all } n \geq n_a} \geq 1-\delta'$. 
    Thus,~\eqref{eq_Zn below epsa} holds, and the proof is complete.
\end{IEEEproof}
The parameters $n_a$ and $\eta$ found in the above proof depend on the process $Z_n$ only through $\kappa$. 
Thus, they universally apply to any process for which~\eqref{eq_bhattacharyya recursion} holds. 
In particular, one can set in advance a universal length $\hat{N}$ for the polar code in the fast stage.

The values of $n_a$ and $\eta$ are not optimized in the above proof, and the actual required length of the fast stage is expected to be shorter in
practice. 
When designing a universal polar code, one can try out several small values of $\eta$ and numerically run the recursion~\eqref{eq_Znbar recursion} until
$\bar{Z}_n$ is sufficiently small for most indices. 
The above proof implies that if $\eta$ is small enough and we run the recursion for sufficiently long, we are ensured that most indices will polarize fast.

\section{Auxiliary Proofs for \Cref{sec_probabilistic model with memory}} \label{ap_auxiliary proofs for sec3A}
We denote $T_j = (X_j,Y_j)$, $j \in\mathbb{Z}$, with realization $t_j$, and $T_M^N = (X_M^N,Y_M^N)$ with realization $t_M^N$. For brevity, we denote  
$\prrv{T_M^N}{} = \prrv{T_M^N}{t_M^N}$, and similarly $\prrv{S_N}{} = \prrv{S_N}{s_N}$.

\begin{IEEEproof}[Proof of \Cref{lem_FAIM is psi mixing}]
    Although~\eqref{eq_psi mixing} was already proved in~\cite[Lemma 5]{Shuval_Tal_Memory_2017}, we provide a proof here for completeness. 

    We will prove that~\eqref{eq_psiphi mixing} holds with
    \begin{equation}
        \label{eq_psi_k}
        \psi_k = 
        \begin{dcases} 
            \max_{s,\sigma} \frac{\Prob{S_0=s,S_k=\sigma}}{\Prob{S_0 = s} \Prob{S_k = \sigma}}, & k > 0,  \\ 
            \max_s \frac{1}{\Prob{S_0 = s}}, & k =0 
        \end{dcases}
    \end{equation} 
    and 
    \begin{equation} 
        \label{eq_phi_k}
        \phi_k = 
        \begin{dcases}
            \min_{s,\sigma} \frac{\Prob{S_0=s,S_k=\sigma}}{\Prob{S_0 = s} \Prob{S_k = \sigma}}, & k > 0,  \\ 
            0, & k = 0.  
        \end{dcases}
    \end{equation} 
    Recall that by stationarity, $P_{S_0} = P_{S_k}$ for any $k$. 
    Further, observe that by Bayes' law, 
    \[ 
        \frac{\Prob{S_0=s,S_k=\sigma}}{\Prob{S_0 = s} \Prob{S_k = \sigma}} = \frac{\Prob{S_k=\sigma|S_0=s}}{\Prob{S_k = \sigma}}. 
    \]

    To prove~\eqref{eq_psiphi mixing}, we first consider the case $M>L$. 
Denote by $a,b,c,d$ the \emph{values} of states $S_0,S_L,S_M,$ and $S_N$, respectively (see \Cref{fig_two blocks LM}). Then,  
\begin{figure}[t]
\begin{center}
\begin{tikzpicture}
	\node[draw, thick, minimum width = 3cm, minimum height = 0.9 cm] (box1) at (0,0) {$T_1^L = (X_1^L,Y_1^L)$}; 
	\node[draw, thick, minimum width = 4cm, minimum height = 0.9 cm, right = 1 cm of box1] (box2) {$T_{M+1}^N= (X_{M+1}^N,Y_{M+1}^N)$}; 
	\node[above = 0.1 of box1.north west] {$S_0$};
	\node[below = 0.4 of box1.south west, anchor = base] {$a$};
	\node[above = 0.1 of box1.north east] {$S_{L}$};
	\node[below = 0.4 of box1.south east, anchor = base] {$b$};
	\node[above = 0.1 of box2.north west] {$S_M$};
	\node[below = 0.4 of box2.south west, anchor = base] {$c$};
	\node[above = 0.1 of box2.north east] {$S_{N}$};
	\node[below = 0.4 of box2.south east, anchor = base] {$d$};
\end{tikzpicture}
\end{center}
\caption{Two blocks of a FAIM process, not necessarily of the same length. The state $S_0$, just before the first block, assumes value $a
\in\mathcal{S}$. The final state of the first block, $S_L$, assumes value $b \in\mathcal{S}$. The state $S_M$, just before the second block, assumes value $c \in\mathcal{S}$. The final state of the second block, $S_N$, assumes value $d \in\mathcal{S}$.}\label{fig_two blocks LM}
\end{figure}

\begin{align*}
	\prrv{T_1^L,T_{M+1}^N}{} 
	&= \sol{\sum_{t_{L+1}^M}} \prrv{T_1^L,T_{L+1}^M,T_{M+1}^N}{} \\
				  &= \sol[l]{\sum_{t_{L+1}^M}} \sol[r]{\sum_{d,a}} \prrv{T_1^L,T_{L+1}^M,T_{M+1}^N,S_N|S_0}{} \prrv{S_0}{}\\
				  &= \sol[l]{\sum_{\substack{d,c, \\ b, a}}} \sum_{t_{L+1}^M} \prrv{T_{M+1}^N,S_N|S_M}{}\prrv{T_{L+1}^M,S_M|S_L}{} \prrv{T_1^L,S_L|S_0}{} \prrv{S_0}{}\\
				  &=\sol{\sum_{\substack{d,c, \\ b, a}}}  \prrv{T_{M+1}^N,S_N|S_M}{}\left(\sum_{t_{L+1}^M} \prrv{T_{L+1}^M,S_M|S_L}{}\right) \prrv{T_1^L,S_L|S_0}{} \prrv{S_0}{}\\
				  &= \sol{\sum_{\substack{d,c, \\ b, a}}}  \prrv{T_{M+1}^N,S_N|S_M}{} \prrv{S_M|S_L}{} \prrv{T_1^L,S_L|S_0}{} \prrv{S_0}{} \\
				  &= \sol{\sum_{\substack{d,c, \\ b, a}}}  \prrv{T_{M+1}^N,S_N|S_M}{} \prrv{S_M}{} \frac{\prrv{S_M|S_L}{}}{\prrv{S_M}{}} \prrv{T_1^L,S_L|S_0}{} \prrv{S_0}{} \\
                  &\eqann[\leq]{a} \psi_{M-L} \left( \sol{\sum_{d,c}}  \prrv{T_{M+1}^N,S_N|S_M}{} \prrv{S_M}{}\right) \left( \sol{\sum_{b,a}}\prrv{T_1^L,S_L|S_0}{}
                      \prrv{S_0}{}\right) \\
				  &= \psi_{M-L} \prrv{T_1^L}{} \prrv{T_{M+1}^N}{}, 
				  \end{align*}
                  where~\eqannref{a}  follows from the definition of $\psi_k$. 
This shows~\eqref{eq_psi mixing}. 
To see~\eqref{eq_phi mixing} we follow the exact steps above up to just before inequality \eqannref{a}, and proceed with
\begin{align*}
	\prrv{T_1^L,T_{M+1}^N}{} 
    &\geq \phi_{M-L} \left( \sol{\sum_{d,c}}  \prrv{T_{M+1}^N,S_N|S_M}{} \prrv{S_M}{}\right) \left( \sol{\sum_{b,a}}\prrv{T_1^L,S_L|S_0}{}
    \prrv{S_0}{}\right) \\
    &= \phi_{M-L} \prrv{T_1^L}{} \prrv{T_{M+1}^N}{}. 
\end{align*}
Again, the inequality follows from the definition of $\phi_k$.

For the case $M=L$, we need only establish~\eqref{eq_psi mixing}, as~\eqref{eq_phi mixing} is trivially true for $M=L$. Again, $a$ and $d$ represent the \emph{values} of states $S_0$ and $S_N$. Both $b$ and $b'$ represent values of state $S_L$; this distinction is to distinguish the summation variables of two different sums over values of $S_L$. Thus,  
\begin{align*}
	\prrv{T_1^L,T_{L+1}^N}{} 
			   &= \sol{\sum_{\substack{a,b, \\ d}}} \prrv{T_{L+1}^N,S_N|S_L}{} \frac{\prrv{S_L}{}}{\prrv{S_L}{}} \prrv{T_1^L,S_L|S_0}{} \prrv{S_0}{}\\
               &\leq \psi_{0} \sol{\sum_{d,b}} \prrv{T_{L+1}^N,S_N|S_L}{} \prrv{S_L}{} \cdot \left(\sum_{b',a}\prrv{T_1^L,S_L|S_0}{} \prrv{S_0}{}\right)\\
               &= \psi_{0} \prrv{T_1^L}{} \prrv{T_{L+1}^N}{};
				  \end{align*}
where the inequality is by the definition of $\psi_0$ and because $\prrv{T_1^L,S_L|S_0}{} \leq \sum_{b'} \prrv{T_1^L,S_L|S_0}{}$.

To see that that $\psi_k$ is nonincreasing, observe that for any $s, \sigma \in \mathcal{S}$:
\begin{align*}
    P_{S_{k+1},S_0}(\sigma,s) & = \sol{\sum_{a\in\mathcal{S}}} P_{S_{k+1}|S_k}(\sigma|a)\cdot P_{S_k,S_0}(a,s) \\ 
    & \leq \psi_k \sol{\sum_{a\in\mathcal{S}}} P_{S_{k+1}|S_k}(\sigma|a)\cdot P_{S_k}(a) P_{S_0}(s) \\ 
    & =\psi_k P_{S_{k+1}}(\sigma)P_{S_0}(s).
\end{align*}
Therefore, we must have $\psi_{k+1} \leq \psi_k$. 
The proof that $\phi_k$ is nondecreasing is similar, with ``$\leq \psi_k$'' replaced with ``$\geq \phi_k$''. 

Finally, the asymptotic properties of $\phi_k$ and $\psi_k$ are due to $S_j$ being an aperiodic and irreducible stationary finite-state Markov chain. 
For in this case there exist $\gamma<1$ and $0<\alpha<\infty$ such that for any $s,\sigma \in \mathcal{S}$ and $k \geq 0$,  
\[ 
    |P_{S_k|S_0}(\sigma|s) - P_{S_k}(\sigma)| \leq \alpha\cdot \gamma^k,
\] 
see~\cite[Theorem 4.3]{Iosifescu} for a proof. 
Rearranging and observing that $\psi_0 < \infty$, we obtain that 
\[
    \left| \frac{\Prob{S_0 = s, S_k = \sigma}}{\Prob{S_0=s} \Prob{S_k=\sigma}} -1 \right| \leq \psi_0 \cdot\alpha\cdot \gamma^k \xrightarrow[k\to\infty]{} 0.
    \] 
Hence, both $\psi_k$ and $\phi_k$ must tend to $1$ exponentially fast as $k \to \infty$. 
\end{IEEEproof}

\begin{IEEEproof}[Proof of \Cref{lem_information for other klm}]
    We will prove~\eqref{eq_memory length inequality general b}. The proof of~\eqref{eq_memory length inequality general a} is identical, with the replacement of $Y_a^b$ with $X_a^b, Y_a^b$ throughout for any $a$ and $b$. 

    The process $(S_j, X_j, Y_j)$, $j\in \mathbb{Z}$ is FAIM, so it satisfies the Markov property~\eqref{eq_markov property of FAIM}. The proof follows from the following chain of inequalities. 
    \begin{align*}
        I(S_1 ; S_{\lambda} | Y_1^{\lambda})  
                 &\eqann[\geq]{a} I(S_1 ; (Y_{\lambda+1}^k, S_m) | Y_1^{\lambda}) \\ 
                 & = I(S_1 ; Y_{\lambda+1}^k | Y_1^{\lambda}) + I(S_1; S_m | Y_1^k) \\ 
                 &\eqann[\geq]{b} I(S_1; S_m | Y_1^k) \\ 
                 &\eqann[\geq]{c} I(S_{\ell} ; S_m | Y_1^k).
    \end{align*}
    We now justify the inequalities: 
    \begin{itemize}
        \item \eqannref{a} is by~\eqref{eq_DPI}, noting that since $m \geq k \geq \lambda \geq 1$,~\eqref{eq_markov property of FAIM} implies 
        \[
            S_1 \markov (S_{\lambda},Y_1^{\lambda}) \markov (Y_{\lambda+1}^k, S_m);
        \]
        \item \eqannref{b} is because mutual information is nonnegative; 
        \item \eqannref{c} is by~\eqref{eq_DPI}, noting that since $\ell \leq 1$,~\eqref{eq_markov property of FAIM} implies 
        \[ 
            S_m \markov (S_1, Y_{1}^{k}) \markov S_{\ell}
        \]  
        (observe that this is a Markov chain in reverse order of time). 
    \end{itemize}
    This completes the proof. 
\end{IEEEproof}

\begin{IEEEproof}[Proof of \Cref{lem_finite memory for HMM}]
    The FAIM process is forgetful, so we let $\memlength$ be the $\epsilon$-recollection of the process. 
    For this $\memlength$,~\eqref{eq_memory length inequality} is satisfied. 

    By the chain rule for mutual information, 
    \begin{IEEEeqnarray}{l} 
        I(S_0; S_{-k},S_k | X_{-\ell}^{-1},Y_{-\ell}^m) \qquad\qquad \IEEEnonumber
        \\\qquad = I(S_0; S_{k} | X_{-\ell}^{-1},Y_{-\ell}^m) + I(S_0;S_{-k}|X_{-\ell}^{-1},Y_{-\ell}^m, S_{k}). \IEEEeqnarraynumspace \IEEEyesnumber\label{eq_chain rule for two states} 
    \end{IEEEeqnarray} 
    We will upper-bound each of the terms on the right-hand side of~\eqref{eq_chain rule for two states} by $\epsilon$, yielding the desired result.

    For any $m,\ell,k$ such that $\min\{m,\ell\} \geq k \geq \memlength$ we have 
    \begin{align*}
        \epsilon &\eqann[\geq]{a} I(S_0; S_k | Y_0^k) \\ 
                 &\eqann[\geq]{b} I(S_0; (S_k,Y_{k+1}^m) | Y_0^k) \\ 
                 &= I(S_0; Y_{k+1}^m | Y_0^k) +  I(S_0; S_k | Y_0^m)\\ 
                 &\eqann[\geq]{c}   I(S_0; S_k | Y_0^m)\\ 
                 &\eqann[\geq]{d} I((S_0,X_{-\ell}^{-1}, Y_{-\ell}^{-1}); S_k | Y_0^m)\\ 
                 &= I(X_{-\ell}^{-1}, Y_{-\ell}^{-1}; S_k | Y_0^m) +  I(S_0; S_k | X_{-\ell}^{-1},Y_{-\ell}^m)\\ 
                 &\eqann[\geq]{e} I(S_0; S_k | X_{-\ell}^{-1},Y_{-\ell}^m).
    \end{align*}
    We now justify the inequalities:\footnote{We remark that in~\eqannref{b} and~\eqannref{d} we can replace the inequalities with equalities.} 
    \begin{itemize}
        \item \eqannref{a} is by~\eqref{eq_memory length inequality Y} and stationarity. 
        \item \eqannref{b} is by~\eqref{eq_DPI}, noting that~\eqref{eq_markov property of FAIM} implies 
        \[
            S_0 \markov (S_k,Y_0^k) \markov (S_k,Y_{k+1}^m);
        \]
        \item \eqannref{c} is because mutual information is nonnegative; 
        \item \eqannref{d} is by~\eqref{eq_DPI}, noting that~\eqref{eq_markov property of FAIM} implies 
        \[ 
            S_k \markov (S_0, Y_{0}^{m}) \markov (S_0, X_{-\ell}^{-1}, Y_{-\ell}^{-1})
        \]  
        (observe that $X_{-\ell}^{-1}, Y_{-\ell}^{-1}$ is ``in the past'' whereas $Y_0^m$ is ``in the future,'' and the state $S_0$ is in between);
        \item \eqannref{e} is because mutual information is nonnegative.
    \end{itemize}

    The derivation for the second term in the right-hand side of~\eqref{eq_chain rule for two states} 
    is similar. 
    For any $m,\ell,k$ such that $\min\{m,\ell\} \geq k \geq \memlength$ we have 
    \begin{align*}
        \epsilon &\eqann[\geq]{a} I(S_0; S_{-k} | X_{-k}^{-1}, Y_{-k}^{-1}) \\ 
                 &\eqann[\geq]{b} I(S_0; (S_{-k},X_{-\ell}^{-k-1}, Y_{-\ell}^{-k-1}) | X_{-k}^{-1}, Y_{-k}^{-1}) \\
                 &\eqann[\geq]{c} I(S_0; S_{-k} | X_{-\ell}^{-1}, Y_{-\ell}^{-1}) \\ 
                 &\eqann[\geq]{d} I((S_0, Y_{0}^{m}, S_k); S_{-k} | X_{-\ell}^{-1}, Y_{-\ell}^{-1})\\ 
                 &\eqann[\geq]{e} I(S_0; S_{-k} | X_{-\ell}^{-1},Y_{-\ell}^m, S_k).
    \end{align*}
    Again, we justify the inequalities: 
    \begin{itemize}
        \item \eqannref{a} is by~\eqref{eq_memory length inequality XY 0} and stationarity. 
        \item \eqannref{b} is by~\eqref{eq_DPI}, noting that~\eqref{eq_markov property of FAIM} implies 
        \[
            S_0 \markov (S_{-k},X_{-k}^{-1}, Y_{-k}^{-1}) \markov (S_{-k},X_{-\ell}^{-k-1},Y_{-\ell}^{-k-1});
        \]
        \item \eqannref{c} is by the chain rule for mutual information
        \item \eqannref{d} is by~\eqref{eq_DPI}, noting that~\eqref{eq_markov property of FAIM} implies 
        \[ 
            S_{-k} \markov (S_0, X_{-\ell}^{-1}, Y_{-\ell}^{-1}) \markov (S_0, Y_{0}^{m}, S_k); 
        \]  
        \item \eqannref{e} is by the chain rule for mutual information. 
    \end{itemize}
    This completes the proof.
\end{IEEEproof}

\begin{IEEEproof}[Proof of \Cref{cor_forgetting multi blocks}] 
    The FAIM process is forgetful, so we set $\memlength$ as the $\epsilon$-recollection of the process. 
    The corollary holds for $k=1$ by \Cref{lem_finite memory for HMM}. 
    We proceed by induction. 
    Assume that the corollary holds for $k-1 \geq 1$, and we will show it holds for $k$.

    Let 
    \begin{align*}
        \bv{i}' &= \begin{bmatrix} i_1 & i_2 & \cdots & i_{k-1} \end{bmatrix}, \\ 
        \bv{i}  &= \begin{bmatrix} i_1 & i_2 & \cdots & i_{k-1} & i_k \end{bmatrix}
        = \begin{bmatrix} \bv{i}' & i_{k} \end{bmatrix}.
    \end{align*}
    For brevity, denote
    \[ 
        C_i = (X_{i-L_0}^{i-1}, Y_{i-L_0}^{i+L_0}). 
    \] 
    Our goal is thus to show that 
    \begin{IEEEeqnarray*}{l}
        I(S_{\bv{i}} ; S_{\bv{i}-L_0}, S_{\bv{i}+L_0} | C_{\bv{i}} ) \\ 
             \qquad   =
        I(S_{\bv{i}'}, S_{i_k}; S_{\bv{i}'-L_0},  S_{\bv{i}'+L_0}, S_{i_k - L_0}, S_{i_k + L_0} | 
        C_{\bv{i}'}, C_{i_k}) \leq k\cdot 2\epsilon.
    \end{IEEEeqnarray*}
    Indeed, 
    \begin{align*}
        &I(S_{\bv{i}'}, S_{i_k}; S_{\bv{i}'-L_0},  S_{\bv{i}'+L_0}, S_{i_k - L_0}, S_{i_k + L_0} | 
        C_{\bv{i}'}, C_{i_k})\\ 
        &\quad= I(S_{\bv{i}'} ;  S_{\bv{i}'-L_0},S_{\bv{i}'+L_0}, S_{i_k-L_0},S_{i_k+L_0}| C_{\bv{i}'}, C_{i_k}) \\ 
        &\qquad+ I(S_{i_k}; S_{\bv{i}'-L_0},S_{\bv{i}'+L_0}, S_{i_k-L_0},S_{i_k+L_0} | S_{\bv{i}'}, C_{\bv{i}'}, C_{i_k})\\ 
        &\quad\eqann[\leq]{a} I(S_{\bv{i}'} ;  S_{\bv{i}'-L_0},S_{\bv{i}'+L_0},
        (S_{i_k-L_0},S_{i_k+L_0}, C_{i_k})| C_{\bv{i}'}) \\ 
        &\qquad+ I(S_{i_k}; S_{i_k-L_0},S_{i_k+L_0}, (S_{\bv{i}'}, S_{\bv{i}'-L_0},S_{\bv{i}'+L_0}, C_{\bv{i}'}) | C_{i_k})\\ 
        &\quad\eqann[\leq]{b} I(S_{\bv{i}'}; S_{\bv{i}'-L_0},S_{\bv{i}'+L_0} | C_{\bv{i}'})
        + I(S_{i_k}; S_{i_k-L_0},S_{i_k+L_0} | C_{i_k}) \\ 
        &\quad\eqann[\leq]{c} (k-1)\cdot 2\epsilon + 2\epsilon \\
        &\quad= k \cdot 2\epsilon, 
    \end{align*}
    where \eqannref{a} is by the chain rule; ~\eqannref{b} is by~\eqref{eq_DPI} and~\eqref{eq_markov
    property of FAIM}, used for the Markov chains (See \Cref{fig_timeline} for an illustration):
    \[
        S_{\bv{i}'} \markov ( S_{\bv{i}'-L_0},S_{\bv{i}'+L_0}, C_{\bv{i}'}) \markov
        (S_{\bv{i}'-L_0},S_{\bv{i}'+L_0}, S_{i_k-L_0},S_{i_k+L_0}, C_{i_k}),
    \]
    which holds because $i_{k-1} \leq i_{k-1} + L_0 \leq i_k - L_0$ so
    $(S_{i_k-L_0},S_{i_k+L_0},C_{i_k})$ are independent of $S_{\bv{i}'}$ given
        $S_{i_{k-1} + L_0}$, which is part of $S_{\bv{i}'+L_0}$, and 
        \begin{align*} 
            S_{i_k}  &\markov ( S_{i_k-L_0},S_{i_k+L_0}, C_{i_k})\\ &\markov ( S_{i_k-L_0},S_{i_k+L_0}, S_{\bv{i}'},
        S_{\bv{i}'-L_0},S_{\bv{i}'+L_0}, C_{\bv{i}'}), 
    \end{align*}
    which again holds because $i_{k-1}+L_0 \leq i_k - L_0 \leq i_k$, so $(S_{\bv{i}'}, S_{\bv{i}'-L_0},S_{\bv{i}'+L_0}, C_{\bv{i}'})$ are independent of $S_{i_k}$ given
    $S_{i_{k-L_0}}$; 
    finally,~\eqannref{c} is because $I(S_{\bv{i}'}; S_{\bv{i}'-L_0},S_{\bv{i}'+L_0} | C_{\bv{i}'}) \leq (k-1)\cdot 2\epsilon$ by the induction hypothesis 
    and $I(S_{i_k}; S_{i_k-L_0},S_{i_k+L_0}|C_{i_k}) \leq 2\epsilon$ by \Cref{lem_finite memory for HMM}. 
    This completes the proof. 
    \begin{figure}[t]
        \begin{center}
            \begin{tikzpicture}[>=latex]
                \node[rectangle, draw, minimum height = 0.2cm, minimum width = 1cm, fill = red!25!white] (R1) at (1,0) {};  
                \node[rectangle, draw, minimum height = 0.2cm, minimum width = 1cm, fill = green!25!white] (R2) at (3,0) {};  
                \node[rectangle, draw, minimum height = 0.2cm, minimum width = 1cm, fill = blue!25!white] (R3) at (5,0) {};  
                \node[rectangle, draw, minimum height = 0.2cm, minimum width = 1cm, fill = orange!25!white] (R4) at (7,0) {};  
                
                \draw[|-|] (0,0)  -- (8,0); 

                \foreach \x/\y/\z in {1/red/0.5,2/green!50!black/0.5,3/blue/0.5,4/orange/0.5}
                {
                \draw[->, inner sep = 1pt, \y] ($(R\x.north west) + (0,0.45)$) node[above] {\tiny $i_\x - \ell$} --(R\x.north west); 
                \draw[->, inner sep = 1pt, \y] ($(R\x.north east) + (0,0.45)$) node[above] {\tiny $i_\x + \ell$} --(R\x.north east); 
                \draw[->, inner sep = 1pt, \y] ($(R\x.north east)!\z!(R\x.north west) + (0,0.45)$) node[above] {\tiny $i_\x$} --($(R\x.north
                    east)!\z!(R\x.north west)$); 
                }

                \draw[dashed] (6.5,0.125) -- (6.5,-0.7) node[right=0.44cm, pos=0.67,anchor=base] (A) {\scriptsize ``future''}; 
                    \draw[dashed] (5.5,0.125) -- (5.5,-0.7) node[left=0.34cm, pos = 0.67,anchor=base](B){ \scriptsize ``past''}
                                                            node[right=0.5cm, pos = 0.67,anchor=base](C){ \scriptsize ``present''}; 
            \end{tikzpicture}
        \end{center}
        \caption{Illustration of the timeline for $k=4$. Given $S_{i_4-\ell}$, the
        ``future'' is independent of the ``present'' and ``past.'' Given $S_{i_3+\ell}$, the ``past'' is independent of the ``present'' and
    ``future.''}
        \label{fig_timeline}
    \end{figure}
\end{IEEEproof}

\section{Auxiliary Proofs for \Cref{sec_otbst}} \label{ap_auxiliary proofs for otbst}

Recall from~\eqref{eq_def of h2} that the binary entropy function $h_2:[0,1] \to [0,1]$ is defined by
\[ 
    h_2(x) = -x \log x -(1-x) \log(1-x). 
\] 
This is a concave-$\cap$ function that satisfies $h_2(x) = h_2(1-x)$ for any $x \in [0,1]$,  and is monotone increasing over $[0,1/2]$. 
The inverse of the binary entropy function is $h_2^{-1}:[0,1] \to [0,1/2]$. 
The following three technical lemmas will be used to prove \Cref{lem_binary rvs}. 

\begin{lemma} 
    For any $0 \leq x \leq 1/2$, 
    \begin{equation} \label{eq_inequality for 1-h2}
        1-h_2(x) \geq \frac{2}{\ln 2}\left(\frac{1}{2}-x\right)^2. 
    \end{equation} 
\end{lemma}
\begin{IEEEproof}
    Denote $g(x) = 1-h_2(x)$. 
    Clearly, $1 = g(0) > 1/(2\ln2) \approx 0.721$. 
    For any $\epsilon>0$, the function $g(x)$ is $4$ times continuously differentiable over $[\epsilon,1/2]$. 
    Therefore, by Taylor's formula with remainder~\cite[Theorem 5.19]{apostol}, for any $x \in [\epsilon,1/2]$, there exists $y \in [x,1/2]$ such that
    \[ 
        g(x) = \frac{2}{\ln 2} \left(\frac{1}{2} -x\right)^2
        + \frac{g^{(4)}(y)}{4!}\left(\frac{1}{2}-x\right)^4.
    \] 
    However, $g^{(4)}(y) = 2(y^{-3}+(1-y)^{-3})/\ln 2 > 0$ for any $y \in [\epsilon,1/2]$. 
    Hence, $1-h_2(x) \geq 2(1/2-x)^2/(\ln2)$ for any $0 \leq x \leq 1/2$ as well. 
\end{IEEEproof}
\begin{lemma}
    For any $0 \leq y \leq x \leq 1/2$, 
    \begin{equation} \label{eq_inequality for h2 difference}
        h_2(x) - h_2(y) \geq \frac{1}{\ln 2}(x-y)\left(1-2y\right). 
    \end{equation}
\end{lemma}

\begin{IEEEproof}
    There is nothing to prove if $x = y$, so we assume that $y < x$. 
    Due to the concavity of $h_2(x)$, for any $x_1 \leq x_2 \leq x_3$ we have 
    \begin{equation}        \label{eq_concavity of h2}
        (x_3 - x_1)(h_2(x_2)-h_2(x_1)) \geq (x_2 - x_1)(h_2(x_3) - h_2(x_1))
    \end{equation} 
    (see, for example, \cite[Section 1.4.3]{Mitrinovic}, or~\cite[Exercise 6.17]{Steele}). 
    Setting $x_1 = y, x_2 = x, x_3 = 1/2$ in~\eqref{eq_concavity of h2} we obtain 
    \[ 
        \left(\frac{1}{2} - y\right) (h_2(x) - h_2(y)) \geq (x-y)(1-h_2(y)). 
    \] 
    Since $y<x \leq 1/2$ by assumption, $1/2-y>0$. Therefore, we rearrange the above inequality and
    obtain 
    \[ 
        h_2(x) - h_2(y) \geq (x-y)\frac{1-h_2(y)}{1/2-y} \geq \frac{1}{\ln 2}
        (x-y)\left(1-2y\right),
    \]
    where the rightmost inequality is by~\eqref{eq_inequality for 1-h2}. 
\end{IEEEproof}
            
\begin{lemma} \label{lem_prop of h2 diff}
    For any $ x,y \in (0, 1/2)$, the function
    \begin{equation} \label{eq_def of fxy}
        f(x,y) = h_2(h_2^{-1}(x) * h_2^{-1}(y)) - y
    \end{equation}
    is increasing in $x$ and decreasing in $y$. 
\end{lemma}

\begin{IEEEproof}
    Denote, for $x,y \in (0, 1/2)$,  
    \[ 
        g(x,y) = h_2(x * y) - h_2(y). 
    \] 
    Then, $f(x,y) = g(h_2^{-1}(x), h_2^{-1}(y))$. 
    The function $h_2(x)$ is monotone increasing over $[0,1/2]$, so $h_2^{-1}(x)$ is also monotone increasing over $[0,1/2]$. 
    Therefore, the claim will be true once we establish that $g(x,y)$ is increasing in $x$ and decreasing in $y$. 

    To this end, recall the function 
    \[ 
        \arctanh(x) = \frac{1}{2}\ln \left( \frac{1+x}{1-x} \right),
    \] 
    defined for $x \in [0,1]$. 
    This is an increasing function of $x$ (since its derivative is $(1-x^2)^{-1}$, which is positive). Moreover, $\arctanh(x) > 0 $ for $x> 0$. 
    
    Now, 
    \[ 
        \frac{\partial g(x,y)}{\partial x} = \frac{2}{\ln 2} (1-2y) \arctanh\big((1-2x)(1-2y)\big). 
    \] 
    This is positive since $\arctanh(z)>0$ for $z>0$, and both $(1-2x) > 0$ and $(1-2y)>0$.
    Thus, $g(x,y)$, and by proxy $f(x,y)$, is increasing in $x$. 
    Next, 
    \begin{align*} 
        &\frac{\partial g(x,y)}{\partial y}\\
        &= \frac{2}{\ln 2}\Big( (1-2x) \arctanh\big((1-2x)(1-2y)\big) - \arctanh(1-2y) \Big) \\ 
        &\leq \frac{2}{\ln 2}\Big( (1-2x) \arctanh(1-2y) - \arctanh(1-2y) \Big)  \\ 
        &=  \frac{2}{\ln 2}\left( (1-2x) - 1 \right)\cdot\arctanh(1-2y)  \\ 
        & < 0, 
    \end{align*} 
    where the first inequality is because $(1-2x)(1-2y) < (1-2y)$ and $\arctanh(\cdot)$ is increasing. 
    Thus, $g(x,y)$, and by proxy $f(x,y)$, is decreasing in $y$. 
\end{IEEEproof}

\begin{IEEEproof}  [Proof of \Cref{lem_binary rvs}] 
    It was shown in \cite[Lemma 2.1]{sasoglu_thesis} that 
    \[ 
    \sum_{a,b} p_a q_b h_2(\alpha_a * \beta_b) \geq h_2(h_2^{-1}(A)* h_2^{-1}(B)),
    \] 
    where
    \[ 
        A = \sum_a p_a h_2(\alpha_a), \quad B = \sum_b q_b h_2(\beta_b).
    \] 
    Therefore, 
    \begin{align*}
        \sum_{a,b} p_a q_b \left( h_2(\alpha_a * \beta_b) - h_2(\beta_b) \right)   
        &\geq h_2(h_2^{-1}(A)* h_2^{-1}(B)) - B \\ 
        &= f(A,B),
    \end{align*}
    where $f(\cdot,\cdot)$ was defined in~\eqref{eq_def of fxy}. 
    By~\eqref{eq_max min entropy}, $A \geq \xi_1$ and $B \leq \xi_2$. 
    Since, by \Cref{lem_prop of h2 diff}, $f(A,B)$ is increasing in $A$ and decreasing in $B$, we conclude that
    \[ 
        \sum_{a,b} p_a q_b \left( h_2(\alpha_a * \beta_b) - h_2(\beta_b) \right) \geq 
            h_2(h_2^{-1}(\xi_1) * h_2^{-1}(\xi_2)) - \xi_2.
        \] 
    Define, therefore,  
    \begin{equation} \label{eq_def of Delta}
         \Delta(\xi_1,\xi_2) \triangleq h_2(h_2^{-1}(\xi_1) * h_2^{-1}(\xi_2)) - \xi_2.
     \end{equation} 
    It remains to show that $ \Delta(\xi_1, \xi_2) > 0$. 

    To this end, observe that for any $x,y \in (0, 1/2)$,  
    \begin{align*} 
        h_2(x*y) - h_2(y)  &\eqann[\geq]{a} \frac{1}{\ln 2}(x*y-y)\cdot(1-2y) \\ 
        &= \frac{1}{\ln 2} x (1-2y)^2. 
    \end{align*} 
    where~\eqannref{a} is by~\eqref{eq_inequality for h2 difference}. 
    Therefore, 
    \[ 
        \Delta(\xi_1, \xi_2) \geq  \frac{1}{\ln 2} h_2^{-1}(\xi_1) \left(1-2h_2^{-1}(\xi_2) \right)^2 > 0.  \IEEEQEDhereeqn
    \] 
\end{IEEEproof}
We note in passing that the expression for $\Delta(\xi_1,\xi_2)$ derived here (or its lower bound) may be used to obtain a tighter lower bound than that of \cite[Lemma 11]{sasoglu_Tal_mem}.

\section{Auxiliary Proofs for \Cref{sec_monopolarization for FAIM derived processes}} \label{ap_auxiliary proofs for sec3B}

\begin{IEEEproof}[Proof of \Cref{lem_entropy difference if distributions close}]
    Denote $F = f(A)$, $\tilde{F} = f(\tilde{A})$, $G= g(A)$, and $\tilde{G} = g(\tilde{A})$. 
    For any $f_0 \in \{0,1\}$, $g_0 \in \mathcal{G}$, we abuse notation and write  
    \begin{IEEEeqnarray}{rCl} \IEEEyesnumber \label{eq_pf0g0 and qf0g0}
        p(f_0,g_0) &\triangleq \Prob{F = f_0, G = g_0} = \sol{\sum_{\substack{a: f(a) = f_0, \\ \hphantom{a:{}}g(a) = g_0}}} p(a), \IEEEyessubnumber
        \label{eq_pf0g0} \\[0.1cm] 
        q(f_0,g_0) &\triangleq \Prob{\tilde{F} = f_0, \tilde{G} = g_0} = \sol{\sum_{\substack{a: f(a) = f_0, \\\hphantom{a:{}} g(a) = g_0}}} q(a).
        \IEEEyessubnumber \label{eq_qf0g0} 
    \end{IEEEeqnarray} 
    With this notation we also have $p(g_0) = \Prob{G = g_0}$ and $p(f_0|g_0) = \Prob{F = f_0 | G = g_0}$. The distributions $q(g_0), q(f_0|g_0)$ are
    similarly defined. 
    By~\eqref{eq_pq inequality} and~\eqref{eq_pf0g0 and qf0g0} we have for all $f_0 \in \{0,1\}$ and $g_0 \in \mathcal{G}$, 
    \begin{IEEEeqnarray}{rCcCl}
        (1-\varepsilon)q(f_0,g_0) &\leq& p(f_0,g_0) &\leq& (1+\varepsilon)q(f_0,g_0), \IEEEnonumber \\ 
        (1-\varepsilon)q(g_0) &\leq& p(g_0) &\leq& (1+\varepsilon)q(g_0). \IEEEyesnumber \label{eq_pg0qg0}
    \end{IEEEeqnarray}
    Therefore, 
    \[ 
        \frac{1-\varepsilon}{1+\varepsilon}\cdot q(f_0|g_0) \leq p(f_0|g_0) \leq \frac{1+\varepsilon}{1-\varepsilon}\cdot q(f_0|g_0). 
    \] 
    When $0 \leq \varepsilon \leq \frac{1}{3}$, we have $(1+\varepsilon)/(1-\varepsilon) \leq 1+3\varepsilon$ and $(1-\varepsilon)/(1+\varepsilon) \geq 1-3\varepsilon
    \geq 0$ by straightforward algebra. 
    Hence, for any $f_0 \in \{0,1\}$ and $g_0 \in \mathcal{G}$, 
    \[ 
        (1-3\varepsilon) q(f_0|g_0) \leq p(f_0|g_0) \leq (1+3\varepsilon) q(f_0|g_0), 
    \] 
    by which $ |p(f_0|g_0) - q(f_0|g_0)|\leq 3\varepsilon\cdot q(f_0|g_0)$. 
   Thus, for any $g_0 \in \mathcal{G}$, since $\varepsilon < \frac{1}{6}$ by assumption, 
   \[ 
        d(g_0) \triangleq \sum_{f_0 =0}^1|p(f_0|g_0) - q(f_0 | g_0)| 
        \leq 3\varepsilon \sum_{f_0=0}^1 q(f_0|g_0) 
        = 3\varepsilon < \frac{1}{2}.
    \]
Since $F$ and $\tilde{F}$ are binary,   we conclude from \cite[Theorem 17.3.3]{cover_thomas} that for any $g_0 \in \mathcal{G}$, 
    \begin{align} 
            \big| H(F|G=g_0) - H(\tilde{F} | \tilde{G}=g_0) \big| &\leq - d(g_0) \log \frac{d(g_0)}{2} \nonumber \\ 
            &\eqann[\leq]{a} -3 \varepsilon \log \frac{3 \varepsilon}{2}.  \label{eq_HFG tilde}
    \end{align} 
    Inequality~\eqannref{a} is true because $x \mapsto -x \log \frac{x}{2}$ is increasing for $0\leq x < \frac{2}{e} \approx 0.736$, and $0\leq d(g_0)
    \leq 3\varepsilon < \frac{1}{2} < \frac{2}{e}$ by assumption.

    Let $\sum\nolimits^+$ denote summation over all $g_0 \in \mathcal{G}$ for which $p(g_0) \geq q(g_0)$, and $\sum\nolimits^-$ denote summation
    over all $g_0 \in \mathcal{G}$ for which $p(g_0) < q(g_0)$. 
    Since $\sum_{g_0} p(g_0) = \sum_{g_0} q(g_0) = 1$, we have 
    \begin{align*} 
        \sum\nolimits^+ (p(g_0) - q(g_0)) &= -\sum\nolimits^-(p(g_0)- q(g_0))\\
        &= \frac{1}{2} \sum_{g_0} |p(g_0) - q(g_0)| \\ 
        &\leq \frac{\varepsilon}{2} \sum_{g_0} q(g_0) = \frac{\varepsilon}{2}, 
    \end{align*} 
    where the inequality is by~\eqref{eq_pg0qg0}. 
    Hence, for any nonnegative function $h:\mathcal{G} \to \mathbb{R}^+$,
    \begin{align} 
        &\sum_{g_0} (p(g_0) - q(g_0))h(g_0)\nonumber \\ 
        &\quad= \sum\nolimits^+ |p(g_0) - q(g_0)| h(g_0) - \sum\nolimits^-|p(g_0)-q(g_0)|h(g_0) \nonumber \\  
        &\quad\leq \Big(\sup_{g_0} h(g_0) - \inf_{g_0} h(g_0)\Big) \cdot \frac{1}{2}\sum_{g_0}|p(g_0) - q(g_0)| \nonumber \\ 
        &\quad\leq \Big(\sup_{g_0} h(g_0) - \inf_{g_0} h(g_0)\Big)\cdot \frac{\varepsilon}{2}.    \label{eq_pgh markov ineq}
    \end{align}
    Therefore, 
    \begin{align*}
        &H(F|G) - H(\tilde{F} | \tilde{G}) \\ 
        &= \sum_{g_0} p(g_0) H(F|G=g_0) - \sum_{g_0} q(g_0) H(\tilde{F} | \tilde{G}=g_0) \\ 
        &\eqann[\leq]{a} \sum_{g_0} p(g_0) \Big(H(\tilde{F}|\tilde{G}=g_0)-3\varepsilon \log \frac{3\varepsilon}{2}\Big) \\ 
        &\quad -\sum_{g_0} q(g_0) H(\tilde{F} | \tilde{G}=g_0) \\ 
        &=- 3\varepsilon \log \frac{3\varepsilon}{2} +  \sum_{g_0} (p(g_0) - q(g_0)) H(\tilde{F}|\tilde{G}=g_0)  \\ 
        &\eqann[\leq]{b} - 3\varepsilon \log \frac{3\varepsilon}{2} + \Big(\max_{g_0}
        H(\tilde{F}|\tilde{G}=g_0)-\min_{g_0}H(\tilde{F}|\tilde{G}=g_0)\Big)\cdot \frac{\varepsilon}{2} \\ 
        &\eqann[\leq]{c} \frac{\varepsilon}{2} - 3\varepsilon \log \frac{3 \varepsilon}{2}, 
    \end{align*}
    where~\eqannref{a} is by~\eqref{eq_HFG tilde},~\eqannref{b} is by~\eqref{eq_pgh markov ineq}, and~\eqannref{c} is because the entropy of a binary
    random variable assumes values between $0$ and $1$. 

    Similarly, 
    \[ 
        \sum_{g_0} (p(g_0) - q(g_0))h(g_0)\geq -\Big(\sup_{g_0} h(g_0) - \inf_{g_0} h(g_0)\Big)\cdot \frac{\varepsilon}{2}, 
    \] 
    by which 
    \[ 
        H(F|G) - H(\tilde{F} | \tilde{G}) \geq - \left( \frac{\varepsilon}{2} - 3\varepsilon \log \frac{3 \varepsilon}{2} \right).  
    \] 
    Thus, we have shown that 
    \[ 
        \big| H(F|G) - H(\tilde{F} | \tilde{G}) \big| \leq \frac{\varepsilon}{2} - 3\varepsilon \log \frac{3\varepsilon}{2}. 
    \] 
    By \Cref{lem_ylny} below and some algebra, we obtain that 
    \[
        \frac{\varepsilon}{2} - 3 \varepsilon \log \frac{3\varepsilon}{2} \leq \sqrt{8} \cdot \frac{ 2^{1/12} \sqrt{3}}{e \cdot \ln 2} \sqrt{\varepsilon}
        < \sqrt{8 \varepsilon},
    \] 
    which completes the proof. 
\end{IEEEproof}

\begin{lemma}\label{lem_ylny} 
    For any $y > 0$, we have
    \[ 
        y(2-\ln y) \leq 2\sqrt{y}.
    \] 
\end{lemma}
\begin{IEEEproof}
    This inequality is illustrated in \Cref{fig_ylny sqrty inequality}. 
    A formal proof follows. 
    The Fenchel dual of $f(x) = e^x$ \cite[p. 105]{rockafellar} is 
    \[ 
        f^*(y) = \sup_x (x y - e^x) = \begin{cases} 
            y \ln y - y, & y > 0, \\
            0, & y=0, \\ 
            \infty, & \text{otherwise}. 
        \end{cases}
    \]
    Therefore, for any $x\in \mathbb{R}$ and $y>0$ we have $x y - e^x \leq y\ln y - y$.
    Now, set $x = \frac{1}{2}\ln y$ and rearrange to yield 
    $y(2-\ln y)  \leq 2 \sqrt{y}$ as desired. 
    \begin{figure}
        \begin{center}
\begin{tikzpicture}[domain=0.00001:1.4, samples =300, xscale = 5, >=latex]
\draw[thin,color=gray, dotted, xstep = 0.2, ystep = 0.5] (-0.1/5,-0.1) grid (1.4,2.5);
  \draw[->] (-0.2/5,0) -- (1.5,0) node[right] {$y$};
  \draw[->] (0,-0.2) -- (0,2.8); 
  \draw[color=red,smooth] plot (\x,{(\x)*(2-ln(\x))} ); 
  \draw[color=blue,smooth] plot (\x,{2*sqrt(\x)} ); 

  \node[red] at (0.38,0.6) {\scriptsize $y(2-\ln y)$}; 
  \node[blue] at (0.9,2.15) {\scriptsize $2\sqrt{y}$}; 

  \foreach \x in {0.2,0.4,0.6,0.8,1,1.2,1.4}
    \node at (\x, -0.2) {\tiny $\x$}; 
  \foreach \y in {0.5,1,1.5,2,2.5}
    \node at (-0.3/5,\y) {\tiny $\y$}; 
\end{tikzpicture}
\end{center}
\caption{Illustration of the inequality $y(2-\ln y) \leq 2 \sqrt{y}$.} 
\label{fig_ylny sqrty inequality}
\end{figure}
\end{IEEEproof}

\section{Equivalence of the Deterministic and Probabilistic Formulations of Hidden Markov Models}
\label{ap_markov model equivalence}
Recall that in a FAIM process, the observations are a \emph{probabilistic} function of the state, see~\eqref{eq_markov property of FAIM}. 
However, in \Cref{sec_HMM}, we defined the observations of a hidden Markov model as a \emph{deterministic} function of the state. 
Seemingly, the deterministic model is less general than the probabilistic FAIM model.
As in~\cite{Hochwald_Jelenkovic_Markov_1999} and~\cite{kaijser1975}, we now show that the deterministic and probabilistic models are equivalent. 

Using the notation of \Cref{sec_HMM}, a hidden Markov model consists of a Markov state $A_n$ and an observation $B_n$. 
In the deterministic model, $B_n = f(A_n)$, where $f$ is a deterministic function. 
In the probabilistic model, there exists a distribution $q$ such that
    \begin{equation} \label{eq_observation in probablistic case}
    \Prob{B_n = b\, | \, A_n =j, B_1^{n-1}, A_1^{n-1}} =  \Prob{B_n = b | A_n = j} = q(b|j).
\end{equation}

One direction of the equivalence is  easy: any deterministic model can be thought of a probabilistic model with $q(\cdot|j)$ assuming only the values $0$ and $1$. 
To cast the probabilistic model as a deterministic one, observe that by the Markov property and~\eqref{eq_observation in probablistic case}, we have
\begin{align*}
    &\Prob{B_n = b, A_n = j \, | \, A_{n-1} = i, A_1^{n-2}, B_1^{n-1}} \\ 
    &\quad= \Prob{B_n = b, A_n = j \, | \, A_{n-1} = i} \\
    &\quad= \Prob{A_n = j \, | \, A_{n-1} =i} \cdot \Prob{B_n = b \, | \, A_n = j} \\ 
    &\quad= p(j|i)q(b|j). 
\end{align*}

We call a pair $(j,b)$, $j \in \mathcal{A}$, $b \in \mathcal{B}$, \emph{viable} if $q(b|j) > 0$. 
Define a new Markov chain $C_n$ with states $(j,b)$ whenever $(j,b)$ is a viable pair,\footnote{States for which $q(b|j) = 0$ can never appear
with positive probability and are therefore removed.} and whose transition probability function for any two states $(j,b)$ and $(i,k)$ is
$\Prob{C_n = (j,b)  \, | \, C_{n-1} = (i,k) } = p(j|i)q(b|j)$. 
Set $f: \mathcal{A} \times \mathcal{B} \to \mathcal{B}$ as the deterministic function that outputs its second argument. 
That is, $f(a,b) = b$. This model is deterministic, and is equivalent to the probabilistic one. 
  
We are now almost done; all that remains is to show that $C_n$ is regular (aperiodic and irreducible) if and only if $A_n$ is. 

\begin{lemma}
    Let $A_n$ be a finite-state  homogeneous Markov chain and let $B_n$ be a probabilistic observation of $A_n$, as in~\eqref{eq_observation in probablistic case}. 
    Then, $A_n$ is aperiodic and irreducible  if and only if $C_n = (A_n, B_n)$ as defined above is aperiodic and irreducible. 
\end{lemma}
\begin{IEEEproof}
    Recall that a finite-state homogeneous Markov chain is aperiodic and irreducible if and only if its transition matrix is primitive. 
    That is, if and only if there exists an integer $m$ such that the $m$-step transition probability from state $i$ to state $j$ is positive for any $i,j$
    \cite[Theorem 1.4 and Section 4.2]{Seneta}, also~\cite[Section 4.1]{Iosifescu}. 

    Assume first that $A_n$ is aperiodic and irreducible. 
    Hence, there exists $m$ such that $\Prob{A_n = j | A_{n-m} = i} > 0$ for all $i$, $j$, and $n$.
    Therefore, for any viable pairs $(j,b)$ and $(i,k)$, 
    \begin{equation*} 
        \Prob{C_n = (j,b) | C_{n-m} = (i,k)} = q(b|j) \Prob{A_n = j | A_{n-m} = i} > 0.
    \end{equation*} 
    Since the states of $C_n$ consist only of viable pairs, we conclude that $C_n$ is aperiodic and irreducible. 

    Next, assume that $C_n$ is aperiodic and irreducible. 
    Then, there exists $m$ such that $\Prob{C_n = (j,b) | C_{n-m} = (i,k)} > 0$ for any two viable pairs (states) $(j,b)$ and $(i,k)$, and all $n$. 
    Therefore, for any $k$ such that $(i,k)$ is viable (at least one such $k$ must exist), 
    \begin{equation*} 
        \Prob{A_n = j | A_{n-m} = i} = \sum_b \Prob{C_n = (j,b) | C_{n-m} = (i,k)} > 0.
    \end{equation*} 
    Hence, $A_n$ is aperiodic and irreducible.
\end{IEEEproof}
\begin{example}
    \label{ex_M for Gilbert Elliott}
    The Gilbert-Elliott channel~\cite{mushkin_ge_channel} is a classic example of a channel with memory.
    It is defined as follows.
    The channel may be at one of two states, \emph{good} and \emph{bad}.
    In the good state, the channel is a binary symmetric channel (BSC) with crossover probability $\gamma$ and in the bad state, the channel is a BSC with
    crossover probability $\beta$.
    The probability of transitioning from the good state to the bad state is $p$, and the probability of transitioning from the bad state to the good
    state is $q$. 

    Assuming a symmetric channel input, we construct a deterministic model $C_n = (S_n, X_n, Y_n)$ with states 
    \begin{align*} 
                1 &= (\mathrm{good},0,0), & 5 &= (\mathrm{bad},0,0), \\  
                2 &= (\mathrm{good},0,1), & 6 &= (\mathrm{bad},0,1),\\ 
                3 &= (\mathrm{good},1,0), & 7 &= (\mathrm{bad},1,0),  \\ 
                4 &= (\mathrm{good},1,1), & 8 &= (\mathrm{bad},1,1). 
            \end{align*}
    For brevity, for a number $x \in [0,1]$ we denote $\bar{x} = 1-x$. 
    The transition probability matrix of $C_n$ is
    \[
        \mat{M} = \frac{1}{2}
        \begin{bmatrix} 
            \bar{p}\bar{\gamma} & \bar{p}\gamma & \bar{p}\gamma & \bar{p}\bar{\gamma} & p\bar{\beta}       & p\beta       & p\beta       & p\bar{\beta} \\ 
            \bar{p}\bar{\gamma} & \bar{p}\gamma & \bar{p}\gamma & \bar{p}\bar{\gamma} & p\bar{\beta}       & p\beta       & p\beta       & p\bar{\beta} \\ 
            \bar{p}\bar{\gamma} & \bar{p}\gamma & \bar{p}\gamma & \bar{p}\bar{\gamma} & p\bar{\beta}       & p\beta       & p\beta       & p\bar{\beta} \\ 
            \bar{p}\bar{\gamma} & \bar{p}\gamma & \bar{p}\gamma & \bar{p}\bar{\gamma} & p\bar{\beta}       & p\beta       & p\beta       & p\bar{\beta} \\ 
            q\bar{\gamma}       & q\gamma       & q\gamma       & q\bar{\gamma}       & \bar{q}\bar{\beta} & \bar{q}\beta & \bar{q}\beta & \bar{q}\bar{\beta} \\ 
            q\bar{\gamma}       & q\gamma       & q\gamma       & q\bar{\gamma}       & \bar{q}\bar{\beta} & \bar{q}\beta & \bar{q}\beta & \bar{q}\bar{\beta} \\ 
            q\bar{\gamma}       & q\gamma       & q\gamma       & q\bar{\gamma}       & \bar{q}\bar{\beta} & \bar{q}\beta & \bar{q}\beta & \bar{q}\bar{\beta} \\ 
            q\bar{\gamma}       & q\gamma       & q\gamma       & q\bar{\gamma}       & \bar{q}\bar{\beta} & \bar{q}\beta & \bar{q}\beta & \bar{q}\bar{\beta}  
        \end{bmatrix}. 
    \]
    The possible observations $(X,Y)$ are $(0,0), (0,1), (1,0),$ and $(1,1)$. 
    The matrices $\mat{M}(b)$, $b \in \{ (0,0), (0,1), (1,0), (1,1)\}$ are obtained from $\mat{M}$ by replacing all but two columns of $\mat{M}$ with zeros. 
    Namely, in $\mat{M}(0,0)$, all but columns $1$ and $5$ are replaced with zeros; in $\mat{M}(0,1)$ all but columns $2$ and $6$ are replaced with zeros; in
    $\mat{M}(1,0)$ all but columns $3$ and $7$ are replaced with zeros; and in $\mat{M}(1,1)$ all but columns $4$ and $8$ are replaced with zeros.  
\end{example}

\section*{Acknowledgment}
The authors are grateful to Prof.~Rami Atar and Dr.~Lele Wang for helpful discussions at the preliminary stages of this work. 

\bstctlcite{BIBctrl}
\bibliographystyle{IEEEtran} 
\bibliography{mybib.bib} 

\end{document}